\documentclass[11pt]{article}
\usepackage{dsfont}

\usepackage{amsmath,amssymb,graphicx}
\usepackage{hyperref}
\usepackage{cite}

\usepackage{slashed}

\DeclareFontFamily{U}{matha}{\hyphenchar\font45}
\DeclareFontShape{U}{matha}{m}{n}{
      <5> <6> <7> <8> <9> <10> gen * matha
      <10.95> matha10 <12> <14.4> <17.28> <20.74> <24.88> matha12
      }{}
\DeclareSymbolFont{matha}{U}{matha}{m}{n}

\DeclareMathSymbol{\oleft}{2}{matha}{"68}
\DeclareMathSymbol{\oright}{2}{matha}{"69}
\usepackage{multicol,color,longtable}

\usepackage[T1]{fontenc}
\usepackage{hyphenat}
\usepackage{amsfonts}
\usepackage{mathrsfs}
\usepackage{mathdots}
\usepackage{multicol,color,longtable}
\definecolor{darkred}{rgb}{0.65,0.15,0}
\hypersetup{pdfborder={0 0 0},colorlinks=true,urlcolor=blue,citecolor=blue,linkcolor=darkred,linktocpage=true}

\usepackage{array}
\newcolumntype{P}[1]{>{\centering\arraybackslash}p{#1}}
\usepackage{lscape}
\usepackage{multirow}

\usepackage{empheq}

\usepackage{comment}

\newcommand{\be}{\begin{equation}}
\newcommand{\ee}{\end{equation}}
\newcommand{\bea}{\setlength\arraycolsep{2pt} \begin{eqnarray}}
\newcommand{\eea}{\end{eqnarray}}

\newcommand{\dsusy}{\delta^{\rm \scriptscriptstyle susy}}

\newcommand{\ord}[1]{{\scriptscriptstyle (#1)}}

\newcommand{\nn}{\nonumber}

\newcommand{\lin}{\text{\tiny{(lin.)}}}

\newcommand{\ints}{\mathds{Z}}
\newcommand{\reals}{\mathds{R}}

\newcommand{\cM}{\mathcal{M}}
\newcommand{\cV}{\mathcal{V}}
\newcommand{\cE}{\mathcal{E}}
\newcommand{\cT}{\mathcal{T}}

\newcommand{\lb}{\left[}
\newcommand{\rb}{\right]}

\newcommand{\mf}[1]{{\mathfrak{#1}}}

\newcommand{\eprint}[1]{{\href{http://arxiv.org/abs/#1}{[\texttt{#1}]}}}
\newcommand{\eprintN}[1]{{\href{http://arxiv.org/abs/#1}{[\texttt{#1 [hep-th]}]}}}
\newcommand{\eprintNT}[1]{{\href{http://arxiv.org/abs/#1}{[\texttt{#1 [math.NT]}]}}}
\newcommand{\eprintGR}[1]{{\href{http://arxiv.org/abs/#1}{[\texttt{#1 [math.GR]}]}}}
\newcommand{\doi}[2]{{\hypersetup{urlcolor=darkred}\href{http://dx.doi.org/#2}{#1}\hypersetup{urlcolor=blue}}}

\def\CR{\nonumber \\}
\newcommand{\w}[1]{\\[0.#1cm]}
\newcommand{\cU}{\mathcal{U}}
\newcommand{\cJ}{\mathcal{J}}
\def\ie{{i.e.}}
\def\eg{{ e.g.}}

\newcommand{\beps}{\bar\epsilon}

\newcommand{\T}{\Theta}
\def\cD{\mathcal{D}}

\newcommand{\Fs}[1]{{}^{\scalebox{0.6}{$(\frac{#1}{2})$}}{}\hspace{-0.6mm}F}

\newcommand{\scal}[1]{\bigl ({#1} \bigr )}
\newcommand{\lsharp}{\text{\raisebox{-0.45ex}{\Large\guilsinglleft}}}
\newcommand{\rsharp}{\text{\raisebox{-0.45ex}{\Large\guilsinglright}}}

\makeatletter

\@addtoreset{equation}{section}
\makeatother

\begin{document}

 {\flushright  CPHT-RR043.072019 \\MI-TH-192\\[15mm]}

\begin{center}

{\LARGE \bf \sc  On supersymmetric $E_{11}$ exceptional field theory}\\[5mm]

\vspace{6mm}
\normalsize
{\large  Guillaume Bossard${}^{1}$, Axel Kleinschmidt${}^{2,3}$ and Ergin Sezgin${}^4$}

\vspace{10mm}
${}^1${\it Centre de Physique Th\'eorique, CNRS,  Institut Polytechnique de Paris\\
91128 Palaiseau cedex, France}
\vskip 1 em
${}^2${\it Max-Planck-Institut f\"{u}r Gravitationsphysik (Albert-Einstein-Institut)\\
Am M\"{u}hlenberg 1, DE-14476 Potsdam, Germany}
\vskip 1 em
${}^3${\it International Solvay Institutes\\
ULB-Campus Plaine CP231, BE-1050 Brussels, Belgium}
\vskip 1 em
${}^4${\it Mitchell Institute for Fundamental Physics and Astronomy\\ Texas A\&M University
College Station, TX 77843, USA}

\vspace{10mm}

\hrule

\vspace{5mm}

 \begin{tabular}{p{14cm}}

We construct an infinite system of non-linear duality equations, including fermions, that are invariant under global $E_{11}$ and gauge invariant under generalised diffeomorphisms upon the imposition of a suitable section constraint. We use finite-dimensional fermionic representations of the R-symmetry $K(E_{11})$ to describe the fermionic contributions to the duality equations.  These duality equations reduce to the known equations of $E_8$ exceptional field theory or eleven-dimensional supergravity for appropriate (partial) solutions of the section constraint. Of key importance in the construction is an indecomposable representation of $E_{11}$ that entails extra non-dynamical fields beyond those predicted by $E_{11}$ alone, generalising the known constrained $p$-forms of exceptional field theories. The construction hinges on the tensor hierarchy algebra extension of $\mathfrak{e}_{11}$,  both for the bosonic theory and its supersymmetric extension.
\end{tabular}

\vspace{6mm}
\hrule
\end{center}

\newpage

\setcounter{tocdepth}{2}
\tableofcontents

\section{Introduction}

Exceptional field theories  \cite{Berman:2010is,Hohm:2013vpa,Hohm:2013uia,Hohm:2014fxa} are based on generalised exceptional geometries in which 
diffeomorphisms are unified with tensor gauge transformations in such a way that the closure of the local  transformations require constraints on the fields, known as section constraints~\cite{Berman:2010is,Berman:2011cg,Berman:2012vc}. These theories live on a space which is locally a direct product of $D$-dimensional `external' space-time with an `internal' space whose coordinates are in a representation of a split real form of the exceptional group $E_{n}$ and are subject to the $E_n$ covariant section constraint. Here, $E_n$ is the usual hidden Cremmer--Julia symmetry group of ungauged maximal supergravity in $D=11-n$ space-time dimensions \cite{Cremmer:1979up}, which also governs gauged maximal supergravity through the embedding tensor formalism~\cite{Nicolai:2000sc,deWit:2002vt} and the associated tensor hierarchy~\cite{deWit:2005hv,deWit:2008ta}. The tensor hierarchy fields play a central role in constructing exceptional field theory.  Solving the section constraint amounts to restricting the dependence on the extra coordinates so that the dynamics of an appropriate supergravity theory emerges. 

The study of exceptional field theories is interesting for several reasons. Besides providing a unified description of supergravity theories that are related by duality transformations (like $D=11$ and type IIB supergravity~\cite{Hohm:2013vpa}), they allow for the derivation of uplift formul\ae{} for solutions of gauged supergravity~\cite{Godazgar:2013nma,Hohm:2013pua,Varela:2015ywx,Kruger:2016agp} and the construction of gauged supergravities via a generalised Scherk--Schwarz mechanism~\cite{Berman:2012uy,Musaev:2013rq,Aldazabal:2013mya,Godazgar:2013oba,Lee:2014mla,Hohm:2014qga}. They are also instrumental in studying non-geometric string theory solutions~\cite{Shelton:2005cf,Dabholkar:2005ve,Hull:2006va,Hull:2007zu,Pacheco:2008ps,Andriot:2011uh,Andriot:2012wx}. Further aspects of exceptional field theory have been discussed in the recent overview~\cite{Hohm:2019bba}.

It is a remarkable fact that the bosonic sector of exceptional field theory is  completely determined by generalised diffeomorphisms without the use of supersymmetry. Another key property is that these theories typically require extra $p$-forms of rank $p\ge D-2$ \textit{beyond} those present in the usual tensor hierarchy of $D$-dimensional maximal gauged supergravity. These obey constraints that are similar to the section constraints. They are related to the physical fields by first-order equations and  do not themselves describe new physical degrees of freedom. 

So far, $E_n$ exceptional field theories have been constructed explicitly for $n=6,7,8$ in\cite{Hohm:2013vpa,Hohm:2013uia,Hohm:2014fxa} and in~\cite{Berman:2010is,Hohm:2015xna,Abzalov:2015ega,Musaev:2015ces,Berman:2015rcc} for smaller $n$. The cases beyond $n=8$ involve infinite dimensional groups and bring in formidable new challenges. A dent has been made recently in the case of $E_9$ \cite{Bossard:2018utw}. The present paper studies the case of $E_{11}$.

It has been proposed by West long ago and prior to the development of exceptional field theory that the $D=11$ supergravity equations of motion should emerge from an $E_{11}$ invariant theory formulated in the framework of a non-linear realisation of $E_{11}$ with coordinates in the  `vector' representation, such that the dynamics would follow from an $E_{11}$ invariant set of duality equations~\cite{West:2001as,West:2011mm,West:2014eza}.\footnote{The idea of extended Kac--Moody symmetries, in particular in connection with low-dimensional gravitational systems, was first expressed in~\cite{Julia}. An explicit trace of $E_{10}$ symmetry was found in a Belinskii--Khalatnikov--Lifshitz  analysis of eleven-dimensional supergravity~\cite{Damour:2000hv} and was later generalised to a cosmological $E_{10}$ model~\cite{Damour:2002cu}.}
It has been realised recently that these first order duality equations can only hold modulo certain equivalence relations~\cite{West:2014qoa,Tumanov:2016abm,Tumanov:2016dxc}. These ambiguities are argued to be liftable by passing to equations of motion that are eventually of arbitrarily high order in derivatives. These equivalence relations may potentially be interpreted as arising from additional gauge symmetries, although their precise form has not been determined. The section constraint was not used in those references in connection with the gauge invariance of the equations of motion, but only in connection with the description of $1/2$-BPS states~\cite{West:2012qm}. $E_{11}$ does capture the supergravity tensor hierarchy field content in $D$ dimensions \cite{Riccioni:2007au,Bergshoeff:2007qi}, but not the extra constrained $p$-forms of the exceptional field theories mentioned above.

In \cite{Bossard:2017wxl}, it was explained that constructing linearised gauge invariant first order field equations with $E_{11}$ symmetry requires the fields to satisfy the section constraint as well as the introduction of additional fields that do not appear in the $E_{11}$ coset space. This construction is based on an infinite-dimensional super-algebra $\cT(\mf{e}_{11})$,  that includes $\mf{e}_{11}$ as a subalgebra and that generalises the tensor hierarchy algebra $\cT(\mf{e}_{n})$ introduced in~\cite{Palmkvist:2013vya} for $n\le 8$ to the Kac--Moody case. The tensor hierarchy algebra then includes  a non-semi-simple extension $\cT_0(\mf{e}_{11})$ of the algebra $\mf{e}_{11}$ that entails the introduction of extra fields already in the linearised theory.  A gauge invariant linearised duality equation can be written in this formulation for a field strength that transforms covariantly under $E_{11}$ provided one introduces these extra fields in the corresponding indecomposable representation. The extra fields are necessary to write a gauge invariant duality equation for the graviton in eleven dimensions  \cite{Bossard:2017wxl}. We exhibit here that our $E_{11}$ duality equation, including the extra fields, gives in the linearised approximation an infinite tower of gauge invariant duality equations of the type described in  \cite{Riccioni:2006az}. Gauge invariance of these equations as described in \cite{Boulanger:2012df,Boulanger:2015mka} is only satisfied in the presence of the extra fields.

The primary goal of this paper is to construct the $E_{11}$ and gauge invariant \textit{non-linear} duality equation that captures all the duality equations of all $E_n$ exceptional fields theories. We will show that one can generalise the duality equation constructed in \cite{Bossard:2017wxl} to a non-linear equation invariant under generalised diffeomorphism. The key observation that facilitates this construction is that the derivative of the extra fields found in \cite{Bossard:2017wxl} at the linearised level are the cohomologically trivial part of extra fields that turn out to underlie the extra constrained $p$-forms fields mentioned above in the $GL(11-n,\reals)\times E_n$ decomposition and subsequent truncation of the $E_{11}$ invariant theory. In analogy with what happens in lower-dimensional exceptional field theories, and as mentioned above, these extra $p$-forms are related to the propagating fields by first order equations but they do not themselves describe new physical degrees of freedom. These first order equations for the constrained fields are sourced by bilinear terms in the derivatives of the fields parametrizing $E_{11}$. Therefore they cannot follow from the $E_{11}$ variations of the duality equation we construct in this paper, and they must be derived separately by requiring gauge invariance and integrability of the equations. We shall not attempt to determine these first order equations in this paper, and will only make some comments on their expected structure.

Besides the investigation of a non-linear bosonic theory based on the tensor hierarchy algebra, an important part of the present paper is the study of its supersymmetric extension. Fermions are introduced here ---~as for maximal supergravity and other exceptional field theories~--- as representations of (the double cover of) the involution invariant subgroup $K(E_{11})$ that plays the role of a generalised R-symmetry group. As noticed in~\cite{Nicolai:2004nv,deBuyl:2005zy,Damour:2005zs,deBuyl:2005sch,Damour:2006xu}, this subgroup admits finite-dimensional (a.k.a. unfaithful) spinor representations in the case of Kac--Moody groups $K(E_n)$ with $n\geq 9$. In particular, these representations were constructed for the gravitino and supersymmetry parameter of $K(E_{11})$ in~\cite{Kleinschmidt:2006tm} with beginnings of the supersymmetry parameter representation already given in~\cite{West:2003fc}. The compatibility of local $K(E_n)$ symmetry with supersymmetry, \ie\ whether the supersymmetry generator transforms correctly as a spinor under $K(E_n)$ was investigated in~\cite{Damour:2006xu,Henneaux:2008nr,Kleinschmidt:2014uwa} for $n=10$ where it was found that there was an inconsistency in the transformation arising for the bosonic fields beyond the six-form, \ie\ starting from the so-called dual graviton. Based on~\cite{West:2003fc} and~\cite{Damour:2005zs,deBuyl:2005sch,Damour:2006xu}, fermions in $K(E_{11})$ were introduced in~\cite{Steele:2010tk} and a similar calculation was carried out up to the level of the six-form.

In the present paper, we resolve this inconsistency starting from the dual graviton by considering not only $\mathfrak{e}_{11}$ but its non-semi-simple extension $\overline{\cT}_0(\mf{e}_{11})$ that appears in the tensor hierarchy algebra. As already emphasised above, one important consequence of the tensor hierarchy algebra is that it introduces additional fields into the theory beyond those of the standard $E_{11}/K(E_{11})$ symmetric space. These fields will resolve the inconsistencies with  the supersymmetry transformations, because the supersymmetry transformation of fields in $\overline{\cT}_0(\mf{e}_{11}) \ominus K(\mathfrak{e}_{11})$ can be written consistently with $K(E_{11})$. We shall use this construction to write linearised supersymmetry transformation rules and equations of motion for the (unfaithful) gravitino field. We also show that one obtains a closed supersymmetry algebra at linearised order. These results will be derived explicitly at low levels, including the dual graviton. At present, we do not have a complete algebraic proof to all levels.

After establishing the linearised supersymmetry and the equations of motion for the Fermi fields, we investigate their non-linear extension and their compatibility with the non-linear duality equation  proposed in Section~\ref{sec:NLDual}. We present some first steps in this direction by introducing the non-linear $K(E_{11})$ connection and a Pauli coupling to the $E_{11}$ field strength.   Although we do not have the complete expression of the non-linear equations, the first few levels exhibit promising cancellations that lead to the desired couplings of eleven-dimensional supergravity.

\subsection*{Structure of the paper and summary of main results}

Given the length of the paper we here give a telegraphic summary of our main results for the reader's convenience.
\begin{itemize}
\item Inspired by the structure of the tensor hierarchy algebra given in Section~\ref{sec:e11tha}, we propose non-linear bosonic field strengths that transform covariantly under $E_{11}$ in an infinite-dimensional representation that generalises the embedding tensor representation of gauged supergravity for finite-dimensional $E_n$ and that is neither highest nor lowest weight. Labelling its component by $I$ we show in~\eqref{eq:NLFS} that the following definition is $E_{11}$ covariant:
\begin{align}
F^I = C^{IM}{}_\alpha J_M^\alpha + C^{IM}{}_{\tilde\alpha} \chi_M{}^{\tilde\alpha} + \ldots\,.
\end{align}
Here, $J_M^\alpha$ is the non-linear $\mf{e}_{11}$ Lie-algebra valued current constructed out of the $E_{11}/K(E_{11})$ representative $\cM$ using $\cM^{-1} \partial_M \cM$ with $\partial_M$ denoting the derivative with respect to the infinitely many coordinates of the $R(\Lambda_1)$ representation of $E_{11}$ subject to a section constraint. The fields $\chi_M{}^{\tilde\alpha}$ are constrained fields, i.e. they are (section) constrained in the $M$ index in the same way as the partial derivative $\partial_M$, and $\tilde\alpha$ labels the representation ${R(\Lambda_2)}$ of $E_{11}$. However, the indecomposability of $\cT_0(\mf{e}_{11})$ is importantly such that they form an \textit{indecomposable} representation together with the adjoint current components in such a way that the structure constants $C^{IM}{}_\alpha$ and $C^{IM}{}_{\tilde\alpha}$ appearing in the expressions above ensure $E_{11}$ covariance of the field strengths $F^I$, whereas $ C^{IM}{}_\alpha J_M^\alpha$ alone would not be covariant. The dots indicate additional constrained fields discussed in Sections~\ref{sec:e11tha} and~\ref{sec:NLDual}.

\item The representation-theoretic content of the tensor hierarchy algebra  permits writing a non-linear duality equation~\eqref{DualityEquation} for the non-linear field strengths:
\begin{align}
\label{eq:DEintro}
F^I - \cM^{IK}\Omega_{KJ} F^J = 0 \,.
\end{align}
The tensor hierarchy algebra ensures the existence of a symplectic form $\Omega_{IJ}$ that acts on the field strengths $F^I$. The above first-order equation is a vast generalisation of (twisted) duality equations that have appeared elsewhere in the literature~\cite{Cremmer:1998px,West:2001as} and covers both the matter and the gravitational sector. As we analyse in Section~\ref{sec:fulldyn}, the duality equation is not sufficient to determine the dynamics of the constrained fields $\chi_M{}^{\tilde\alpha}$, just as \eg \ for $E_7$ exceptional field theory~\cite{Hohm:2013uia}. Assuming integrability conditions at linearised order, we relate the constrained fields and their dynamics to our previously studied model in~\cite{Bossard:2017wxl}.

\item We propose non-linear gauge transformations of all fields in~\eqref{eq:gt4} and~\eqref{eq:gtt} and show that the duality equation~\eqref{eq:DEintro} is gauge invariant under these gauge transformations if a certain group-theoretic identity~\eqref{MasterAssum2} holds. This identity is then verified at low levels in decompositions of $E_{11}$ under its $GL(11)$ and $GL(3)\times E_8$ subgroups in Sections~\ref{sec:gl11bos} and~\ref{sec:E8bos}, respectively. We also write explicitly the duality equation \eqref{eq:DEintro} in components in the corresponding parametrisations, and exhibit that it reproduces the known duality equations of eleven-dimensional supergravity and of  $E_8$ exceptional field theory.  We exhibit in particular in Section \ref{sec:dualfields} the gauge invariance of the infinite tower of linearised duality equations underlying eleven-dimensional supergravity \cite{Riccioni:2006az,Boulanger:2012df,Boulanger:2015mka}.

\item Starting from Section~\ref{sec:susyALG}, we study the fermionic extension of the model. Given the unfaithful spinors $\Psi$ and $\epsilon$ of $\widetilde{K}(E_{11})$, we show how their bilinears relate to the tensor hierarchy algebra and how this can be used to define supersymmetry transformation rules and a consistent supersymmetry algebra. We show that this consistency also connects to the reducible gauge structure of the $E_{11}$ generalised Lie derivative and introduces yet more additional bosonic fields into the theory in order to make all symmetries manifest. 

\item We establish a linearised, $\widetilde{K}(E_{11})$ covariant equation of motion for the gravitino field in Section~\ref{sec:susyEOM} that reads (see~\eqref{eq:LRSeq})
\begin{align}
\label{eq:RSintro}
G^{a;b M}\partial_M \psi_b = 0 \,,
\end{align}
where $\psi_b$ are the components of $\Psi$ in a $Spin(1,10)$ basis and $G^{a;bM}$ are $\widetilde{K}(E_{11})$ invariant tensors that are constructed out of $Spin(1,10)$ gamma matrices and Kronecker symbols. We show how this gravitino equation of motion is consistent with the bosonic dynamics under supersymmetry. This requires also introducing gravitino bilinears in the non-linear duality equation~\eqref{eq:DEintro} in the form (see~\eqref{DEwO}) 
\begin{align}
F^I - \cM^{IK}\Omega_{KJ} F^J  = \mathcal{V}^{-1 I}{}_{\underline{I}} O^{\underline{I}}\,,
\end{align}
where $O^{\underline{I}}\sim (\Psi \Psi)^{\underline{I}}$ denotes fermion bilinears transforming in the $K(E_{11})$ representation of the field strength equation of motion. The underlined index $\underline{I}$ here indicates a `local' $K(E_{11})$ index that is converted into a `global' $E_{11}$ index $I$ by means of the inverse generalised vielbein $\mathcal{V}^{-1}$  in $E_{11}/K(E_{11})$. The possibility of making this fermionic modification of the first-order duality equation rests on a non-trivial relation between the unfaithful spinors and the tensor hierarchy algebra that we demonstrate at low levels. In this way we obtain a supersymmetric non-linear duality equation for the bosons including the non-linear fermionic terms.

\item When studying the supersymmetry algebra and gauge algebra it is important to also study the generalised diffeomophisms on the fermions. We provide a general expression for this in~\eqref{eq:liePsi} that involves the compensating $K(\mf{e}_{11})$ transformation arising due to the gauge-fixed generalised vielbein. We verify in Appendix~\ref{E7susy} that our formula, when restricted to $E_7$ exceptional field theory, agrees with previous results in the literature.

\item  In Section~\ref{sec:NLferm}, we study also the extension of the linearised fermionic equation of motion~\eqref{eq:RSintro} and supersymmetry variations to the non-linear level.

\end{itemize}

\section{\texorpdfstring{$E_{11}$}{E11} and tensor hierarchy algebra  }
\label{sec:e11tha}

In this section, we shall review elements of the group $E_{11}$ with the underlying Lie algebra $\mf e_{11}$, and the tensor hierarchy algebra $\cT(\mf{e}_{11})$ that will be needed in the construction of the $E_{11}$ invariant duality equations.

\subsection{\texorpdfstring{$E_{11}$}{E11} and its Lie algebra}

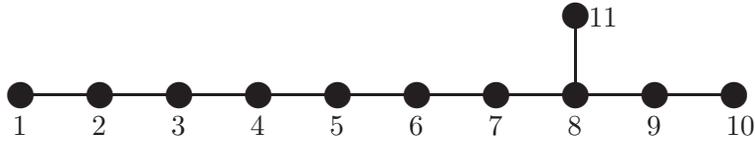
\begin{figure}[t!]
\centering
\begin{picture}(300,50)
\thicklines
\multiput(10,10)(30,0){10}{\circle*{10}}
\put(10,10){\line(1,0){270}}
\put(220,40){\circle*{10}}
\put(220,10){\line(0,1){30}}
\put(7,-5){$1$}
\put(37,-5){$2$}
\put(67,-5){$3$}
\put(97,-5){$4$}
\put(127,-5){$5$}
\put(157,-5){$6$}
\put(187,-5){$7$}
\put(217,-5){$8$}
\put(247,-5){$9$}
\put(277,-5){$10$}
\put(225,36){$11$}
\end{picture}
\caption{\label{fig:e11dynk}\textit{Dynkin diagram of $E_{11}$ with labelling of nodes used in the text.}}
\end{figure}

The Lie algebra $\mf e_{11}$ is an infinite-dimensional Lorentzian Kac--Moody algebra with Dynkin diagram shown in Figure~\ref{fig:e11dynk}. For a detailed description of the algebra see for example~\cite{Kac,West:2001as,Kleinschmidt:2003mf}. We will denote a representation with highest weight $\Lambda$ by $R(\Lambda)$ where $\Lambda=\sum_i p^i \Lambda_i$, with $\Lambda_i$ denoting the fundamental weights and $p^i$ are the Dynkin labels. For example, $R(\Lambda_1)$ refers to  representation with Dynkin labels $(1,0,...,0)$. We will use the notation $R(\Lambda)$ to also refer to the module associated with the corresponding representation. The dual representation of $R(\Lambda)$ will be denoted by $\overline{R(\Lambda)}$ and it is a lowest weight representation.

A convenient way of organising the generators of $\mf e_{11}$ is by decomposing the adjoint representation of $\mf{e}_{11}$ under its $\mf{gl}(11)$ subalgebra obtained by removing node $11$ from the diagram. Defining the $\mf{gl}(11)$ level $\ell$ as the eigenvalue of the generator $\frac13 K^m{}_m$, where $K^m{}_n$ for $m,n=0,\ldots,10$ denotes the generators of the $\mf{gl}(11)$, levels $0\leq \ell\leq 4$ of the $\mf{gl}(11)$ decomposition of the adjoint $\mf{e}_{11}$ are given in Table~\ref{tab:e11adj} in Appendix~\ref{app:e11}. The appendix also contains more details on the $\mf{gl}(11)$ algebra in \eqref{gl11} and similar decompositions of some other representations of $\mf{e}_{11}$ that play a role in this work.  At levels $0$ and $1$, the generators have the same index structure as the graviton and the $3$-form field of 11D supergravity, respectively, and their dual $6$-form and the dual graviton appear at levels $2$ and $3$.

The highest weight representation $R(\Lambda_1)$ plays an important role in the dynamical description of the $E_{11}$ exceptional field theory as it gives the representation structure of the $E_{11}$ space-time coordinates~\cite{West:2003fc}. Its dual lowest weight representation $\overline{R(\Lambda_1)}$ is the representation denoted by $\ell_1$ in~\cite{West:2003fc} that can be used to contract the coordinates when forming a generalised translation group element. The level decomposition of $\overline{R(\Lambda_1)}$ under $\mf{gl}(11)$ is displayed in Table~\ref{tab:e11l1} in Appendix~\ref{app:e11}. The names of the generators there already anticipate their roles as central charge type coordinates in a $D=11$ interpretation and associated translation generators and gauge parameters. 

We shall also need to make use of tensor products of $\mf{e}_{11}$ representations. The tensor product of highest weight (respectively lowest weight) representations is completely reducible into infinitely many highest (respectively lowest) weight representations. By contrast tensor products of highest with lowest representations fall outside what is called category $\mathcal{O}$ and there are no complete reducibility results~\cite{Kac}.  It is known nonetheless that the tensor product of a representation and its dual contains the adjoint representation. The following decompositions of tensor products will prove to be useful 
\begin{align}
\label{HighestWeightDcompo} 
\left(R(\Lambda_1) \otimes R(\Lambda_1)\right)_{\rm sym} &= R(2\Lambda_1) \oplus [ R(\Lambda_{10}) \oplus \ldots]\ ,
\nn\w2
\left(R(\Lambda_1) \otimes R(\Lambda_1)\right)_{\rm antisym} &= R(\Lambda_2) \oplus [ R(\Lambda_4) \oplus \ldots]\ ,
\nn\w2
\overline{R(\Lambda_1)} \otimes \overline{R(\Lambda_2)} &= \overline{R(\Lambda_1+\Lambda_2)} \oplus \overline{R(\Lambda_3)} \oplus \ldots 
\end{align}
The representation $[R(\Lambda_{10})\oplus \ldots]$ encodes the weak section constraint in the $E_{11}$ exceptional field theory which will be described in the next section and the low levels of its $\mf{gl}(11)$ decomposition is given in Table~\ref{tab:e11l10}. The representation $[R(\Lambda_4) \oplus\ldots]$ completes this to the strong section constraint. 

The Kac--Moody group $E_{11}$ should be properly defined either as the minimal (or small) group generated by products of real root generators~\cite{KacPeterson} or as the completed group that is obtained from the minimal definition by completion with respect to the building topology~\cite{Carbone:2003}. For the purposes of this paper, we will consider the completed group $E_{11}$ as formal exponentials of $\mf{e}_{11}$ Lie algebra elements completed in the positive Borel direction. A more detailed discussion of the Kac--Moody symmetric space and possible coordinates on it will be given in Section~\ref{sec:Vgauge}.

\subsection{Tensor hierarchy algebra}
\label{sec:THA}

For any $\mf e_{n}$ algebra, the tensor hierarchy algebra $\mathcal{T}(\mf{e}_n)$ is a super-algebra extension of $\mf e_{n}$~\cite{Palmkvist:2013vya,Bossard:2017wxl}. It admits generally a $\mathds{Z}$-grading $\mathcal{T}(\mf e_n) = \bigoplus_p \mathcal{T}_p(\mf e_n)  $ consistent with the Grassmann $\mathds{Z}_2$ grading  (\ie\ such that $\bigoplus_k \mathcal{T}_{2k}(\mf e_n) $ is the bosonic subalgebra). For $n\le 8$, $\mathcal{T}_p(\mf{e}_n)$ for $0\leq p \leq 11-n$ corresponds to the $\mf{e}_n$ representation of the $p$-forms in $(11-n)$-dimensional maximal supergravity. In particular one has $\mathcal{T}_0(\mf{e}_n)= \mf{e}_n$ corresponding to the Cremmer--Julia hidden symmetry of the scalar sector that extends to the $p$-form sector. The tensor hierarchy algebra is not symmetric under $p \leftrightarrow -p$, meaning $\cT_{p} \ncong \overline{\cT_{-p}}$. The component $\mathcal{T}_{-1}(\mf e_n)$ is the so-called embedding tensor representation \cite{Palmkvist:2013vya} which is used for describing gaugings of supergravity~\cite{Nicolai:2000sc,deWit:2002vt}. The tensor hierarchy algebra was constructed in \cite{Bossard:2017wxl} for $n\ge 9$ as the quotient of the superalgebra generated by a local superalgebra by its maximal ideal, using the construction of~\cite{Kac2}. This construction is very similar to the one of a Kac--Moody algebra, for which the local algebra is defined by the Chevalley generators associated to each simple root, and the maximal ideal is defined by the Serre relations. For the tensor hierarchy algebra, the local superalgebra can be described explicitly but the maximal ideal does not admit a closed-form definition generalising the Serre relations. In the following we shall simply use $\mathcal{T} = \oplus_{p\in \mathds{Z}} \mathcal{T}_p$ when we refer to the tensor hierarchy algebra extension of $\mf{e}_{11}$.

The important difference between $\mathcal{T}$ and the tensor hierarchy algebras associated to $\mf e_{n}$ for $n\le 8$ is that $\mathcal{T}_0 \supsetneq \mf e_{11}$ and is a non-simple extension of $\mf e_{11}$ that decomposes as an $\mf e_{11}$ module as follows 
\be 
\mathcal{T}_0 \cong \mf{e}_{11}\oleft R(\Lambda_2)\oplus R(\Lambda_{10}) \oplus \cdots 
\ee
where the notation $\mf{e}_{11}\oleft R(\Lambda_2)$ indicates that it is not the direct sum of two modules, but rather that $\mf{e}_{11}$ is a submodule and  $[\mf{e}_{11}\oleft R(\Lambda_2)] / \mf{e}_{11}$ is the highest weight module  $R(\Lambda_2)$ as a quotient only. Thus $\mf{e}_{11}\oleft R(\Lambda_2)$ forms an indecomposable representation. It was shown in~\cite{Bossard:2017wxl} that the next term $R(\Lambda_{10})$ forms a direct sum with this space but the full module structure contained in the dots is presently not known. The other degrees $\mathcal{T}_p$ have a similar structure. 

The tensor hierarchy algebra admits an antisymmetric bilinear form such that   $ \mathcal{T}_p \cong \overline{\mathcal{T}_{-2-p}}$. The components $\mathcal{T}_p$ for $p\ge 1$ are highest weight modules of $\mf e_{11}$, and therefore lowest weight for $p\le -3$. $\mathcal{T}_{-1} $ is a symplectic representation of $\mf e_{11}$, but  very little is known about its reducibility, since it is neither a highest/lowest weight representation of $\mathfrak{e}_{11}$ nor an extension of the adjoint itself. We will therefore refer to this representation as $\mathcal{T}_{-1} $, both as an $\mf e_{11}$ module and as the component of the tensor hierarchy algebra. The known structure of $\mathcal{T}_p$ for $ -3\le p\le 2$ is summarised in Table \ref{tha2} where also notation for the corresponding generators is introduced. 

\begin{table}
\begin{center}
\begin{tabular}{c|c|c}
level & $E_{11}$ rep & index notation\\
\hline
&&\\
$p=2$ & $R(\Lambda_{10})\oplus \cdots $ & $ P^{\widehat\Lambda} = ( P^\Lambda, \ldots) $
\\
&&\\
$p=1$ & $R(\Lambda_1) \oplus R(\Lambda_1+\Lambda_{10}) \oplus R(\Lambda_{11}) \oplus \cdots $ & $P^{\widehat M}= (P^M, \tilde P^{M\Lambda} , \ldots)$
\\
&&\\
$p=0$ & $\mf{e}_{11}\oleft R(\Lambda_2)\oplus R(\Lambda_{10}) \oplus \cdots $  & $ t^{\widehat\alpha} = (t^\alpha, \tilde{t}^{\tilde\alpha},  {\tilde{t}}^{\Lambda}, \ldots)$ \\
&&\\
$p=-1$ &  $ \mathcal{T}_{-1}$ & $t_I$
\\
&&\\
$p=-2$ & $\mathfrak{e}_{11}\oright \overline{R(\Lambda_2)}\oplus \overline{R(\Lambda_{10})} \oplus \cdots$ & $\bar t_{\widehat\alpha} = (\bar{t}_\alpha$, $\bar{\tilde{t}}_{\tilde\alpha}$,  $\bar{\tilde{t}}_{\Lambda}, \ldots) $, 
\\
&&\\
$p=-3$ & $\overline{{R}(\Lambda_1)}\oplus \overline{R(\Lambda_1+\Lambda_{10})} \oplus \overline{R(\Lambda_{11})} \oplus \cdots $ & $\bar{P}_{\widehat M} = (\bar P_M, \bar {\tilde{P}}_{M\Lambda},\ldots )$
\end{tabular}
\end{center}
\caption{\label{tha2}\textit{$E_{11}$ representations arising at level $ -3\le p\le 2$ elements of the tensor hierarchy algebra. In particular, $P^M$ and $\bar P_M$ denote the representations $R(\Lambda_1)$ and $\overline{R(\Lambda_1)}$, respectively. }}
\end{table}

The indecomposability of $\mathcal{T}_0$ is reflected in the commutation relations of the level $p=0$ generators $t^{\widehat{\alpha}}= ( t^\alpha, \tilde{t}^{\tilde\alpha}, \tilde{t}^\Lambda,\ldots)$ as 
\be
\lb t^\alpha , t^\beta \rb = f^{\alpha\beta}{}_{\gamma} t^\gamma\ ,
\qquad
\lb t^\alpha , \tilde{t}^{\tilde\beta} \rb =  -T^{\alpha\tilde\beta}{}_{\tilde\gamma}\tilde{t}^{\tilde\gamma} - K^{\alpha \tilde{\beta}}{}_{\gamma} t^{\gamma}\ ,
\qquad
\lb t^\alpha , \tilde{t}^\Lambda \rb = -T^{\alpha\Lambda}{}_{\Xi} \tilde{t}^{\Xi}\ . \label{E11Commute} 
\ee
The presence of the non-trivial structure constant $K^{\alpha \tilde{\beta}}{}_\gamma$ in the middle equation is due to the indecomposability, showing that there is commutator of $\mf{e}_{11}$ with $R(\Lambda_2)$ going back to $\mf{e}_{11}$. The action of the group $E_{11}$ on $\mf{e}_{11}\oleft R(\Lambda_2)$ is defined by 
\begin{align}
\label{eq:indact}
g^{-1} t^\alpha g = g^\alpha{}_\beta t^\beta\,,\quad g^{-1} \tilde{t}^{\tilde\alpha} g = g^{\tilde\alpha}{}_{\tilde\beta} \tilde{t}^{\tilde\beta} + \omega^{\tilde\alpha}_\beta(g) t^\beta\,,
\end{align}
such that $g_1^\alpha{}_\beta g_2^\beta{}_\gamma = (g_1g_2)^\alpha{}_\gamma$ and where $\omega^{\tilde\alpha}_\beta(g)$ is a group 1-cocycle satisfying
\begin{align}
\omega^{\tilde\alpha}_\beta(g_1g_2) = \omega^{\tilde\alpha}_\gamma(g_1) g_2^\gamma{}_\beta + g_1^{\tilde\alpha}{}_{\tilde\gamma} \omega^{\tilde\gamma}_\beta(g_2)
\end{align}
and that linearises to $\omega^{\tilde\alpha}_\beta( e^{\Lambda_\gamma t^\gamma}) = \Lambda_\gamma K^{\gamma\tilde\alpha}{}_\beta + O(\Lambda^2)$ consistently with the commutation relation~\eqref{E11Commute}. Finding an explicit form of the cocycle $\omega^{\tilde\alpha}_\beta(g)$ for the tensor hierarchy algebra $\cT(\mf{e}_{11})$ seems to be a formidable task, although one can write $\omega^{\tilde\alpha}_\beta(e^{X})$ as a formal power series in $X$, see~\cite{Bossard:2018utw} for formulas in the case of $\mf{e}_9$.

The action of $\mf{e}_{11}$ on the other levels $\mathcal{T}_p$ is given by
\be 
\label{eq:E11rep}
\begin{split} [ t^\alpha , P^M] &= -T^{\alpha M}{}_N P^N  \; ,  \\
[ t^\alpha , \bar t_\beta ]  &= - f^{\alpha\gamma}{}_\beta { \bar t}_\gamma +K^{\alpha \tilde \gamma}{}_\beta \bar{\tilde{t}}_{\tilde \gamma}\; , 
\end{split} 
\quad 
\begin{split} [ t^\alpha , P^\Lambda] &= -T^{\alpha \Lambda}{}_\Xi P^\Xi \\
 [ t^\alpha , \bar{\tilde{t}}_{\tilde \beta} ] &=  T^{\alpha\tilde\gamma}{}_{\tilde \beta}  \bar{\tilde{t}}_{\tilde \gamma} 
\end{split} \qquad 
\begin{split}
[ t^\alpha , t_I ] &= T^{\alpha J}{}_I t_J \; , \\
[ t^\alpha , \bar{\tilde{t}}_\Lambda ]  &=  T^{\alpha\Xi}{}_{\Lambda} \bar{\tilde{t}}_{\Xi}\ . 
\end{split}
\ee
Since $p=-2$ is the dual representation to $p=0$, the indecomposability is now in $\mf{e}_{11}^*\oright \overline{R(\Lambda_2)}$, such that $K^{\alpha\tilde{\gamma}}{}_{\beta}$ now appears in the commutator of $t^\alpha$ with the element $\bar{t}_\beta$ of the co-adjoint $\mf{e}_{11}^*$.
The convention for the $\mf{e}_{11}$ representation matrices is such that 
\be 
T^{\alpha M}{}_P T^{\beta P}{}_N - T^{\beta M}{}_P T^{\alpha P}{}_N =  f^{\alpha\beta}{}_\gamma T^{\gamma M}{}_N \,,\quad \textrm{etc.} 
\ee
We shall also use the notation $f^{\widehat{\alpha}\widehat{\beta}}{}_{\widehat{\gamma}}$ for the complete $\mathcal{T}_0$ structure coefficients, such that $f^{\alpha\widehat{\beta}}{}_{\widehat{\gamma}} = - T^{\alpha\widehat{\beta}}{}_{\widehat{\gamma}}$, $f^{\alpha\widehat{\beta}}{}_{\gamma} = - K^{\alpha\widehat{\beta}}{}_{\gamma}$ for example.

Further (anti)commutators that will be needed later are given by 
\bea
\lb P^M, \bar{t}_\alpha \rb &=& C^{I M}{}_{\alpha} t_I\ ,
\qquad
\lb P^M, \bar{\tilde{t}}_{\tilde\alpha} \rb =   C^{I M}{}_{\tilde\alpha} t_I\ ,
\qquad
\lb P^M, \bar{\tilde{t}}_\Lambda \rb =   C^{I M}{}_\Lambda t_I\ ,
\nn\w2
\left\{ P^M, \bar{P}_N \right \} &=&  T^{\alpha M}{}_N \bar{t}_\alpha + T^{{\tilde\alpha}M}{}_N \bar{\tilde{t}}_{\tilde\alpha}+ T^{\Lambda M}{}_N \bar{\tilde{t}}_{\Lambda} \ ,
\qquad
\left\{ P^M, P^N \right\} = \Pi_\Lambda{}^{M N} P^\Lambda\ ,
 \nn\\
\lb P^M, t_I \rb  &=&- \Omega_{IJ} C^{J M}{}_{\alpha} t^\alpha -\Omega_{IJ} C^{J M}{}_{\tilde{\alpha}} t^{\tilde{\alpha}} -\Omega_{IJ} C^{J M}{}_{\Lambda} t^\Lambda 
\ ,
\eea
where the coefficients are $E_{11}$ invariant tensors, except for $K^{\alpha \tilde{\beta}}{}_{\gamma}$, $C^{IM}{}_\alpha$ and $T^{\tilde{\alpha}M}{}_N$ that mix with the indecomposable structure, although the complete tensors $C^{IM}{}_{\hat{\alpha}}$ and $T^{\hat{\alpha} M}{}_N$ are invariant tensors in the indecomposable representation $\mathcal{T}_0$. In particular for $E_{11}$ group elements one has 
\be
g^I{}_J\,g^{M}{}_N\, g^{-1 \beta}{}_{{\alpha}}\, C^{J N}{}_{{\beta}}  =C^{IM}{}_{{\alpha}}- g^I{}_J\,g^{M}{}_N\,\omega^{\tilde\beta}_{\alpha}(g^{-1})\, C^{J N}{}_{ \tilde{\beta}} \ .
\label{sc1}
\ee 

If, as in~\cite{Bossard:2017wxl}, we take the fields of the theory to be in $\mathcal{T}_{-2}$ such that they are of the form  $\phi^{\widehat{\alpha}} = (\phi^\alpha  , X^{\tilde{\alpha}} , Y^\Lambda,\ldots)$ and defining their $\mf{e}_{11}$ variation by $\delta_\Lambda \phi^{\widehat{\alpha}} \bar{t}_{\widehat{\alpha}} = \Lambda_\alpha [ t^\alpha  , \phi^{\widehat{\alpha}} \bar{t}_{\widehat{\alpha}} ]$, the commutation relations \eqref{E11Commute} yield
\bea
\delta_\Lambda \phi^\alpha &=& - \Lambda_\gamma f^{\gamma\alpha}{}_\beta \phi^\beta  \ ,
\w2
\delta_\Lambda X^{\tilde{\alpha}} &=&  \Lambda_\gamma  T^{\gamma\tilde{\alpha}}{}_{\tilde{\beta}} X^{\tilde{\beta}}  + \Lambda_\gamma K^{\gamma\tilde\alpha}{}_\beta  \phi^\beta \  , 
\eea
which gives after exponentiation 
\bea \phi^\alpha &\rightarrow& \exp( -\Lambda_\gamma f^\gamma)^\alpha{}_\beta \phi^\beta  \ ,
\w2
X^{\tilde{\alpha}} &\rightarrow&\exp( \Lambda_\gamma T^\gamma)^{\tilde\alpha}{}_{\tilde\beta} X^{\tilde\beta}   + \Lambda_\gamma  \sum_{n=0}^\infty \frac{1}{n!} \sum_{k=0}^n [ (\Lambda_\epsilon T^\epsilon)^k K^{\gamma} (-\Lambda_\delta f^\delta)^{n-k} ]^{\tilde \alpha}{}_\beta   \phi^\beta \  .  
\eea
The above transformations satisfy the $E_{11}$ algebra.

\section{Non-linear field strengths and duality equations}
\label{sec:NLDual}

In this section, we shall construct non-linear field strengths for the bosonic fields and propose a duality equation invariant under rigid $E_{11}$ and non-linear local gauge transformations in the spirit of the generalised diffeomorphisms that are encountered in the $E_n$ exceptional field theory formulation of maximal supergravity theories in $D=11-n$ `external' space-time dimensions~\cite{Berman:2010is,Hohm:2013pua,Hohm:2013vpa,Hohm:2013uia,Hohm:2014fxa}. This section constitutes the first central result of the paper and some of the general formul\ae{} given here will be tested in various examples in the following sections.

\subsection{Preliminaries for general exceptional field theories}

Exceptional field theories for $E_n$ are formulated in an extended space-time in which the extra (internal) coordinates transform in a representation of the duality group $E_{n}$~\cite{Berman:2010is,Hohm:2013pua,Hohm:2013vpa,Hohm:2013uia,Hohm:2014fxa}. Furthermore, there exists a generalised diffeomorphism symmetry which closes on the fields satisfying the section constraint for $n\le 7$. This section constraint restricts the dependence of all fields and parameters on the extra coordinates such that they can depend at most on $n$ independent coordinates on the neighbourhood of each point. For $n\ge 8$, the algebra of generalised diffeomorphism needs to be extended to include not only diffeomorphism preserving the $E_n$ structure, but also additional gauge transformations that involve constrained parameters \cite{Hohm:2014fxa,Cederwall:2017fjm,Bossard:2017aae}. 

The formulation of $E_n$ exceptional field theory extends that of gauged maximal supergravity. In particular, the various fields in $E_n$ representations appearing in the tensor hierarchy of gauged supergravity~\cite{deWit:2005hv,deWit:2008ta} are also involved in exceptional field theory. For instance, the external scalar fields parametrizing the coset $E_n/K(E_n)$ play a significant role in the construction of the field equations in the form of a generalised metric.

A key feature of the $E_n$ exceptional field theories is that they typically require extra $p$-forms of rank $p\ge D-2$ (in some specific representations of 
$E_n$) beyond the ones present in the tensor hierarchies of maximal supergravity theories. These extra fields obey extra constraints related to the section constraint mentioned above and do not represent new dynamical degrees of freedom. Their first order field strengths are determined algebraically by source terms quadratic in the original fields. 
Finally, the $E_n$ exceptional field theory possesses an R-symmetry group which is the maximal compact subgroup of $E_n$, and that we denote by $K(E_n)$.\footnote{For fermions one has to consider the double cover $\widetilde{K}(E_n)$ that we shall also encounter for $E_{11}$ when we discuss coupling to fermions starting from Section~\ref{sec:susyALG}.} 

In view of the picture outline above for the $E_n$ exceptional field theories, it is natural to consider $E_{11}$ as the duality group and take the extended space-time to be parametrized by a vector in the fundamental representation $z^M t_M \in \overline{R(\Lambda_1)}$~\cite{West:2003fc}, such that the coordinates themselves $z^M$ transform in $R(\Lambda_1)$. In the following we will mostly refer to the representations of the coordinates, rather than the representations of the vectors.  It is also natural to introduce the coset $E_{11}/K(E_{11})$ where $K(E_{11})$ is a maximal subgroup of $E_{11}$ defined by being invariant under the (temporal) Cartan involution, as was considered long ago by West \cite{West:2001as,Englert:2003py}. 
Indeed, it is known that the $GL(11-n,\mathds{R}) \times E_n$ decomposition of the fields that parametrize the coset $E_{11}/K(E_{11})$ does contain all the  supergravity fields \cite{Riccioni:2007au,Bergshoeff:2007qi}, see also Appendix~\ref{app:e11} for the cases $n=0$ and $n=8$. This remarkable fact is encouraging but the extra constrained $p$-forms discussed above, which play a key role in the description of the exceptional field theories, are absent in this picture. As such, an $E_{11}$ exceptional field theory similar to the $E_n$ theories cannot be formulated only in terms of fields valued in $E_{11}/K(E_{11})$, which depend on coordinates $z^M$. However, the tensor hierarchy algebra introduced in Section~\ref{sec:e11tha} provides the required additional building blocks to tackle this problem. $E_{11}$ invariant consistent field equations were obtained at the linearised level in \cite{Bossard:2017wxl} by employing building blocks provided by the tensor hierarchy algebra~\cite{Palmkvist:2013vya}. Here we are going to reconsider these equations and show that they are in fact invariant under non-linear generalised diffeomorphisms provided one defines appropriately the gauge transformations of the constrained fields.

\subsection{Differential complex from \texorpdfstring{$\cT(\mf{e}_{11})$}{TE11} and the section constraint }

As observed in \cite{Bossard:2017wxl}, the tensor hierarchy algebra $\cT=\oplus_{p} \cT_p$ defines a differential complex of functions depending on coordinates $z^{M}$  that transform as $P^M \in R(\Lambda_1) \subset \cT_1$. The differential is defined through the adjoint action of the basis elements $P^{M}$ in $\cT_1$ and thus shift the degree $p$ by $1$ in the complex. Acting on any function in the complex we let\footnote{Here, we assume that $\cT_1$ as an $E_{11}$ module admits $R(\Lambda_1)$ as an irreducible submodule. We have checked that the components $R(\Lambda_{1}+\Lambda_{10})$ and $R(\Lambda_{11})$ can consistently be set to zero~\cite{Bossard:2017wxl}, but we do not have complete proof that $R(\Lambda_1)$ is indeed an irreducible submodule in $\cT_1$.}
\begin{align}
d = ({\rm ad}\,P^{M} )\,\partial_{M}\ . 
\label{Dext}
\end{align}
For this differential to square to zero one needs 
\begin{align}
d^2 = ({\rm ad}\,P^M)\,({\rm ad}\,P^N) \,\partial_M \partial_N = \Pi_\Lambda{}^{MN}{}\, ({\rm ad}\,P^\Lambda) \,\partial_M \partial_N = 0\ ,
\end{align}
which is equivalent to the condition that any field $\Phi(z)$ in the complex satisfies
\be 
\label{eq:SC}
\Pi_\Lambda{}^{MN}\ \partial_M  \partial_N \Phi (z) = 0 \ .
\ee
This is nothing but the weak section constraint and $P^\Lambda$ is the $\cT_2$ generator introduced in Table~\ref{tha2}. Its strong version (acting on arbitrary products of fields) can be written as \cite{Bossard:2017aae}
\begin{align}
\kappa_{\alpha\beta} T^{\alpha P}{}_M T^{\beta Q}{}_N \partial_P \otimes \partial_Q = -\frac12 \partial_M\otimes \partial_N + \partial_N \otimes \partial_M\ ,
\label{msc}
\end{align}
using the $E_{11}$ generators in the $R(\Lambda_1)$ representation and the inverse $E_{11}$ Killing metric $\kappa_{\alpha\beta}$ in the adjoint of $\mf{e}_{11}$. $\kappa_{\alpha\beta}$ is $E_{11}$ invariant and non-degenerate~\cite{Kac}.

The differential complex defined in this way serves as a basis for the construction of the field equations, such that the degree $p=-3$ supports the gauge parameters, $p=-2$ the potentials, $p=-1$ the field strengths, and $p=0$ the Bianchi identities, as can be anticipated from Table~\ref{tha2}. Note that the potentials belong to a module in the co-adjoint representation of
$\cT_0$ residing at level $p=-2$ rather than level $p=0$. Because $\cT_0$ is not reductive, the co-adjoint $\cT_0^*\cong \cT_{-2}$ is not an algebra. Therefore one cannot define a non-linear theory from a putative Maurer--Cartan form in $\cT_{0}^*$ alone. Moreover, within the tensor hierarchy algebra, $\cT_{-2}$ generates arbitrarily negative levels. This problem will be resolved by treating the fields in $\mf{e}_{11}\cong \mf{e}_{11}^*$ differently from the fields in the complement.

At the linearised level, the differential complex introduced above provides the following explicit expression for the field strengths  at  $p=-1$ given by the exterior differential of the potentials $\phi^{\widehat\alpha}$ at $p=-2$ via\footnote{In analogy with the role played by $\Theta^I$ for $E_n$ exceptional field theories, we could also call this the `embedding tensor representation'. For example, for $E_7$ this is the $(912+56)$-dimensional representation of $E_7$.}
\begin{align} 
\label{eq:Flin}
 \Theta^I t_I   = d (\phi^{\widehat{\alpha}} \bar{t}_{\widehat{\alpha}}  )= C^{IM}{}_{\widehat{\alpha}} \partial_M \phi^{\widehat{\alpha}} t_I
= \left( C^{IM}{}_\alpha \partial_M \phi^\alpha + C^{IM}{}_{\tilde\alpha} \partial_M X^{\tilde\alpha} + C^{IM}{}_\Lambda \partial_M Y^\Lambda +\ldots \right) t_I \, . 
\end{align}
We recall that the well-definedness of the $\mf{e}_{11}$-representation $\cT_{-1}$ of the field strengths follows from the tensor hierachy algebra although $\cT_{-1}$ is not a highest or lowest weight representation.

By virtue of the (weak) section condition~\eqref{eq:SC} this field strength is gauge invariant under the linearised gauge transformation
\begin{align} 
\label{eq:glold}
\delta_\xi \phi^{\widehat{\alpha}}  \bar t_{\widehat{\alpha}} \equiv   d (\xi^{{M}} \bar P_{{M}})  = T^{\widehat{\alpha}M}{}_{{N}} \partial_M \xi^{{N}}\bar t_{\widehat{\alpha}}   \ .
\end{align}
Here, the fields $\phi^{\widehat\alpha}$ in $\cT_{-2}$ are valued in the full representation as indicated by the index $\widehat{\alpha}$. One can divide $\cT_{-2}$ into the co-adjoint of $\mf{e}_{11}$ and the dual of the extending representations. The fields $\phi^\alpha$ associated with the dual $\mf{e}_{11}^*$ can be thought of as the usual fields also arising in the coset $E_{11}/K(E_{11})$ while the remaining fields $(X^{\tilde\alpha}, \zeta^\Lambda,\ldots)$ will be those related to the extra constrained fields needed in the formulation of $E_{11}$ exceptional field theory. As we can see in~\eqref{eq:Flin}, the object $ \Theta^I t_I$ has an explicit derivative $\partial_M$ and therefore satisfies constraints due to the section condition. Now, it turns out that $ \Theta^I \equiv \Theta^I(\phi^\alpha, X^{\tilde{\alpha}}, Y^\Lambda,\ldots)$ can be used to construct a field strength $F^I$ at the linearised level which will provide a building block for the linearised duality equations. We define the linearised field strength on the coset fields by imposing the projection on $\phi^\alpha$ to be in the coset $\mathfrak{e}_{11} \ominus K(\mathfrak{e}_{11})$~\cite{Bossard:2017wxl}
\begin{align}
\label{FieldStrengthDef} 
 F^I_\lin =  \T^I \left( \phi +\eta\phi^\dagger \eta , X, Y,\ldots \right)\ ,
\end{align}
where $ \phi + \eta \phi^\dagger \eta$ is short for $ \phi^\alpha T_{\alpha}{}^M{}_N + \phi^\alpha  \eta^{MQ} \eta_{NP} T_{\alpha}{}^P{}_Q$, where  $\eta_{MN}$ is the $K(E_{11})$ invariant metric  on the $R(\Lambda_1)$ module and $\eta^{MN}$ its inverse. This projection ensures at the linearised level that $\phi^\alpha$ can be shifted by an arbitrary $K(\mathfrak{e}_{11})$ element without modifying the field strength $F^I_\lin $. Note that the field strengths $F^I_\lin$ in~\eqref{FieldStrengthDef} are only $K(E_{11})$ covariant while the $\Theta^I$ are $E_{11}$ covariant. Moreover, this 
additional term violates gauge invariance~\cite{Bossard:2017wxl}. We shall see nonetheless that one can accommodate the gauge transformation of the fields $X$ and $Y$ such that the duality equations described below are gauge invariant. 

 With that understood, and due to the symplectic structure of $p=-1$ one can write down the duality equation
 \begin{align} \label{EtaLinear}
 F^I_\lin = \eta^{IJ} \Omega_{JK} F^K_\lin\,,
\end{align}
where $\eta^{IJ}$  is a symmetric non-degenerate $K(E_{11})$ invariant bi-linear form on $\mathcal{T}_{-1}$ and $\Omega_{JK}$ the $E_{11}$ invariant symplectic form on  $\mathcal{T}_{-1}$.\footnote{As we do not know whether $\cT_{-1}$ is completely reducible, it is possible that $\eta^{IJ}$ is only well-defined and non-degenerate on some (maximal) completely reducible submodule. For simplicity, we shall only refer to $\cT_{-1}$ as this (maximal) completely reducible submodule.}

In what follows we propose a non-linear extension of this duality equation. One important step will be to  replace the partial derivative of the extra fields $(X^{\tilde\alpha}, \zeta^{\Lambda},\ldots)$ by constrained  fields that are familiar from exceptional field theory, in particular for $E_9$~\cite{Bossard:2018utw}, in which case the tensor hierarchy algebra gives rise to the non-semi-simple Virasoro extension $\mathfrak{e}_9 \oleft \langle L_{-1}\rangle $ of $\mathfrak{e}_9$.

\subsection{Proposal for the non-linear duality equations}

In this section we shall argue that the construction of a  non-linear $E_{11}$ exceptional field theory can be achieved by defining the non-linear duality equation 
\be
\label{eq:guessDE}
F^I = \cM^{IJ} \Omega_{JK} F^K\ ,
\ee
where $\cM_{IJ}$ is the exceptional metric, a function of the fields in $E_{11} / K(E_{11})$, in the field strength representation, and $F^I$ is a non-linear field strength whose definition needs to incorporate the extra constrained fields that are expected to arise from what we already know from the structure of the $E_n$ exceptional field theories for lower $n$. In constructing this field strength, we shall use the tensor hierarchy algebra extension of $\mf{e}_{11}$ introduced in Section~\ref{sec:THA}. 

Before defining the non-linear $F^I$ we shall give more details on the definition of the exceptional metric $\cM_{IJ}$. Let $\cV(z)$ be a coset representative of $E_{11}/K(E_{11})$ transforming as $\cV(z) \to k(z) \cV(g^{-1} z) g$ with $g\in E_{11}$ a global element and $k(z)$ a local $K(E_{11})$ element. `Local' here refers to the dependence on the extended space-time with coordinates $z^M$ where $M$ labels the $R(\Lambda_1)$ representation of $E_{11}$ occurring at level $p=+1$ in the tensor hierarchy algebra. The subgroup $K(E_{11})$ is defined as the subgroup of elements $k$ that preserve a non-Euclidean metric $\eta$ such that $k^\dagger \eta k = \eta$ in a suitable highest (or lowest) weight representation where the Hermitian conjugate can be defined~\cite{Kac}. The non-Euclidean nature means for example that $K(E_{11})\cap GL(11) = SO(1,10)$ where $GL(11)$ denotes the regular $GL(11)$ subgroup of $E_{11}$ that appears in the level decomposition relevant for describing $D=11$ supergravity. In other words, $\eta$ is the standard Minkowski metric of eleven-dimensional space-time extended to the whole extended space-time.\footnote{As shown in~\cite{Keurentjes:2004bv}, there are other $D=11$ signatures embedded in $K(E_{11})$ that relate to so-called exotic forms of supergravity~\cite{Hull:1998ym} that have more than one time-like direction.} 

As usual in exceptional field theory, it is convenient to work with the exceptional metric in order to avoid the introduction of the  $K(E_{11})$ gauge invariance and its gauge-fixing,
\be
\label{eq:Mdef}
\cM(z)= \cV(z)^\dagger \eta \cV(z)\ \quad \to \quad g^\dagger \cM(g^{-1} z) g\ .
\ee
This definition is in complete analogy with non-linear realisations of finite-dimensional groups but requires some care in the case of infinite-dimensional Kac--Moody groups. We shall make more comments on this subtlety when we discuss the vielbein and its gauge transformation in section~\ref{sec:Vgauge}.

A key building block for the duality equations is the current defined as
\begin{align}
J_M{}^\alpha \kappa_{\alpha\beta}  t^\beta  = \cM^{-1} \partial_M \cM\quad\quad \in  \overline{R(\Lambda_1)}\otimes \mf{e}_{11}\ .
\label{J}
\end{align}
This expression makes sense for the so-called small group in any integrable module \cite{DeMedts:2009}. Since $\cM$ only involves $E_{11}/K(E_{11})$, the combination $\cM^{-1}\partial_M \cM$ can be expanded in the adjoint of $E_{11}$ and transforms covariantly in the tensor product of $\overline{R(\Lambda_1)}$ with the adjoint with, where the former factor is due to the partial derivative.

We use the current~\eqref{J} to define the non-linear field strength $F^I$ by
\begin{align}
\label{eq:NLFS}
 F^I = C^{I M}{}_ \alpha J_M{}^\alpha+ C^{I M}{}_{\tilde\alpha} \chi_M{}^{\tilde\alpha}+ C^{I M}{}_\Lambda  \zeta_M{}^\Lambda +\dots 
\end{align}
where the structure coefficients are the same as in \eqref{eq:Flin}, but $\partial_M \phi^\alpha$ has been promoted to the non-linear current $J_M{}^\alpha$, while the partial derivative $\partial_M X^{\tilde{\alpha}},\, \partial_M Y^\Lambda,\, \dots $ are promoted to constrained fields $\chi_M{}^{\tilde{\alpha}},\, \zeta_M{}^{\Lambda},\, \dots $ that are not total derivatives, but satisfy the section constraint~\eqref{msc} on their index $M$. It is important to stress that the structure coefficients are defined such that the potential $\phi^{\hat{\alpha}} \bar{t}_{\hat{\alpha}}$ is in $\mathcal{T}_{-2}$ and not in $\mathcal{T}_0$, such that one cannot simply extend $J_{M \alpha} t^\alpha$ to a current in the extended algebra $\mathcal{T}_0$. Nonetheless, because $\mathfrak{e}_{11}$ is simple one can raise the index of the $E_{11}/K(E_{11})$ current to get $J_M{}^\alpha \bar t_\alpha$. The indecomposable structure of the module $\mathcal{T}_{-2}$ implies that the field strength must necessarily involve an additional field in $\overline{R(\Lambda_1)}\otimes R(\Lambda_2)$, and because $J_M{}^\alpha  \bar t_\alpha$ is not a total derivative, this additional field $\chi_M{}^{\tilde{\alpha}}$ cannot be a total derivative either. It is nevertheless consistent with the indecomposable representation to require that it satisfies the strong section constraint. We shall therefore introduce the \textit{constrained fields}  $\chi_M{}^{\tilde{\alpha}},\, \zeta_M{}^{\Lambda},\, \dots $  so that they transform under $E_{11}$ according to the indecomposable representation and  the field strength $F^I$ is indeed an $E_{11}$ tensor in $\mathcal{T}_{-1}$. Even though the additional field $\zeta_M{}^{\Lambda}$ in $\overline{R(\Lambda_1)}\otimes R(\Lambda_{10})$ is not required by $E_{11}$ covariance, we shall see that both $\chi_M{}^{\tilde{\alpha}}$ and $ \zeta_M{}^{\Lambda}$ are necessary to write down a twisted selfduality equation \eqref{eq:guessDE} covariant under generalised diffeomorphisms.  In the current paper we assume implicitly that we can consistently truncate to this known part of $\cT_{-2}$ but in principle an extension to additional modules in $\cT_{-2}$ can be envisaged as indicated by the ellipsis, which will be dropped for short in the following.\footnote{The field $\zeta_M{}^\Lambda$ associated with $\overline{R(\Lambda_{10})}$ was not considered in the linearised analysis in~\cite{Bossard:2017wxl}. Including it here has the benefit of making the first order duality equation gauge invariant while~\cite{Bossard:2017wxl} only had gauge invariant second order equations.}

We now describe in more detail why~\eqref{eq:NLFS} defines an $E_{11}$ covariant object due to the indecomposable structure of the module $\cT_{-2}$. Under rigid $E_{11}$ transformations one has as in~\eqref{eq:indact}
\vskip -4mm
\begin{subequations}
\label{eq:JchiG}
\begin{align}  
J_M{}^\alpha &\rightarrow   g^{-1N}{}_M\, g^\alpha{}_\beta\,  J_N{}^\beta \ , 
\\
\chi_M{}^{\tilde\alpha} &\rightarrow   g^{-1N}{}_M \bigl( g^{\tilde\alpha}{}_{\tilde\beta}  \chi_N{}^{\tilde\beta}  + \omega^{\tilde\alpha}_\beta       (g) J_N{}^\beta \bigr)\ .
\end{align}
\end{subequations}
Recalling \eqref{sc1}, it follows that
\be 
C^{IM}{}_\alpha J_M{}^\alpha + C^{IM}{}_{\tilde\alpha} \chi_M{}^{\tilde \alpha} \rightarrow g^{I}{}_J \bigl(  C^{JM}{}_\alpha J_M{}^\alpha + C^{JM}{}_{\tilde\alpha}\chi_M{}^{\tilde \alpha} \bigr) 
\ee
transforms covariantly. The cocycle appearing in the indecomposable representation is crucial in this calculation. The infinitesimal transformations under  $g = \exp(\Lambda_\alpha t^\alpha)$ corresponding to~\eqref{eq:JchiG} are
\vskip -6mm
\begin{subequations}
\begin{align}
\delta_\Lambda J_{M}{}^\alpha &=  \Lambda_\gamma \bigl(- f^{\gamma\alpha}{}_\beta J_M{}^\beta - T^{\gamma N}{}_M J_N{}^\alpha \bigr) \ ,
\\
\delta_\Lambda \chi_M{}^{\tilde{\alpha}} &=  \Lambda_\gamma \bigl( T^{\gamma\tilde{\alpha}}{}_{\tilde{\beta}} \chi_M{}^{\tilde{\beta}} - T^{\gamma N}{}_M \chi_ N{}^{\tilde\alpha} \bigr) + \Lambda_\gamma K^{\gamma\tilde\alpha}{}_\beta  J_M{}^\beta \  .  
\label{nw1}
\end{align}
\end{subequations}

\subsection{Gauge invariance of the non-linear field equations}

Having introduced an $E_{11}$ covariant tensor field strength $F^I$, the next step is to compute how it transforms under $E_{11}$ generalised diffeomorphism. The gauge parameter $\xi^M$ of the generalised Lie derivative transforms also in $R(\Lambda_1)$  just as the coordinates $z^M$. Note that additional gauge transformations with constrained gauge parameters are required for the closure of the algebra of generalised Lie derivatives for $\mf{e}_n$ with  $n\geq 8$~\cite{Hohm:2014fxa,Cederwall:2015ica,Bossard:2017aae,Cederwall:2017fjm}. We shall not check the closure of the algebra of generalised Lie derivatives for $\mf{e}_{11}$, but we will comment  on these additional transformations in Section~\ref{sec:fulldyn}.

The dynamical degrees of freedom of the theory appear through the representative $\cM$. Therefore we start by defining the generalised Lie derivative with parameter $\xi$ acting on $\cM$. The formula, as for all exceptional field theories,  can be defined as (see e.g.~\cite{Bossard:2018utw})
\begin{align}
\delta_\xi \cM = \xi^M \partial_M \cM + \kappa_{\alpha\beta} T^{\alpha M}{}_N \partial_M \xi^N \left( \cM t^\beta + t^{\beta\dagger} \cM\right)\ .
\label{gt1}
\end{align}
This formula reproduces the unique linearised gauge transformation studied in~\cite{West:2014eza,Bossard:2017wxl} in the linearised approximation and provides a non-linear extension of it.

Combining the definition of $J_M{}^\alpha$ in \eqref{J} and the transformation \eqref{gt1} it follows that 
\begin{multline}
\delta_\xi J_M{}^\alpha = \xi^N \partial_N J_M{}^\alpha -T^{\beta N}{}_P \partial_N \xi^P  f_{\beta\gamma}{}^\alpha J_M^\gamma + \partial_M\xi^N J_N{}^\alpha
\\
 + T^{\alpha N}{}_P \left( \partial_M\partial_N \xi^P + \cM_{NQ} \cM^{PR} \partial_M\partial_R\xi^Q\right)\ ,
\end{multline}
where the  third term in the first line originates from the derivative in the current acting on the generalised diffeomorphism parameter in the variation of $\cM$. Using the section constraint~\eqref{msc} on this term, one can recognise the first line as the expected generalised Lie derivative, including the transport term, an infinitesimal $\mathfrak{e}_{11}$ transformation plus a weight term, \ie\;  introducing the  notation $T_\alpha{}^M{}_N \equiv \kappa_{\alpha\beta} T^{\beta M}{}_N$
\begin{multline}
\delta_\xi J_M{}^\alpha = \xi^N \partial_N J_M{}^\alpha + T_\beta ^N{}_P \partial_N \xi^P  f^{\beta\alpha}{}_\gamma J_M{}^\gamma +T_\beta{}^P{}_Q \partial_P \xi^Q T^\beta{}^N{}_M J_N{}^\alpha +\frac12 \partial_N \xi^N J_M{}^\alpha  
\\
 +T^{\alpha N}{}_P \left( \partial_M\partial_N \xi^P + \cM_{NQ} \cM^{PR} \partial_M\partial_R\xi^Q\right)\ .
\label{gt1J}
\end{multline}
The  inhomogeneous terms in the second line are non-covariant variations that resemble the linearised gauge transformation of the linearised current. At this level the variation of the current is identical in structure to what one would obtain for any $E_n$ exceptional field theory.

In order to obtain a consistent transformation of the field strength $F^I$, it is necessary that all the components in $\cT_{-2}$ transform according to the indecomposable representation of $E_{11}$. To this end, it is useful to introduce the notation $J_M{}^{\widehat{\alpha}} \equiv ( J_M{}^\alpha , \chi_M{}^{\tilde{\alpha}} , \zeta_M{}^{\Lambda}, \ldots)$ that includes the additional constrained fields in a single object transforming as an element $J_M{}^{\widehat{\alpha}} \bar t_{\hat{\alpha}}$ of $\cT_{-2}$. In this way one can define the following ansatz for the gauge transformation of $J_M{}^{\widehat{\alpha}} $ that is manifestly consistent with the indecomposable representation 
\begin{align}
\label{eq:gt4}
\delta_\xi J_M{}^{\widehat{\alpha}} &= \xi^N \partial_N J_M{}^{\widehat{\alpha}} +T_{\beta}{}^{N}{}_P \partial_N \xi^P  f^{\beta\widehat{\alpha}}{}_{\widehat{\gamma}} J_M{}^{\widehat{\gamma}} +T_\beta{}^P{}_Q \partial_P \xi^Q T^\beta{}^N{}_M J_N{}^{\widehat{\alpha}} +\frac12 \partial_N \xi^N J_M{}^{\widehat{\alpha}}  
\nn\\
&\quad + T^{\widehat{\alpha} N}{}_P \left( \partial_M\partial_N \xi^P + \cM_{NQ} \cM^{PR} \partial_M\partial_R\xi^Q\right)+{\Pi^{\widehat\alpha}{}_{ Q P }} \cM^{NQ} \partial_M \partial_N \xi^P\,.
\end{align}
This formula extends~\eqref{gt1J} to all components $\widehat{\alpha}$ with structure coefficients invariant under $E_{11}$ for the indecomposable module, and we have added one extra term at the very end of the equation with the understanding that ${\Pi^{\alpha}{}_{ Q P }}=0$, while ${\Pi^{\tilde\alpha}{}_{ Q P }} $ and ${\Pi^{\Lambda}{}_{ Q P }} $ are the highest weight projectors from $R(\Lambda_1)\otimes R(\Lambda_1)$ to $R(\Lambda_2)$ and $R(\Lambda_{10})$, respectively. These exist due to \eqref{HighestWeightDcompo}. We again restrict only to $R(\Lambda_2)$ and $R(\Lambda_{10})$; additional $E_{11}$ representations in $\cT_{-2}$ could be accommodated in the same way, as long as they would appear in the tensor product $R(\Lambda_1)\otimes R(\Lambda_1)$.

Assuming the uniform gauge transformation~\eqref{eq:gt4} we can read off the gauge transformations of the constrained fields to be 
\begin{subequations}
\label{eq:gtt}
\begin{align}
\delta_\xi \chi_M{}^{\tilde\alpha} &= \xi^N \partial_N \chi_M{}^{\tilde\alpha} -T_\alpha{}^N{}_P \partial_N \xi^P T^{\alpha\tilde\alpha}{}_{\tilde\beta} \chi_M{}^{\tilde\beta}  + \partial_M \xi^N \chi_{N}{}^{\tilde\alpha} -T_\alpha{}^ N{}_P \partial_N \xi^P K^{\alpha\tilde\alpha}{}_{\beta} J_M{}^\beta 
\nn\\
&\quad + T^{\tilde\alpha N}{}_P \left( \partial_M\partial_N \xi^P + \cM_{NQ} \cM^{PR} \partial_M\partial_R\xi^Q\right)+ {\Pi^{\tilde\alpha}{}_{ Q P }} \cM^{NQ} \partial_M \partial_N \xi^P \,, 
\label{gt2}\w2
\delta_\xi \zeta_{M}{}^\Lambda &= \xi^N \partial_N \zeta_{M}{}^\Lambda - T_\alpha{}^N{}_P \partial_N \xi^P T^{\alpha \Lambda}{}_{\Xi}\zeta_{M}{}^\Xi  + \partial_M \xi^N \zeta_{N}{}^\Lambda  \nn\\
&\quad +  T^{\Lambda N}{}_P\left( \partial_M\partial_N \xi^P + \cM_{NQ} \cM^{PR} \partial_M\partial_R\xi^Q\right)+ {\Pi^\Lambda{}_{Q P }} \cM^{NQ} \partial_M \partial_N \xi^P \ .
\label{gt3}
\end{align}
\end{subequations}
As explained above, this form is by construction in agreement with the indecomposable structure of the module: The gauge transformation of $\chi_M{}^{\tilde{\alpha}}$ must include the same gauge transformation as $J_M{}^\alpha$ with the index $\tilde{\alpha}$ instead, so that the gauge transformation of $J_M{}^{\widehat{\alpha}} $ is written in terms of invariant tensors. The indecomposable structure is such that one has the freedom to add any transformation of $\chi_M{}^{\tilde{\alpha}}$ in the $R(\Lambda_2)$ module, which gives the freedom to add the term involving the projector ${\Pi^{\tilde\alpha}{}_{ Q P }}$. It will turn out that this additional transformation is indeed necessary for the duality equations to be gauge invariant under generalised diffeomorphisms. 

Equipped with the formulas above, we can now compute the variation of the non-linear field strength~\eqref{eq:NLFS} under gauge transformations and find 
\begin{align}
\label{eq:lieF}
\delta_\xi F^I &=  \xi^M \partial_M F^I -  T_\alpha{}^N{}_M \partial_N \xi^M T^\alpha{}^I{}_J F^J +\frac12 \partial_M\xi^M F^I
\nn\\
&\quad  + C^{IM}{}_{\widehat{\alpha}}\Bigl(  T^{\widehat{\alpha} N}{}_P \left( \partial_M\partial_N \xi^P + \cM_{NQ} \cM^{PR} \partial_M\partial_R\xi^Q\right)+{\Pi^{\widehat\alpha}{}_{ Q P }} \cM^{NQ} \partial_M \partial_N \xi^P \Bigr)  \nn\\
&=  \xi^M \partial_M F^I -  T_\alpha{}^N{}_M \partial_N \xi^M T^\alpha{}^I{}_J F^J +\frac12 \partial_M\xi^M F^I
\nn\\
&\quad  + \Bigl(\bigl( C^{IM}{}_ \alpha T^{\alpha R}{}_Q+ C^{IM}{}_{ \tilde\alpha} T^{\tilde\alpha R}{}_Q+ C^{IM}{}_\Lambda  T^{\Lambda R}{}_Q \bigr)  \cM^{QN}\cM_{RP} \CR
& \hspace{40mm} + C^{IM}{}_{\tilde\alpha} \Pi^{\tilde\alpha}{}_{QP}\cM^{QN}+ C^{IM}{}_\Lambda \Pi^\Lambda{}_{QP}\cM^{QN} \Bigr)   \partial_M \partial_N \xi^P \,.
\end{align}
Here, we have recombined the terms into the generalised Lie derivative of $F^I$ in the first lines of the two equations. The transport term and the $\mathfrak{e}_{11}$ transformation term recombine by invariance of the structure coefficients $C^{IM}{}_{\widehat{\alpha}}$. The last term of the first lines determines the weight of $F^I$. The inhomogeneous terms that do not involve conjugation, $C^{IM}{}_{\widehat{\alpha}}  T^{\widehat{\alpha} N}{}_P  \partial_M\partial_N \xi^P$,  combine into the section constraint by the Jacobi identity and have been removed when going to the second step. The remaining inhomogeneous terms at the end show that the field strength $F^I$ is thus \textit{not}  gauge covariant. 

The non-covariance of $F^I$ is not a problem since we are only interested in constructing a gauge invariant dynamics. More specifically, we only demand that the duality equation 
\be \label{DualityEquation}
\cE^I \equiv F^I - \cM^{IK} \Omega_{KJ} F^J =0 
\ee
transforms into itself under generalised diffeomorphism. 

Performing its gauge transformation, we find
\begin{align}
\delta_\xi \cE^I &= \xi^M \partial_M \cE^I - T_\alpha{}^N{}_M \partial_N \xi^M  T^{\alpha I}{}_J \cE^J + \frac12 \partial_M\xi^M  \cE^I+ \left(\delta_J^I - \cM^{IK}\Omega_{KJ}\right)\partial_M \partial_N \xi^P 
\\
&\quad \times \left( C^{JM}{}_{\tilde\alpha} \Pi^{\tilde\alpha}{}_{QP} \cM^{QN} +C^{JM}{}_\Lambda \Pi^\Lambda{}_{QP} \cM^{QN} - \cM^{JL} \Omega_{LK'}  C^{K' M}{}_{\hat{\alpha}} T^{\hat{\alpha}R}{}_Q \cM^{QN} \cM_{RP}\right) \ ,
\nn
\end{align}
where we used $(\cM\Omega)^2 = 1$. Because one derivative comes contracted with $\cM^{MN} \partial_N$ and the other derivative does not, one cannot use the section constraint to cancel the inhomogeneous term. For gauge covariance of the duality equation we must therefore require that 
\be
\label{MasterAssum}  
 C^{JM}{}_{\tilde\alpha} \Pi^{\tilde\alpha}{}_{QP} \cM^{QN} +C^{JM}{}_\Lambda \Pi^\Lambda{}_{QP} \cM^{QN} - \cM^{JL} \Omega_{LK'}  C^{K' M}{}_{\hat{\alpha}} T^{\hat{\alpha} R}{}_Q \cM^{QN} \cM_{RP}  =  0 \ .
\ee
By construction, the conjugation by $\cM$ in various representations allows us to define various conjugate invariant tensors\footnote{We note that $\eta$ is not $E_{11}$ invariant, only $K(E_{11})$ invariant. The construction above is similar to tensor representations of $\mf{sl}(n)$: even though $\delta_{ab}$ is not an invariant tensor of $\mf{sl}(n)$ but only of $\mf{so}(n)$, it can be used to relate a tensor representation to its dual. Abstractly, $\eta$ is also an automorphism of the $E_{11}$ algebra relating the highest weight module to the lowest weight module. In particular $\eta^{\alpha\delta} \eta^{\beta\epsilon} \eta_{\gamma\vartheta} f_{\delta\epsilon}{}^{\vartheta} =- f^{\alpha\beta}{}_\gamma$.}
\bea \label{ConjugateConstants} 
\Pi_{\tilde{\alpha}}{}^{MN}  &=& \cM^{MP} \cM^{NQ} \cM_{\tilde{\alpha}\tilde{\beta}} \Pi^{\tilde{\beta}}{}_{PQ}  = \eta^{MP} \eta^{NQ} \eta_{\tilde{\alpha}\tilde{\beta}} \Pi^{\tilde{\beta}}{}_{PQ} \ , \nn\\
\Pi_\Lambda{}^{MN}  &=& \cM^{MP} \cM^{NQ} \cM_{\Lambda\Xi} \Pi^{\Xi}{}_{PQ}= \eta^{MP} \eta^{NQ} \eta_{\Lambda\Xi} \Pi^{\Xi}{}_{PQ}\  , \eea
 and similarly for $C_{IM}{}^{\widehat{\alpha}}$ assuming the existence of the matrix $\eta_{IJ}$ defining a symmetric $K(E_{11})$ non-degenerate bilinear form on $\cT_{-1}$.  Multiplying \eqref{MasterAssum} by $\cM_{NR}$ to remove the explicit scalar matrix dependence and using  \eqref{ConjugateConstants}, one obtains the following necessary and sufficient algebraic condition for gauge invariance in terms of $E_{11}$ structure constants:  
\be 
\label{MasterAssum2}  
C_{I P}{}^{\tilde\alpha} \Pi_{\tilde\alpha}{}^{MN} +C_{IP}{}^\Lambda \Pi_\Lambda{}^{MN}  \stackrel{!}{=} \Omega_{IJ}   C^{J M}{}_ {\hat{\alpha}} T^{\hat{\alpha} N}{}_P \ .   
\ee
If this condition is satisfied, the first order duality equation~\eqref{DualityEquation} is gauge invariant. We do not have a general proof of this central condition. In the next sections, we will verify condition~\eqref{MasterAssum2} in the $GL(11)$ and the $GL(3)\times E_8$ level decompositions at low levels. This will provide non-trivial support that this condition is satisfied.

Let us now try to give some heuristic argument why~\eqref{MasterAssum2} is plausible.
Taking the symmetric part in $M$ and $N$ of the equation, the right-hand-side becomes the Jacobi identity $2[ P^{(M} , \{ P^{N)} , \bar P_P\}] = [\{ P^M , P^N\} , \bar P_P]$ in the tensor hierarchy algebra. Therefore the symmetrized condition reduces to 
\be  
[ P^\Lambda , \bar P_N ] =- 2  \Omega^{IJ} C_{JP}{}^\Lambda  t_I  \: . 
\ee
That the coefficient $C_{JP}{}^\Lambda$ occurring through the condition~\eqref{MasterAssum2} is the same as this corresponding structure coefficient in the tensor hierarchy algebra is not guaranteed a priori. But since both are $E_{11}$ invariant tensors, they must be proportional to each other such that up to a conventional factor this identity must be true. Since we do not assume that the modules in $\cT(\mf{e}_{11})$ must necessarily be irreducible, meaning that $R(\Lambda_2)$ would be extended to the complete anti-symmetric tensor product and $R(\Lambda_{10})$ to the complete symmetric product minus the highest weight module $R(2\Lambda_1)$, the identity \eqref{MasterAssum2} requires that one can choose the coefficients defining $\Pi_{\tilde\alpha}{}^{MN} $ and $\Pi_\Lambda{}^{MN} $ such that the identity holds, which is simply the statement that there is no component in $R(2\Lambda_1)$ by the Jacobi identity in the tensor hierarchy algebra. 

\subsection{Gauge transformation of the vielbein and compensators}
\label{sec:Vgauge}

In the discussion above, we have relied on the `generalised metric' element $\cM$ defined in~\eqref{eq:Mdef} from the `generalised vielbein' $\cV\in E_{11}/K(E_{11})$. These objects have to be treated with care since the Lie algebra $\mf{e}_{11}$ is infinite-dimensional and one first has to define what group one associates with it. A standard building block is to consider the one-parameter subgroups of the form $e^{t E_\alpha}$ ($t\in \reals$) for generators $E_{\alpha}$ associated to \textit{real} roots. The group built from taking finite products of such real one-parameter subgroups generates what is called the \textit{small Kac--Moody group}~\cite{KacPeterson,Marquis}. The action of this small Kac--Moody group is completely under control for so-called integrable representations of the algebra $\mf{e}_{11}$ where all real root generators are locally nilpotent, meaning that the repeated action of $E_{\alpha}$ on any element of the representation space terminates after a finite number of repetitions. Thus $e^{t E_{\alpha}}$ is effectively represented by a polynomial and multiplying finitely many polynomials gives a well-defined polynomial again without having to worry about convergence or similar matters. All highest and lowest weight representations of $\mf{e}_{11}$ are integrable and so is the adjoint representation~\cite{Kac}, and so $\cT_{-1}$ although it is neither highest nor lowest weight.  One can also associate the matrices $\cM$ and $\cM^{-1}$ with the small Kac--Moody group in the so-called `group model' (generalising the Cartan embedding) of the Kac--Moody symmetric space~\cite{Freyn:2017}. However, the current $\cJ = \cM^{-1} d\cM$ is not evidently meaningful since the continuous map from the small Kac--Moody group to the algebra and back cannot be defined.

An alternative model of the symmetric space can be obtained by using the Iwasawa decomposition, leading to what is sometimes called the `Kostant model'. Elements of the (small) Kac--Moody group have an Iwasawa decomposition $E_{11}= K(E_{11}) B$, where $B$ is the (upper) Borel subgroup~\cite{DeMedts:2009}. This setting also allows for considering the so-called \textit{completed Kac--Moody group} where one completes the Borel subgroup with respect to the topology of an associated building~\cite{Carbone:2003,Marquis}. A representative of the thus completed Kac--Moody symmetric space\footnote{We do not distinguish the completed Kac--Moody group from the uncompleted one in terms of notation since we shall always be using the completed group implicitly.} $E_{11}/K(E_{11})$ can then be chosen in standard form by products of exponentials of \textit{all} generators of $B$~\cite{Marquis}, including the positive imaginary ones. In this construction, it is only the positive Borel subgroup that is completed; the group $K(E_{11})$ is not changed.

Generalising this approach slightly, we can also consider a parabolic gauge with
\begin{align}
\label{eq:mpg}
\cV = v U\,,
\end{align}
where $v$ belongs to a finite-dimensional Levi subgroup of $E_{11}$ (such as $GL(11)$) and $U$ belongs to the unipotent subgroup associated with this parabolic subalgebra of $\mf{e}_{11}$. $U$ lies in a unipotent subgroup of the Borel subgroup $B$. If the parabolic subalgebra is associated with a level $\ell$, the generators appearing in $U$ correspond to level $\ell>0$ while the generators appearing in $v$ correspond to $\ell=0$. Explicitly, we can write
\begin{align}
U = \prod_{\ell>0} \exp \left( A_{\ell} \cdot E_{\ell} \right)\,,
\end{align}
in the completed group and the factors are ordered with smaller levels appearing to the left. Here, $E_{\ell}$ denotes all (finitely many) generators on level $\ell>0$ and $A_\ell$ are the coefficient of a general Lie algebra element on that level. The $A_{\ell}$ correspond to the fields and depend on the extended space-time coordinates $z^M$. 

With a parabolic parametrisation~\eqref{eq:mpg} of $\cV$ one can work out the Maurer--Cartan derivative $d\cV \cV^{-1}\in \mf{e}_{11}$ in a meaningful manner since every generator $E_{\ell}$ is multiplied by a polynomial in $A_m$ and $dA_m$ for $1<m\leq \ell$ (and dressed by the Levi vielbein $v$).\footnote{We note, however, that if one wanted to use this parametrisation to define a metric on the symmetric space from the invariant bilinear form on $\mf{e}_{11}$, every $dA_\ell$ would be multiplied by a infinite series of other fields and the convergence of this expression is doubtful and would at least required a completion of the Lie algebra as well.} Writing $d\cV \cV^{-1}= \mathcal{P} - \mathcal{Q}$ in the usual symmetric space decomposition, the current $\cJ$ in this Kostant model becomes $\cJ = 2\cV^{-1} \mathcal{P} \cV$ and is defined by the adjoint action of the completed Kac--Moody group on the Lie algebra. Formally, this results again in infinite series expressions for the components $J_{M}{}^\alpha$.
If we are interested in only obtaining polynomial expressions in the fields, we are therefore led to working with the components of $\mathcal{P}$. In lowest/highest weight representations one can also make sense of the matrix components of $\cV$ and $\cV^{-1}$, whether this remains true in unbounded representations like the field strength representation is not clear to us.

As is clear from the above discussion, it is typically better to consider the completed group and write the vielbein in a gauge-fixed form using a (maximal) parabolic gauge. Examples of such maximal parabolic gauges use the Levi groups are $GL(11-n)\times E_n\subset E_{11}$ with an associated level decomposition.\footnote{Note that only $GL(11-n)\times E_n\subset E_{11}$, with $GL(11-n)$ the linear group with positive determinant, so we always understand that $GL(11-n,\reals)$ is the connected component of the linear group, and does not include negative determinant elements.}
Since we have fixed a gauge the action of a generalised diffeomorphism on the vielbein $\cV$ will be accompanied by a compensating $K(E_{11})$ rotation that ensures that the gauge is maintained:
\begin{align}
\label{eq:lieV}
\delta_\xi \cV = \xi^N \partial_N \cV + \kappa_{\alpha\beta} T^{\alpha M}{}_N \partial_M\xi^N  \cV t^\beta + X\cV\,,
\end{align}
where the compensating transformation $X$ is an element of the Lie algebra of $K(E_{11})$ that acts on $\cV$ in the chosen representation. The same mechanism has been discussed also in~\cite{West:2014eza,Bossard:2017wxl}.

The compensator $X$ can be written more explicitly in a level decomposition associated with a maximal parabolic subgroup. Since the original $\cV$ is made out of generators at levels $\ell\geq 0$ by definition of the maximal parabolic gauge, only those $t^\beta$ that are associated with negative levels violate the gauge in the middle rotation term. By the Killing form they are paired with positive level generators $T^{\alpha M}{}_N$ contracting the rotation parameter $\partial_M\xi^N$, where it is important that the derivative $\partial_M$ is subject to the section constraint~\eqref{msc}.

Associated with a parabolic decomposition~\eqref{eq:mpg} is also a decomposition of the representation $R(\Lambda_1)$ of the derivatives $\partial_M$. In the case $GL(11-n)\times E_n$, the derivatives decompose into
\begin{align}
\label{eq:derdec}
\partial_M \to (\partial_\mu, \partial_A,\ldots)\,. 
\end{align}
The index $A$ here labels the coordinate representation of $E_n$ exceptional field theory. We can choose a partial solution to the $E_{11}$ section condition by keeping only these two lowest levels of derivatives, \ie\ setting to zero the ellipses in this decomposition. This solution to the section condition is only partial as the derivatives $\partial_A$ still have to satisfy the $E_n$ section condition. In connection to usual exceptional field theory the $\partial_\mu$ are called external derivatives and the $\partial_A$ internal derivatives. 

In such a partial solution to the section constraint there is only one generator that can arise in~\eqref{eq:lieV} and that needs to be compensated. It is the one mapping $\partial_\mu$ to $\partial_A$. All other positive level generators map to zero on this choice of section. The corresponding compensator then can be written explicitly as
\begin{align}
\label{eq:kmpg}
X = V_{\underline{A}}{}^A e_\mu^a \partial_A \xi^\mu (E^{\underline{A}}_a -  \eta_{ab} \delta^{\underline{A}\underline{B}} F_{\underline{B}}^b )\,,
\end{align}
where $V_{\underline{A}}{}^A$ is the $E_n / K(E_n)$ coset representative and $e_\mu{}^a$ the $GL(11-n)$ vielbein; together they form the $\ell=0$ part $v$ in~\eqref{eq:mpg}. The generators $E^{\underline{A}}_a$ and $ F_{\underline{B}}^b$ are the first level generators that are conjugates of each other. 

In particular, we see that formula~\eqref{eq:lieV} provides a completely well-defined expression for the generalised diffeomorphism action on the vielbein $\cV$. The compensating transformation will also be crucial when we consider fermions starting from Section~\ref{sec:susyALG}.

A final comment on the relation between the vielbein and metric formalism here concerns the issue of connection.
It is well-known that in exceptional field theory it is not possible to fix the affine or spin connection completely by the requirement of metric compatibility and torsion-freeness~\cite{Coimbra:2011nw,Coimbra:2012af,Cederwall:2013naa,Hohm:2013pua,Godazgar:2013dma,Godazgar:2014sla}, even though this arbitrariness drops out in the supergravity equations derived in generalised geometry~\cite{Coimbra:2011nw,Coimbra:2012af}. The definition of a (spin-)connection for $E_{11}$, as would be needed in the formulation \cite{West:2014eza} is a complicated open problem that we shall not address in this paper. 
The formulation we are using here avoids the problem of defining a (spin-)connection as we have defined the generalised Lie derivative acting on all objects in the theory.

\subsection{Comments on the extended dynamics and linearised field equations}
\label{sec:fulldyn}

Under the assumption~\eqref{MasterAssum2}, the duality equation \eqref{DualityEquation} is non-linear and  invariant under generalised diffeomorphisms. However, it is not sufficient to describe the whole dynamics. In the following sections where we consider specific solutions to the section constraint, we shall see in detail that most of the duality equation components only fix the constrained fields $\chi_M{}^{\tilde{\alpha}}$, $\zeta_M{}^\Lambda$ etc.  in terms of the current $J_M{}^\alpha$, and do not give any dynamical equation. The additional constrained fields  $\chi_M{}^{\tilde{\alpha}}$, $\zeta_M{}^\Lambda$ indeed appear algebraically in the duality equation~\eqref{DualityEquation}, so for a given solution to the section constraint, one can simply solve a large part of the duality equations in components by expressing the non-vanishing  $\chi_M{}^{\tilde{\alpha}}$, $\zeta_M{}^\Lambda$ as functions of the current. We shall see that among the infinitely many components of the duality equations, only a finite number remains non-trivial and gives rise to dynamical constraints on the fields parametrizing the $E_{11}/ K(E_{11})$ coset. In particular, for a decomposition of the type $GL(11- n) \times E_n\subset E_{11}$, it seems that the only remaining dynamical duality equations are the ones involving $GL(11-n)$ $p$-forms, while the dual graviton equation and the higher rank mixed symmetry equivalents involve the constrained fields in a way that trivialises the dynamics. This should not be so much a surprise since the necessity to introduce extra auxiliary fields trivialising the dynamics seems unavoidable in defining dual gravity at the non-linear level~\cite{West:2002jj,Boulanger:2008nd}. The non-vanishing constrained fields with the $M$ index along $GL(11-n)$ seem then to play the role of the St\"uckelberg fields introduced in~\cite{Boulanger:2008nd}, and their generalisation to all duality equations. 

What is lacking in order to recover the full dynamics of the theory, are additional first order equations for the constrained fields $\chi_M{}^{\tilde{\alpha}}$ and $\zeta_M{}^\Lambda$. 
To re-obtain the dynamics of~\cite{Bossard:2017wxl} in the linearised approximation, we expect that all the constrained fields are enforced to be total derivatives by a curl-free condition
\begin{subequations}
\label{dc}
\begin{align}
&&\partial_{[M} \chi_{N]}{}^{\tilde \alpha} &= 0 \quad\quad\quad \Rightarrow & \chi_M{}^{\tilde{\alpha}}  &= \partial_M X^{\tilde{\alpha}}  \ ,&&&
\\
&&\partial_{[M} \zeta_{N]}{}^{\Lambda} &= 0 \quad\quad\quad \Rightarrow  &\zeta_M{}^{\Lambda}  &= \partial_M Y^{\Lambda}  \ , &&
\end{align}
\end{subequations}
where the solutions are defined up to gauge transformations of the type~\eqref{eq:newgauge}.  The fields $X^{\tilde\alpha}$ are the fields belonging to $\cT_{-2}$ that appeared in the linearised analysis of~\cite{Bossard:2017wxl}. The field $Y$ did not appear there but this did not effect the conclusion that while the duality equations are no longer gauge invariant, the second order integrability equations that are derived from them are indeed gauge invariant on section.

Considering the non-linear system proposed here and parametrizing formally the fundamental $\cM = \exp(\phi)$, where $\phi_\alpha T^{\alpha M}{}_N = \phi_\alpha  \eta^{MQ} \eta_{NP} T^{\alpha P}{}_Q$ due to $\cM^\dagger=\cM$, one obtains with~\eqref{dc} the linearised field strength 
\be 
F^I = C^{IM}{}_{\alpha} \partial_M \phi^\alpha +  C^{IM}{}_{\tilde{\alpha}} \partial_M X^{\tilde{\alpha}} +  C^{IM}{}_{\Lambda} \partial_M Y^\Lambda+ \mathcal{O}(\phi^2)  = C^{IM}{}_{\hat{\alpha}} \partial_M \phi^{\hat{\alpha}}+ \mathcal{O}(\phi^2)\ ,  
\label{LF}
\ee
with the linearised gauge transformations
\be 
\delta_\xi \phi^{\widehat{\alpha}} = T^{\widehat{\alpha} N}{}_P \left( \partial_N \xi^P + \eta_{NQ} \eta^{PR} \partial_R\xi^Q\right)+{\Pi^{\widehat\alpha}{}_{ Q P }} \eta^{NQ}  \partial_N \xi^P \ ,
\label{LG}
\ee
with $\Pi^\alpha{}_{PQ} = 0$. The linearised duality equation 
\be 
F^I - \eta^{IK} \Omega_{KJ} F^J =0 
\label{LD}
\ee
is gauge invariant under these linearised gauge transformations. This agrees with the proposal of~\cite{Bossard:2017wxl}, up to the presence of the additional field $Y^\Lambda$ that does not affect the analysis of the linearised duality equation in \cite{Bossard:2017wxl} at the level it was considered, and up to the additional term in the linearised gauge transformation 
\be \label{deltaxiX} 
\delta_\xi X^{\tilde{\alpha}} = T^{\tilde{\alpha} N}{}_P \left( \partial_N \xi^P + \eta_{NQ} \eta^{PR} \partial_R\xi^Q\right)+{\Pi^{\tilde\alpha}{}_{ Q P }} \eta^{NQ}  \partial_N \xi^P \ ,
\ee
which permits to make the duality equation gauge invariant. 

Thus, the question at hand is how to find the full field equation for the constrained fields $\chi^{\tilde\alpha}$ and $\zeta^\Lambda$ such that upon linearisation it yields \eqref{dc}. To this end, we shall offer  some speculations on how these equations could be obtained. Analogy with exceptional field theories for finite dimensional exceptional group $E_n$ suggests that there should be a pseudo-Lagrangian $\mathcal{L}$ invariant under generalised diffeormorphisms (up to a total derivative) whose equations of motion are obtained by the variation
\be \delta \mathcal{L} = \langle \cE_{\alpha}  t^\alpha , \cM  \delta \cM^{-1} \rangle  + \cE_{\tilde{\alpha}}^M \delta \chi_M{}^{\tilde{\alpha}}  + \cE_{\Lambda}^M \delta \zeta_M{}^{\Lambda} + \partial_M ( \dots )\ ,
 \ee 
with $\langle t^\alpha , t^\beta \rangle = \kappa^{\alpha\beta}$. The general structure of exceptional field theory shows that the equations of motion of the constrained fields gives the first-order duality equations of the supergravity fields \cite{Hohm:2013uia,Hohm:2014fxa}. Assuming the same structure and invariance of the pseudo-Lagrangian under $E_{11}$, one concludes that the equations of motion are of the schematic form 
\be 
\cE_\alpha  = C^{IM}{}_{\alpha} \partial_M( \cM_{IJ} \cE^J) + \widetilde{\cE}_\alpha \, , \quad \cE^M_{\tilde{\alpha}}  = C^{IM}{}_{\tilde{\alpha}} \cM_{IJ} \cE^J \, , \quad \cE^M_{\Lambda}  = C^{IM}{}_{\Lambda} \cM_{IJ} \cE^J \, . 
\label{hatE}
\ee
The (second order) equation $\cE_\alpha$ for the adjoint scalars does not only involve the integrability of the duality equation $\cE^I$ of~\eqref{DualityEquation} but also an additional piece $\widetilde{\cE}_\alpha$ in the adjoint representation of $E_{11}$. It is this piece that should imply the first order equations for the constrained fields $\chi_M{}^{\tilde{\alpha}}$ and $\zeta_M{}^\Lambda$ if the duality equations are satisfied. 
Equations \eqref{dc} should follow from  $\widetilde{\cE}_\alpha$ in the linearised approximation.  However, without this peudo-Lagrangian at hand, we cannot currently propose the general form of $\widetilde{\cE}_\alpha$.

As mentioned above, we also know that the algebra of generalised diffeomorphism of parameter $\xi^M$ will not close and that one must introduce additional gauge transformations to obtain a closed algebra of gauge transformations. The Lagrangian must also be gauge invariant under these transformations. These transformations will at least include the additional parameter $\Sigma_M{}^{N\tilde{\alpha}}$  in $\overline{R(\Lambda_1)} \otimes R(\Lambda_3)$, with a tracelessness condition $\Sigma_M{}^{M\tilde{\alpha}}=0$.\footnote{It is conceivable that there are more constrained parameters. We note that for the case of $E_n$ with $n\leq 9$  a single new parameter suffices~\cite{Bossard:2017aae}. } The gauge transformation of the first constrained scalar under this parameter is
\be 
\label{eq:newgauge}
\delta \chi_M{}^{\tilde{\alpha}} = \partial_N \Sigma_M{}^{N\tilde{\alpha}}   + \dots 
\ee

Turning to \eqref{dc}, we note that while the Bianchi identity for the non-linear field strength contains the curl of $\chi_M{}^{\tilde\alpha}$ and $\zeta_M{}^\Lambda$, it does not provide the desired non-linear field equation for these fields. The $\cT(\mf{e}_{11})$ algebra Jacobi identity $ 2\{ P^{(M} , [ P^{N)} , t_{\hat{\alpha}} ]\}  = [\{P^M , P^N\} , t_{\hat{\alpha}}]$ implies 
\be 
\Omega_{IJ} C^{IM}{}_{\hat{\alpha}} C^{JN}{}_{\hat{\beta}} \partial_{(M} A \partial_{N)} B  = 0 
\ee
on section. Using this we can find the Bianchi identity
\bea  
\Omega_{IJ} C^{IM}{}_\alpha \partial_M F^J &=& \Omega_{IJ} C^{IM}{}_\alpha \Bigl( C^{JN}{}_\beta \partial_{M}  J_N{}^\beta + C^{JN}{}_{\tilde\beta} \partial_{M} \chi_N{}^{\tilde\beta} + C^{JN}{}_\Lambda  \partial_{M} \zeta_{N}{}^\Lambda \Bigr) 
\\
&=&\Omega_{IJ} C^{IM}{}_\alpha \Bigl( C^{JN}{}_\beta \partial_{[M}  J_{N]}{}^\beta 
+ C^{JN}{}_{\tilde\beta} \partial_{[M} \chi_{N]}{}^{\tilde\beta} +C^{JN}{}_\Lambda  \partial_{[M} \zeta_{N]}{}^\Lambda \Bigr) 
\nn\\
&=&\Omega_{IJ} C^{IM}{}_\alpha \Bigl( - \tfrac12 C^{JN}{}_\beta  f_{\gamma\delta}{}^\beta J_M{}^\gamma J_N{}^\delta+ C^{JN}{}_{\tilde\beta}  \partial_{[M} \chi_{N]}{}^{\tilde\beta} + C^{JN}{}_\Lambda  \partial_{[M} \zeta_{N]}{}^\Lambda \Bigr)\ .
\nn \label{Bianchi}
\eea
In the first step one uses the definition of the field strength and the section constraint, and in the second the Maurer--Cartan equation for the current. The corresponding integrability condition on the duality equation~\eqref{DualityEquation} relates then the second-order supergravity fields equations to the curl of the constrained fields, but it does not determine them. A similar situation arises already in $E_n$ exceptional field theory in which the equations for the constrained fields  follow from the Lagrangian, but cannot be obtained from the duality equations of the supergravity fields by $E_n$ symmetry. This is discussed for instance for $E_7$ in~\cite[Eq.~(3.17)]{Hohm:2013uia}.

\section{\texorpdfstring{$GL(11)$}{GL11} decomposition}
\label{sec:gl11bos}

In this section, we analyse the proposed duality equation~\eqref{DualityEquation} in a $GL(11)$ level decomposition of the tensor hierarchy algebra at low levels.  We begin with an analysis of the tensor hierarchy algebra decomposition under $\mf{gl}(11)$ to derive the transformations of the fields under $E_{11}$ and the expression of the field strength and the gauge transformations. Then we study the non-linear duality equation and its gauge invariance. We shall concentrate in particular on the crucial condition~\eqref{MasterAssum2} in $GL(11)$ level decomposition.

\subsection{The field strength representation and rigid \texorpdfstring{$E_{11}$}{E11} transformations} 
\label{SubFieldStrengthGL11} 

\begin{table}
\centering
\renewcommand{\arraystretch}{1.5}
\begin{tabular}{c||c|c|c|c|c|c}
 & $q=-3$ & $q=-2$ & $q=-1$ & $q=0$ & $q=1$ & $q=2$ \\\hline\hline
$p=1$ & $P_8$,\, $P_{7,1}$ & $P_5$ & $P_2$ & $P^1$ &&\\\hline
\multirow{2}{*}{$p=0$} & $F_{8,1}$ & {$F_6$} & {$F_3$} & {$K^1{}_1$} & {$E^3$} & {$E^6$}  
\\& $F_9$ & & & & &\\\hline
$p=-1$ & $K^2{}_1$ & $K^4$ & $K^7$ & $K^{9,1}$,\, $K^{10}$ & $K_1{}^3$,\, $K^{1,1}$  &$\cdots$ \\\hline
\multirow{2}{*}{$p=-2$} & {$\tilde{K}^{1}{}_1$} & {$\tilde{E}^3$} & {$\tilde{E}^6$} & {$\tilde{E}^{8,1}$}  & $\tilde{E}^{9,3}$,\, $\tilde{E}^{10,1,1}$,\, $\tilde{E}^{11,1}$ & $\cdots$\\
&  & & & $\tilde{E}^9$ & $\tilde{E}^{10,2}$,\, $\tilde{E}^{11,1 \prime}$,\, $\tilde{E}^{11,1 \prime\prime}$ & $\cdots$ \\\hline 
$p=-3$ & $P_1$ & $P^2$ & $P^5$ & $P^{7,1}$,\, $P^8$ & $\cdots$ & $\cdots$
\end{tabular}

\begin{picture}(0,0)
\put(-55,76){$\times$}
\end{picture}
\caption{\label{tab:thaGL11}\textit{Part of the tensor hierarchy algebra $\cT$ in $\mf{gl}(11)$ level decomposition. $p$ denotes the general $\mathds{Z}$-grading of the tensor hierarchy algebra and the additional grading $q$ is related to the $\mf{gl}(11)$ level $\ell$ by $\ell=q-\tfrac32p$. The usefulness of $q$ is that the involution of $\cT$ acts on $(p,q)$ by sending it to $(-2-p,-3-q)$ and can be represented in this table by a point reflection about the place marked with a cross. The involution includes mapping $\mf{gl}(11)$ representations to their duals and for $p<0$ we have explicitly dualised all representations. For $p=0$ we have explicitly separated $\cT_0$ into $\mf{e}_{11}$ and the additional generators from the tensor hierarchy algebra extension. The additional generators are given in the second line. Since $\cT_{-2} \cong \cT_0^{*}$, we have performed the same line split for $p=-2$. The fields appear at $p=-2$ and we see explicitly the first additional generator $\tilde{E}^9$ at $(-2,0)$ that is related to the first extra field $X_{n_1\ldots n_9}$. For $(-2,1)$ there are several extra fields arising, with degeneracies in their $\mf{gl}(11)$ tensor structure. The general notation for $\mf{gl}(11)$ tensors here is such that comma separated indices indicate Young-irreducible blocks of antisymmetric indices. If a tensor has both upper and lower indices, it has by definition non-vanishing traces and is thus reducible. As in~\cite[Table~3]{Bossard:2017wxl}, some of the generators have been dualised using the eleven-dimensional Levi--Civita symbol.}}
\end{table}

The tensor hierarchy algebra decomposes under $GL(11)$ as indicated in Table~\ref{tab:thaGL11}. Besides the general $\ints$-grading $\cT= \sum_p \cT_p$, the subalgebra $\mf{gl}(11)\subset \mf{e}_{11}$ introduces another $\ints$-grading that is denoted $q$ in the table. The adjoint of $\mf{e}_{11}$ at level $p=0$ contains the $\mf{gl}(11)$ generators $K^m{}_n$ at level $q=0$ and $\mf{e}_{11}$ is generated by commutators of the elements 
\begin{align}
\label{eq:L1}
\frac1{3!} f^{n_1n_2n_3} F_{n_1n_2n_3} + \frac1{3!} e_{n_1n_2n_3} E^{n_1n_2n_3}\ , 
\end{align}
where $E^{n_1n_2n_3}$ and $F_{n_1n_2n_3}$ are the generators of $\mf e_{11}$ sitting at $q=1$ and $q=-1$, respectively, while $e_{n_1n_2n_3}$ and $f^{n_1n_2n_3}$ are the corresponding constant parameters. The transformations of the tensor fields under rigid $E_{11}$ is determined by their transformations under these generators, and so we shall only display these transformations.  The level $\ell$ that is determined by the action of the trace $K^m{}_m$ that counts the number of upper minus the number of lower indices. The relation between $\ell$ and $(p,q)$ is $q=\ell+\tfrac32p$.

In order to display the field strength representation at level $p=-1$, we introduce components $F^I$ dual to the generators $t_I$ that decompose under $GL(11)$ as indicated in Table~\ref{tab:thaGL11}, \eg, for $K^7\equiv K^{m_1\ldots m_7}$ at $(p,q)=(-1,-1)$ we introduce at field strength $F_{m_1\ldots m_7}$ etc. The tensor hierarchy algebra fixes these components to transform under the rigid $E_{11}$ generator~\eqref{eq:L1} as follows~\cite[Eq.~(4.37)]{Bossard:2017wxl}:
\begin{subequations}
\label{evF}
\begin{align}
\delta F^{m,n} &= \frac{1}{2} f^{p_1p_2(m} F_{p_1p_2}{}^{n)}  - \frac{1}{6} e_{p_1p_2p_3} F^{p_1p_2p_3(m,n)}\ , 
\\
\delta F_{m}{}^{n_1n_2n_3} &= -3 f^{p[n_1n_2} F_{mp}{}^{n_3]} + \frac{3}{4} f^{p_1p_2[n_1} \delta_m^{n_2} F_{p_1p_2}{}^{n_3]} 
- \frac{1}{6}e_{p_1p_2p_3} F_m{}^{n_1n_2n_3p_1p_2p_3}
\nn\\
&\quad  - e_{mpq} F^{n_1n_2n_3p,q} + \frac{3}{8} \delta_m^{[n_1} e_{p_1p_2q} F^{n_2n_3]p_1p_2,q} \ ,
\\
\delta F_{n_1n_2}{}^m &= e_{p_1p_2[n_1} F_{n_2]}{}^{mp_1p_2} - \frac{1}{9} e_{p_1p_2p_3}\delta^m_{[n_1} F_{n_2]}{}^{p_1p_2p_3} +  e_{pn_1n_2} F^{m,p} 
\nn\\
&\quad- \frac{1}{2} f^{mp_1p_2} F_{n_1n_2p_1p_2} - \frac{1}{9} f^{p_1p_2p_3} \delta^m_{[n_1} F_{n_2]p_1p_2p_3} \ ,
\\
\delta F_{n_1n_2n_3n_4} &= - 6 e_{p[n_1n_2} F_{n_3n_4]}{}^p - \frac{1}{6} f^{p_1p_2p_3} F_{n_1n_2n_3n_4p_1p_2p_3} \ ,
\\
\delta F_{n_1\ldots n_7}& = -35 e_{[n_1n_2n_3} F_{n_4...n_7]} - \frac{1}{2} f^{p_1p_2p_3} ( F_{n_1\dots n_7p_1p_2,p_3} - F_{n_1\dots n_7p_1p_2p_3})\ ,
\label{evF3}
\\
\delta F_{n_1\ldots n_9,m} &=-\frac{42\times 3}{5} \left(e_{[n_1n_2n_3} F_{n_4\ldots n_9]m} + e_{m[n_1n_2} F_{n_3\ldots n_9]}\right) 
-\frac1{18} f^{p_1p_2p_3} F_{p_1\langle n_1\dots n_9,m\rangle p_2p_3}  \ ,
\label{evF1}
\\
\delta F_{n_1\ldots n_{10}} &= 4 e_{[n_1n_2n_3} F_{n_4\ldots n_{10}]}+\frac1{18} f^{p_1p_2p_3} F_{p_1[n_1\dots n_9,n_{10}]p_2p_3} \,.
\label{evF2}
\end{align}
\end{subequations}
Here, we have added the $(9,3)$-form terms in \eqref{evF1} and \eqref{evF2} compare to~\cite{Bossard:2017wxl}.

{\allowdisplaybreaks
This is the transformation of the field strength in~\eqref{eq:NLFS} that we recall is composed out of the current $J_M{}^\alpha$ and the constrained fields $\chi_M{}^{\tilde\alpha}$ and $\zeta_M{}^\Lambda$ that transform respectively as components of $R(\Lambda_1)\otimes\overline{R(\Lambda_2)}$, embedded in the indecomposable representation, and $R(\Lambda_1)\otimes \overline{R(\Lambda_{10})}$. This fixes their rigid $E_{11}$ transformation that, for the first components in the $GL(11)$ decomposition, takes the form
\begin{subequations}
\label{eq:chivar}
\begin{align}
\delta \chi_{m;n_1\ldots n_9} &= \frac12 e_{m p_1p_2} \chi^{p_1p_2}{}_{n_1\ldots n_9} -  \tfrac12 f^{p_1p_2p_3}  \chi_{m;n_1\ldots n_9p_1,p_2p_3}\nn\\
&\quad\quad + f^{p_1p_2p_3}  \chi_{m;n_1\ldots n_9p_1p_2,p_3}+\ldots  \ ,
\\
\delta \chi^{pq}{}_{n_1\ldots n_{10}, rs} &= f^{pqt} \chi_{t;n_1...n_{10},rs} +
 6e_{rs[n_1} \chi^{pq}{}_{n_2\ldots n_{10}]} - 9 e_{r[n_1n_2} \chi^{pq}{}_{n_3\ldots n_{10}] s} 
\nn\\
 &\quad\quad +9 e_{s[n_1n_2} \chi^{pq}{}_{n_3\ldots n_{10}] r} +\ldots\ ,
\w2
\delta \chi^{pq}{}_{n_1\ldots n_{11},m} &=  f^{pqr} \chi_{r;n_1...n_{11},m} + 11 e_{m[n_1n_2} \chi^{pq}{}_{n_3\ldots n_{11}]} +\ldots \ ,
\w2
\delta\zeta^{p_1\ldots p_5}{}_{n_1\ldots n_7p_1\ldots p_4,p_5} &= 10 f^{[p_1p_2p_3} \zeta^{p_4p_5]}{}_{n_1\ldots n_7p_1\ldots p_4,p_5}  + \ldots\ , 
\end{align}
\end{subequations}
where we have not displayed the terms involving the $E_{11}$ current, that would appear because of the indecomposable representation. The transformations~\eqref{eq:chivar} will be sufficient for the checks we shall perform. The transformations above can be deduced by combining~\eqref{E11adjoint} with the rigid $E_{11}$ transformation of the derivative index that follows from $\partial_M$ in~\cite{Bossard:2017wxl}
\begin{align}
\delta  \partial_m &= \frac{1}{2} e_{mp_1p_2} \partial^{p_1p_2} \ ,
\nn\\
\delta  \partial^{n_1n_2}  &= f^{n_1n_2p}\partial_p + \frac16 e_{p_1p_2p_3} \partial^{n_1n_2p_1p_2p_3} \ ,
\nn\\
\delta  \partial^{n_1n_2n_3n_4 n_5}  &= 10 f^{[n_1n_2n_3} \partial^{n_4n_5]} + \cdots  \ .
\label{eq:KE11der}
\end{align}
}

We shall next derive an explicit form of the non-linear field strength by choosing an explicit parametrisation of the $E_{11}/K(E_{11})$ fields appearing in the current $J_M{}^\alpha$. The matrix $\cM$ appearing in the definition~\eqref{J} is formal and involved intricate infinite sums. To write a meaningful equation in the parabolic decomposition one must resort to the  coset representative $\cV$ of $E_{11}/K(E_{11})$ in a maximal parabolic gauge. Such a maximal parabolic gauge decomposes $\cV$ into the Levi factor  $v \in GL(11) / SO(1,10)$ and the unipotent component $\cU$  in the unipotent subgroup of positive $GL(11)$ levels as (cf.~\eqref{eq:mpg})
\begin{align}
\cV =  v \, \cU \ .
\end{align}
We take the unipotent element concretely of the form
\begin{align}
\label{eq:Upar11}
\cU =   \exp\left( \tfrac1{3!} A_{n_1n_2n_3} E^{n_1n_2n_3} \right)  \exp\left( \tfrac1{6!} A_{n_1\dots n_6} E^{n_1\dots n_6} \right) \exp\left( \tfrac1{8!} h_{n_1\dots n_8,m} E^{n_1\dots n_8,m} \right) \cdots \,.
\end{align}
With the $GL(11)$ metric
\begin{align}
\label{mMGL11} m = v^\dagger \eta v \quad\quad \Rightarrow \quad j=m^{-1} dm\ ,
\end{align}
one has $\cM = \cV^\dagger \eta \cV = \cU^\dagger  m\,  \cU$ and
\begin{align}
\label{CurrentDecompo}  
\cJ_M = \cM^{-1} \partial_M \cM = \cU^{-1}  \bigl( j_M + \partial_M \cU\, \cU^{-1}   + m^{-1} (  \partial_M \cU\, \cU^{-1}  )^\dagger m \bigr)  \cU \,.
\end{align}
Although $\cJ_M$ is a formal expression in this representation involving infinite sums, it is conjugate under the  unipotent element $\cU$ to 
\begin{align}
\tilde{\cJ}_M = \cU \cJ_M \cU^{-1}=j_M + \partial_M \cU\, \cU^{-1}   + m^{-1} (   \partial_M \cU\, \cU^{-1}  )^\dagger m\ ,
\end{align}
that does admit a well-defined expansion in fields. It satisfies  $\tilde{\cJ}_M = m^{-1}\tilde{\cJ}_M^\dagger m$. 
We shall refer to such a $\tilde{\cJ}_M$ as a `semi-flattened' current as the conjugation by $U$ makes all indices associated with positive levels flat while keeping curved indices on level 0. We can therefore still use the metric formalism associated with the Levi subgroup and its `metric' $m$. 
We similarly define a semi-flattened version of the constrained fields according to
\begin{align}
\label{ChiGL11} 
\chi_{M}{}^{\tilde{\alpha}} = \mathcal{U}^{-1 \tilde{\alpha}}{}_{\tilde{\beta}}  \tilde{\chi}_{M}{}^{\tilde{\beta}} + \omega^{\tilde{\alpha}}_\beta(\cU^{-1}) \tilde{\cJ}_{M}{}^\beta  \ , \qquad \zeta_{M}{}^\Lambda =    \mathcal{U}^{-1 \Lambda}{}_{\Xi}  \tilde{\zeta}_{M}{}^\Xi  \ , 
\end{align}
where we see again the indecomposable structure of the representation structure in $\cT$. The associated semi-flattened field strength is then $\tilde{F}^I = \mathcal{U}^{ I}{}_{J} {F}^J $ with
\begin{align}
\label{FieldStrengthFGL11} 
\tilde{F}^I = C^{I M }{}_{\alpha}   \cU^{-1}_M{}^N \tilde{\cJ}_{N}{}^{\alpha} +C^{I M}{}_{ \tilde\alpha}  \cU^{-1}_M{}^N  \tilde{\chi}_{N}{}^{\tilde\alpha}+ C^{I M}{}_\Lambda  \cU^{-1}_M{}^N  \tilde{\zeta}_{N}{}^\Lambda \ . 
\end{align}
Written in terms of $\tilde{F}^I$, the first order non-linear duality equation~\eqref{DualityEquation} only involves the matrix $m$ rather than $\cM$, 
\begin{align}  
\label{DualitySemiFrame} \tilde{F}^I =  m^{IK} \Omega_{KJ} \tilde{F}^J  \ . 
\end{align}
The purpose of this construction is that $m$ acts diagonally in the level decomposition and just expresses the raising/lowering of the indices by the  metric $g_{mn}$, with a multiplication by the density term  $\sqrt{-\det g_{mn}}$. Moreover, the current $\tilde{\cJ}$ written in this way is a well-defined finite expression in the fields of the theory, level by level, whereas the current $\cJ$ itself would involve formal entangled infinite sums.  
 
Writing out the  non-linear field strengths~\eqref{FieldStrengthFGL11} in the parametrisation~\eqref{eq:Upar11} one finds when restricting to the $D=11$ solution of the section constraint
\begin{subequations}
\label{eq:tildeF}
 \begin{align}
 \tilde{F}_{n_1n_2}{}^m &= 2 g^{mp} \partial_{[n_1} g_{n_2]p}   \,, \\
 \tilde{F}_{n_1\dots n_4} &= 4 \partial_{[n_1} A_{n_2n_3n_4]} \,, \\
  \tilde{F}_{n_1\dots n_7} &= 7 \partial_{[n_1} A_{n_2n_3\dots n_7]}+ 70 A_{[n_1n_2n_3} \partial_{n_4} A_{n_5n_6n_7]} \,, \\
  \tilde{F}_{n_1\dots n_9,m}-\tilde{F}_{n_1\dots n_9m}  &= 9 \partial_{[n_1} h_{n_2n_3\dots n_9],m}  + 840 A_{m[n_1n_2} A_{n_3n_4n_5} \partial_{n_6} A_{n_7n_8n_9]} \, \\
  &\quad + 168 \bigl( A_{[n_1n_2n_3} \partial_{n_4} A_{n_5\ldots n_9]m} + A_{m[n_1n_2}  \partial_{n_3} A_{n_4\ldots n_9]}\bigr)  +\tilde  \chi_{m;n_1\dots n_9} \,.\nn
\end{align}
\end{subequations}
The restriction to the $D=11$ solution of the section constraint means that we are here only retaining the derivatives $\partial_m$. Note that choosing a solution to the section condition breaks the global $E_{11}$ symmetry to a (finite-dimensional) subgroup.

The non-linear duality equations~\eqref{DualityEquation} take the following  form
\begin{subequations}
\label{de}
\begin{align}
\tilde{F}_{n_1\ldots n_4} &=-\frac1{7! \sqrt{-g}} \varepsilon_{n_1\ldots n_4 m_1\ldots m_7} \tilde{F}^{m_1\ldots m_7}\ ,
\label{de1}\\
\tilde{F}_{n_1n_2}{}^m + \frac15 \delta^m_{[n_1} \tilde{F}_{n_2]p}{}^p &= \frac1{9!\sqrt{-g}} g_{n_1q_1}g_{n_2q_2} \varepsilon^{q_1q_2p_1\ldots p_9} g^{mq}\tilde{F}_{p_1\ldots p_9,q}\ ,
\label{de2}\\
\tilde{F}_{np}{}^p &=\frac{1}{9!\sqrt{-g}} g_{nq}\varepsilon^{qp_1\ldots p_{10}} \tilde{F}_{p_1\ldots p_{10}}\,.
\label{de3}
\end{align}
\end{subequations}
Equation~\eqref{de2} is tracefree while~\eqref{de3} is the pure trace part. We see that~\eqref{de1} represents the standard type of duality of the four-form field strength and the seven-form field strength in $D=11$ supergravity and, moreover, $\tilde{F}_4$ and $\tilde{F}_7$ are free of the extra constrained field $\chi_{m}{}^{\tilde\alpha}$. We stress that the equations~\eqref{de} do not depend on the explicit parametrisation of $\cU$.

Unlike~\eqref{de1}, equations~\eqref{de2} and~\eqref{de3} do contain the field $\chi_{m}{}^{\tilde\alpha}$ algebraically and can be seen as just determining it in terms of the values of the other fields. Put differently,~\eqref{de2} and~\eqref{de3} are not strong enough to determine the dynamics of the graviton and the dual graviton unless one has further extra equations that determine $\chi_{m}{}^{\tilde\alpha}$. This is exactly the phenomenon discussed in Section~\ref{sec:fulldyn} where these extra equations were called $\widetilde{\cE}_\alpha$. In the absence of these additional equations,  the self-duality equation --- though consistent and gauge invariant ---  is not sufficient to fully determine the non-linear dynamics. The equation of motion of $\tilde  \chi_{m;n_1\dots n_9}  $ in particular should reproduce the same mechanism as depicted in \cite{West:2002jj,Boulanger:2008nd} for the St\"{u}ckelberg field  to restore Einstein equation.

\subsection{Gauge invariance}

We now discuss in more detail the non-linear gauge invariance of the duality equation~\eqref{DualityEquation} in the $GL(11)$ decomposition. For doing this, we begin with the linearised analysis.

{\allowdisplaybreaks
The non-linear field strengths defined in~\eqref{eq:tildeF} linearise to
\begin{subequations}
\label{F}
\begin{align}
\Fs{-5}^{n_1,n_2}_\lin &=  \partial^{q(n_1} h_q{}^{n_2)} + \frac{1}{6!} \partial^{p_1p_2p_3p_4p_5p_6(n_1,n_2)} A_{p_1p_2p_3p_4p_5p_6}  +\dots\ , 
\\
\Fs{-5}_m{}^{n_1n_2n_3}_\lin &= - \partial_m A^{n_1n_2n_3} + 3 \partial^{[n_1n_2} h_m{}^{n_3]} + \frac{1}{2}\partial^{n_1n_2n_3p_1p_2} A_{mp_1p_2} 
\nn\\
 &\quad + \frac{1}{4!}    \partial^{n_1n_2n_3p_1p_2p_3p_4,q} A_{mp_1p_2p_3p_4q} - \frac{1}{5!} \partial^{n_1n_2n_3p_1p_2p_3p_4p_5} A_{mp_1p_2p_3p_4p_5}
\nn\\
&\quad + \frac{3}{2} \delta_m^{[n_1} \Bigl( \partial^{n_2|q} h_q{}^{n_3]} 
- \frac{1}{6} \partial^{n_2n_3]p_1p_2p_3} A_{p_1p_2p_3} - \frac{3}{2\cdot 5!}  \partial^{n_2n_3]p_1\dots p_5,q} A_{p_1\dots p_5q}\nn\\
& \hspace{40mm}  
+\frac{1}{6!}  \partial^{n_2n_3]p_1\dots p_6} A_{p_1\dots p_6} \Bigr) +\dots \ , 
\\
\Fs{-3}^\lin_{n_1n_2}{}^m &= 2 \partial_{[n_1} h_{n_2]}{}^m + \partial^{mp} A_{n_1n_2p} + \frac13\delta_{[n_1}^m  \partial^{p_1p_2} A_{n_2]p_1p_2} +\ldots\ ,
\\
\Fs{-1}_{m_1\ldots m_4}^\lin &= 4 \partial_{[m_1} A_{m_2m_3m_4]} - \frac12 \partial^{n_1n_2} A_{m_1\ldots m_4 n_1n_2} - \frac1{24} \partial^{n_1\ldots n_5} h_{m_1\ldots m_4n_1\ldots n_4,n_5}
\nn\\
&\quad + \frac1{5!} \partial^{n_1\ldots n_5} X_{m_1\ldots m_4n_1\ldots n_5} +\ldots \ ,
\\
\label{eq:F7imp}
\Fs{1}_{m_1\ldots m_7}^\lin &= 7 \partial_{[m_1} A_{m_2\ldots m_7]} + \partial^{n_1n_2} h_{m_1\ldots m_7n_1,n_2} - \frac12\partial^{n_1n_2} X_{m_1\ldots m_7n_1n_2} 
\nn\\
&\quad -\frac1{12} \partial^{n_1\ldots n_5} X_{m_1\ldots m_7n_1n_2n_3,n_4n_5} 
 - \frac1{24} \partial^{n_1\ldots n_5} X_{m_1\ldots m_7n_1\ldots n_4,n_5}   
 \nn\\
 &\quad - \frac1{24} \partial^{n_1\ldots n_5} Y_{m_1\ldots m_7n_1\ldots n_4,n_5} +\ldots \ ,
\\
\Fs{3}_{m_1\ldots m_{10}}^\lin &= \partial_{[m_1} X_{m_2\ldots m_{10}]} - \frac1{15}\partial^{pq} X_{m_1 \ldots m_{10} ,pq} -\frac1{30} \partial^{pq} X_{m_1 \ldots m_{10} p,q}\nn\\
&\quad -\frac7{30} \partial^{pq} Y_{m_1 \ldots m_{10} p,q} +\ldots \ ,\\
\Fs{3}_{n_1\ldots n_9,m}^\lin &= 9 \partial_{[n_1} h_{n_2\ldots n_9],m}  +\partial_m X_{n_1\ldots n_9} - \partial_{[m} X_{n_1\ldots n_9]}+ \frac12 \partial^{p_1p_2} A_{n_1\dots n_9,mp_1p_2}
\nn\\*
& \quad +\frac{9}{10} \left(\partial^{pq} X_{p n_1\ldots n_9,m q}+\partial^{pq} X_{pm[ n_1\ldots n_8,n_9] q}\right) \nn\\*
&\quad 
 + \frac{27}{20} \left(\partial^{pq} X_{pq n_1\ldots n_9, m} +\partial^{pq} X_{pqm [n_1\ldots n_8,n_9]}\right)\nn\\*
&\quad+ \frac{9}{20}  \left(\partial^{pq} Y_{pq n_1\ldots n_9, m} +\partial^{pq} Y_{pqm [n_1\ldots n_8,n_9]}\right)+\ldots \ ,\\
\Fs{5}_{m_1\ldots m_{10},n_1n_2n_3}^\lin &= 10 \partial_{[m_1} A_{m_2\dots m_{10}],n_1n_2n_3} + \dots 
\end{align}
\end{subequations}
For later reference, we have explicitly given the $\mf{gl}(11)$ levels for the various components of the field strengths, but we shall often suppress them for simplicity of notation when there is no risk of confusion. In the expressions above we also have implemented several things. First, we removed the tildes on $F$ since the semi-flattening has no effect in the linearised approximation. Moreover, we have reinstated the terms with partial derivatives beyond $\partial_m$, thus not enforcing the $D=11$ solution to the section constraint as this would break $K(E_{11})$ covariance and gauge invariance. And finally, the extra constrained fields at linearised order are expressed as
\begin{align}
\label{eq:chiX}
\chi_M{}^{\tilde\alpha} =\partial_M X^{\tilde\alpha}\quad\textrm{and} \quad \zeta_M{}^\Lambda= \partial_M Y^\Lambda\,,
\end{align}
where we have also used explicitly that the first few extra fields are the additional potentials $X_{n_1...n_{10},rs}$, $X_{n_1...n_{11},r}$ and $Y_{n_1...n_{11},r}$ coming from $\mf{gl}(11)$ level $\ell=4$ in $\cT_{-2}$ with $X^{\hat{\alpha}}$ in $\overline{R(\Lambda_2)}$ and $Y^\Lambda$ in $\overline{R(\Lambda_{10})}$, see Table~\ref{tab:thaGL11}. The additional potentials $X_{n_1...n_{10},rs}$, $X_{n_1...n_{11},r}$ and $Y_{n_1...n_{11},r}$ are dual to generators for $(p,q)=(-2,1)$ in that table, where $q= \ell + \tfrac32 p$ in relation to the $\mf{gl}(11)$ level $\ell$.

We will now write the decomposition of the gauge transformation~\eqref{eq:gt4}. under $\mathfrak{gl}(11)$. For many of the fields this was already carried out in~\cite{Bossard:2017wxl}, but not for the crucial inhomogeneous terms involving the invariant tensor  $\Pi^{\widehat{\alpha}}{}_{QP}$ in the gauge transformation of the constrained fields in~\eqref{gt2} and~\eqref{gt3}. Here we concentrate on the central condition~\eqref{MasterAssum2}, which would imply the gauge invariance of the duality equations. An important observation is that, as the condition~\eqref{MasterAssum2} is a condition on invariant tensors, it suffices to verify it at linear order in order to deduce gauge invariance of the non-linear duality equation.

One obtains for the first level fields in the  $GL(11)$  decomposition that the gauge transformations~\eqref{eq:gt4}  give in the linearised approximation 
\begin{subequations}
\label{xt}
\begin{align}
\delta_\xi h_{n}{}^m &=  (\partial_n \xi^m -\partial^{mp} \lambda_{np} +\frac16 \delta_n^m \partial^{p_1p_2} \lambda_{p_1p_2}) + \partial^m \xi_n - \partial_{np} \lambda^{mp} +\frac16\delta_n^m \partial_{pq} \lambda^{pq} + \ldots\ ,
\label{xt1}\\
\delta_\xi A_{n_1n_2n_3} &= (3\partial_{[n_1} \lambda_{n_2n_3]} + \frac12 \partial^{p_1p_2} \lambda_{n_1n_2n_3p_1p_2} ) +  3 \partial_{[n_1n_2} \xi_{n_3]} +\ldots\ ,
\label{xt2}\w2
\delta_\xi  A_{n_1\cdots n_6} &=  (6 \partial_{[n_1} \lambda_{n_2\cdots n_6]} 
- \partial^{p_1p_2} \xi_{n_1\cdots n_6p_1,p_2} 
+  \partial^{p_1p_2} \lambda_{n_1\cdots n_6p_1p_2})+\cdots \ ,
\label{xt22}\w2
\delta_\xi h_{n_1\cdots n_8,m} &= ( 8 \partial_{[n_1} \xi_{n_2\cdots n_8],m}
+24 \partial_{\!\lsharp \! n_1} \lambda_{n_2\cdots n_8 ,m\!\rsharp}) +\cdots \ ,
\label{xt23}\w2
\delta_\xi \chi_{M;n_1\ldots n_9} &=    24 \partial_M \partial_{[n_1} \lambda_{n_2\cdots n_9]}     - \varepsilon_{n_1\ldots n_9pq}  \partial_M \partial^{p} \xi^q +\ldots\ ,
\label{xt3}\w2
\delta_\xi \chi_{M;\, m_1\ldots m_{\scalebox{0.6}{$10$}},n_1n_2} &=  \varepsilon_{m_1\ldots m_{\scalebox{0.6}{$10$}}p} \partial_M (  \partial^p \lambda_{n_1n_2} + \tfrac15 \delta^p_{[n_1} \partial^q \lambda_{n_2]q} - \partial_{n_1n_2} \xi^p - \tfrac15 \delta^p_{[n_1} \partial_{n_2]q} \xi^q ) +\ldots\ , 
\label{xt4}\\
\delta_\xi \chi_{M;\, m_1\ldots m_{\scalebox{0.6}{$11$}},n} &=  \frac1{10} \varepsilon_{m_1\ldots m_{\scalebox{0.6}{$11$}} } \partial_M ( \partial^p \lambda_{np} -  \partial_{np} \xi^p ) +\ldots \ ,
\label{xt5}\\
\delta_\xi \zeta_{M;\, m_1\ldots m_{\scalebox{0.6}{$11$}},n} &=  -\frac1{2} \varepsilon_{m_1\ldots m_{\scalebox{0.6}{$11$}} } \partial_M (  \partial^p \lambda_{np} + \partial_{np} \xi^p ) +\ldots
\label{xt6}
\end{align}
\end{subequations}
The terms corresponding to the non-trivial tensors $\Pi^{\tilde\alpha}{}_{QP}$ are the terms containing $\varepsilon_{11}$ in the gauge variations~\eqref{xt3}, \eqref{xt4}, \eqref{xt5} and~\eqref{xt6} of the constrained fields. The corresponding coefficients are determined by $E_{11}$ invariance up to an overall coefficient that is fixed by the terms in~\eqref{xt3} and~\eqref{xt6}. These overall coefficients will be determined below by requiring gauge invariance of the duality equation~\eqref{de2} at linear order.

Using the gauge transformations above one can derive the intermediate result that the linearised field strengths transform as 
\begin{subequations}
\begin{align}
\delta_\xi F_{n_1n_2n_3n_4}^\lin &= 12 \partial_{[n_1} \partial_{n_2n_3} \xi_{n_4]} + 2 \partial_{[n_1} \partial_{n_2n_3n_4]p_1p_2} \lambda^{p_1p_2} + 3 \partial^{p_1p_2} \partial_{[n_1\dots n_4p_1} \xi_{p_2]} \nonumber \\  & \qquad - \frac1{5!}\varepsilon_{n_1n_2n_3n_4p_1\dots p_7} \partial^{p_1p_2p_3p_4p_5} \partial^{p_6} \xi^{p_7} \ ,\\ 
\delta_\xi F^\lin_{n_1\dots n_7} &= - 42 \partial_{[n_1} \partial_{n_2\dots n_6} \xi_{n_7]} + \frac{1}{2} \varepsilon_{n_1\dots n_7p_1p_2p_3p_4} \partial^{p_1p_2} \partial^{p_3} \xi^{p_4} \\ 
& \qquad + \frac{1}{24} \varepsilon_{n_1\dots n_7p_1p_2p_3p_4}  \bigl( 2 \partial^{p_1}\partial^{p_2p_3 p_4q_1q_2}  \lambda_{q_1q_2} +3 \partial^{[p_1p_2p_3 p_4q_1}  \partial_{q_1q_2} \xi^{q_2]} \bigr)    \ , \nonumber\\
\delta_\xi \bigl( F^\lin_{n_1n_2}{}^m \!+\! \frac15 \delta^m_{[n_1} F^\lin_{n_2]p}{}^p \bigr) &= 2 \partial_{[n_1} \partial^m \xi_{n_2]} + \frac15 \delta^m_{[n_1} (\partial_{n_2]} \partial^p \xi_p - \partial_p \partial^p \xi_{n_2]})
\\
&\quad - 2 \partial_{[n_1} \partial_{n_2]p} \lambda^{mp} - \frac15 \delta^m_{[n_1} ( \partial_{n_2]p}  \partial_q \lambda^{pq}+ \partial_{n_2]}\partial_{pq} \lambda^{pq}) 
\nn\\
&\quad + \partial^{mp} \partial_{n_1n_2} \xi_p + 2 \partial^{mp} \partial_{p[n_1} \xi_{n_2]} + \frac15 \delta^m_{[n_1} \partial^{pq} ( \partial_{pq} \xi_{n_2]}+ 2 \partial_{n_2]p} \xi_q ) \ ,
\nn\\
\delta_\xi F^\lin_{np}{}^p &= \partial_n \partial^p \xi_p - \partial_p \partial^p \xi_n + \frac23 \partial_n \partial_{pq} \lambda^{pq}  + \partial_p \partial_{nq} \lambda^{pq}- 2 \partial^{pq}\partial_{[pq} \xi_{n]}  \, , \\
\delta_\xi F^\lin_{n_1\dots n_9,m} &= - \frac12 \varepsilon_{n_1\dots n_9p_1p_2} \delta_\xi \bigl( F^{\lin p_1p_2}{}_m  + \frac15 \delta^{p_1}_m F^{\lin p_2q}{}_q  \bigr) \; , \\
\delta_\xi F^\lin_{n_1\dots n_{\scalebox{0.6}{$10$}}} &= - \frac1{10} \varepsilon_{n_1\dots \dots n_{\scalebox{0.6}{$10$}} p} \delta_\xi \bigl(F^{\lin pq}{}_q  \bigr) \ . 
\end{align} 
\end{subequations}
This shows again that the (linearised) field strengths are not gauge invariant, however, one checks that all the terms cancel in  the duality equation~\eqref{DualityEquation}, proving gauge invariance of the self-duality equation at this level. One finds that it fixes all coefficients in the transformation of the constrained fields consistently with $E_{11}$ covariance. For instance the term in~\eqref{xt3} is needed to cancel the gauge variation of the form  $\partial_1\partial^1\xi_1$ in the dual graviton equation~\eqref{de2} as well as the gauge variation of the form $ \partial_1 \partial_2 \xi_1$ in the 7-form duality equation \eqref{de1}. Note that the field $\zeta_{M;m_1\ldots m_{11},n}$ is necessary here to ensure gauge invariance.

To end this section, we come back to the additional gauge transformations mentioned at the end of Section~\ref{sec:fulldyn}. These additional gauge transformations are needed in order to closed the algebra of generalised diffeomorphisms and the first such parameter is one with parameter $\Sigma_M{}^{N\tilde\alpha}$ where the $M$ index obeys the same section constraint as a partial derivative $\partial_M$. We do not know the full sequence of additional gauge transformations, but demanding actual invariance of the linearised field strengths under $\Sigma_M{}^{N\tilde\alpha}$ one can derive that
\begin{subequations}
\label{SigmaGaugeTranformations}
\begin{align} 
\delta_\Sigma A_{n_1n_2n_3} &=  \frac{1}{5!} \Sigma^{p_1\dots p_5}{}_{n_1n_2n_3 p_1\dots p_5} \, ,  \\
\delta_\Sigma A_{n_1\dots n_6} &= - \frac{1}{2} \Sigma^{p_1p_2}{}_{n_1\dots n_6p_1p_2} \ ,  \\
\delta_\Sigma h_{n_1\dots n_8;m} &= \Sigma_{m;n_1\dots n_8} \; ,  \\
\delta_\Sigma \chi_{M; n_1\dots n_9} &=- 9 \partial_{[n_1} \Sigma_{M;|n_2\dots n_9]} \ .
\end{align}
\end{subequations}
These transformations will play a role when we discuss the closure of the supersymmetry algebra.}

\subsection{Propagating free fields at all levels}
\label{sec:dualfields}

We shall now describe in more detail the tower of duality relations at the linearised level. To understand the dynamical content of the field equations in eleven dimensions, it is convenient to consider the structure of the tensor hierarchy algebra at positive levels $q$ (see Table \ref{tab:thaGL11}). For $q\ge 1$, the generators have at most level $p=0$, in which case they are the positive $q$ level generators $E^{\alpha_+}$ of $\mathfrak{e}_{11}$, with $\alpha_+$ the adjoint index restricted to positive $\mathfrak{gl}(11)$ level. At a given $\mathfrak{gl}(11)$ level $\ell$, $E^{\alpha_+}$ includes various irreducible representations associated to Young tableaux with $3\ell$ boxes, including possibly columns of $11$ boxes~\cite{West:2002jj,Nicolai:2003fw,Kleinschmidt:2003mf}. The $\mf{e}_{11}$ generators are the top-form components  $E^{11;\alpha_+}$  of supermultiplets of generators transforming under the  ($q=0$ component of $\mathcal{T}(\mathfrak{e}_{11})$) superalgebra  $W(11)$ of vector fields in eleven Grassmann variables $\vartheta_m$ as a superfield of representations $E^{11+p;\alpha_+}$ for $-11\le p \le 0$, in the tensor product representation of the $(11+p)$-form with the reducible representation associated to $\alpha_+$ at a given level $\ell$ \cite{Bossard:2017wxl}. 

As an example, consider the case $q=1$ in Table~\ref{tab:thaGL11} that has the top component $E^3$ at $p=0$, corresponding to the familiar $3$-form generator of $E_{11}$. This can also be written as $E^{11;3}$ using the $\varepsilon$-tensor of $\mf{sl}(11)$. The next generator for $p=-1$ and $q=1$ is $K_1{}^3$ arising as the tenth order term in $\vartheta_m$ in the expansion of  the superfield with top component $E^{3}$. After dualisation this can be written as the reducible tensor $E^{10;3}$. At $p=-2$ one gets the reducible generator $\tilde{E}^{9;3}$, which decomposes into $\tilde{E}^{9,3}$, $\tilde{E}^{10,2}$ and $\tilde{E}^{11,1}$ in Table~\ref{tab:thaGL11}. We recall that comma separated indices denote irreducible representations  
while the full set of indices separated by a semi-colon denotes the reducible representation obtained by the tensor product of the irreducible representations on each side of the semi-colon.

There are additional superfields of generators starting at $p\le-1$, but they involve at least one $11$-form at $p=-1$ (\ie\ one column of $11$ boxes) and  one 10-form at $p=-2$ (\ie\ one column of $10$ boxes). If one considers only fields involving at most a 9-form components in $\mathfrak{e}_{11}$ (Young tableaux with no column of more than nine boxes), the corresponding fields are all of the type $A_{9^n,3}$ at level $\ell=3n+1$,  $A_{9^n,6}$ at level $\ell=3n+2$, $h_{9^n,8,1}$ at level $\ell=3n+3$ \cite{Riccioni:2006az}, as can be seen by consistency with the $\mathfrak{e}_9$ subalgebra. 

These potentials at level $\ell = 3(n+1)+k$ appear in the tensor hierarchy algebra $\mathcal{T}(\mathfrak{e}_{11})$ at level $p=-2$ and $q=3n+k$, such that the last 9-form comes from the $W(11)$ 9-form components. They are part of the reducible representations given by the tensor products 
\begin{align} 
[9]\otimes [9^n,3k] &=  [9^{n+1},3k]\oplus [10,9^{n},3k-1]\oplus [11,9^{n},3k-2] \\
&\hspace{21mm}\oplus [10,9^{n-1},8,3k]\oplus [11,9^{n-1},8,3k-1]\oplus [11,9^{n-1},7,3k]\nn
\end{align}
for $k=1$ or $2$ and 
\begin{align} 
[9]\otimes [9^n,8,1] &=  [9^{n+1},8,1]\oplus [10,9^{n},7,1]\oplus [10,9^{n},8]\oplus [11,9^{n},6,1] \oplus [11,9^{n},7] \\
&\hspace{22mm}\oplus [10,9^{n-1},8^2,1]\oplus [11,9^{n-1},8^2]\oplus [11,9^{n-1},8,7,1] \nn
\end{align}
for $k=3$. The corresponding fields in the irreducible representations $ [9^{n+1},3]$, $ [9^{n+1},6]$ and  $ [9^{n+1},8,1]$ are the $E_{11}$ potentials, while the fields corresponding to the other irreducible representations always involve the constrained fields $\chi =\partial X $  for $10$-forms and $11$-forms, and possibly the constrained  fields $\zeta=\partial Y$ for $11$-forms. In the combination that appears naturally in the $W(11)$ superfield, $\mathcal{A}_{9;9^{n},3k} = A_{9^{n+1},3k} + 3 X_{10,9^{n},3k-1} + \dots $, the  $\mathcal{T}(\mathfrak{e}_{11})$ exterior derivative simply acts as the exterior derivative on the last $9$-form component \cite[Appendix B]{Bossard:2017wxl}, 
\begin{align} 
F_{n_1\dots n_{10};9^n,3k} = 10 \partial_{[n_1}  \mathcal{A}_{n_2\dots n_{10}];9^{n},3k}  \; , \quad  F_{n_1\dots n_{10};9^n,8,1} = 10 \partial_{[n_1}  \mathcal{A}_{n_2\dots n_{10}];9^{n},8,1}  \; . 
\end{align}
In this way the corresponding linearised field strength restricted to the eleven-dimensional solution to the section constraint (with fields only depending on eleven coordinates) is manifestly invariant under the gauge transformation 
\be
\delta \mathcal{A}^{\widehat{\alpha}_+} = T^{\widehat{\alpha}_+ m}{}_N  \partial_m \xi^N  \quad \Rightarrow \quad \begin{cases}  \delta \mathcal{A}_{n_1\dots n_9;9^{n},3k}  =  9 \partial_{[n_1} \lambda_{n_2\dots n_9];9^n,3k}  \hspace{3mm} &\mbox{for } k=1,\; 2 \\
 \delta \mathcal{A}_{n_1\dots n_9;9^{n},8,1}  =  9 \partial_{[n_1} \lambda_{n_2\dots n_9];9^n,8,1}   \hspace{2mm} &\mbox{for } k=3
 \end{cases} \; .
\ee
In eleven dimensions, assuming the fields depend on the eleven coordinates only, the dual field strength only involves the ordinary derivative of a potential so that the duality equation becomes 
\begin{align}
F_{n_1 \dots n_{10},9^n,3k}  = \bigl[  F_{n_1 \dots n_{10};9^n,3k} \bigr]_{10,9^n,3k} &= - \bigl[ \varepsilon_{n_1\dots n_{10} m} \eta^{mp} \partial_p  A_{9^n,3k} \bigr]_{10,9^n,3k} \; , \nn\\
F_{n_1 \dots n_{10},9^n,8,1}  = \bigl[  F_{n_1 \dots n_{10};9^n,8,1} \bigr]_{10,9^n,8,1} &= - \bigl[ \varepsilon_{n_1\dots n_{10} m} \eta^{mp} \partial_p  A_{9^n,8,1} \bigr]_{10,9^n,8,1} \; ,
\end{align}
where we take the projection to the irreducible $SL(11)$ representation indicated.  The right-hand side is not gauge invariant, but it follows from \eqref{MasterAssum2} that the lack of gauge invariance is compensated by the gauge transformation 
\begin{align}
\label{deltaX11D} 
\delta \mathcal{A}^{\widehat{\alpha}_+} = {\Pi^{\widehat{\alpha}_+}{}_{ mP  }} \eta^{mn}  \partial_n \xi^P  \; , 
\end{align}
on the left-hand side, such that the duality equation is indeed gauge invariant under the total linearised gauge transformation 
\begin{align}
\delta \mathcal{A}^{\widehat{\alpha}_+} =T^{\widehat{\alpha}_+ m}{}_N  \partial_m \xi^N + {\Pi^{\widehat{\alpha}_+}{}_{ mP  }} \eta^{mn}  \partial_n \xi^P  \; . 
\end{align}

In order to understand these duality equations better, it is useful to compare them with the equations derived in \cite{Boulanger:2015mka}. The first instance uses the potentials in the reducible $(9;3)$ representation composed out of $E_{11}$ potentials and additional fields as
\begin{align} 
\mathcal{A}_{n_1\dots n_9;p_1p_2p_3} = A_{n_1\dots n_9;p_1p_2p_3} + 3 X_{n_1\dots n_9[p_1,p_2p_3]} - 24 X^\prime_{n_1\dots n_9[p_1p_2,p_3]} \,.
\end{align}
The duality equation 
\begin{align}
\varepsilon_{n_1\dots n_{10} m} \eta^{mp} \partial_p  A_{p_1p_2p_3}  &=  F_{n_1\dots n_{10};p_1p_2p_3} + \frac{3}8 \bigl(  10 F_{n_1\dots n_9] [p_1;p_2p_3][n_{10}}- F_{n_1\dots n_{10};p_1p_2p_3} \bigr)  \\
&=  10 \partial_{[n_1} A_{n_2\dots n_{10}],p_1p_2p_3} - 3 \partial_{[p_1} X_{n_1\dots n_{10},|p_2p_3]} -  6  \partial_{[p_1} X^\prime_{n_1\dots n_{10}|p_2,p_3]} \ \nn
\end{align}
is gauge invariant under~\eqref{xt}. Here, we have defined for a free coefficient $c$ the combination
\begin{align}
X_{11,1}^\prime = (1+5 c) X_{11,1} +  c Y_{11,1} \ . 
\end{align}
The duality equation above is gauge invariant in eleven dimensions for any $c$, but $E_{11}$ covariance determines its value. 

At level $\ell = 3n+7$ one has similarly  
\begin{align} 
& \mathcal{A}_{n_1\dots n_9;9^n,p_1\dots p_9,q_1q_2q_3} \\
=&  \Bigl[ {A}_{n_1\dots n_9,9^n,p_1\dots p_9,q_1q_2q_3} + 3 {X}_{n_1\dots n_9q_1,9^n,p_1\dots p_9,q_2q_3}  + 9 {X}_{n_1\dots n_9p_1,9^n,p_2\dots p_9,q_1q_2q_3} \nn\\
&  -6  {X}^\prime_{n_1\dots n_9q_1q_2,9^n,p_1\dots p_9,q_3} -27  {X}^\prime_{n_1\dots n_9p_1q_1,9^n,p_2\dots p_9,q_2q_3} -72  {X}^\prime_{n_1\dots n_9p_1p_2,9^n,p_3\dots p_9,q_1q_2q_3} \Bigr]_{9;9^{n+1},3} \; . \nn
\end{align}
 For each irreducible field strength component, each field can only appear with a unique tensor structure, and in particular 
\begin{align} 
\label{UnfoldingStrength}  
& F_{n_1\dots n_{10},9^n,p_1\dots p_9,q_1q_2q_3} \\
=&  \bigl[ \scalebox{0.9}{$ 10 \partial_{[n_1} (   {A}_{n_2\dots n_{10}],9^n,p_1\dots p_9,q_1q_2q_3} 
 + 3 {X}_{n_2\dots n_{10}]q_1,9^n,p_1\dots p_9,q_2q_3}  + 9 {X}_{n_2\dots n_{10}]p_1,9^n,p_2\dots p_9,q_1q_2q_3} ) $}\bigr]_{10,9^{n+1},3} \nn  \\
= &  \bigl[ \scalebox{0.9}{$ 10 \partial_{[n_1}   {A}_{n_2\dots n_{10}],9^n,p_1\dots p_9,q_1q_2q_3}  -  3 \partial_{[q_1|} {X}_{n_1\dots n_{10},9^n,p_1\dots p_9,|q_2q_3]}  - 9 \partial_{[p_1|} {X}_{n_1\dots n_{10},9^n,|p_2\dots p_9],q_1q_2q_3}$} \bigr]_{10,9^{n+1},3} \; , \nn
  \end{align} 
 only involves the $X$ fields and not the $X^\prime$ fields that include an $11$-form. 
The duality equation
\begin{align}
\bigl[F_{n_1\dots n_{10};9^n,p_1\dots p_9,q_1q_2q_3}\bigr]_{10,9^{n+1},3}      = - \bigl[ \varepsilon_{n_1\dots n_{10}m} \eta^{mr} \partial_r A_{9^n,p_1\dots p_9,q_1q_2q_3}\bigr]_{10,9^{n+1},3}    \; ,
\end{align}
is gauge invariant provided \eqref{deltaX11D}  gives
\begin{align} \label{PiXHigher} 
\delta X_{n_1\dots n_{10},9^{n+1},2}  &= \bigl[ \varepsilon_{n_1\dots n_{10}m} \eta^{mp} \partial_p \lambda_{9^{n+1},2} \bigr]_{10,9^{n+1},2} \; , \nn\\
  \delta X_{n_1\dots n_{10},9^{n},8,3}  &= \bigl[ \varepsilon_{n_1\dots n_{10}m} \eta^{mp} \partial_p \lambda_{9^{n},8,3} \bigr]_{10,9^{n},8,2} \; . 
\end{align}
This is indeed consistent with the structure of the $E_{11}$ invariant tensor $\Pi^{\tilde{\alpha}}{}_{MN}$. Recall that  $\Pi^{\tilde{\alpha}}{}_{MN}$ defines the projection from the exterior product of two $R(\Lambda_1)$ representations onto $R(\Lambda_2)$, see~\eqref{HighestWeightDcompo}. The subspace that is annihilated by this projector starts with $R(\Lambda_4)$ at $\mathfrak{gl}(11)$ level $6$ and contains as lowest component a $(11,7)$ form. All the components of the kernel of the projector are associated to Young tableaux with at least one 11-form.  Therefore, the component  $\Pi^{\tilde{\alpha}}{}_{mN} \eta^{mp} \partial_p \xi^N$ for components of $\xi^N$ different of $\xi^n$ and with a Young-tableau partition $\mathcal{P}(N)$ that does not include columns of more than 9 boxes, are simply the projections of $ \varepsilon_{n_1\dots n_{10}m} \eta^{mp} \partial_p \lambda_{\mathcal{P}(N)}$ to the irreducible representation of partition $(10,\mathcal{P}(N))$. Assuming that the fields $X$ are canonically normalised with respect to this structure, one obtains indeed  the formula \eqref{PiXHigher} above, consistently with~\eqref{xt4} for $n=-1$.

The field strength \eqref{UnfoldingStrength} has the right structure for the $(n+3)^{\rm th}$ order field strength in the irreducible representation $[10^{n+2},4]$ 
\begin{align}
\bigl[ \partial_1^{\; n} \partial_{p_1} \partial_{q_1}  F_{n_1\dots n_{10},9^n,p_2\dots p_{10} ,q_2q_3q_4}\bigr]_{[10^{n+2},4]}     = - \bigl[ \varepsilon_{n_1\dots n_{10}m} \eta^{mr} \partial_r \partial_1^{\; n} \partial_{p_1} \partial_{q_1}  A_{9^n,p_2\dots p_{10},q_2q_3q_4}]_{[10^{n+2},4]}  \; ,
\end{align}
to not depend on the extra fields $X$, and then reproduces the duality equations  described  in  \cite{Boulanger:2015mka}. Conversely, according to the generalised Poincar\'e lemma proved in  \cite{Bekaert:2002dt}, this duality equation is locally equivalent to the first order duality equation satisfied by \eqref{UnfoldingStrength}, where the $X$ fields parametrise the ambiguities in integration described in \cite{Boulanger:2012df}. The propagating degrees of freedom in this equation are therefore the fields $ {A}_{9^{n},3}$ of the $E_{11} / K(E_{11})$ coset \cite{Riccioni:2006az}, but the additional fields $X$ are necessary to write a gauge invariant first-order duality equation. The same argument goes through at level $\ell = 3n+8$ and $\ell = 3n+9$, with 
 \begin{multline}   
 \bigl[F_{n_1\dots n_{10};9^n,p_1\dots p_9,q_1\dots q_8,m}\bigr]_{10,9^{n+1},8,1}   =  \Bigl[ 10 \partial_{[n_1} {h}_{n_2\dots n_{10}],9^n,p_1\dots p_9,q_1\dots q_8,m}\\  - 8 \partial_{[q_1} {X}_{n_1\dots n_{10},9^n,p_1\dots p_9,|q_2\dots q_8],m}  - \partial_m X_{n_1\dots n_{10},9^n,p_1\dots p_9,q_1\dots q_8}  \\ - 9 \partial_{[p_1} {X}_{n_1\dots n_{10},9^n,|p_2\dots p_9],q_1\dots q_8,m} \Bigr]_{10,9^{n+1},8,1}\; ,
 \end{multline}
 and
 \begin{align} \label{PiXGraHigher} 
 \delta h_{n_1\dots n_9,9^{n+1},q_1\dots q_8,m} &= \bigl[ 9 \partial_{n_1} \lambda_{n_2\dots n_9;9^{n+1},q_1\dots q_8,m} \bigr]_{9^{n+2},8,1} \;  \\
  &\hspace{-30mm} = \bigl[  8 \partial_{[q_1}  \lambda_{n_1\dots n_9,9^{n+1},|q_2\dots q_8],m}  + \partial_m \lambda_{n_1\dots n_9,9^{n+1},q_1\dots q_8}+9 \partial_{[n_1} \lambda_{9^{n+1},|n_2\dots n_9],q_1\dots q_8,m} \bigr]_{9^{n+2},8,1} \; , \nn \\
\delta X_{n_1\dots n_{10},9^{n+1},q_1\dots q_7,m}  &= \bigl[ 30 \partial_{[n_1} \lambda_{n_2\dots n_9|;9^{n+1},|n_{10}]q_1\dots q_7,m} + \varepsilon_{n_1\dots n_{10}m} \eta^{mp} \partial_p \lambda_{9^{n+1},q_1\dots q_7,m} \bigr]_{10,9^{n+1},7,1} \, , \nn\\
\delta X_{n_1\dots n_{10},9^{n+1},8}  &= \bigl[ 9 \partial_{[n_1} \lambda_{n_2\dots n_9|;9^{n+1},8,|n_{10}]} + \varepsilon_{n_1\dots n_{10}m} \eta^{mp} \partial_p \lambda_{9^{n+1},8} \bigr]_{10,9^{n+1},8} \, , \nn\\
\delta X_{n_1\dots n_{10},9^{n},q_1\dots q_8,8,m}  &= \bigl[ 45 \partial_{[n_1} \lambda_{n_2\dots n_9|;9^{n},|n_{10}]q_1\dots q_8,8,m} + \varepsilon_{n_1\dots n_{10}m} \eta^{mp} \partial_p \lambda_{9^{n},q_1\dots q_8,8,m} \bigr]_{10,9^{n},8,8,1} \, , \nn
\end{align}
The duality equation~\eqref{eq:guessDE} therefore relates all the higher level dual fields to the ones of the graviton and the $3$-form in order to give the expected bosonic degrees of freedom of $D=11$ supergravity, as already discussed in~\cite{Riccioni:2006az,Boulanger:2012df,Boulanger:2015mka}. The gauge invariance of the system presented here requires the introduction of the extra constrained fields.

\section{\texorpdfstring{$E_{8}$}{E8} decomposition}
\label{sec:E8bos}
 
 In this section, we consider the decomposition of the tensor hierarchy algebra $\cT$ under the subgroup $GL(3)\times E_{8}$ and study the duality equation~\eqref{DualityEquation} in this decomposition. More details on this decomposition and the construction of the tensor hierarchy algebra in the $E_8$ basis are given in Appendix~\ref{app:e8tha}. In Table~\ref{tab:THA3}, we present a part of the tensor hierarchy algebra in this decomposition for reference in this section.
 
\subsection{Fields, field strengths and rigid transformations}
 
\begin{table}[t!]
\setlength{\arraycolsep}{3.5pt}
\renewcommand{\arraystretch}{1.5}
\begin{tabular}{r||c|c|c|c|c}
&
$q=-2$ & $q=-1$ & $q=0$ & $q=1$ & $q=2$ \\
\hline\hline
$p=2$
&  $F^{AB}$\phantom{, $G_{ab}$} & &&& \\ \hline
$p=1$
&   $F_\mu^{AB}$, $G_\mu^A$ &  $F^A$\phantom{, $G_{ab}$}   & $ P^\mu$\phantom{, $t_{AB}$} && \\ \hline
$p=0$ 
 & $F_{\mu_1\mu_2}^{AB}$, $G_{\nu;\mu}^A$ & $F_\mu^A$, $G_\mu$ & $K^\mu{}_\nu{}$, $t^A$ & $E^{\mu A}$\phantom{, $E^{AD}$} &
$E^{\mu_1\mu_2 AB}$ , $E^{\mu,\nu A}$\phantom{, $\dots$} \\ \hline
$p=-1$ 
 & $F_{3}^{AB}$, $G_{\nu_1\nu_2;\mu}^A$ & $F_{\mu_1\mu_2}^A$, $G_{\nu;\mu}$ & $P^\mu{}_{\nu_1\nu_2}$, $t_\nu^A$ & $E_{\nu}^{\mu A}$,  $E^{AB}$ &
$E^{\mu_1\mu_2 AB}_\nu$, $E^{\mu,\nu A}_{\sigma}$, $\dots$  \\ \hline
$p=-2$ 
 &  \phantom{$F_{\mu_1\mu_2}$,} $G_{3;\mu}^A$ & $F_{3}^A$, $G_{\nu_1\nu_2;\mu}$ & $P^\mu{}_{3}$, $t_{\nu_1\nu_2}^A$ & $E_{\nu_1\nu_2}^{\mu A}$,  $E^{AB}_\mu$ &
$E^{\mu_1\mu_2 AB}_{\nu_1\nu_2}$, $E^{\mu,\nu A}_{\sigma_1\sigma_2}$, $\dots$ \\ \hline
$p=-3$
 &  &\phantom{$F_{\mu_1\mu_2}$, } $G_{3;\mu}$ &  \phantom{$P_{abc}$,} $t_{3}^A$& $E_{3}^{\mu A}$, $E^{AB}_{\mu_1\mu_2}$ &
$E^{\mu_1\mu_2AB}_{3}$, $E^{\mu,\nu A}_{3}$, $\dots$ \\ \hline
$p=-4$ 
  &  & & &\phantom{$E_{ab}$, } $E^{AB}_{3}$ &
\end{tabular}

\begin{picture}(0,0)
\put(188.5,77){$\times$}
\end{picture}

\caption{\label{tab:THA3}\textit{Part of the tensor hierarchy algebra $\cT$ decomposed under $\mathfrak{gl}(3)\oplus \mf e_{8}$. 
The subalgebra at $p=0$ is the extension of $\mathfrak{e}_{11}$. The components of fixed $p$ (the rows) are in $\mf{e}_{11}$ representations. 
The cross marks the fixed point of a reflection symmetry explained in the text. The notation and structure here is similar to that of Table~\ref{tab:thaGL11}.
 }}
\end{table}

We begin with the fields parametrising $E_{11}/ K(E_{11})$ before proceeding to the tensor hierarchy algebra. Under $GL(3) \times E_{8(8)}\subset E_{11}$ the coset $E_{11}/ K(E_{11})$ can be parametrized by~\cite{Riccioni:2007au,Bergshoeff:2007qi}
\begin{align}
 ( g_{\mu\nu} , V; A_\mu^A ; B_{\mu\nu}^{AB} , h_{\mu,\nu}^A ; \dots ) \; , 
\end{align}
where the semi-colon in this list separates different levels given by the central $GL(1)\subset GL(3)$, see Appendix~\ref{app:e8tha}. These first four fields can be identified with the supergravity fields as follows. At level zero $g_{\mu\nu}$ is the metric, $V\in E_8/(Spin(16)/ \mathds{Z}_2)$ is parametrised by the $D=3$ scalar fields. At level one $A_\mu^A$ in the ${\bf 248}$ of $E_{8}$ are the vector fields dual to the scalars. At level two  the 2-form $B_{\mu\nu}^{AB}$ is symmetric in $AB$  and belongs to the reducible representation  ${\bf 1} \oplus {\bf 3875}$, so that a potential ${\bf  27000}$ component in $\textrm{Sym}^2( {\bf 248})$ is understood to vanish. It is dual to the embedding tensor constants. The symmetric tensor  $h_{\mu,\nu}^A$  in the ${\bf 248}$ of $E_{8}$ at level two does not appear in the supergravity tensor hierarchy~\cite{deWit:2008ta} nor in $E_8$ exceptional field theory and is the gradient dual to $A_\mu^A$ similar to the construction in \cite{Boulanger:2015mka} as will be discussed later. It extends to an infinite sequence of rank $n$ symmetric tensors in the adjoint representation at level $n$, that provide all the fields related by duality to the propagating scalars. This infinite sequence is associated to the affine subgroup $E_9 \subset E_{11}$. The two-forms $B_{\mu\nu}^{AB}$ do not carry any propagating degrees of freedom and we note also that there is no analogue of the usual dual graviton in $D=3$ since gravity is non-propagating.

When needed we will decompose the reducible tensor $B_{\mu\nu}^{AB} = \acute{B}_{\mu\nu}^{AB} + \kappa^{AB} B_{\mu\nu}$ in terms of $\acute{B}_{\mu\nu}^{AB} = P^{AB}{}_{CD} B_{\mu\nu}^{CD}$ in the ${\bf 3875}$ and the $E_8$ singlet $B_{\mu\nu} = \frac1{248} \kappa_{AB} B_{\mu\nu}^{AB}$, $\kappa_{AB}$ being the Killing--Cartan metric on $\mathfrak{e}_{8}$. We use the conventions of \cite{Koepsell:1999uj}, in which the projector $P^{AB}{}_{CD}$ to the ${\bf 3875}$ is 
\begin{align}
14 P^{AB}{}_{CD} = \delta^A_C \delta^B_D + \delta^A_D \delta^B_C  + f^{AE}{}_{(C} f_{D)E}{}^B - \tfrac14 \kappa^{AB} \kappa_{CD} \ . 
\end{align}

We shall also need the fields in the first components of $\chi_M{}^{\tilde{\alpha}}$ in  the $R(\Lambda_1)\otimes \overline{R(\Lambda_2)}$ module of $E_{11}$ under  $GL(3)\times E_8$ 
\begin{align}
(\chi_{M;\nu} ; \chi_{M;}{}^A_{\nu\sigma}  ; \dots ) \ ,
\end{align}
such that the $(1,1)$-form $\chi_{M;\nu}$ is at level $1$ and the $(1,2)$-form $\chi_{M;}{}^A_{\nu\sigma} $ in the ${\bf 248}$ is at level $2$. The first component of $\zeta_M{}^\Lambda \in R(\Lambda_1) \otimes \overline{R(\Lambda_{10})}$ only appears at level $3$ and will not be considered in this section.  The linearised field $X^{\tilde\alpha}$ in $\overline{R(\Lambda_2)}$ are obtained by removing the derivative $M$ index as before.
  
The derivatives in $\overline{R(\Lambda_1)}$ decompose as 
\begin{align}
\label{eq:delE8}
  (\partial_\mu; \partial_A; \partial^\mu_{AB},\partial^\mu_A; \dots ) \,,
\end{align}
where at level $-\frac12$, $\partial_\mu$ is the external space-time derivative, at level $-\frac32$, $\partial_A$ is the internal derivative of exceptional field theory \cite{Hohm:2014fxa}. The additional derivative at level $- \frac52$, $ \partial^\mu_{AB} = \acute{\partial}^\mu_{AB} + \kappa_{AB} \partial^\mu$ and $\partial^\mu_A$  are respectively in the ${\bf 3875}\oplus{\bf 1}$ and the ${\bf 248}$. 
The components of the section constraints that will be relevant in this section are
 \begin{align}
 7 P^{CD}{}_{AB} \partial_C \partial_D = \acute{\partial}_{AB}^\mu \partial_\mu \ , \qquad \kappa^{AB} \partial_A \partial_B = 8 \partial^\mu \partial_\mu \ ,
 \end{align}
 or equivalently 
 \begin{align} 
 2 \partial_A \partial_B + f_{EA}{}^C f^{ED}{}_B \partial_C \partial_D = 2 \partial_{AB}^\mu \partial_\mu \ . 
 \end{align}

The derivatives~\eqref{eq:delE8} transform under the generator \eqref{eq:phi11trm} of $\mathfrak{e}_{11}$ as 
\begin{subequations}
 \begin{align}
 \delta \partial_\mu &= e_\mu^A \partial_A \,,\\
 \delta \partial_A &= f_A^\mu \partial_\mu + e_\mu^B \partial^\mu_{AB} + f_{AB}{}^C e_\mu^B \partial^\mu_C \,,\\
 \delta \partial^\mu_{AB} &= 14 P^{CD}{}_{AB}   f^\mu_C \partial_D + \tfrac14 \kappa_{AB} \kappa^{CD} f^{\mu}_C \partial_D   \,, \\
 \delta \partial^\mu_A &= \tfrac12 f^{BC}{}_A f_B^\mu \partial_C \ . 
 \end{align}
 \end{subequations}
 
The field strengths in the representation $\mathcal{T}_{-1}$ decompose in components that transform as
\begin{subequations}
 \begin{align}
  \delta F_{\mu A} &= e_\mu^B F_{AB} + e_\nu^B f_{AB}{}^C F_{\mu }{}^\nu_C - f_{AB}{}^C f_C^\nu F_{\mu\nu}^B + f^\nu_A F_{\nu;\mu}  \,,\\
\delta F_{\mu\nu}{}^\sigma &= - 2 e_{[\mu}^A  F_{\nu]}{}^\sigma_A + 2 e_\rho^A \delta^\sigma_{[\mu} F_{\nu]}{}^\rho_A - f_A^\sigma F_{\mu\nu}^A - 2 f_A^\rho \delta^\sigma_{[\mu} F_{\nu]\rho}^A \,,\\
\delta F_{\mu\nu}^A &= - e_\sigma^A F_{\mu\nu}{}^\sigma + 2 f^{AB}{}_C e_{[\mu}^C F_{\nu]B}  + f^B_\mu ( \dots) \,,\\
\delta F_{\mu;\nu} &=  e_\nu^A  F_{\mu A} + f^\sigma_A ( \dots ) \,,\\
\delta F_{AB} &= \bigl( 14 P^{CD}{}_{AB} + \tfrac14 \kappa_{AB} \kappa^{CD} \bigr) f_C^\mu F_{\mu D} +  e_\mu^C  (\dots )  \,,\\ 
\delta F_{\mu}{}^\nu_A  &= f_A^\sigma F_{\mu\sigma}{}^\nu +f^{BC}{}_A \bigl( f_B^\nu F_{\mu C} - \tfrac12 \delta_\mu^\nu f_B^\sigma F_{\sigma C} \bigr)  +e_\sigma^B \bigl(  \dots  \bigr)   \ , \qquad  
\end{align}
\end{subequations}
where we have left out some components that are irrelevant for the discussion in this section. The representations that appear with their levels are recorded in Table~\ref{tab:THA3} in the appendix.

Explicit expressions for the field strength components can be obtained from~\eqref{eq:NLFS} by following the same procedure as in Section \ref{SubFieldStrengthGL11}, but now using instead the parabolic gauge 
 \begin{align}
 \cV =  v \, \cU \ , 
 \end{align}
 in which $v \in GL(3) \times E_{8(8)}$, and $\cU$ is in the unipotent subgroup
\begin{align} 
\cU =   \exp( A_\mu^A E_A^\mu )  \exp( \tfrac1{28} \acute{B}_{\mu\nu}^{AB} \acute{E}_{AB}^{\mu\nu}  +2 B_{\mu\nu} E^{\mu\nu} +  h_{\mu,\nu}^A E_A^{\mu,\nu}  ) \cdots 
\end{align}
 with in particular
 \begin{align}
 [E_A^\mu , E_B^\nu ] = -E_{AB}^{\mu\nu} + f_{AB}{}^C E_C^{\mu,\nu}=-\acute{E}_{AB}^{\mu\nu} - \kappa_{AB} E^{\mu\nu} + f_{AB}{}^C E_C^{\mu,\nu} \ . 
\end{align}
 One uses as in Section \ref{SubFieldStrengthGL11} the definition \eqref{mMGL11} with $m = v^\dagger \eta v$ and similarly the `semi-flattened' objects \eqref{CurrentDecompo}, \eqref{ChiGL11}, \eqref{FieldStrengthFGL11} and \eqref{DualitySemiFrame}. The element $m$ corresponds to the metric $g_{\mu\nu}$ of Minkowski signature and the symmetric $E_{8}$ matrix $M_{AB}$. 
 
 \subsection{Linearised field strengths and gauge transformations}
 
Unlike in Section~\ref{sec:gl11bos}, we begin with the linearised field strengths and the gauge invariance of the duality equations to exhibit that~\eqref{MasterAssum2} holds also in the $E_8$ decomposition to the level checked. We define the linearised theory around Minkowski space and around the origin of $E_8$ moduli space
\begin{align}
g_{\mu\nu} = \eta_{\mu\nu} + h_{\mu\nu} \,,\quad
M_{AB} = \delta_{AB} + \delta_{C(A} f^{CD}{}_{B)} \Phi_D\,.
\end{align}
The linearised field strengths are
 \begin{align}
  \Fs{-1}^\lin_{\mu\nu}{}^\sigma &= 2 \partial_{[\mu} h_{\nu]}{}^\sigma +  \tfrac1{14} \acute{\partial}^{\sigma}_{AB} \acute{B}_{\mu\nu}^{AB} +4\partial^\sigma B_{\mu\nu} +2 \partial^\sigma_A  X_{\mu\nu}^{A}  +2\delta^\sigma_{[\mu} \bigl(\partial_A A_{\nu]}^A  +\tfrac17 \acute{\partial}^\rho_{AB} \acute{B}_{\nu]\rho}^{AB} + 8 \partial^\rho B_{\nu]\rho}   ) \,,\nn\\
 \Fs{-1}^\lin_{\mu A}  &=  \partial_\mu \Phi_A + f_{AB}{}^C \partial_C A_\mu^B + \partial_A X_\mu  \nn\\
 &\quad + \tfrac17 f_{AB}{}^C {\partial}^\nu_{CD} {B}^{BD}_{\mu\nu} - \partial^\nu_{AB} ( h^B_{\mu,\nu} + X^B_{\mu\nu}) - f_{AB}{}^C \partial_C^\nu (h^B_{\mu,\nu} - X^B_{\mu\nu})+  \dots 
 \end{align}
 at level $-1/2$ and
 \begin{align}
  \Fs{1}^\lin_{\mu\nu}{}^A&= 2 \partial_{[\mu} A_{\nu]}^A - \partial_B B_{\mu\nu}^{AB} - f^{AB}{}_{C} \partial_B X_{\mu\nu}^C \ ,\nn\\
 \Fs{1}^\lin_{\mu;\nu} &= \partial_\mu X_\nu + \partial_A X^A_{\mu\nu} - \partial_A h^A_{\mu,\nu}  
 \end{align}
 at level $1/2$, where the second is the analogue of the dual graviton field strength that only involves the extra field $\chi_{\mu;\nu}  \sim \partial_{\mu} X_{\nu}$ in that case. At level $-3/2$, one gets 
\begin{align}
 \Fs{-3}^\lin_{AB} &= \bigl( 14 P^{CD}{}_{AB}  + \tfrac14 \kappa^{CD} \kappa_{AB} \bigr) \partial_C \Phi_D \nn = 2 \partial_{(A} \Phi_{B)} + f^{EC}{}_{A} f_{EB}{}^{D} \partial_{(C} \Phi_{D)}  \,,\\
\Fs{-3}^\lin_{\mu}{}^\nu_A  &=  \partial_\mu A^{\nu}_A - \partial_A h_\mu{}^\nu - \partial^\nu_{AB} A_\mu^B - f_{AB}{}^C \partial_C^\nu A_\mu^A - \tfrac12 \delta_\mu^\nu f^{BC}{}_A \partial_B \Phi_C + \delta_\mu^\nu \partial^\sigma_{AB} A_\sigma^B  \ .\quad  
\end{align}
 
The linearised gauge transformations can be derived using the structure constants given in~\eqref{GaugeStrucCoeff} as 
\begin{subequations}
  \label{GaugeTransfE8} 
\begin{align}
  \delta_\xi  h_\mu{}^\nu &= \partial_\mu \xi^\nu - \tfrac1{14} \acute{\partial}^\nu_{AB} \acute{\lambda}_\mu^{AB} - 4 \partial^\nu \lambda_\mu + 2 \partial^\nu_A \xi_\mu^A + \delta_\mu^\nu \Bigl( \partial_A \lambda^A+ \tfrac17  \acute{\partial}^\sigma_{AB} \acute{\lambda}_\sigma^{AB}+8 \partial^\sigma \lambda_\sigma   \Bigr) \nn \\
  &\quad + \eta_{\mu\sigma} \eta^{\nu\rho} \Bigl( \partial_\rho \xi^\sigma - \tfrac1{14} \acute{\partial}^\sigma_{AB} \acute{\lambda}_\rho^{AB} - 4 \partial^\sigma \lambda_\rho + 2 \partial^\sigma_A \xi_\rho^A  \Bigr) + \dots\,, \\
  \delta_\xi  \Phi_A &=  ( \delta_A^E+ \kappa_{AF} \delta^{EF} ) \Bigl( - f_{EB}{}^C \partial_C \lambda^B- \tfrac17 f_{EB}{}^C \partial_{CD}^\mu \lambda_\mu^{BD} + 2 f_{EB}{}^C \partial_C^\mu \xi_\mu^B  + \dots\Bigr)  \,,\\
  \delta_\xi A_\mu{}^A &= \partial_\mu \lambda^A + \partial_B \lambda^{AB}_\mu - f^{AB}{}_C \partial_B \xi_\mu^C  + \eta_{\mu\nu} \delta^{AB} \Bigl( \partial_B \xi^\nu  + \partial_{BC}^\nu \lambda^C +f_{BC}{}^D \partial_D^\nu \lambda^C \Bigr) + \dots \,,\\
  \delta_\xi B_{\mu\nu}^{AB} &= 2 \partial_{[\mu} \lambda_{\nu]}^{AB}  - 2  \eta_{\mu\sigma} \eta_{\nu\rho} \delta^{AC} \delta^{BD}\partial_{CD}^{[\sigma}  \xi^{\rho]}+   \dots \,,\\
  \delta_\xi  h_{\mu,\nu}^{A} &= 2 \partial_{(\mu} \xi^A_{\nu)}  + 2 \eta_{\mu\sigma} \eta_{\nu\rho} \delta^{AB}  \partial^{(\sigma}_B \xi^{\rho)} +  \dots
 \end{align}
 \end{subequations}
The linearised gauge transformation of the fields in $R(\Lambda_2)$ includes similarly the three contributions from \eqref{deltaxiX} as
\begin{subequations}
\begin{align}
  \delta_\xi X_\mu &= 2 \partial_A \xi^A_\mu+ 2  \eta_{\mu\nu} \partial_A^\mu \lambda^A - \varepsilon_{\mu\nu\sigma}  \eta^{\nu\rho}  \partial_\rho \xi^\sigma \, , \\
\delta_\xi X_{\mu\nu}^A &=- 2  \partial_{[\mu} \xi^A_{\nu]}  +2  \eta_{\mu\sigma} \eta_{\nu\rho} \delta^{AB} \partial^{[\sigma}_B \xi^{\rho]}  - \varepsilon_{\mu\nu\sigma} ( \eta^{\sigma\rho} \partial_\rho \lambda^A - \delta^{AB} \partial_B   \xi^\sigma)  \ . 
\end{align}
\end{subequations}
The terms involving the Levi--Civita $\varepsilon$ symbol correspond to the contributions from $\Pi^{\tilde\alpha}{}_{QP}$ in the general formula~\eqref{deltaxiX}.
One straightforwardly checks that the linearised duality equations 
\begin{subequations}
\begin{align}
 \cE_{\mu\nu}^{^\lin A} &= F^\lin_{\mu\nu}{}^A - \varepsilon_{\mu\nu\sigma}  \eta^{\sigma\rho} \delta^{AB}  F^\lin_{\rho B}  = 0 \ , \\
\cE^\lin_{\mu;\nu} &= F^\lin_{\mu;\nu} + \frac{1}{2} \eta_{\mu\sigma}  \varepsilon^{\sigma\rho\lambda}  F^\lin_{\rho\lambda}{}^\nu + \eta_{\mu\sigma}  \eta_{\nu\rho} \varepsilon^{\sigma\rho\lambda}  F^\lin_{\lambda\kappa}{}^\kappa = 0 \ 
\end{align}
\end{subequations}
 are gauge invariant modulo the section constraint within this level truncation. This provides an additional check of equation \eqref{MasterAssum2}.

\subsection{Non-linear field strengths}
 
We also provide a formula for the field strengths at the non-linear level and compare with the known $E_8$ exceptional field theory~\cite{Hohm:2014fxa}.
To obtain the formula for the field strength at the non-linear level, one needs to replace the derivative of the field in $\mathfrak{e}_{11}$ by the components of the semi-flattened current 
\begin{multline} \label{E8current}
\tilde J_\alpha {t}^\alpha  = g^{\nu\sigma} d g_{\mu\sigma}   {K}^\mu{}_\nu + J_{A} {t}^A   +  d A_\mu^A ( {E}^\mu_A  + g^{\mu\nu} M_{AB} F_\nu^B)\\
 +\tfrac1{28} \bigl(d \acute{B}_{\mu\nu}^{AB}- 14  A_{[\mu}^A d A_{\nu]}^B \bigr)  ( \acute{{E}}_{AB}^{\mu\nu}+ g^{\mu\sigma}g^{\nu\rho} M_{AC} M_{BD} \acute{F}_{\sigma\rho}^{CD}  )   \\
+2 \bigl( d B_{\mu\nu}+ \tfrac{1}{4} \kappa_{AB} A_{[\mu}^A d A_{\nu]}^B \bigr)(  {E}^{\mu\nu}  + g^{\mu\sigma} g^{\nu\rho} F_{\nu\rho}) \\+ \bigl( d h_{\mu,\nu}^{A} + f_{BC}{}^A  A_{(\mu}^B d A_{\nu)}^C \bigr) (   {E}_A^{\mu\nu}  + g^{\mu\sigma}g^{\nu\rho} M_{AB}  F_{\sigma\rho}^{B} ) + \dots 
\end{multline}
where $J_A$ is the $E_{8(8)}$ current 
\begin{align}
  J_{A} f^{AB}{}_C = -M^{BD} d M_{CD}  \ .  
\end{align}
It satisfies 
\begin{align}
  \kappa^{AB} J_B  = M^{AB} J_B \  . 
\end{align}
 
For simplicity we shall consider the solution to the section constraint for which only the derivatives $\partial_\mu$ and $\partial_A$ are non-zero, and $\partial_A$ satisfies the $E_{8}$ section constraint~\cite{Berman:2012vc}
 \begin{align}
   P^{CD}{}_{AB} \partial_C \otimes \partial_D = 0  \ , \qquad \kappa^{AB} \partial_A \otimes \partial_B = 0 \ , \quad f^{AB}{}_C \partial_A \otimes \partial_B = 0  \; .  
\end{align}
One obtains then that the only non-trivial components of the derivative $ \cU^{-1}_M{}^N  \partial_N $  are
 \begin{align}
  \tilde{\partial}_\mu = \cU^{-1}_\mu{}^M \partial_M = \partial_\mu - A_\mu^A \partial_A \; , \qquad  \cU^{-1}_A{}^M \partial_M = \partial_A \ . 
  \end{align}
 The formulae for the semi-flattened field strengths $\tilde{F}^I$ can be obtained directly from the linear expressions by substituting these currents to the linear derivatives, with the derivative modified according to the formula above. 
One obtains in this way 
\begin{align}
 \tilde{F}_{\mu\nu}{}^\sigma &= 2 g^{\sigma\rho} (\partial_{[\mu}  - A_{[\mu}^A \partial_A ) g_{\nu]\rho} +2 \delta^\sigma_{[\mu} \partial_A A_{\nu]}^A  \,, \nn\\
 \tilde{F}_{\mu A} &= \tilde{J}_{\mu A}   +  f_{AC}{}^D  \partial_D A_\mu^C +  \chi_{A;\mu}  \,,
\end{align}
 where $\tilde{J}_{\mu A}  $ includes the transport term in the derivative, 
 \begin{align}
   \tilde{J}_{\mu A} f^{AB}{}_C = -M^{BD} ( \partial_\mu - A_\mu^E \partial_E) M_{CD} \; ,
\end{align}
and 
 \begin{align}
  \tilde{F}_{\mu\nu}^A&=2 \partial_{[\mu} A_{\nu]}^A- 2 A_{[\mu}^B \partial_B A_{\nu]}^A 
 - \partial_B B_{\mu\nu}^{AB} +\bigl(  14 P^{AB}{}_{CD} + \tfrac{1}{4} \kappa^{AB} \kappa_{CD} \bigr) A_{[\mu}^C \partial_B A_{\nu]}^D-f^{AB}{}_{C} \tilde{\chi}_{B;}{}_{\mu\nu}^C \,,\nn\\
 \tilde{F}_{\mu;\nu} &=- \partial_A h^A_{\mu,\nu}  - \tfrac12 f_{BC}{}^A A_{(\mu}^B  \partial_A A_{\nu)}^C  +\tilde{\chi}_{\mu;\nu} -A_{(\mu}^A \chi_{A;\nu)} + \tilde{\chi}_{A;}{}^A_{\mu\nu}\,.
 \end{align}

 The structure of these field strengths and their dependence in the constrained field $\tilde\chi_{M,\tilde\alpha}$ can be compared to \cite{Hohm:2014fxa} with the identification $\tilde{\chi}_{A;\mu} = B_{\mu A}$ and 
\be  
\label{HohmChi} \tilde{\chi}_{B;}{}_{\mu\nu}^A= C_{\mu\nu B}{}^A+\tfrac12 f^A{}_{KL} A_{[\mu}^K \partial_B A_{\nu]}^L +\frac{1}{\sqrt{-g}} g_{\mu\sigma} g_{\nu\rho} \varepsilon^{\sigma\rho\lambda} \partial_B A^A_\lambda \ , \ee
such that the $E_{11}$ duality equation gives
\be \label{DualYMscalar} \tilde{F}_{\mu\nu}^A = \frac{1}{\sqrt{-g}} g_{\mu\sigma} g_{\nu\rho} \varepsilon^{\sigma\rho\lambda} M^{AB}\tilde{F}_{\lambda B}\ , 
\ee
which coincides with  \cite{Hohm:2014fxa} up to moving the last term in \eqref{HohmChi} to the right-hand-side such as to reproduce the covariant current  of the $E_8$ exceptional field theory
\begin{align}
 j_{\mu}{}^A &= M^{AB} \tilde{F}_{\mu B} + \kappa^{AB} \bigl( f_{BC}{}^D  \partial_D A_\mu^C +  \chi_{B;\mu}  \bigr) \nn\\
&= \kappa^{AB} \tilde{J}_{\mu B}   + (\kappa^{AB} + M^{AB} ) (f_{BC}{}^D  \partial_D A_\mu^C +  \chi_{B;\mu})  \ ,
 \end{align}
up to the term in $\kappa^{AB} \chi_{B;\mu}$ that does not appear in \eqref{DualYMscalar}. Note however that this equation in  \cite{Hohm:2014fxa} is only satisfied up to a trivial parameter since it is the equation of motion of $B_{\mu A}= \chi_{A;\mu}$. This additional term can be produced by adding to the action in  \cite{Hohm:2014fxa}   a term in $\sqrt{-g} g^{\mu\nu} \eta^{AB} B_{\mu A} B_{\nu B} $ that vanishes on section. 

The other equation at this order is the dual graviton equation 
\be 
\tilde{F}_{\mu;\nu} = - \frac{1}{2\sqrt{-g}} g_{\mu\sigma}  g_{\nu\kappa}  \varepsilon^{\sigma\rho\lambda}  \tilde{F}_{\rho\lambda}{}^\kappa - \frac{1}{\sqrt{-g}} g_{\mu\sigma}  g_{\nu\rho} \varepsilon^{\sigma\rho\lambda}  \tilde{F}_{\lambda\kappa}{}^\kappa \ . 
\ee
Similar to the $GL(11)$ decomposition, this equation is not dynamical by itself and only determines the field $\tilde{\chi}_{\mu;\nu}$ algebraically. The integrability condition for the  Einstein equation \cite{Hohm:2014fxa} to be satisfied determines the first order equation for $\tilde{\chi}_{\mu;\nu}$. 

\section{Supersymmetry transformations and algebra}
\label{sec:susyALG}

In the remainder of the paper we study aspects of the supersymmetric extension of the model we have developed in the preceding sections that is achieved through the inclusion of an unfaithful (vector-)spinor $\Psi$ that transform under $\widetilde{K}(E_{11})$, the double cover of $K(E_{11})$.\footnote{It was shown recently that this double cover of the group $K(E_{11})$ is the universal cover~\cite{Harring:2019}.} We note that these results do \textit{not} depend on the speculative full dynamics discussed in Section~\ref{sec:fulldyn}. One of the key results we establish here is that certain bilinears in $\Psi$ take values in the anti-selfdual subspace of the $\cT_{-1}$ part of the tensor hierarchy algebra and therefore terms of the form $\Psi \Psi$ can be added to the first order self-duality equation~\eqref{DualityEquation}. We shall also see how to define supersymmetry transformation rules for all the fields, including the constrained fields and how to write down a $\widetilde{K}(E_{11})$ covariant equation of motion for the vector-spinor $\Psi$. The compatibility of supersymmetry with Kac--Moody symmetries was previously investigated in~\cite{Damour:2005zs,deBuyl:2005sch,Damour:2006xu} for $K(E_{10})$ and in~\cite{Steele:2010tk} for $K(E_{11})$. We shall see explicitly how our inclusion of the extra constrained fields is crucial for resolving apparent inconsistencies in the supersymmetry algebra observed in~\cite{Damour:2006xu}.

\subsection{Spinors of \texorpdfstring{$\widetilde{K}(E_{11})$}{K(E11)}}

We begin with the description of the spinors for $\widetilde{K}(E_{11})$. The existence of an unfaithful vector-spinor $\Psi$ for $\widetilde{K}(E_{11})$ was deduced in~\cite{Kleinschmidt:2006tm}, relying heavily on previous results for vector-spinors for $\widetilde{K}(E_{10})$~\cite{Damour:2005zs,deBuyl:2005sch}. In this section we write $\widetilde{K}(E_{11})$ for the double cover of the maximal subgroup of $E_{11}$ defined by the Cartan involution. Unlike $E_{11}$, the subgroup $K(E_{11})$ is not a Kac--Moody group and its general representation theory is unknown. However, one can demonstrate the existence of an unfaithful Dirac-spinor $\epsilon$~\cite{West:2003fc,deBuyl:2005zy,Damour:2005zs,Kleinschmidt:2006tm} and of an unfaithful vector-spinor $\Psi$~\cite{Damour:2005zs,deBuyl:2005sch,Kleinschmidt:2006tm}. This is possible because $\widetilde{K}(E_{11})$ is not a reductive group, and contains ideals. Finite dimensional unfaithful representations of $\widetilde{K}(E_{11})$ exist for which the ideal $\mathcal{I}$ acting trivially on the finite-dimensional vector space are such that $\widetilde{K}(E_{11})/ \mathcal{I}$ is a finite-dimensional group. In the case of the Dirac-spinor, $\widetilde{K}(E_{11})/ \mathcal{I}_\epsilon \cong SL(32)$, a result anticipated in~\cite{West:2003fc}.  The dimension of the vector-spinor representation is $352$. 

The representations can be succinctly described in terms of the $Spin(1,10)$ Lorentz subgroup of $\widetilde{K}(E_{11})$. Under this subgroup the $32$-component Dirac-spinor $\epsilon$ is irreducible and becomes the standard Majorana spinor in $D=11$ dimensions up to a rescaling. The vector-spinor $\Psi$ is reducible under this subgroup and decomposes into a gamma-traceless vector-spinor of $Spin(1,10)$ and a simple spinor that can be viewed as the gamma trace. We shall combine the two and write the $\widetilde{K}(E_{11})$ vector-spinor $\Psi$ as $\psi_a$ when we think of it as a (reducible) $Spin(1,10)$ representation. Here, $a=0,\ldots,10$ is a Lorentz tangent index. The use of $Spin(1,10)$ here parallels the $GL(11)$ decomposition studied in Section~\ref{sec:gl11bos}. 

The $\widetilde{K}(E_{11})$ transformations of the spinors are completely determined by giving the transformation under $Spin(1,10)$ and under the combination
\begin{align}
\frac1{3!} \Lambda_{a_1a_2a_3} \left( E^{a_1a_2a_3} - \eta^{a_1b_1} \eta^{a_2b_2} \eta^{a_3b_3} F_{b_1b_2b_3}\right) \in K(\mf{e}_{11})\,.
\end{align}
This combination is invariant under the involution defining $K(E_{11})$ and the occurrence of the Minkowski metric $\eta^{ab}$ is due to the signature of the involution. When we write $SO(1,10)$ tensors we shall use $\eta^{ab}$ freely to raise and lower indices.

The $Spin(1,10)$ transformations of $\epsilon$ and $\psi_a$ are implicit in their index structure and the result of~\cite{Kleinschmidt:2006tm} is that the transformations
\begin{subequations}
\label{LV}
\begin{align}
\delta \psi_a  &= - \frac{1}{12} \Lambda_{bcd} \Gamma^{bcd}  \psi_a - \frac{2}{3} \Lambda_{abc} \Gamma^b \psi^c +\frac16  \Lambda_{bcd} \Gamma_a{}^{bc} \psi^d \ ,
\label{LV1}\w2
\delta \epsilon &= -\frac1{12} \Lambda_{abc} \Gamma^{abc} \epsilon\ ,
\label{LV2}
\end{align}
\end{subequations}
define consistent unfaithful spinors of $\widetilde{K}(E_{11})$. The overall signs in these transformations were chosen to match the commutators of the generators under $K(\mf{e}_{11})\subset \mf{e}_{11}$.  Here, the conjugate Majorana spinor is defined as $\bar{\psi}_a = \psi_a^{\intercal} \mathcal{C}$ with $\mathcal{C}=i\Gamma^0$ and we use gamma matrices satisfying $\Gamma^{a_1\ldots a_{11}} = -\varepsilon^{a_1\ldots a_{11}}$. The rules~\eqref{LV} are sufficient to study the transformation of any polynomial in these fermions.

\subsection{Coset scalar fields supersymmetry transformations}
\label{sec:susybos}

In $D=11$ supergravity one has the (linearised) supersymmetry transformation rules for the bosonic fields given by
\begin{subequations}
\label{eq:susy11}
\begin{align}
\dsusy h_{ab} &= -\bar{\epsilon} \Gamma_{(a} \psi_{b)}\,,\\
\dsusy A_{a_1a_2a_3} &= \frac32 \bar{\epsilon} \Gamma_{[a_1a_2} \psi_{a_3]} \,,\\
\dsusy A_{a_1\ldots a_6} &= 3\bar{\epsilon} \Gamma_{[a_1\ldots a_5} \psi_{a_6]} \,,
\end{align}
\end{subequations}
where we have extended the transformations of~\cite{Cremmer:1978km} to also include the dual six-form potential~\cite{Damour:2006xu} and written the expression in tangent space. In order to define the supersymmetry variation of the scalar fields parametrising $E_{11}  / K(E_{11})$, one needs a priori to show that the bilinear $\epsilon \Psi$ includes a representation of  $K(E_{11})$ that can be consistently embeded in $\mf{e}_{11} \ominus K(\mf{e}_{11})$. Starting from \eqref{eq:susy11}, we shall therefore study the representation $\epsilon\Psi$. With the definitions 
\begin{subequations}
\label{Xi}
\begin{align}
\Xi_{ab} &= - \bar\epsilon \Gamma_{(a} \psi_{b)} \,,
\\
\Xi_{a_1a_2a_3} &=  \frac32 \bar\epsilon \Gamma_{[a_1a_2} \psi_{a_3]} \,,
\\
\Xi_{a_1\dots a_6} &= 3 \bar\epsilon \Gamma_{[a_1\dots a_5} \psi_{a_6]} \,,
\\
\Xi_{a_1\dots a_8,b} &= \bar\epsilon \Gamma_{\langle a_1 \dots a_8,} \psi_{b\rangle}+ \frac1{12} \varepsilon_{a_1\dots a_8}{}^{c_1c_2c_3} \bar\epsilon \Gamma_{c_1c_2c_3bd } \psi^d \; ,  \\
\Xi_{a_1\dots a_9,}{}^{b_1b_2b_3} &= 1512 \delta^{b_1}_{[a_1} \delta^{b_2}_{a_2} \delta^{b_3}_{a_3}   \bar \epsilon \Gamma_{a_4\dots a_8} \psi_{a_9]} \nn\\
& \qquad  -9  \delta^{[b_1}_{[a_1} \delta^{b_2}_{a_2} \varepsilon_{a_3\dots a_9]c_1\dots c_4} ( \bar \epsilon \Gamma^{c_1\dots c_4} \psi^{|b_3]} +\tfrac12 \eta^{b_3]c_1} \bar \epsilon \Gamma^{c_2\dots c_4d} \psi_{d})  \ ,
\end{align}
\end{subequations}
one computes using the formulae \eqref{E11adjoint} that the $\Xi$-bilinears transform into each other under the $K(\mf e_{11})$ transformations~\eqref{LV} according to\footnote{A similar calculation was done up to the level of the six-form~\eqref{eq:ds6} for $K(\mf{e}_{11})$ in~\cite{Steele:2010tk} where fermions and $\mf{e}_{11}$ were also considered. For $K(\mf{e}_{10})$, it had been done previously to higher level in~\cite{Damour:2006xu,Henneaux:2008nr} where also the extra nine-form $\Upsilon^*_9$ of~\eqref{eq:dx9} had been found.}
\begin{subequations}
\label{XiKE11} 
\begin{align}
\delta \Xi_{ab} &= \Lambda_{(a}{}^{c_1c_2} \Xi_{b)c_1c_2} - \frac19 \eta_{ab} \Lambda^{c_1c_2c_3} \Xi_{c_1c_2c_3}\,,\\
\label{eq:ds6}
\delta \Xi_{a_1a_2a_3} &= - 3 \Lambda_{[a_1a_2}{}^b \Xi_{a_3]b} +\frac16  \Lambda^{c_1c_2c_3} \Xi_{a_1a_2a_3c_1c_2c_3}\,,\\
\label{eq:dx9}
\delta \Xi_{a_1\dots a_6} &= 20 \Lambda_{[a_1a_2a_3} \Xi_{a_4a_5a_6]}   + \frac12 \Lambda^{b_1b_2c} ( \Xi_{a_1\dots a_6b_1b_2,c} + \Upsilon^\star_{a_1\dots a_6b_1b_2c} ) \ ,  \\
\delta \Xi_{a_1\dots a_8,b} &= 56 \Lambda_{\langle a_1a_2a_3} \Xi_{a_4\dots a_8,b\rangle} + \frac{1}{2} \Lambda^{c_1c_2c_3} \Xi_{c_1\langle a_1\dots a_8,b\rangle c_2c_3} + \dots 
\end{align}
\end{subequations}
up to the introduction of an additional new bilinear  $\Upsilon^\star_9$ appearing in the transformation of $\Xi_{a_1...a_6}$, 
\begin{align}
\Upsilon^\star_{a_1\dots a_9} = \bar\epsilon \Gamma_{[a_1 \dots a_8} \psi_{a_9]} =  \frac1{18} \varepsilon_{a_1\dots a_9b_1b_2} \Upsilon^{b_1b_2} \,.
\end{align}
Such a nine-form is not present in the coset $\mf{e}_{11} \ominus K(\mf{e}_{11})$ since the dual graviton is in the irreducible $(8,1)$ representation and this signals a potential inconsistency between $\mf{e}_{11}$ and supersymmetry. However, considering the tensor hierarchy algebra extension $\cT_0\supset \mf{e}_{11}$, the nine-form is consistent with the $K(\mf e_{11})$ representation of a field parametrising the degree zero component of the tensor hierarchy algebra $\mathcal{T}_0\ominus K(\mf{e}_{11})$. In fact we shall see that we need to consider the conjugate algebra $\overline{\mathcal{T}}_0\ominus K(\mf{e}_{11})$, extending $\mf{e}_{11}$ with generators in the conjugate representation $\overline{R(\Lambda_2)}$.

{\allowdisplaybreaks
The indecomposable $E_{11}$ module $\mf{e}_{11}\oleft \overline{R(\Lambda_2)}\subset \overline{\mathcal{T}}_0$ induces the dual indecomposable structure on its components $(\Xi,\Upsilon) \in (\mf{e}_{11} \ominus K(\mf e_{11})) \oright {R(\Lambda_2)}$, such that $\Xi$ transforms into $\Xi$ and $\Upsilon$, whereas $\Upsilon$ transforms into itself under $K(\mf{e}_{11})$. To further check this property one computes that the element $\Upsilon^{a_1a_2}$ generates a $K(\mf e_{11})$ module that is indeed consistent with the structure of  ${R(\Lambda_2)}$. For this purpose one defines 
\begin{subequations}
\label{Upsilon}
\begin{align} 
\Upsilon^{b_1b_2} &=  \bar\epsilon \Gamma^{b_1b_2 c} \psi_c\,,
\w2
\Upsilon^{a_1a_2;b} &= - \bar\epsilon \Gamma^{a_1a_2bc} \psi_{c} - 4 \bar \epsilon \Gamma^{b[a_1} \psi^{a_2]} - 2 \eta^{b[a_1} \bar \epsilon \psi^{a_2]}  \ , 
\w2
\Upsilon^{a_1a_2a_3a_4a_5;b}  &= \bar\epsilon \Gamma^{a_1a_2a_3a_4a_5} \psi^b +4 \bar\epsilon \Gamma^{[a_1a_2a_3a_4a_5} \psi^{b]} +10 \eta^{b[a_1} \bar\epsilon \Gamma^{a_2a_3a_4} \psi^{a_5]} \ ,
\nn\\*
&\quad -\frac1{48} \varepsilon^{a_1a_2a_3a_4a_5bc_1c_2c_3c_4c_5} \bar\epsilon \Gamma_{c_1c_2c_3c_4} \psi_{c_5} \ , 
\w2
\Upsilon^{a_1a_2a_3,b} &= \bar\epsilon \Gamma^{a_1a_2a_3} \psi^b - \bar\epsilon \Gamma^{[a_1a_2a_3} \psi^{b]} + 6 \eta^{b[a_1} \bar\epsilon \Gamma^{a_2} \psi^{a_3]}\ .
\end{align}
\end{subequations}
and one checks that they transform into each other according to 
\begin{subequations}
\label{variations}
\begin{align}
\label{eq:dU2}
\delta \Upsilon^{a_1a_2} &= \Lambda^{b_1b_2[a_1} \Upsilon_{b_1b_2;}{}^{a_2]} \ , 
\w2
\delta \Upsilon^{c(a;b)} &= -\frac13 \Lambda_{def} \Upsilon^{defc(a,b)}+2 \Lambda_{de}{}^{(a} \Upsilon^{b)cd,e}\ ,  
\w2
\delta \Upsilon^{a_1a_2a_3} &= 3 \Lambda_b{}^{[a_1a_2} \Upsilon^{a_3]b} - \frac13 \Lambda_{b_1b_2b_3}  \Upsilon^{a_1a_2a_3b_1b_2;b_3} - 2 \Lambda_{bc}{}^{[a_1} \Upsilon^{a_2a_3]b,c} \,.
\end{align}
\end{subequations}
These transformations are consistent with the structure of the tensor hierarchy algebra that is described in~\eqref{E11adjoint}. In particular, the transformation~\eqref{eq:dU2} shows that the new nine-form that appears for the extra fields in ${R(\Lambda_2)}$  does not transform back into the $\Xi$ components in $\mathfrak{e}_{11}\ominus K(\mathfrak{e}_{11})$. Moreover, demanding that the $K(E_{11})$ transformation commute with the supersymmetry transformation, one can determine the linearised supersymmetry variation of fields belonging to the $R(\Lambda_2)$ module as
\begin{subequations}
\label{eq:susyRL2}
\begin{align}
\dsusy h_{a_1\dots a_9} & =\frac1{18} \varepsilon_{a_1\dots a_9b_1b_2} \Upsilon^{b_1b_2}  =  \bar\epsilon \Gamma_{[a_1 \dots a_8} \psi_{a_9]} \,,\\
\dsusy A_{a_1\dots a_{10};b_1b_2} &= \frac1{18} \varepsilon_{a_1\dots a_{10}c} \, \Upsilon_{b_1b_2;}{}^c   \,,\\
\dsusy A_{a_1\ldots a_{10};b_1\dots b_5} &=  \frac1{18} \varepsilon_{a_1\dots a_{10}c}  \Upsilon_{b_1\dots b_5;}{}^c \; , \qquad \dsusy B_{a_1\dots a_{11},b_1b_2b_3,c} =   \frac1{18} \varepsilon_{a_1\dots a_{11}} \Upsilon^{b_1b_2b_3,c} \,,
\end{align}
\end{subequations}
consistently with the decomposition of the module ${R(\Lambda_2)}$ under $GL(11)$. }

Defining $G(\overline{\mathcal{T}}_0)$ as the group associated to $\overline{\mathcal{T}}_0$ (the algebra defined from $\mf{e}_{11}$ with the additional generators in the conjugate representation $\overline{R(\Lambda_2)}$), we conclude that one can define the linearised supersymmetry transformation of an extended field $\cV \in G(\overline{\mathcal{T}}_0) / K(E_{11})$ consistently with the $K(E_{11})$ representation of $\overline{\mathcal{T}}_0$ as an indecomposable module. Defining $T(\Xi,\Upsilon)$ as the $\overline{\mathcal{T}}_0$ element of parameter $(\Xi,\Upsilon)$ defined above, one can write the supersymmetry transformation of   $\cV \in G(\overline{\mathcal{T}}_0) / K(E_{11})$ as
\be 
\label{NonLinsusy}  \dsusy \cV  = T(\Xi,\Upsilon) \cV \ , 
\ee
such that supersymmetry commutes with the action of $\widetilde{K}(E_{11})$ in the standard way, at the price of introducing additional fields into the theory. We stress here that the additional fields, that we denote by $h^{\tilde{\alpha}}$, parametrising $G(\overline{\mathcal{T}}_0)$ must not be confused with the constrained fields $\chi_M{}^{\tilde{\alpha}}$ that transform instead as components of the co-adjoint module $\mathcal{T}_{-2}$.  In order to have well-defined expressions, one will choose in practice a given parametrization of the coset $G(\overline{\mathcal{T}}_0) / K(E_{11})$, so one will define the non-linear supersymmetry transformation from \eqref{NonLinsusy} with the compensating local $\widetilde{K}(E_{11})$ transformation on the left.  

Let us compare the situation to that in $E_9$ exceptional field theory \cite{Bossard:2018utw}. In this case, $\overline{\cT}_{0}(\mf{e}_9)$ is the extension of $\mf{e}_9$ by a single Virasoro generator $L_{-1}$.  
There is a single additional field $\tilde\rho$ on top of the $E_9/K(E_9)$ coset fields that is dual to the dilaton such that all fields together parametrise $G(\overline{\cT}_0(\mf{e}_9))/K(E_9)$.
In addition, the indecomposable representation of the co-adjoint  $\cT_0(\mf{e}_9)^*$ entails a single constrained field $\chi_M$. 
In order to have manifestly $K(E_9)$ covariant supersymmetry rules one has to introduce $\tilde\rho$ with a non-trivial supersymmetry transformation~\cite{Nicolai:1988jb}. This is also what we see above for $K(E_{11})$ where we have to extend $\mf{e}_{11}$ to $\overline{\cT}_0$. Moreover, the dual dilaton $\tilde{\rho}$ is associated with an additional gauge symmetry shifting $\tilde\rho$ whose parameter is called $\Sigma$ in~\cite{Bossard:2018utw}  and which is generally needed for the closure of the algebra of generalised Lie derivatives. In the present case, there is a similar additional gauge parameter $\Sigma_M{}^{N\tilde{\alpha}}$ in $\overline{R(\Lambda_1)} \otimes R(\Lambda_3)$ as discussed in Section \ref{sec:fulldyn}:
\begin{align}
  \label{ExtendedSigmaGauge}
\delta A_{n_1n_2n_3} &=  \frac{1}{5!} \Sigma^{p_1\dots p_5}{}_{n_1n_2n_3 p_1\dots p_5} \, , \nn \\
\delta A_{n_1\dots n_6} &= - \frac{1}{2} \Sigma^{p_1p_2}{}_{n_1\dots n_6p_1p_2} \ , \nn \\
\delta h_{n_1\dots n_8;m} &= \Sigma_{m;n_1\dots n_8} \; , \nn \\
\delta \chi_{M; n_1\dots n_9} &=- 9 \partial_{[n_1} \Sigma_{M;|n_2\dots n_9]} \ ,
\end{align}
where, compared to~\eqref{SigmaGaugeTranformations}, we have relaxed the condition that $ \Sigma_{[m;n_1\dots n_8]}=0$. This condition is the vanishing of the first component of the trace of $\Sigma_M{}^{N\tilde\alpha}$. The gauge symmetry of the traced parameter $\Sigma_N{}^{N\tilde{\alpha}}$ fits precisely with the representation of the additional fields in ${R(\Lambda_2)}$, and should allow to gauge-fix them at the price of making the supersymmetry realised non-linearly with a compensating $\Sigma_N{}^{N\tilde{\alpha}}$ transformation breaking $\widetilde{K}(E_{11})$ invariance in the linearised approximation. 

In the $GL(11)$ decomposition, we would like to think of these additional gauge symmetries as being related to local Lorentz transformations, such that 
\begin{subequations}
 \label{ExtendedFields} 
\begin{align} h_{a_1\dots a_8;b} &\equiv  h_{a_1\dots a_8,b} + h_{a_1\dots a_8b}\; ,  \\
A_{a_1\dots a_9;b_1b_2b_3} &\equiv  A_{a_1\dots a_9,b_1b_2b_3} + A_{a_1\dots a_9[b_1;b_2b_3]} \; , 
\end{align}
\end{subequations}
could be thought of as the dual graviton and the 3-form gradient dual in the vielbein formulation. Recall that the semi-colon indicates the general tensor product, whereas the comma implies instead that this is an irreducible $GL(11)$ tensor.  This justifies the notation for these additional fields, which should not be confused with the fields $X\in R(\Lambda_2)$ with the transformation rules \eqref{E11adjoint}. 

Later on in Section~\ref{supercovariantF} we shall introduce extended field strengths including the additional fields \eqref{ExtendedFields} in \eqref{FieldStrengthsT0}. As shown there, the extended field strengths are indeed invariant under the transformations \eqref{ExtendedSigmaGauge}. 

We conclude this section by giving a few consequences of the above supersymmetry transformations for further reference. Combining the linearised supersymmetry transformations of the irreducible fields, one concludes that the reducible field $h_{8;1}$ transforms as
\begin{align}
\dsusy h_{a_1\dots a_9;b} &= \bar\epsilon \Gamma_{a_1 \dots a_8,} \psi_{b}+ \frac1{12} \varepsilon_{a_1\dots a_8}{}^{c_1c_2c_3} \bar\epsilon \Gamma_{c_1c_2c_3bd } \psi^d \,.
 \end{align}
The coset field component $A_{9,3}$ transforms as
\begin{align}
 \dsusy A_{a_1\dots a_9,b_1b_2b_3} &= \Xi_{a_1\dots a_9,b_1b_2b_3} \ .
 \end{align}

We also do similar checks for the $GL(4) \times E_7$ decomposition in Appendix \ref{E7susy}, and find prefect agreement with the supersymmetry transformations of $E_7$ exceptional field theory \cite{Godazgar:2014nqa}.

\subsection{Vector-spinor field transformations and supersymmetry algebra}

The supersymmetry variation of the gravitino in  $D=11$ supergravity in the linearised approximation is $\dsusy \psi_a = \partial_a \epsilon$. We shall now extend this transformation to a $\widetilde{K}(E_{11})$ covariant supersymmetry variation. This can be done by making an ansatz involving the higher level  derivatives and fixing the free coefficients by $\widetilde{K}(E_{11})$ covariance, such that $\dsusy \psi_a$ transforms as $\psi_a$ under  ${K}(\mf{e}_{11})$. The result is
\begin{align} 
\dsusy \psi_a &= \partial_a \epsilon + \frac23  \Gamma^b \partial_{ab}\epsilon - \frac16 \Gamma_{abc} \partial^{bc} \epsilon 
- \frac1{3\cdot 4!}  \Gamma^{b_1b_2b_3b_4} \partial_{ab_1b_2b_3b_4}\epsilon  
\nn\\
&\quad + \frac{2}{3 \cdot 5!} \Gamma_{ab_1\dots b_5} \partial^{b_1\dots b_5} \epsilon 
+\frac{2}{7!} \Gamma^{a_1...a_7} \partial_{a_1...a_7,a}\epsilon -\frac{8}{7!} \Gamma_{aa_1...a_6} \eta_{bc}\partial^{a_1...a_6 b,c}  \epsilon
\nn\\
&\quad
+\frac{2}{3\cdot 7!} \Gamma_{aa_1...a_8} \partial^{a_1...a_8} \epsilon -\frac{2}{3\cdot 7!}  \Gamma^{a_1...a_7}\partial_{aa_1...a_7}\epsilon
 +\ldots 
\label{dpsi}
\end{align}
We have verified this expression for $\dsusy \psi_a $ to be covariant under $K(\mf e_{11})$ transformations including all terms varying into $\partial_1, \partial_2$ and $\partial_5$ derivatives.

We can now verify that the linearised supersymmetry transformations are consistent with the closure of the supersymmetry algebra on the bosonic fields. As usual in supergravity, one expects the algebra of local supersymmetry to only close modulo the equations of motion and gauge transformations with parameters that are bilinear in the supersymmetry spinor parameters $\epsilon$. In the linearised approximation, the closure of the supersymmetry algebra on the bosonic fields does not depend on the fields, and therefore cannot involve  the equations of motion. In this approximation one expects to simply get a bosonic gauge transformation of parameter bilinear in the spinor $\epsilon$. We recall that the bosonic theory, without fermions $\Psi$, is not only invariant under generalised diffeomorphism of parameter $\xi^M$ in $R(\Lambda_1)$, but also under gauge transformation of constrained parameters $\Sigma_{M}{}^{N \tilde{\alpha}}$ in $\overline{R(\Lambda_1)}\otimes R(\Lambda_3)$ as discussed in Section \ref{sec:fulldyn} and at the end of the preceding section. As the closure of two supersymmetries generally produces all gauge symmetries, we therefore expect both of them to appear in the supersymmetry algebra. For simplicity we consider a supersymmetry transformation of commuting parameter $\epsilon$, such that the algebra is obtained by applying twice the same variation. One computes straightforwardly (neglecting derivatives $\partial_{a_1a_2a_3a_4a_5}$ and those of higher level)
\begin{align}
  \label{SusyAlgebra}  
  (\dsusy{})^2 h_{ab} &= \dsusy( - \bar \epsilon \Gamma_{(a} \psi_{b)} ) = - \frac12 \partial_{(a} ( \bar \epsilon \Gamma_{b)} \epsilon) - \frac12 \partial_{(a}{}^c ( \bar \epsilon \Gamma_{b)c} \epsilon) + \frac1{12} \eta_{ab} \partial^{cd}( \bar \epsilon \Gamma_{cd} \epsilon) \ ,
   \\
(\dsusy{})^2 A_{a_1a_2a_3} &= \dsusy(\tfrac32 \bar \epsilon \Gamma_{[a_1a_2} \psi_{a_3]} )  = \frac34 \partial_{[a_1} ( \bar \epsilon \Gamma_{a_2a_3]} \epsilon) - \frac34 \partial_{[a_1a_2} ( \bar \epsilon \Gamma_{a_3]} \epsilon) - \frac18 \partial^{b_1b_2} ( \bar \epsilon \Gamma_{a_1a_2a_3b_1b_2} \epsilon) \ ,
 \nn \\
 (\dsusy{})^2 A_{a_1\dots a_6} &= \dsusy(3 \bar \epsilon \Gamma_{[a_1\dots a_5} \psi_{a_6]} )  = - \frac32 \partial_{[a_1} ( \bar \epsilon \Gamma_{a_2\dots a_6]} \epsilon) + \frac14 \partial^{bc} ( \tfrac{7}{5!} \eta_{c[a_1} \varepsilon_{a_2\dots a_6b]d_1\dots d_5} \bar \epsilon \Gamma^{d_1\dots d_5} \epsilon ) \nn\\
 & \hspace{50mm}- \frac12 \bar \epsilon \Gamma_{a_1\dots a_6b_1b_2} \partial^{b_1b_2} \epsilon  \; . \nn
\end{align}
Apart from the last line in the variation of $A_{a_1\dots a_6}$, all these terms are total derivatives and can be rewritten as generalised Lie derivative gauge transformations \eqref{xt} of parameter $- \frac14 \xi^M$, with the components of $\xi^M$ given by
\begin{subequations}
 \label{GaugeSusyClose}
 \begin{align}
\xi^a &= \bar \epsilon\,  \Gamma^a \epsilon\ , 
\nn\\
\lambda_{ab} &= - \bar \epsilon\,  \Gamma_{ab} \epsilon \ ,
\nn\\
\lambda_{a_1a_2a_3a_4a_5} &= \bar \epsilon\,  \Gamma_{a_1a_2a_3a_4a_5} \epsilon \ ,
\nn\\
\xi_{a_1\dots a_7,b} &= - \frac{7}{5!} \eta_{b[a_1} \varepsilon_{a_2\dots a_7]c_1c_2c_3c_4c_5} \bar \epsilon\,  \Gamma^{c_1c_2c_3c_4c_5} \epsilon \ . 
\end{align}
\end{subequations}

The last line in \eqref{SusyAlgebra}, however, is not a total derivative and must be the component of the parameter $\Sigma_{M}{}^{N \tilde{\alpha}}$  in the gauge transformation  \eqref{ExtendedSigmaGauge}. One can indeed cast it in the form
\begin{align}
 \label{SigmaFermions} \Sigma_{M; a_1\dots a_8} = \bar \epsilon \Gamma_{a_1\dots a_8} \partial_M \epsilon \ , 
 \end{align}
which is not a total derivative, but  satisfies the strong section constraint as necessary for the parameter $\Sigma_M{}^{N\tilde\alpha}$.

Up to this level truncation in the higher level derivatives, and obtains therefore that the supersymmetry algebra closes up to the expected gauge transformations of the theory. This relies on the fact that the symmetric bilinear $\epsilon \epsilon$ can be consistently embedded in the representation $R(\Lambda_1)$. The antisymmetric bilinear $\epsilon_1\wedge  \epsilon_2$ can in turn be embedded consistently in the representation  $R(\Lambda_3)$. This is necessary for \eqref{SigmaFermions} to extend to a well-defined $\Sigma_M{}^{N\tilde\alpha}$ parameter in $\overline{R(\Lambda_1) }\otimes R(\Lambda_3)$. One checks that the low level truncation exhibits that this is indeed possible, in particular we discuss the case of the symmetric bilinear in more detail in Appendix \ref{GaugeParaSpinor}.\footnote{For the antisymmetric product, one finds that the 3-form, the 4-form and the scalar bilinear form the antisymmetric rank 2 irreducible representation of $SL(32) \cong \widetilde{K}(E_{11})/ \mathcal{I}_{\epsilon}$. Indeed $R(\Lambda_3)$ first gives a 3-form, then a 4-form, a (3,1)-form and a (2,2)-form. The 4-form and the double trace of the (2,2)-form are represented by the four-form and the scalar bilinears, while the other components vanish. All these three bilinear representations appear repeatedly then at each level.}

\subsection{Constrained scalar fields}

The definition of the $E_{11}$ exceptional field theory also requires the introduction of the additional  field $\chi_M{}^{\tilde{\alpha}}$ that transforms instead in the representation associated to $\mathcal{T}_{-2}$, so in order to construct a supersymmetric theory we also need to extend the module $(\Xi,\Upsilon) \in (\mf{e}_{11} \ominus K(\mf e_{11})) \oright {R(\Lambda_2)}$ to include the component $R(\Lambda_2)$ of $\mathcal{T}_{-2}$. We shall argue first that there is an indecomposable module with the structure 
\begin{align}
  (X,\Xi,\Upsilon)\in R(\Lambda_2) \oright (\mf{e}_{11} \ominus K(\mf e_{11})) \oright {R(\Lambda_2)} \ . \end{align}
This module seems to exist as a restriction of an $E_{11}$ module (which is not a submodule of $\cT$)
\begin{align}
 M_{-2} \cong R(\Lambda_2) \oright \mf{e}_{11}  \oright R(\Lambda_2) \ ,
 \end{align}
extending the module of components of $\mathcal{T}_{-2}$ such that $M_{-2} / \mathcal{I}_{R(\Lambda_2)} \cong \mathcal{T}_{-2}$.\footnote{The existence of the module $M_{-2}$ can be checked at low level in the $GL(3) \times E_8$ decomposition, with the additional component $A_\mu,B_{\mu\nu}^M,\dots$ in $R(\Lambda_2)$ with respect to $\mathcal{T}_{-2}$,
 \begin{align}  
\delta h_\mu{}^\nu &= e_\mu^M \bar A_M^\nu - f^\nu_M A_\mu^M - \delta_\mu^\nu \bigl( e_ \sigma^M \bar A_M^\sigma - f^\sigma_M A^M_\sigma \bigr)\ , 
\nn \\ 
 \delta \Phi_M &= f_{MN}{}^P e_\mu^N \bar A^\mu_P - f_{MN}{}^P f^\mu_P A_\mu^N + f^\mu_M A_\mu\ ,
  \nn\\
  \delta A_\mu^M &= - e_\nu^M h_\mu{}^\nu + f^{PM}{}_N e_\mu^N \Phi_P - f^\nu_N B_{\mu\nu}^{MN} - f^{MN}{}_P f^\nu_N h_{\mu\nu}^P   + \frac12 f^{MN}{}_P f_N^\nu B_{\mu\nu}^P\ ,
  \nn\\
    \delta B_{\mu\nu}^{MN}  &= 28 P^{MN}{}_{PQ}  e_{[\mu}^P A_{\nu]}^Q + \tfrac12 \eta^{MN} \eta_{PQ}  e_{[\mu}^P A_{\nu]}^Q\ ,    \nn\\
         \delta h_{\mu\nu}^{M}  &= -  f_{NP}{}^M e_{(\mu}^N A_{\nu)}^P + e_{(\mu}^M A_{\nu)}\ , 
\nn
 \end{align}
and
  \begin{align}
  \delta X_\mu &= f^\nu_M ( X_{\mu\nu}^M + h_{\mu\nu}^M ) +  e_\mu^M \Phi_M- \frac12 f^\nu_M B_{\mu\nu}^M \; , &  \delta A_\mu &= f^\nu_M B_{\mu\nu}^M \; ,
  \nn\\ 
 \delta  X_{\mu\nu}^M &= - 2 e_{[\mu}^M X_{\nu]} -  f^M{}_{NP} e_{[\mu}^N A_{\nu]}^P +  e_{[\mu}^M A_{\nu]} \ , &
 \delta  B_{\mu\nu}^M &= - 2 e_{[\mu}^M A_{\nu]}  \ . \nn
 \end{align}
The important feature is that one cannot avoid that $\delta X^{\tilde{\alpha}}$ transforms back into $\Phi^{\tilde{\alpha}}$ parametrised here by $A_\mu,B_{\mu\nu}^M,\dots $. }

We consider the representative of the nine-form in  $R(\Lambda_2)$
\begin{align}
\label{susyX9} X_{a_1\dots a_9} = - \frac12  \varepsilon_{a_1\dots a_9b_1b_2} \bar \epsilon  \Gamma^{b_1} \psi^{b_2} \ . 
\end{align}
Its $K(\mf e_{11})$   variation gives according to  Appendix \ref{BilinearVariations}
\begin{align} 
\delta \left( \bar \epsilon  \Gamma_{[a} \psi_{b]}\right) = - \frac12 \Lambda_{cd[a}   \bar \epsilon \Gamma^{cd} \psi_{b]}   + \frac13   \Lambda_{cd[a}  \bar \epsilon \Gamma_{b]}{}^c \psi^d  + \frac23 \Lambda_{abc}\bar \epsilon \psi^c + \frac16 \Lambda_{cde} \bar \epsilon \Gamma_{ab}{}^{cd} \psi^e \ , 
\end{align}
consistently with the assumption that 
\begin{align}
\delta X_{a_1\dots a_9} = - 28 \Lambda_{[a_1a_2a_3}  \Xi_{a_4\dots a_9]} + \frac1{18} \Lambda^{b_1b_2b_3} \Xi_{a_1\dots a_9,b_1b_2b_3} + \dots 
\end{align}
according to \eqref{E11adjoint}, where the dots stand for terms in $\bar \epsilon \Gamma_2 \psi_a$ and $\bar \epsilon \psi_a$ that would appear in the other fields (\ie\   $\Xi_{11,1},\;  X_{10,2},\; X_{11,1}$) that we disregard here.

We conjecture therefore that the set of bilinear $\epsilon \Psi$ can be identified as a $K(\mf{e}_{11})$ module with successive  quotients defined from $E_{11}$ modules: 
\bea (X,\Xi,\Upsilon) &\in & [R(\Lambda_2) \oright (\mf{e}_{11} \ominus K(\mf e_{11})) \oright {R(\Lambda_2)} ] / \mathcal{I}_{ (X,\Xi,\Upsilon)} \; , \\
( \Xi,\Upsilon) &\in&  [ (\mf{e}_{11} \ominus K(\mf e_{11})) \oright {R(\Lambda_2)}]  / \mathcal{I}_{( \Xi,\Upsilon)}  \; , \\
 \Upsilon &\in&   {R(\Lambda_2)}  / \mathcal{I}_{\Upsilon}  \; . \eea

By construction of the module, the supersymmetry transformation of the field $\chi_M{}^{\tilde{\alpha}}$ must include $\partial_M X^{\tilde{\alpha}}$ at linearised order. But because $\Upsilon^{\tilde{\alpha}}$ is in a submodule ${R}(\Lambda_2)$, it can also appear in the supersymmetry variation of $\chi_M{}^{\tilde{\alpha}}$ with the derivative $\partial_M$ acting either on $\epsilon$ or $\psi$ while preserving the strong section constraint. We shall now find that there is a unique definition of the supersymmetry variation of $\chi_M{}^{\tilde{\alpha}}$ that is consistent with the supersymmetry algebra in the linearised approximation. 

According to the discussion above, $K(E_{11})$ imposes that the linearised supersymmetry variation of the constrained field is of the form
\begin{align}
\dsusy \chi_{M;a_1\dots a_9} =\frac12  \varepsilon_{a_1\dots a_9 b_1 b_2 } \Bigl(    \alpha \partial_{M} \bar \epsilon \Gamma^{b_1b_2c} \psi_{c} - \beta \bar \epsilon \Gamma^{b_1b_2c} \partial_{M} \psi_{c}   -  \partial_{M}(  \bar \epsilon \Gamma^{b_1} \psi^{b_2} ) \Bigr)\ , \nn\\
\end{align}
with free coefficients $\alpha$ and $\beta$ that are not determined by  $\widetilde{K}(E_{11})$ covariance, but will be fixed by closure of the supersymmetry algebra momentarily. We have indeed seen that the third term in \eqref{susyX9} is fixed by $\widetilde{K}(E_{11})$ through the indecomposable structure of the representation, while the only other possible terms must be obtained from the bilinear $\Upsilon^{a_1a_2}\in R(\Lambda_2)$ with a derivative on either $\epsilon$ or $\psi_a$.   Note that the section constraint implies that the index $M$ of $\chi_{M}{}^{\tilde{\alpha}}$ must be attached to a derivative, but not necessarily to a total derivative. The closure of the supersymmetry algebra implies that 
\begin{align}
(\dsusy)^2 \chi_{M; a_1 \dots a_9}  = \frac{1}{4} \varepsilon_{a_1\dots a_9b_1b_2} \partial_M \partial^{b_1} ( \bar \epsilon \Gamma^{b_2} \epsilon) - 9 \partial_{[a_1} ( \bar \epsilon \Gamma_{a_2\dots a_8]} \partial_M \epsilon )\ , 
\end{align}
where the first term is the gauge transformation of parameter $\xi^a$ in \eqref{GaugeSusyClose} and the second the gauge transformation of parameter $\Sigma_{M; a_1\dots a_8} $ with \eqref{SigmaFermions}.  One computes that this is the case if and only if $\alpha=\beta=1$. We conclude that
\begin{align} 
\dsusy \chi_{M;a_1\dots a_9} &=\frac12  \varepsilon_{a_1\dots a_9 b_1 b_2 } \Bigl(     \partial_{M} \bar \epsilon \Gamma^{b_1b_2c} \psi_{c} - \bar \epsilon \Gamma^{b_1b_2c} \partial_{M} \psi_{c}   -  \partial_{M}(  \bar \epsilon \Gamma^{b_1} \psi^{b_2} ) \Bigr)\ , \nn\\
&= 9 \partial_M \bar\epsilon \Gamma_{[a_1\dots a_8} \psi_{a_9]} - 9 \bar\epsilon \Gamma_{[a_1\dots a_8} \partial_M \psi_{a_9]} - \frac12  \varepsilon_{a_1\dots a_9 b_1 b_2 } \partial_{M}(  \bar \epsilon \Gamma^{b_1} \psi^{b_2} ) \ .
\label{Ga}  
\end{align}
Note that this transformation is in agreement with the supersymmetry transformation of the constrained 2-form found in $E_7$ exceptional field theory~\cite[Eq.~(3.33)]{Godazgar:2014nqa} with the identification of $\chi_{M;a_1a_245678910} = B_{a_1a_2 M}$. From this ansatz, one extrapolates that $\widetilde{K}(E_{11})$  will fix the linearised supersymmetry transformation of $\chi_{M}{}^{\tilde{\alpha}}$ to be in general of the form
\be 
\dsusy \chi_{M}{}^{\underline{\tilde{\alpha}}}  = \Upsilon^{\underline{\tilde{\alpha}}}( \partial_M \epsilon , \Psi) -\Upsilon^{\underline{\tilde{\alpha}}}(  \epsilon , \partial_M \Psi) + \partial_M X^{\underline{\tilde{\alpha}}}(\epsilon, \Psi)  \ ,
\label{susychi} 
\ee
where $\Upsilon^{\tilde{\alpha}}(\epsilon,\Psi)$ and $X^{\tilde{\alpha}}(\epsilon, \Psi)$ are the fermion bilinears in $R(\Lambda_2)$ introduced in this section.

\section{Supersymmetry of the field equations}
\label{sec:susyEOM}

Having established the linearised supersymmetry rules for all fields such that the supersymmetry algebra closes, we now turn to studying the supersymmetry of the field equations. In a first step, we determine the linearised Rarita--Schwinger equation for the vector-spinor $\Psi$ through $\widetilde{K}(E_{11})$  covariance. Next, we turn our attention to the duality equations, and show that suitably constructed bilinears in fermions can be utilised to supercovariantise them.

\subsection{Linearised Rarita--Schwinger equation and its supersymmetry} 

If the vector-spinor equation follows from the variation of a Lagrangian, the equation of motion for $\psi_a$ should transform in the $\widetilde{K}(E_{11})$ representation conjugate to that of $\psi_a$. For $\widetilde{K}(E_{11})$, the conjugate representation is given by $\rho^a$ with
\begin{align}
\label{VaryRarita} 
\delta \rho^a = \frac1{12} \Lambda_{bcd} \Gamma^{bcd} \rho^a + \frac{2}{3} \Lambda^{abc} \Gamma_b \rho_c - \frac1 6 \Lambda^{abc} \Gamma_{bcd} \rho^d \ ,
\end{align}
such that $\delta ( \rho^a \psi_a ) =0$ under $K(\mathfrak e_{11})$. It is important here that the index contraction in the last term differs from that in~\eqref{LV1}. Note that because of the non-existence of an invariant bilinear form on the $\widetilde{K}(E_{11})$ vector-spinor, the conjugate representation cannot be obtained by applying such a bilinear form. This is different from the situation for $K(E_{10})$ where $ \Gamma^{ab} \psi_b$ is conjugate to $\psi_a$~\cite{Damour:2006xu}. 

The starting point for a $\widetilde{K}(E_{11})$ invariant Rarita--Schwinger equation is the usual $Spin(1,10)$ covariant linearised Rarita--Schwinger equation of eleven-dimensional supergravity: $\Gamma^{abc} \partial_b \psi_c=0$. As the partial derivative $\partial_a$ transforms into the other derivatives in $\overline{R(\Lambda_1)}$ according to~\eqref{eq:KE11der}, one needs to extend this equation by the other derivatives in order to ensure $\widetilde{K}(E_{11})$ covariance. Making an ansatz for the extended derivatives and requiring the Rarita--Schwinger equation to transform as in~\eqref{VaryRarita} leads to
\begin{align}
\label{eq:RSeq}
\rho^a &= \Gamma^{abc} \partial_b \psi_c +5  \partial^{ab} \psi_b +2 \Gamma^{ab} \partial_{bc} \psi^c +2 \Gamma_{bc} \partial^{ab} \psi^c +\frac12 \Gamma^a{}_{bcd} \partial^{bc} \psi^d 
 + \frac56 \Gamma_{b_1b_2b_3} \partial^{ab_1b_2b_3c} \psi_c
 \nn\\
 &\quad + \frac{1}{12} \Gamma^{ab_1b_2b_3b_4} \partial_{b_1b_2b_3b_4c} \psi^c + \frac{1}{12} \Gamma_{b_1b_2b_3b_4c} \partial^{ab_1b_2b_3b_4} \psi^c 
+\frac{1}{120} \Gamma^{ac_1...c_5 b} \partial_{c_1...c_5} \psi_b  +\dots  
\end{align}
up to the higher level derivatives in $\partial^{7;1}$ etc. We have verified that this expression is $\widetilde{K}(E_{11})$ covariant in all terms varying into $\partial_1$ and $\partial^2$ and expect that this structure can be extended recursively to all orders in the derivatives. This will produce a formally infinite set of terms but on section only a finite number of these will be non-zero, so equation \eqref{eq:RSeq} only involves finitely many terms for any given specific solution to the section constraint.

In the linearised approximation, supersymmetry of the Rarita--Schwinger equation~\eqref{eq:RSeq} amounts to its gauge invariance under \eqref{dpsi} for a spinor $\epsilon$ satisfying the section constraint. Up to terms  involving $\partial_2 \partial_5, \partial_5\partial_5$ and higher level derivatives, we find 
\begin{align}
\dsusy \rho^a = 8 \partial^{ab} \partial_b \epsilon + 4 \Gamma^{ab} \partial_{bc} \partial^c \epsilon -\frac16 \left(\Gamma^a{}_{b_1b_2b_3b_4}+8\delta^a_{b_1} \Gamma_{b_2b_3b_4}\right) \bigl(3\partial^{b_1b_2} \partial^{b_3b_4} -\partial^{b_1b_2b_3b_4c} \partial_c \bigr)\epsilon + \dots \ ,
\end{align}
which vanishes by virtue of the section constraints
\begin{align}
\partial^{ab} \partial_b \epsilon =0\ ,\qquad 3\partial^{[a_1a_2} \partial^{a_3a_4]}  \epsilon=\partial^{a_1a_2a_3a_4b} \partial_b \epsilon \ .
\end{align}
Therefore we see that the section constraint is crucial for obtaining equations of motion that are invariant under local supersymmetry.

\subsection{Gauge invariant and supercovariant self-duality equation}
\label{supercovariantF} 

We now study the fermionic modification of the duality equation~\eqref{DualityEquation}. We shall first argue that there is a remarkable representation-theoretic property of the $\widetilde{K}(E_{11})$ spinor $\Psi$ in relation to the field strengths, allowing the addition of fermion bilinears. Then we show that one can define the generalised diffeomorphisms on $\Psi$ such that the modification maintains gauge invariance. Finally we show that the modified duality equation is supercovariant under linearised supersymmetry transformations.

\subsubsection{Embedding of fermion bilinears in field strength representation}
\label{sec:psiO}

We want to argue now that the representation of $K(E_{11})$ defined by the duality equation $\cE^I$ in~\eqref{DualityEquation}, includes an unfaithful representation constructed out of bilinears in the vector-spinor $\Psi$. More precisely, under $K(E_{11})$, the representation of the field strength splits into self-dual and anti-selfdual components $\cT_{-1} \cong S_+ \oplus S_-$, and the field equation \eqref{eq:guessDE} is the statement that $F^{\underline{I}} = \cV^{\underline{I}}{}_J F^J$ belongs to $S_+$. We shall argue that the vector-spinor bilinears $\Psi \Psi$ define an unfaithful representation of $K(E_{11})$ homomorphic to $S_- / \mathcal{I}_{\Psi\Psi}$, where $\mathcal{I}_{\Psi\Psi}$ denotes a certain $K(E_{11})$ invariant subspace in $S_-$.

As shown in Table~\ref{tab:thaGL11} and equation~\eqref{de1}, the central terms of the duality equation of $GL(11)$ weight $\frac12$ involve a four-form field strength and the dual of a seven-form field strength. These can be constructed out of fermion bilinears as
\begin{align}
\bar\psi_{[a_1} \Gamma_{a_2a_3} \psi_{a_4]}  
\quad  \textrm{and} \quad
\frac{1}{5!} \varepsilon_{a_1a_2a_3a_4}{}^{b_1\dots b_7}   \bar\psi_{b_1}  \Gamma_{b_2\dots b_6}  \psi_{b_7}\,.
\end{align}

Using the transformations~\eqref{LV1} one finds that\footnote{A collection of fermion bilinear transformations can be found in Appendix~\ref{BilinearVariations}.}
\begin{subequations}
\begin{align}
& \delta \left(\bar\psi_{[a_1} \Gamma_{a_2 a_3} \psi_{a_4]}\right) = \Lambda_{b[a_1 a_2} \left( \bar\psi_{a_3} \Gamma^b \psi_{a_4]} + 2 \bar\psi^b \Gamma_{a_3} \psi_{a_4]} \right) + \frac{7}{12} \Lambda^{b_1 b_2 b_3} \psi_{[b_1} \Gamma_{b_2 b_2 a_1 a_2 a_3} \psi_{a_4]} 
\nn\\
& \hspace{3cm} - \frac1{12}  \Lambda_{b_1 b_2 b_3} \psi^{b_1} \Gamma_{a_1 a_2 a_3 a_4}{}^{b_2} \psi^{b_3} 
\end{align}
and
\begin{align}
&  \delta \left(\frac{1}{5!} \varepsilon_{a_1a_2a_3a_4}{}^{b_1\dots b_7}   \bar\psi_{b_1} \Gamma_{b_2\dots b_6}  \psi_{b_7}\right)  = -24 \Lambda_{b[a_1a_2} \bar\psi^b \Gamma_{a_3a_4]c} \psi^c -\frac{1}{12} \varepsilon_{a_1a_2a_3a_4}{}^{b_1\dots b_7} \Lambda_{b_1b_2b_3}\bar\psi_{b_4} \Gamma_{b_5b_6} \psi_{b_7} 
\nn\\
& \hspace{6cm}  \ \ + \Lambda_{b_1b_2b_3} \bar\psi^{b_1} \Gamma^{b_2}{}_{a_1a_2a_3a_4} \psi^{b_3} + 6 \Lambda_{[a_1a_2}{}^{b_1} \bar\psi^{b_2} \Gamma_{a_3a_4]b_1b_2b_3} \psi^{b_3}\,.
\end{align}
\end{subequations}
Therefore the variation of the combination
\begin{align} 
\label{eq:O4}
O_{a_1a_2a_3a_4} &= 3 \bar\psi_{[a_1} \Gamma_{a_2a_3} \psi_{a_4]} + \frac{1}{480} \varepsilon_{a_1a_2a_3a_4}{}^{b_1b_2b_3b_4b_5b_6b_7} \bar\psi_{b_1} \Gamma_{b_2b_3b_4b_5b_6} \psi_{b_7} \nn\\
&= 3 \bar\psi_{[a_1} \Gamma_{a_2a_3} \psi_{a_4]} +\frac14 \bar\psi^{b_1} \Gamma_{a_1\ldots a_4b_1b_2} \psi^{b_2}
\end{align}
does not contain the term $\bar\psi^{[b_1} \Gamma^{b_2}{}_{a_1a_2a_3a_4} \psi^{b_3]}$ and satisfies
\begin{align}
\label{eq:deltaO4}
\delta O_{a_1a_2a_3a_4} = - 6 \Lambda_{b[a_1a_2} O_{a_3a_4]}{}^b - \frac{1}{144} \Lambda^{b_1b_2b_3} \varepsilon_{b_1b_2b_3a_1a_2a_3a_4}{}^{c_1c_2c_3c_4} O_{c_1c_2c_3c_4} \ , 
\end{align}
where
\begin{align}
\label{eq:O21}
O_{a_1a_2}{}^b =- \frac12 \bar\psi_{[a_1} \Gamma^b \psi_{a_2]} -\bar\psi^b \Gamma_{[a_1} \psi_{a_2]} - \frac14 \bar\psi_{c_1} \Gamma_{a_1a_2}{}^{bc_1c_2} \psi_{c_2} + \bar\psi^b \Gamma_{a_1a_2c} \psi^c  \ . 
\end{align}

The transformation~\eqref{eq:deltaO4} is in complete agreement with the $K(E_{11})$ transformation of the combination $F_{a_1a_2a_3a_4}+\frac{1}{7!} \varepsilon_{a_1a_2a_3a_4}{}^{b_1 \dots b_7} F_{b_1 \dots b_7}$ in~\eqref{de1}  as given in \cite[Eq.~(5.58)]{Bossard:2017wxl}. This exhibits that the bilinear $\Psi\Psi$ indeed transforms in the representation $S_-$. At the next level one obtains
\bea
\delta O_{a_1a_2}{}^b &=& \Lambda_{cd [a_1} O_{a_2]}{}^{bcd}  -\frac19 \Lambda_{cde}\delta_{[a_1}^b O_{a_2]}{}^{cde} +\Lambda_{a_1a_2 c} O^{b,c} 
\nn\\
&&
+ \frac12 \Lambda^{bcd} O_{a_1a_2cd} +\frac19 \Lambda^{cde} \delta_{[a_1}^b O_{a_2]cde} \ ,
\eea
where
\bea \label{O13} 
O_a{}^{b_1b_2b_3} &=&   \frac32  \bar\psi^{[b_1}\Gamma^{b_2b_3]} \psi_{a} - \frac{15}{2}  \bar\psi^{[b_1}\Gamma_a{}^{b_2} \psi^{b_3]}
 - 3 \bar \psi_c \Gamma_a{}^{c[b_1b_2} \psi^{b_3]}   +\frac14 \bar\psi_{c_1} \Gamma_a{}^{b_1b_2b_3c_1c_2}\psi_{c_2} 
 \nn\\
 && - \frac32  \delta_a^{[b_1} \bar \psi^{b_2} \Gamma^{b_3]c} \psi_c    \ ,
\w2
O^{a,b} &=& -3\bar\psi^a\psi^b    - \bar\psi^{(a}\Gamma^{b)c}\psi_c  -\frac14 \eta^{ab} \bar  \psi_c \Gamma^{cd} \psi_d \ ,
\eea
consistently with \cite[Eq.~(5.59)]{Bossard:2017wxl}. Further variation under $K(E_{11})$ according to \cite[Eq.~(4.37)]{Bossard:2017wxl}
\begin{align}
\delta O^{a,b} &= -\frac16 \Lambda_{c_1c_2c_3} O^{c_1c_2c_3(a,b)} -\frac12 \Lambda^{c_1c_2(a} O_{c_1c_2}{}^{b)}\,,\\
\delta O_a{}^{b_1b_2b_3} &= 3 \Lambda^{c[b_1b_2} O_{ac}{}^{b_3]} - \frac34 \Lambda^{c_1c_2[b_1} \delta_a^{b_2} O_{c_1c_2}{}^{b_3]} -\frac16 \Lambda_{c_1c_2c_3} O_a{}^{b_1b_2b_3c_1c_2c_3}\nn\\
&\hspace{10mm} - \Lambda_{ac_1c_2} O^{b_1b_2b_3c_1,c_2} + \frac38 \Lambda_{c_1c_2c_3} \delta_a^{[b_1} O^{b_2b_3]c_1c_2,c_3}
\end{align}
gives
\begin{align}
O^{a_1a_2a_3a_4,b} &= - \bar \psi^b \Gamma^{a_1a_2a_3a_4c} \psi_c + \bar \psi^{[b} \Gamma^{a_1a_2a_3a_4]c} \psi_c - 12 \bar \psi^b \Gamma^{[a_1a_2a_3} \psi^{a_4]} \nn\\
& \hspace{10mm} + \eta^{b[a_1} \bar \psi_{c_1} \Gamma^{a_2a_3a_4]c_1c_2} \psi_{c_2} - 12 \eta^{b[a_1} \bar \psi^{a_2}\Gamma^{a_3a_4]c} \psi_c  - 30 \eta^{b[a_1}\bar \psi^{a_2} \Gamma^{a_3} \psi^{a_4]} \ ,
\\
O_a{}^{b_1 \ldots b_6} &= \frac14 \bar{\psi}_{c_1} \Gamma_a{}^{b_1\ldots b_6 c_1c_2} \psi_{c_2} +6 \bar{\psi}^{[b_1} \Gamma_a{}^{b_2\ldots b_6] c} \psi_c +12 \delta_a^{[b_1} \bar \psi^{b_2} \Gamma^{b_3\ldots b_6]c} \psi_c \nn\\
&\hspace{10mm} +3 \bar{\psi}_a \Gamma^{[b_1\ldots b_5} \psi^{b_6]} +\frac{75}2 \bar{\psi}^{[b_1} \Gamma_a{}^{b_2\ldots b_5} \psi^{b_6]}\,.
\label{O16a}
\end{align}
This consistency check includes all field strengths from level $-\frac{7}{2}$ to $\frac{7}{2}$, and therefore takes into account not only the standard field strength for the fields that appear already in $E_n$ exceptional field theory for $n\le 8$, but also the gradient dual 10-form field strengths (through $O_a{}^{b_1b_2b_3} $ and $O_a{}^{b_1\dots b_6}$) that are reminiscent of the affine structure of $E_9$, and even the non-dynamical 11-form field strengths (through $O^{a,b}$ and $O^{a_1a_2a_3a_4,b}$) that only appear in $E_{11}$. 

We also check in Appendix \ref{E7susy} that the fermion bilinear  decomposed under $Spin(1,3) \times SU(8)$ gives consistently the supercovariantisation of the field strengths in $E_7$ exceptional field theory \cite{Godazgar:2014nqa,Butter:2018bkl}.

The proposal, checked here at lowest levels, is therefore that the duality equation~\eqref{DualityEquation} can be extended by fermion bilinears in the form
\begin{align}
\label{DEwO}
\widehat{\cE}^I \equiv F^I - \cM^{IK} \Omega_{KJ} F^J  - \mathcal{V}^{-1 I}{}_{\underline{I}} O^{\underline{I}}  =0 \ . 
\end{align}
Here, $ \mathcal{V}^{-1 I}{}_{\underline{I}} $ is the $E_{11}/K(E_{11})$ vielbein in the field strength representation $\cT_{-1}$ with $\underline{I}$ denoting a local $K(E_{11})$ index in that representation. The bilinears $O^{\underline{I}}$ are the embedding of the unfaithful representation of the $\Psi\Psi$ bilinear into $S_-$,  mentioned at the beginning of the section. Equation~\eqref{DEwO} is an $E_{11}$ invariant extension of the bosonic duality equation $\cE^I$ by fermion bilinears. We will use the symbol hat to denote the supercovariantisation as usual.

\subsubsection{Gauge invariance of modified duality equation}

For discussing gauge invariance below we also need to establish the action of generalised diffeomorphisms on the spinor $\Psi$. As we discussed in Section~\ref{sec:Vgauge} and as is usual for fermions one has to consider the vielbein formalism. Moreover, we consider the vielbein in a maximal parabolic gauge and this entails a compensating transformation $X\in K(\mathfrak{e}_{11})$ in its gauge transformation~\eqref{eq:lieV}. The compensating transformation that appears in the gauge transformation of the spinor is
\begin{align}
\label{eq:liePsi}
\delta_\xi \Psi = \xi^M \partial_M \Psi + \frac14 \partial_M\xi^M \Psi + X \Psi\,.
\end{align}
Thus, $\Psi$ is a scalar density from the point of view of diffeomorphisms but there is a non-trivial induced $K(\mathfrak{e}_{11})$ action due to the compensator. In general, the compensating transformation $X$ involves infinitely many generators of $K(E_{11})$. However, if one chooses a partial solution  to the section constraint associated with the maximal parabolic gauge as explained around~\eqref{eq:derdec}, the compensating transformation $X$ takes the simple form~\eqref{eq:kmpg}. For the case of Levi $GL(11)$, the solution to the section constraint amounts to keeping only the external derivatives $\partial_\mu$ and there is no compensating transformation. For general $GL(11-n)\times E_n$ there is a non-trivial compensating transformation. In Appendix~\ref{E7susy}, we demonstrate that the resulting generalised diffeomorphism on the fermions is consistent with formulas that have appeared in the case of $E_7$ exceptional field theory~\cite{Butter:2018bkl}.

The weight given in~\eqref{eq:liePsi} above is fixed by gauge invariance of~\eqref{DEwO} as follows. As the contribution $O^{\underline{I}}$ is bilinear in fermions the weight of a single fermion should be half the weight of $F^I$ to match the weight of the left-hand side of the equation, recalling that $\cV$ has no weight. As we derived in~\eqref{eq:lieF} that $F^I$ has weight $1/2$, this fixes the weight of $\Psi$ to $1/4$. All these weights can be ultimately traced back to the non-trivial weight of the derivative $\partial_M$ as the vielbein $\cV$ has no weight. We shall see later that the weight $1/4$ is also consistent with a formal Rarita--Schwinger Lagrangian being gauge invariant.

Under a gauge transformation $\delta_\xi$ we now find that the bosonic and fermionic terms of the modified duality equation~\eqref{DEwO} transform in the same way  with respect to the transport and weight terms. The compensating $X$ transformation on the fermion bilinear $O^{\underline{I}}$ gets converted into an $E_{11}$ rotation in the field strength representation by the inverse vielbein $\cV^{-1\, I}{}_{\underline{I}}$ such that $\widehat{\cE}^I$ transforms covariantly under generalised diffeomorphisms and the modified duality equation is gauge invariant.

\subsubsection{Supercovariance of modified duality equation}

According to the discussion in Section \ref{sec:susybos}, the manifestly $\widetilde{K}(E_{11})$ invariant representation of supersymmetry requires to extend the field content such that $\cV \in  G(\overline{\mathcal{T}}_0) / K(E_{11})$. In this formulation one should take the element $\mathcal{V}^{-1 I}{}_{\underline{I}}$ of the group $ G(\overline{\mathcal{T}}_0)$ accordingly in the representation  $\mathcal{T}_{-1}$.  Note that $\mathcal{T}_{-1}$ is by construction a representation of $ G(\overline{\mathcal{T}}_0)$. Nevertheless, we expect that there is a partially gauged fixed version of the theory in which $\cV \in E_{11}/K(E_{11})$, and that these formulae are not modified; see the discussion below \eqref{NonLinsusy}.

Note that there is no notion of superconvariant field strength in $E_{11}$, and only the supercovariant equation \eqref{DEwO} defined above transforms under $K(E_{11})$ into itself in the module $S_-$. Nonetheless, it will be convenient for comparison with eleven-dimensional supergravity to write the supercovariant duality equation  \eqref{DEwO} as $\widehat{\cE}^{\underline{I}} =0$ with $\widehat{\cE}^{\underline{I}} = \widehat{F}^{\underline{I}}  - \eta^{\underline{I} \underline{K}} \Omega_{\underline{K}\underline{J}}\widehat{F}^{\underline{K}}$ for some $\widehat{F}^{\underline{I}} =F^{\underline{I}} + \Psi \Gamma^{\underline{I}} \Psi$ whose components are reminiscent of the supercovariant supergravity field strengths. However, it is important to keep in mind that these  $\widehat{F}^{\underline{I}} $ do not transform into themselves under $K(E_{11})$  and that they are not supercovariant, only the anti-selfdual component $\widehat{\cE}^{\underline{I}}$ belongs to $S_-$ and is supercovariant. 

In $D=11$ supergravity the corresponding supercovariant expressions are in our conventions 
\begin{subequations}
\begin{align}
\widehat{F}_{a_1a_2a_3a_4} &= F_{a_1a_2a_3a_4} - 3 \bar\psi_{[a_1} \Gamma_{a_2a_3} \psi_{a_4]} 
\,, \\
\widehat{F}_{a_1a_2a_3a_4a_5a_6a_7} &= F_{a_1a_2a_3a_4a_5a_6a_7} - \frac{21}{2}  \bar\psi_{[a_1} \Gamma_{a_2a_3a_4a_5a_6} \psi_{a_7]} \,.
\end{align}
\end{subequations}
That $\dsusy \widehat{F}$ is independent of $\partial \epsilon$ when keeping only $\partial_a$ can be checked easily using the transformation laws~\eqref{eq:susy11} and~\eqref{dpsi}. Moreover, this combination is exactly the one that is produced by the extended duality equation~\eqref{DEwO} when taking the terms in $O_{a_1a_2a_3a_4}$ of~\eqref{eq:O4} without the Levi--Civita symbol $\varepsilon_{11}$ into $\widehat{F}_4$ and those with into $\widehat{F}_7$. Similarly, one checks that the supercovariant spin connection $\widehat{\omega}_{ab} = e^c \widehat{\omega}_{c,ab}$ is in agreement with the first three terms in $O_{a_1a_2}{}^b$ given in~\eqref{eq:O21}. The field strength $F_{a_1a_2}{}^b$ is related at the non-linear level to the spin connection as
\be 
F_{a_1a_2}{}^b = - 2\omega^b{}_{,a_1a_2} +  2e^{bm} ( e_{[a_1}{}^n \partial_m e_{n | a_2]} ) \ ,  \qquad \omega_{ab} = - \frac12e_c F_{ab}{}^c + e_{[a}{}^n d e_{n|b]} \; , 
\ee
where the second term is a component of the Maurer--Cartan form in $K(\mf{e}_{11})$ that does not require a supercovariantisation with the supersymmetry realisation in which
\be 
\dsusy  e_m{}^a =- \frac12 \bar\epsilon \Gamma^{(a} \psi^{b)} e_{mb} \ , 
\ee
which is natural in a coset construction. Therefore the supercovariantisation of $F_{ab}{}^c$ must be minus the one of the spin connection
\be 
\widehat{F}_{a_1a_2}{}^b = F_{a_1a_2}{}^b + \frac12 \bar\psi_{[a_1} \Gamma^b \psi_{a_2]} +\bar\psi^b \Gamma_{[a_1} \psi_{a_2]} - \frac14 \bar\psi_{c_1} \Gamma_{a_1a_2}{}^{bc_1c_2} \psi_{c_2} \ . 
\ee
The remaining term $ \psi^b \Gamma_{a_1a_2c} \psi^c$ in~\eqref{eq:O21} is traceless, and will only contribute to the supercovariantisation of $F_{9,a}$ as 
\be 
\widehat{F}_{a_1\dots a_9,b} = {F}_{a_1\dots a_9,b} - 9 \bar\psi_{[a_1} \Gamma_{a_2\dots a_9]} \psi_b \ ,
\ee
which is indeed consistent with the expected supercovariantisation of the dual graviton. 


The discussion above was based on the field strengths of usual $D=11$ supergravity. In the $E_{11}$ model built using the tensor hierarchy the field strengths receive additional contributions from the constrained fields $\chi_M{}^{\tilde\alpha}$ and $\zeta_M{}^\Lambda$ in \eqref{eq:NLFS}.  Moreover, the manifestly $\widetilde{K}(E_{11})$ covariant formulation of supersymmetry requires the additional fields parametrising $ G(\overline{\mathcal{T}}_0)/ K(E_{11})$, so one needs to complete the expressions given in~\eqref{F} for the explicit field strengths in the $GL(11)$ decomposition. They are given by 
\begin{subequations}
\label{FieldStrengthsT0} 
\begin{align}
F_{a_1a_2}{}^b &= 2 \partial_{[a_1} h_{b_2]}{}^c + \partial^{bc} A_{a_1a_2c} + \frac{1}{3}\delta^b_{[a_1}  \partial^{c_1c_2} A_{b_2]c_1c_2}  + \dots \; \ ,
\\ 
F_{a_1a_2a_3a_4} &= 4 \partial_{[n_1} A_{a_2a_3a_4]} 
- \frac{1}{2}\partial^{c_1c_2} A_{a_1a_2a_3a_4c_1c_2} +\dots \ ,
\\
F_{a_1\cdots a_7} &= 7 \partial_{[a_1} A_{a_2\cdots a_7]} + \partial^{b_1b_2} h_{a_1\cdots a_7 b_1;b_2}  
-\frac{1}{2} \chi^{b_1b_2}{}_{;a_1\cdots a_7b_1b_2}+\dots \ ,
\\
F_{a_1\cdots a_9,b} -  F_{a_1\cdots a_9b} &= 9 \partial_{[a_1} h_{a_2\cdots a_9];b} 
+\chi_{{b}; a_1\dots a_9}  +\dots \ ,
\end{align}
\end{subequations}
where $h_{a_1\dots a_8;b} = h_{a_1\dots a_8,b}+h_{a_1\dots a_8b}$ according to the discussion in Section \ref{sec:susybos}. In this section we shall ignore derivatives at levels higher than $\partial_n$ and $\partial^{n_1n_2}$. We checked at first order that these field strengths still transform under $K(\mf e_{11})$ according to 
\bea \delta F_{a_1\dots a_7} &=& 7  \partial_{[a_1} \delta A_{a_2\dots a_7]} + \delta \partial^{bc} h_{a_1\dots a_7b;c} + \dots \nn\\
&=& \frac{9}{2} \Lambda^{b_1b_2c} \partial_{[a_1} h_{a_2\dots a_7b_1b_2];c} + \dots \nn\\
&=& \frac12 \Lambda^{b_1b_2c} ( F_{a_1a_2\dots a_7b_1b_2;c}- F_{a_1a_2\dots a_7b_1b_2c}) + \dots \ ,   \eea
when including the nine-form component $h_{[a_1\dots a_8;b]}$.

The supersymmetry transformation of the field $\chi_M{}^{\tilde{\alpha}}$ was determined in~\eqref{Ga}. Let us now show that this is consistent with the supercovariance of the duality equation 
\begin{align}
 \widehat{\cE}_{a_1a_2a_3a_4} \equiv  \widehat{F}_{a_1a_2a_3a_4}+\frac{1}{7!} \varepsilon_{a_1a_2a_3a_4}{}^{b_1 \dots b_7} \widehat{F}_{b_1 \dots b_7}  = 0 \ . 
 \end{align}
The constrained field $\chi^{2}{}_{9}$ appears in the field strength $\widehat{F}_7$ since $F_7$ contains $\chi_M{}^{\tilde\alpha}$ as written in~\eqref{FieldStrengthsT0}.  The supersymmetry variation is 
\begin{align}
\dsusy \widehat{F}_{a_1\dots a_7} &= - 21 \bar \epsilon \Gamma_{[a_1\dots a_5} \partial_{a_6} \psi_{a_7}  - 21\beps \Gamma_{b[a_1...a_5} \partial_{a_6}{}^b \psi_{a_7]}
 + \frac92  \partial^{b_1b_2} \bar \epsilon \Gamma_{[a_1\dots a_7b_1}  \psi_{b_2]} \nn\\
 &\quad +    \bar \epsilon \Gamma_{a_1\dots a_7b_1} \partial^{b_1b_2} \psi_{b_2} - \frac72 \partial_{b[a_1}\left(  \bar \epsilon \Gamma_{a_2\dots a_7]} \psi^b \right)  - \frac12 \dsusy \chi^{b_1b_2}{}_{a_1\dots a_7b_1b_2} \ ,
 \label{f7s}
\end{align}
while that of $\hat{F}_4$ is given by
\begin{align}
\dsusy \hat{F}_{a_1\dots a_4} &=  6  \bar \epsilon \Gamma_{[a_1a_2} \partial_{a_3} \psi_{a_4]} -\frac12 \partial^{b_1b_2} \bigl( \bar \epsilon \Gamma_{a_1\dots a_4 b_1} \psi_{b_2} \bigr) - \bar \epsilon \Gamma_{b_1b_2[a_1a_2a_3} \partial^{b_1b_2} \psi_{a_4]} +  6 \partial_{[a_1a_2} \bar \epsilon \Gamma_{a_3} \psi_{a_4]}  \  .
\label{f4s}
\end{align}
The supercovariance of $\widehat{\cE}_{a_1a_2a_3a_4}$ determines all the terms in $\dsusy \chi^{b_1b_2}{}_{a_1\dots a_7b_1b_2}$ in~\eqref{Ga} with a derivative on $\epsilon$, and the ones with a derivative on $\psi_a$ are exactly such that 
\be
\dsusy\,\widehat{\cE}_{abcd}= -\frac16 \beps \Gamma_{abcde}\,\rho^e +\frac43 \beps \Gamma_{[abc}\,\rho_{d]}\ .
\label{svde}
\ee
The above shows that the duality equation for the bosons is related by supersymmetry to the Rarita--Schwinger equation as in supergravity. We have checked this relation for all terms containing $\partial_1$ and $\partial_2$ derivatives. The supersymmetry variation of the field $\chi^{b_1b_2}{}_{a_1\ldots a_9}$ given in~\eqref{Ga} also plays a key role for the cancellation of the unwanted terms involving $\partial_2$ derivatives of the supersymmetry parameter and the gravitino, in order to achieve supercovariance and supersymmetry. 

A similar calculation can be done for the dual graviton equation using 
\begin{subequations}
\begin{align} 
\dsusy \widehat{F}_{a_1a_2}{}^b &= - \bar \epsilon \Gamma^b \partial_{[a_1} \psi_{a_2]} + \bar \epsilon \Gamma_{[a_1} ( \partial_{a_2]} \psi^b - \partial^b \psi_{a_2]} ) + \partial^b ( \bar \epsilon \Gamma_{[a_1} \psi_{a_2]} ) \nn\\
&\hspace{28mm}+ \frac12 \partial_{c_1} \bar \epsilon \Gamma_{a_1a_2}{}^{bc_1c_2} \psi_{c_2} \,,\\
\dsusy(  \widehat{F}_{a_1\cdots a_9,b} -  \widehat{F}_{a_1\cdots a_9b} ) &= 9 \bar \epsilon \Gamma_{a_1\dots a_8} ( \partial_{a_9]} \psi^b - \partial^b \psi_{a_9]}) - \tfrac34 \partial_{[a_1} ( \varepsilon_{a_2\dots a_9c_1c_2c_3} \bar \epsilon \Gamma^b{}_{c_1c_2c_3d} \psi^d ) \,,\nn\\
& \hspace{28mm} - \tfrac12 \varepsilon_{a_1\dots a_9c_1c_2} \partial^b ( \bar \epsilon \Gamma^{c_1} \psi^{c_2})\,. 
\end{align}
\end{subequations}
Putting these together one obtains that 
\begin{align}
 & \dsusy \Bigl( \widehat{F}_{a_1a_2}{}^b - \tfrac1{9!} \varepsilon_{a_1a_2}{}^{c_1\dots c_9} (  \widehat{F}_{c_1\cdots c_9,}{}^b -  \widehat{F}_{c_1\cdots c_9}{}^b \bigr) \Bigr) \nn\\
&= - \bar \epsilon \Gamma^b \partial_{[a_1} \psi_{a_2]} +  \bar \epsilon \Gamma_{[a_1} ( \partial_{a_2]} \psi^b - \partial^b \psi_{a_2]} )+2 \bar\epsilon \Gamma_{a_1a_2c} \partial^{[c} \psi^{b]} - \frac12  \bar \epsilon \Gamma_{a_1a_2}{}^{bc_1c_2} \partial_{c_1}  \psi_{c_2}\nn\\
&=\frac12 \bar \epsilon \Gamma_{a_1a_2} \rho^b - \bar \epsilon \Gamma^b{}_{[a_1} \rho_{a_2]} + \frac{2}{9} \delta^b_{[a_1} \bar \epsilon \Gamma_{a_2]} \Gamma^c \rho_c \ . 
\end{align}
We see again that the supersymmetry of the bosonic equations of motion gives the fermionic Rarita--Schwinger equation. Note that in this case the component $\chi_{b;a_1\dots a_9}$ is involved, so this additional field is already necessary to understand the supersymmetry of the linearised dual graviton equation in eleven dimensions. 

{\allowdisplaybreaks
The complete equation will take the form
\be 
\label{SupercovarianceRep} 
\dsusy \widehat{\mathcal{E}}^{\underline{I}} = \bar \epsilon G_a{}^{\underline{I}} \rho^a \  , 
\ee
where $G_a{}^{\underline{I}} $ defines a $\widetilde{K}(E_{11})$ invariant tensor, implying that the multiplet of bilinears in $\epsilon$ and $\rho^a$ is in the $S_-$ module as $\Psi\Psi$. One computes in particular that 
\begin{subequations}
\begin{align} 
\label{SusyEM} 
\dsusy \widehat{\cE}_{a_1a_2a_3a_4} &= \frac{4}{3} \bar \epsilon \Gamma_{[a_1a_2a_3} \rho_{a_4]} - \frac16 \bar \epsilon \Gamma_{a_1a_2a_3a_4b} \rho^b \  ,
\\
 \dsusy \widehat{\cE}_{a_1a_2}{}^b   &= \frac12 \bar \epsilon \Gamma_{a_1a_2} \rho^b - \bar \epsilon \Gamma^b{}_{[a_1} \rho_{a_2]} + \frac29 \delta_{[a_1}^b \bar \epsilon \Gamma_{a_2]} \Gamma^c \rho_c\ ,
 \\ 
 \dsusy \widehat{\cE}_a{}^{b_1b_2b_3} &= \frac16 \bar \epsilon \Gamma_a{}^{b_1b_2b_3c} \rho_c - \bar \epsilon \Gamma_a{}^{[b_1b_2} \rho^{b_3]} - \frac23 \bar \epsilon \Gamma^{b_1b_2b_3} \rho_a + \frac12 \delta_a^{[b_1} \bigl( \bar \epsilon \Gamma^{b_2b_3]c} \rho_c - \bar \epsilon \Gamma^{b_2} \rho^{b_3]}\bigr)  \; , \qquad \quad 
 \\
 \dsusy \widehat{\cE}^{a,b} &= \bar \epsilon \Gamma^{(a} \rho^{b)} - \frac16 \eta^{ab} \bar \epsilon \Gamma^c \rho_c \ , 
 \\
 \dsusy \widehat{\cE}^{a_1a_2a_3a_4,b} &=  \bar\epsilon \Gamma^{a_1a_2a_3a_4} \rho^b  -\bar\epsilon \Gamma^{[a_1a_2a_3a_4} \rho^{b]} + \frac43 \eta^{b[a_1} \bar \epsilon \Gamma^{a_2a_3a_4]c} \rho_c - 2 \eta^{b[a_1} \bar \epsilon \Gamma^{a_2a_3} \rho^{a_4]}  \; ,  \qquad \\
 \dsusy\widehat{\cE}_a{}^{b_1...b_6} &=- \frac{1}{18} \varepsilon_a{}^{b_1\dots b_6c_1c_2c_3c_4} \Bigl( \bar \epsilon \Gamma_{c_1c_2c_3} \rho_{c_4} - \frac18 \bar \epsilon \Gamma_{c_1c_2c_3c_4d} \rho^d \Bigr) \nn\\
 & \hspace{25mm}  + \bar \epsilon \Gamma^{b_1\dots b_6} \rho_a + 2 \delta_a^{[b_1} \bigl( \bar \epsilon \Gamma^{b_2\dots b_6]c} \rho_c - \bar \epsilon \Gamma^{b_2\dots b_5} \rho^{b_6]} \bigr)   
 \end{align}
 \end{subequations}
 transform in the representation of the self-duality equation as they should. }

\section{Non-linear theory with fermions}
\label{sec:NLferm}

In this section, we shall investigate how much of the structure of Sections~\ref{sec:susyALG} and~\ref{sec:susyEOM} can be made non-linear. We  propose that the non-linear self-duality equations including fermions in  \eqref{DEwO} and the non-linear supersymmetry transformations for the bosonic fields \eqref{NonLinsusy}, are the actual equations and fields transformations of the $E_{11}$ exceptional field theory. With this assumption, we shall now attempt to define also the non-linear generalisations of the field equations and supersymmetry transformations of the fermionic fields. 

We consider the generalisation of the Rarita--Schwinger equation~\eqref{eq:RSeq} in Section~\ref{sec:NLRS} and in Section~\ref{sec:NLsusy}, we investigate the non-linear generalisation of the fermionic supersymmetry transformation~\eqref{dpsi}. As we shall see by comparison to $D=11$ supergravity, our non-linear proposals reproduce correctly the structure of the non-linear terms of $D=11$ supergravity, due to remarkable cancellations yielding only gauge invariant combinations of the low level field strengths. However, we also get undesired additional contributions involving higher level fields. While we do not know how to remove these contributions at present, we provide evidence that the structures we write must be part of the complete answer.

\subsection{Non-linear Rarita--Schwinger equation}
\label{sec:NLRS}

In order to study possible non-linear equations of motion for the fermions, we first introduce appropriate covariant derivatives and covariant tensors. We propose a Lagrangian in~\eqref{RSLagrange} to describe the gravitino kinetic term, its Pauli couplings to generalised field strengths and quartic fermion terms. Finally, we investigate the relation of our proposal to $D=11$ supergravity.

\subsubsection{Ingredients of the non-linear fermionic terms}

Equation~\eqref{eq:RSeq} defines a $\widetilde{K}(E_{11})$ covariant linearised equation for the vector-spinor $\Psi$ through 
\begin{align}
\label{eq:LRSeq}
\rho^a= G^{a;b M} \partial_M \psi_b = 0\ ,
\end{align}
where $G^{a;b M}$ is a $\widetilde{K}(E_{11})$ invariant tensor that also acts on the not explicitly shown spinor indices.  This equation is moreover consistent with linearised supersymmetry as defined in~\eqref{dpsi}. We expect the non-linear equation to be defined in a similar way but with the partial derivative being replaced by a covariant derivative, plus additional terms depending on the field strength $F^I$, as well as appropriate cubic terms in the fermions. The natural candidate for a covariant derivative is the one obtained from the  ${K}(\mf{e}_{11})$ component of the Maurer--Cartan form valued in the ${K}(\mf{e}_{11})$  representation of  $\psi_a$.  One defines the covariant derivative from  the Maurer--Cartan derivative
\begin{align}
 \partial_M \cV \cV^{-1} = \mathcal{P}_M - \mathcal{Q}_M\,, \quad \mathcal{Q}_M \in K(\mf{e}_{11})\,,\quad \mathcal{P}_M \in \overline{\mathcal{T}}_0 \ominus K(\mf{e}_{11})\, ,
\end{align}
where here $\cV$ is an element of  $ G(\overline{\mathcal{T}}_0)$. For the terms in the level decomposition we shall consider in this section, there is not yet a distinction between $ G(\overline{\mathcal{T}}_0)$ and $E_{11}$, so the reader may consider as well that $\cV$ is the standard $E_{11} / K(E_{11})$ coset representative for simplicity. The covariant derivative in tangent frame is defined as
\begin{align}
 \mathcal{D}_{\underline{M}}  = \cV^{-1 N}{}_{\underline{M}} ( \partial_N + \mathcal{Q}_N )  \ , \label{cosetDer} 
 \end{align}
where we denote by $\underline{M}$ the tangent frame indices that transform under $K(E_{11})$. We shall also use the notation that 
\begin{align}
 J_{\underline{M}\, \underline{\alpha}} t^{\underline{\alpha}} =  \cV^{-1 N}{}_{\underline{M}}  J_{N \alpha } \cV t^\alpha  \cV^{-1} = 2 \mathcal{P}_{\underline{M}\,\underline{\alpha}} t^{\underline{\alpha}} \ ,
 \end{align}
for the current components in tangent frame.

The $\widetilde{K}(E_{11})$  Rarita--Schwinger equation must reduce to the standard eleven-dimensional supergravity equation upon choosing the solution to the section constraint in which the fields only depend on the eleven coordinates $x^m$. In this case the covariant derivative reduces to 
   \be \cV^{-1 N}{}_{a} ( \partial_N + \mathcal{Q}_N )\Big|_{11}   = e^{\frac12} e_a{}^m ( \partial_m +\mathcal{Q}_m) \ , \ee
where the notation $|_{11}$ indicates that fields only dependent on the eleven coordinates $x^m$ so that all the higher level derivatives can be disregarded, and the additional factor of the vielbein determinant comes from the $GL(11)$ weight of the $R(\Lambda_1)$ module. Note, however, that $\mathcal{Q}_m$ still involves  an infinity of fields as the section constraint only affects the derivative index.

The spin connection can be rewritten in terms of the coset space connection $ \mathcal{Q}|_{11} $ in $\mathfrak{so}(1,10)$ and  the field strength  component  $F_{n_1n_2}{}^m \big|_{11} = 2g^{mp} \partial_{[n_1} g_{n_2]p} $ as 
\be \label{SpinConnection} \omega_{ab} = e_{[a}{}^n d e_{n|b]} - \frac1{2 } e_c F_{ab}{}^c \Big|_{11} \ .  \ee

The vielbein determinant is part of the $E_{11} / K(E_{11})$ coset representative, and as such cannot appear separately without violating $E_{11}$ symmetry. The way it is resolved for the  eleven-dimensional gravitino field $\psi^{\scalebox{0.5}{11D}}_m$, is that it is related to the vector-spinor through 
\be 
\psi_a = e^{\frac14} e_a{}^m \psi^{\scalebox{0.5}{11D}}_m  \ . 
\label{11DF}
\ee 
Note that similar redefinitions were also necessary for $\widetilde{K}(E_{10})$, see~\cite{Damour:2006xu}.

\subsubsection{Non-linear fermionic Lagrangian and Rarita--Schwinger equation}

We shall now investigate the construction of the Lagrangian for the non-linear Rarita--Schwinger equation. We are guided first by $K(E_{11})$ invariance using the ingredients introduced above.  In principle, one would like to also check gauge invariance of the Lagrangian using~\eqref{eq:lieF} and~\eqref{eq:liePsi}. Doing so requires new identities of $K(E_{11})$ tensors that remain to be investigated. We shall only study gauge invariance indirectly below in Section~\ref{sec:NL11} when we analyse the Lagrangian in the $D=11$ decomposition. As we shall see our proposal is incomplete as it requires additional terms in order to reproduce $D=11$ supergravity and these additional terms are also expected to be necessary for gauge invariance.

Using the $K(E_{11})$ invariant tensors we have introduced we can write the following Lagrangian quadratic in fermionic fields
\begin{align} \label{Lzero} 
 \mathcal{L}^{\rm \scriptscriptstyle RS}_0 \sim \bar \psi_a G^{a;b \underline{M}} \mathcal{D}_{\underline{M}} \psi_b + \frac{1}{4} \eta_{\underline{I} \underline{J}} F^{\underline{I}} O^{\underline{J}} \ , 
 \end{align}
where $G^{a;b \underline{M}}$ is defined as in \eqref{eq:LRSeq} and where $O^{\underline{I}}$ is the $\Psi\Psi$ bilinear in the $\mathcal{T}_{-1}$ representation defined in Section \ref{supercovariantF}.  By construction the covariant derivative $\mathcal{D}_{\underline{M}} $ is $\widetilde{K}(E_{11})$ covariant, so the first term is manifestly  $\widetilde{K}(E_{11})$ invariant.  The second term is also manifestly $\widetilde{K}(E_{11})$ invariant, and is non-zero according to the property that on-shell
\be 
\label{eq:DEflat}
F^{\underline{I}}  = \eta^{\underline{I}\underline{K}}\Omega_{\underline{K}\underline{J}} F^{\underline{J}} + O^{\underline{I}} \; , \qquad  O^{\underline{I}} = -  \eta^{\underline{I}\underline{K}}\Omega_{\underline{K}\underline{J}} O^{\underline{J}} \ . 
\ee
The bilinear form $\eta_{\underline{I} \underline{J}} $  is given in  \cite[Eq.~(5.39)]{Bossard:2017wxl} as
\begin{multline}  \eta_{\underline{I} \underline{J}} F^{\underline{I}} O^{\underline{J}} =  \frac{1}{9!} \Fs{3}_{a_1\cdots a_9,b}  O^{a_1\cdots a_9,b} 
-  \frac{1}{8!} \Fs{3}_{a_1\cdots a_{10}} O^{a_1\cdots a_{10}}+ \frac{1}{7!} \Fs{1}_{a_1\cdots a_7}  
O^{a_1\cdots a_7}  + \frac{1}{4!} \Fs{-1}_{a_1\cdots a_4} O^{a_1\cdots a_4}  
\\  +\frac{1}{2} \Fs{-3}_{a_1a_2}{}^b O^{a_1a_2}{}_b -\Fs{-3}_{ab}{}^b O^{ac}{}_c 
+\frac{4}{6} \Fs{-5}^{a_4}{}_{[a_1a_2a_3} O_{a_4]}{}^{a_1a_2a_3} + \Fs{-5}_{a,b} O^{a,b} + \frac{1}{4!} \Fs{-7}_{a_1\cdots a_4,b} O^{a_1\cdots a_4,b}
\\  + \frac{7}{6!} \Fs{-7}^{a_7}{}_{[a_1\cdots a_6} O_{a_7]}{}^{a_1\cdots a_6} + \frac{9}{8!}  \Fs{-9}^{[a_1}{}_{a_2\dots a_9,b} O_{a_1}{}^{a_2\dots a_9],b} - \frac{1}{8!} \Fs{-9}^{b}{}_{a_1\dots a_8,b} O_{c}{}^{a_1\dots a_8,c}+ \dots \end{multline} 
However, this Lagrangian involves infinitely many fields and is formally infinite. We shall now argue that one can partially resolve this problem by exhibiting that infinitely many terms cancel upon using the self-duality equation \eqref{eq:DEflat}, and that the resulting Lagrangian agrees with the eleven-dimensional supergravity Lagrangian at low order in the level truncation. However, this will not yet provide the complete answer. The Lagrangian must not only give rise to a meaningful finite Rarita--Schwinger  equation, but this equation must moreover be gauge invariant. The Lagrangian \eqref{Lzero} is not a priori gauge invariant, since neither the $K(\mathfrak{e}_{11})$ covariant derivative nor the field strength $F^{\underline{I}}$ is covariant under generalised diffeomorphisms. Thus one will need to check gauge invariance separately. We shall see that gauge invariance can also be achieved partially by modifying the corresponding Rarita--Schwinger equation by a term proportional to the bosonic field equation $\widehat{\cE}^I$.

First of all we will introduce a Darboux basis on $\mathcal{T}_{-1}\cong S_+\oplus S_-$  as a $K(E_{11})$ module. To argue why one needs to do this it is useful to recall the case of $\mathcal{N}=8$ supergravity in four dimensions. In this case, the Lagrangian includes two terms that are not invariant under the full R-symmetry group, but only under the subgroup $SO(8) \subset SU(8)$ acting on the 28 vector fields \cite{deWit:1982bul}. One of these two terms is in particular the source of the Pauli coupling $F\Psi\Psi$. For $E_{11}$ we have similarly that  $\mathcal{T}_{-1}$ is a symplectic representation of $E_{11}$, and therefore $S_\pm$ are conjugate unitary representations of  $K(E_{11})$, and one needs to introduce a Lagrangian subspace that further breaks $K(E_{11})$ to a subgroup preserving a quadratic norm $\eta_{\underline{I}\underline{J}}^+$ on $S_+$ and $S_-$. The choice of Lagrangian subspace is not unique, as it is neither for the symplectic frame in four dimensions, but there is a natural choice associated to any maximal standard parabolic subgroup of $E_{11}$. This choice is defined by the positive weight components in the corresponding parabolic subgroup decomposition. Since in this section we want to compare with eleven-dimensional supergravity, we shall use the Lagrangian subspace determined by the $GL(11)$ weight, such that $\eta_{\underline{I}\underline{J}}^+$ is the projection of $\eta_{\underline{I}\underline{J}}$ to the negative weight components along $\underline{I}$, \ie\  for a positive weight component ${}^{\scalebox{0.6}{$(\frac{1}{2}+k)$}}\hspace{-0.6mm}F^{\underline{I}}$ one has $^{\scalebox{0.6}{$(\frac{1}{2}+k)$}}\hspace{-0.6mm}F^{\underline{I}}\eta_{\underline{I}\underline{J}}^+{}=0$, and is non-degenerate on the negative weight ${}^{\scalebox{0.6}{$(-\frac{1}{2}+k)$}}\hspace{-0.6mm}F^{\underline{I}}\eta_{\underline{I}\underline{J}}^+{}\neq 0$. More explicitly, we take  
\begin{multline} 
\label{etaplus}  
\eta^+_{\underline{I} \underline{J}} F^{\underline{I}} O^{\underline{J}} =  \frac{1}{4!} \Fs{-1}_{a_1\cdots a_4} O^{a_1\cdots a_4}   +\frac{1}{2} \Fs{-3}_{a_1a_2}{}^b O^{a_1a_2}{}_b -\Fs{-3}_{ab}{}^b O^{ac}{}_c 
\\+\frac{4}{6} \Fs{-5}^{a_4}{}_{[a_1a_2a_3} O_{a_4]}{}^{a_1a_2a_3} + \Fs{-5}_{a,b} O^{a,b} + \frac{1}{4!} \Fs{-7}_{a_1\cdots a_4,b} O^{a_1\cdots a_4,b}
\\  + \frac{7}{6!} \Fs{-7}^{a_7}{}_{[a_1\cdots a_6} O_{a_7]}{}^{a_1\cdots a_6} + \frac{9}{8!}  \Fs{-9}^{[a_1}{}_{a_2\dots a_9,b} O_{a_1}{}^{a_2\dots a_9],b} - \frac{1}{8!} \Fs{-9}^{b}{}_{a_1\dots a_8,b} O_{c}{}^{a_1\dots a_8,c}+ \dots 
\end{multline} 
 Now we can define 
\begin{align}
\mathcal{L}^{\rm \scriptscriptstyle RS} &= \bar \psi_a G^{a;b \underline{M}} \mathcal{D}_{\underline{M}} \psi_b+ \frac{1}{4} \bigl(  \eta_{\underline{I} \underline{J}} F^{\underline{I}} O^{\underline{J}} +   \eta^+_{\underline{I} \underline{J}} \; : \tfrac12 ( {\cE}^{\underline{I}} +\widehat{\cE}^{\underline{I}} ) O^{\underline{J}} : \bigr)   \ , \nn\\
&= \bar \psi_a G^{a;b \underline{M}} \mathcal{D}_{\underline{M}} \psi_b + \frac{1}{2} \eta^+_{\underline{I} \underline{J}} F^{\underline{I}} O^{\underline{J}} - \frac{1}{8} :   \eta^+_{\underline{I} \underline{J}}   O^{\underline{I}}  O^{\underline{J}} :  \ , 
\label{RSLagrange} 
\end{align}
where the normal ordered product is introduced on the infinite sum of quartic fermions to regularise it. The bilinear terms combine to give a finite set of contributions for each field in the $GL(11)$ decomposition for a chosen solution to the section constraint, so even if there is an infinite set of fields contributing to the Rarita--Schwinger equation, it makes sense as a formal sum over the infinite set of fields. By contrast, $ \eta^+_{\underline{I} \underline{J}}   O^{\underline{I}}  O^{\underline{J}}  $ would involve infinitely many times the same vector-spinor fields and must be replaced by a finite polynomial in the vector-spinor that we write $:   \eta^+_{\underline{I} \underline{J}}   O^{\underline{I}}  O^{\underline{J}} : $. We shall argue below that this polynomial can be determined by $\widetilde{K}(E_{11})$ invariance. The Lagrangian~\eqref{RSLagrange} is not manifestly $\widetilde{K}(E_{11})$ invariant, but the corresponding Rarita--Schwinger equation $\widehat{\rho}^a=0$ obtained by variation via
\be 
\delta \bar \psi_a \widehat{\rho}^a   = \frac12 \delta \mathcal{L}^{\rm \scriptscriptstyle RS}+ \partial_M (\dots )  \ , 
\ee
differs from the manifestly covariant one $\widehat{\rho}^a_0$ defined from  $\mathcal{L}^{\rm \scriptscriptstyle RS}_0$ above by a term proportional to the equations of motion 
\begin{align}
 \widehat{\rho}^a &= G^{a;b \underline{M}} \bigl( \mathcal{D}_{\underline{M}} \psi_b  - \tfrac{1}{2} T^{\underline{\alpha} \underline{N}}{}_{\underline{M}} \mathcal{P}_{\underline{N} \underline{\alpha}}  \psi_b \bigr) + \frac{1}{4} \eta^+_{\underline{I} \underline{J}} {F}^{\underline{I}} \frac{\partial^L  O^{\underline{J}}}{\partial \psi^a } - \frac{1}{8} : \eta^+_{\underline{I} \underline{J}} {O}^{\underline{I}} \frac{\partial^L  O^{\underline{J}}}{\partial \psi^a }  : \nn\\
& = \rho^a_0 + \frac{1}{8} \eta^+_{\underline{I} \underline{J}} \widehat{\mathcal{E}}^{\underline{I}} \frac{\partial^L  O^{\underline{J}}}{\partial \psi^a }  \ , 
\end{align}
so that it is ensured to be covariant under $\widetilde{K}(E_{11})$, modulo term that vanish when the duality equation  $\widehat{\mathcal{E}}^{\underline{I}} =0$ is satisfied. Note that the $\mathcal{N}=8$ Rarita--Schwinger equation is only covariant under $SU(8)$ modulo the twisted self-duality equation for the 28 vector fields   in four dimensions \cite{deWit:1982bul}, so it is to be expected that the same complication must arise in $E_{11}$ exceptional field theory.

The extra term appearing with the covariant derivative can be understood in terms of the current $J_{M \alpha}$ as
\be  
\label{CovVecSpin} \cD_M - \tfrac12 \cV^{\underline{N}}{}_M T^{\underline{\alpha} \underline{P}}{}_{\underline{N}} \mathcal{P}_{\underline{P} \underline{\alpha}}   =  \cD_M - \tfrac14   T^{\alpha N}{}_M J_{\alpha N}  \ , 
\ee
such that it is a weight term that appears at level 0 in the $GL(11)$ decomposition because of the $e^{\frac{1}{4}}$ in the definition of the vector-spinor~\eqref{11DF}.

Note moreover that the $\widetilde{K}(E_{11})$  invariant Lagrangian $ \mathcal{L}^{\rm \scriptscriptstyle RS}_0$ in~\eqref{Lzero} does not include quartic terms in the vector-spinor. Indeed, one can infer from the invariance under the $\widetilde{K}(E_{10})\subset \widetilde{K}(E_{11})$ subgroup that there is no quartic invariant in the vector-spinor. Under $SO(10)$, $\psi_a$ decomposes as $\psi_a$ for $a=1$ to $10$ and $ \lambda = \psi_0 - \Gamma_0 \sum_{a=1}^{10} \Gamma^a \psi_a$, which transform respectively under $K(\mf{e}_{10})$ as a vector-spinor and a spinor \cite{Damour:2006xu}. However, the vector-spinor of $\widetilde{K}(E_{10})$ transforms under a quotient subgroup $SO(32,288)= \widetilde{K}(E_{10}) / \mathcal{I}_\psi$ \cite{Kleinschmidt:2018hdr}, so there is no quartic antisymmetric invariant that can be written. One straightforwardly concludes that there is no quartic invariant under  $\widetilde{K}(E_{11})$. 

The necessity of introducing cubic terms in the fermions would therefore arise when one introduces a Lagrangian subspace to define the Rarita--Schwinger equation. The Lagrangian $\mathcal{L}^{\rm \scriptscriptstyle RS}$ in~\eqref{RSLagrange} is not $\widetilde{K}(E_{11})$ invariant and does have quartic fermion terms. We stress that it is still formal, since it involves a sum over infinitely many fields and infinitely many components. We shall argue that one can make sense of its part that is quadratic in the vector-spinor by expanding in level, such that  cancellations arise for different fields at each level. This answer is nonetheless incomplete, and one will need to add other terms to the Lagrangian. The situation is more complicated for the quartic terms in the vector-spinors, since the naive polynomial does not even make sense formally. The bilinear $\Psi\Psi$ includes $61\, 776$ components that appear infinitely many times in $O^{\underline{I}}$ in the infinite representation $S_-$, so there is no way to directly make sense of the infinite sum of terms appearing in $\eta^+_{\underline{I}\underline{J}} O^{\underline{I}}O^{\underline{J}}$. Instead of defining  $:\eta^+_{\underline{I}\underline{J}} O^{\underline{I}}O^{\underline{J}}:$ through some regularisation scheme, we hope that one could think of $:\eta^+_{\underline{I}\underline{J}} O^{\underline{I}}O^{\underline{J}}:$  as a finite quartic polynomial in the vector-spinor that is determined by (on-shell) $K(\mf{e}_{11})$ covariance of the non-linear Rarita--Schwinger equation. Let us explore this idea in some more detail. 

The  component of the Rarita--Schwinger equation linear in $\psi_a$ is\footnote{The notation $\frac{\partial^L}{\partial \psi^a}$ means the convention to differentiate fermions from the left.}
\begin{align}
\rho^a  = \rho^a_0 + \frac{1}{8} \eta^+_{\underline{I} \underline{J}} {\mathcal{E}}^{\underline{I}} \frac{\partial^L  O^{\underline{J}}}{\partial \psi^a }  \ , 
\end{align}
and transforms by construction under $K(\mf e_{11})$ as 
\bea
&&\delta \rho^a - \frac1{12} \Lambda_{bcd} \Gamma^{bcd} \rho^a - \frac{2}{3} \Lambda^{abc} \Gamma_b \rho_c +\frac1 6 \Lambda^{abc} \Gamma_{bcd} \rho^d  \\
&=& \delta \Bigl(  \frac{1}{8} \eta^+_{\underline{I} \underline{J}} {\mathcal{E}}^{\underline{I}} \frac{\partial^L  O^{\underline{J}}}{\partial \psi^a } \Bigr) 
 - \eta^+_{\underline{I} \underline{J}} {\mathcal{E}}^{\underline{I}} \Bigl(  \frac1{96} \Lambda_{bcd} \Gamma^{bcd} \frac{\partial^L  O^{\underline{J}}}{\partial \psi^a }+ \frac{1}{12} \Lambda^{abc} \Gamma_b   \frac{\partial^L  O^{\underline{J}}}{\partial \psi_c } -\frac1 {48} \Lambda^{abc} \Gamma_{bcd} \frac{\partial^L  O^{\underline{J}}}{\partial \psi_d }  \Bigr) \nn\\
 &=&  \Lambda_{b_1b_2b_3} 
 \cE^{\underline{I}} \mathcal{R}_{a \, \underline{I}}{}^{b_1b_2b_3} \; ,\nn  \eea
for some $\mathcal{R}_{a \, \underline{I}}{}^{b_1b_2b_3}$, since $\cE^{\underline{I}}$ transforms into itself under $K(\mf e_{11})$. When evaluated on section, one may hope that most of the components of  $ \cE^{\underline{I}}$ cancel such that $ \cE^{\underline{I}} \mathcal{R}_{a \, \underline{I}}{}^{b_1b_2b_3} $ would only involve finitely many components of $ \cE^{\underline{I}}$ in a level decomposition. If this were true, the purported regularised quartic term in the fermions would be determined then such that it would transform under $K(\mf e_{11})$ as 
\begin{align}
\label{eq:NOpsi}
\delta \left[ \frac{1}{8} :   \eta^+_{\underline{I} \underline{J}}   O^{\underline{I}}  O^{\underline{J}} : \right] =  \frac12 \Lambda_{b_1b_2b_3} O^{\underline{I}}\bar \psi^a \mathcal{R}_{a \, \underline{I}}{}^{b_1b_2b_3} \,.
\end{align}
The full Rarita--Schwinger equation with cubic fermion terms would then transform under $K(\mf{e}_{11})$ as
\be \delta \widehat{\rho}^a - \frac1{12} \Lambda_{bcd} \Gamma^{bcd} \widehat{\rho}^a - \frac{2}{3} \Lambda^{abc} \Gamma_b \widehat{\rho}_c +\frac1 6 \Lambda^{abc} \Gamma_{bcd} \widehat{\rho}^d  =  \Lambda_{b_1b_2b_3} 
 \widehat{\cE}^{\underline{I}} \mathcal{R}_{a \, \underline{I}}{}^{b_1b_2b_3} \; .  \ee
Whether a regularisation prescription with~\eqref{eq:NOpsi} exists and produces finite expressions, needs to be established.

\subsubsection{Relation to $D=11$ supergravity}
\label{sec:NL11}

Let us finally describe how the conjectured Lagrangian  $\mathcal{L}^{\rm \scriptscriptstyle RS}$ partly reproduces the eleven-dimen\-sional supergravity Lagrangian when the fields only depend on the eleven coordinates $x^m$. In $GL(11)$ parabolic gauge, the Maurer--Cartan form $d \cV \cV^{-1} $ only has components at positive levels 
\begin{align}
\label{eq:MCform}
d \cV \cV^{-1} = e_a{}^m d e_m{}^b K_b{}^a + \sum_{k=1}^\infty J^\ord{k}_{A_k} E^{A_k} =\mathcal{P}-\mathcal{Q} \,,
\end{align}
where
\bea 
 J^\ord{1}_{a;b_1b_2b_3}\Big|_{11}  &=& e^{\frac{1}{2}} e_a{}^m e_{b_1}{}^{n_1}e_{b_2}{}^{n_2}e_{b_3}{}^{n_3} \partial_m A_{n_1n_2n_3}\ ,
  \nn \\
J^\ord{2}_{a;b_1\dots b_6} \Big|_{11}  &=& e^{\frac{1}{2}} e_a{}^m e_{b_1}{}^{n_1}\dots e_{b_6}{}^{n_6}\bigl(  \partial_m A_{n_1\dots n_6} -10 A_{n_1n_2n_3} \partial_m A_{n_4n_5n_6} \bigr) \ , \eea
and similarly for higher $GL(11)$ levels. Here we slight abuse of notation to denote the strictly positive components of the Cartan--Maurer by $J^\ord{k}$ although they only agree with the strictly positive components of the current $\cV\cJ\cV^{-1}$ and not $\cJ$ itself.
Because of the expression of the Maurer--Cartan form~\eqref{eq:MCform}, at non-zero level, the composite $K(\mf e_{11})$ connection and the coset components are both defined by $J^\ord{k}_{A_k}$:
\begin{align}
 \label{cd2}
\mathcal{Q}= e_{[a}{}^m d e_{m|b]} K^{ab} - \frac12 \sum_{k=1}^\infty J^\ord{k}_{A_k} (E^{A_k}-F^{A_k}) \; , 
\quad \mathcal{P} = e_{(a}{}^m d e_{m|b)} K^{ab} + \frac12 \sum_{k=1}^\infty J^\ord{k}_{A_k} (E^{A_k}+F^{A_k})  \ . 
\end{align}
The covariant derivative is not covariant under generalised diffeomorphisms, because of the terms in $J^\ord{k}_{A_k}$  that involve the higher level fields through an ordinary partial derivative and not an exterior derivative. But this is also the case for the field strength at level $\ell \le - \frac32-k$ for $k\ge 1$ evaluated on section, see~\eqref{F}. One finds that most of these field strengths vanish on section, in particular $\Fs{-5}^{a,b} |_{11} = 0$ and  $\Fs{-7}^{a_1a_2a_3a_4,b} |_{11} = 0$. The non-vanishing ones are those that are in the $SO(1,10)$ representation of the field of level $k$ times the standard co-tangent space, and in that case 
\be 
{}^{\scalebox{0.6}{$(-\frac{3}{2}-k)$}}\hspace{-0.6mm}F_{a;A_k} \Big|_{11} =(-1)^k J^\ord{k}_{a;A_k} \; , 
\ee
according to \cite{Bossard:2017wxl}. We observe that the components $J^{\ord{k}}$ include the gauge invariant field strengths ${}^{\scalebox{0.6}{$(-\frac{3}{2}+k)$}}\hspace{-0.6mm}F$ for $k=0,1,2$. In fact, we are going to show that in the Rarita--Schwinger Lagrangian these components $J^\ord{k}$ combine remarkably into the gauge invariant combination ${}^{\scalebox{0.6}{$(-\frac{3}{2}+k)$}}\hspace{-0.6mm}F$. This kind of recombination cannot occur, however, for $k\geq 3$ since ${}^{\scalebox{0.6}{$(-\frac{3}{2}+k)$}}\hspace{-0.6mm}F$ contains the constrained $\chi$ fields that are not present in  $J^\ord{k}$.
 For $k\geq3$, the $J^\ord{k}$ would need to cancel in the Rarita--Schwinger Lagrangian for gauge invariance but preliminary calculations show that they do not.

We now assemble the various pieces for expanding the Lagrangian~\eqref{RSLagrange} in $GL(11)$ decomposition to exhibit the remarkable recombinations mentioned above.
The covariant derivative at level zero $ \cD^\ord{0}_m$ only includes the $SO(1,10)$ connection $\mathcal{Q}_m^\ord{0}$
\be 
\cD^\ord{0}_m \psi_a = \partial_m \psi_a + e_{[a}{}^n \partial_m e_{n|b]} \psi^b  + \frac14 e_{b}{}^n \partial_m e_{n c} \Gamma^{bc} \psi_a \ . 
\ee
The Pauli couplings at level $-\frac12$ and $-\frac32$ give 
\bea 
&&  \frac1{4!}\Fs{-1}_{a_1a_2a_3a_4} O^{a_1a_2a_3a_4}  + \frac12 \Fs{-3}_{a_1a_2}{}^b O^{a_1a_2}{}_b - \Fs{-3}_{ab}{}^b O^{ac}{}_c 
\\
&=& 2 \bar \psi_a \Bigl( - \frac18 \Gamma^{a}{}_{bc} \Gamma^{de}  F_{de}{}^b \psi^c + \frac12 \Gamma^{ab}{}_d F_{bc}{}^d \psi^c \Bigr) + \frac1{8} F_{a_1a_2a_3a_4} \Bigl( \bar \psi^{a_1} \Gamma^{a_2a_3} \psi^{a_4} +\tfrac1{12}\bar \psi_{b_1} \Gamma^{a_1\dots a_4b_1b_2} \psi_{b_2} \Bigr) \nn\\
&& \quad - \frac{3}{2} F_{[a_1a_2;a_3]} \bigl( \tfrac12 \bar \psi^{a_1} \Gamma^{a_2} \psi^{a_3} + \tfrac14 \bar \psi_{c_1} \Gamma^{a_1a_2a_3c_1c_2} \psi_{c_2} - \bar \psi^{a_1} \Gamma^{a_2a_3c} \psi_c \bigr)  \ ,
\nn  
\eea
so, using that $F_{[a_1a_2;a_3]}|_{11}=0$, the corresponding contributions to the Lagrangian combine into 
\bea && \Bigl[ \bar \psi_a \Gamma^{abc}\cD^\ord{0}_b \psi_c +\frac{1}{2} \Bigl(  \frac1{4!}F_{a_1a_2a_3a_4} O^{a_1a_2a_3a_4}  + \frac12 F_{a_1a_2}{}^b O^{a_1a_2}{}_b - F_{ab}{}^b O^{ac}{}_c  \Bigr) \Bigr] \Big|_{11} \nn\\
&=& \bar \psi_a \Gamma^{abc} e_b{}^m \bigl( \partial_m \psi_c + \omega_{m c}{}^d \psi_d + \tfrac14 \omega_{m d_1d_2} \Gamma^{d_1d_2} \psi_c \bigr)  + \frac1{192} F_{a_1a_2a_3a_4} \bar \psi_{b_1} \Gamma^{b_1} \Gamma^{a_1\dots a_4} \Gamma^{b_2} \psi_{b_2} \ . \qquad 
\eea
The first term reproduces the standard covariant derivative of the gravitino field, where one notes that the term in $ \frac14 \bar \psi_a\Gamma^{abc} e^{-1}\partial_b e\, \psi_c =0$ due to the $e^{\frac14}$ rescaling in \eqref{11DF} drops out by symmetry. Note that we drop the explicit level when there is no ambiguity, but we shall keep it for the field strength of level $\ell \le - \frac52$. The second term is the expected Pauli coupling in eleven-dimensional supergravity, but with a factor of one-half.

To exhibit the cancellation of the non gauge invariant terms between $\bar \psi_a \rho^a $ and $ \frac{1}{2} \eta^+_{\underline{I} \underline{J}} F^{\underline{I}} O^{\underline{J}}$ one needs to consider higher levels. The kinetic term expands as
\be 
\bar{\psi}_a \Gamma^{abc}\mathcal{D}_b \psi_c = \bar{\psi}_a \Gamma^{abc}\mathcal{D}^{\ord{0}}_b \psi_c +\sum_{k=1}^\infty \bar{\psi}_a \Gamma^{abc} \delta(\tfrac12J^\ord{k}_b) \psi_c \ ,
\ee
where $\delta(\tfrac12J^\ord{k}_b)$ denotes the $K(\mf{e}_{11})$ action of the corresponding component. At the first level  has the contribution 
\bea 
\bar{\psi}_a \Gamma^{abc} \delta(\tfrac12J^\ord{1}_b) \psi_c = - \frac1{12} J_a{}^{b_1b_2b_3} \bar \psi_{c_1} \Gamma^{c_1ac_2} \bigl( \tfrac12 \Gamma_{b_1b_2b_3} \psi_{c_2} + 4 \eta_{c_2b_1} \Gamma_{b_2} \psi_{b_3} - \Gamma_{c_2b_1b_2} \psi_{b_3} \bigr) 
\nn \\
= \frac12 J_a{}^{b_1b_2b_3} \Bigl( \tfrac1{12} \bar \psi_{c_1} \Gamma^{c_1c_2a}{}_{b_1b_2b_3} \psi_{c_2} - \frac32 \bar \psi_{b_1} \Gamma^a{}_{b_2} \psi_{b_3} - \delta^a_{[b_1} \bar \psi_{b_2} \Gamma_{b_3]c} \psi^c + \tfrac12 \bar \psi_c \Gamma^{ca}{}_{[b_1b_2} \psi_{b_3]}\Bigr)  \ . 
\eea
The Pauli coupling at level $-\frac52$ gives using  \eqref{O13} and \eqref{etaplus} 
\begin{multline}  
\frac{4}{6} O^{a}{}_{[b_1b_2b_3} \Fs{-5}_{a]}{}^{b_1b_2b_3} = \Bigl( \frac{1}{4} \bar \psi^{[b_1} \Gamma^{b_2b_3]} \psi_a - \frac54 \bar \psi^{[b_1} \Gamma_a{}^{b_2} \psi^{b_3]} +\frac12 \bar \psi^c \Gamma_{ca}{}^{[b_1b_2} \psi^{b_3]} \\ +\frac1{24} \bar\psi_{c_1} \Gamma^{c_1c_2a}{}_{b_1b_2b_3} \psi_{c_2} - \delta_a^{[b_1} \bar \psi^{b_2} \Gamma^{b_3]c} \psi_c \Bigr)\Fs{-5}_{a}{}^{b_1b_2b_3}\ .
\end{multline}
Using $\Fs{-5}_a{}^{b_1b_2b_3} |_{11} =- J^\ord{1}_a{}^{b_1b_2b_3} $ when neglecting higher level derivatives, one gets from the previous two results
\be 
\bar{\psi}_a \Gamma^{abc} \delta(\tfrac12J^\ord{1}_b) \psi_c  + \frac12 \frac{4}{6} O^{a}{}_{[b_1b_2b_3} \Fs{-5}_{a]}{}^{b_1b_2b_3} \big|_{11}
= \frac1{192} F_{a_1a_2a_3a_4} \bar \psi_{b_1} \Gamma^{b_1}\Gamma^{a_1\dots a_4} \Gamma^{b_2} \psi_{b_2}  \ ,  
\ee
so that all the non gauge invariant terms disappear. Importantly, one gets an additional contribution to the Pauli coupling and the sum of these first two terms is in full agreement with the supergravity Lagrangian.

At the next level, using the fact that the $K(\mf{e}_{11})$ action on the vector-spinor at level 2 is the same as for $K(\mf e_{10})$  given in \cite{Damour:2006xu}, \ie 
\be 
\delta(\tfrac12J^\ord{2}_b) \psi_c =  \frac{1}{2\cdot 6!} \Bigl( \frac{1}{2} J^\ord{2}_b{}^{a_1\dots a_6} \Gamma_{a_1\dots a_6} \psi_c - 10 J^\ord{2}_{b:c}{}^{a_1\dots a_5} \Gamma_{a_1\dots a_4} \psi_{a_5} + 4 J^\ord{2}_b{}^{a_1\dots a_6} \Gamma_{ca_1\dots a_5} \psi_{a_6} \Bigr)\ , 
\label{magic}
\ee
one finds that 
\begin{multline}
\bar{\psi}_a \Gamma^{abc} \delta(\tfrac12J^\ord{2}_b) \psi_c =  \frac{1}{2\cdot 6!} \Big(  -\frac12\bar\psi^c \Gamma^a{}_{b_1...b_6cd}\psi^d -45\bar\psi_{b_1} \Gamma_{b_2...b_5}{}^ a\psi_{b_6} 
\\
+ 6\bar\psi_{b_1} \Gamma_{b_2...b_6}{}^{ac}\psi_c +30\delta_{b_1}^a \bar\psi^c\Gamma_{cb_2...b_5}\psi_{b_6}\Big) J^\ord{2}_a{}^{b_1...b_6}\ .
\label{P1}
\end{multline}
The Pauli coupling  on the other hand, upon using  \eqref{O16a},  gives
\begin{multline}
 \frac{7}{2\cdot 6!} O^{a}{}_{[b_1\dots b_6} \Fs{-7}_{a]}{}^{b_1\dots b_6} = \frac{1}{2\cdot 6!} \Big(\frac14  \bar\psi^c \Gamma^a{}_{b_1...b_6cd}\psi^d 
 +\frac{75}{2}\bar\psi_{b_1} \Gamma_{b_2...b_5}{}^ a\psi_{b_6} 
+3 \bar\psi^a \Gamma_{b_1...b_5} \Gamma_{b_6} \\
 - 6\bar\psi_{b_1} \Gamma_{b_2...b_6}{}^{ac}\psi_c -30\delta_{b_1}^a \bar\psi^c\Gamma_{cb_2...b_5}\psi_{b_6}\Big)  \Fs{-7}_{a}{}^{b_1\dots b_6}\ .
\label{P2}
\end{multline}
Combining \eqref{P1} with  \eqref{P2} gives
\be 
\label{level2cont} 
\bar{\psi}_a \Gamma^{abc} \delta(\tfrac12J^\ord{2}_b) \psi_c  + \frac12 \frac{7}{6!} O^{a}{}_{[b_1\dots b_6} \Fs{-7}_{a]}{}^{b_1\dots b_6}\big|_{11}  =-\frac1{192} \frac{1}{7!} \varepsilon_{a_1a_2a_3a_4}{}^{b_1\dots b_7}F_{b_1\dots b_7}  \bar \psi_{b_1} \Gamma^{b_1}\Gamma^{a_1\dots a_4} \Gamma^{b_2} \psi_{b_2}   \ ,
\ee
where we note that the last two terms in \eqref{P1} with  \eqref{P2} cancel each other and the remaining terms sum up to an expression that is totally antisymmetric in $[b;b_1...b_6]$, thereby making it possible to use the relation $J^\ord{2}_{[a;b_1...b_6]} |_{11}= \Fs{-7}_{[a;b_1...b_6]}|_{11}= \frac17 \Fs{1}_{ab_1...b_6}|_{11}$. From level $\ell=2$ one gets therefore in total
\begin{align}
\label{eq:F7toomuch}
\bar{\psi}_a \Gamma^{abc} \mathcal{D}_b \psi_c  + \frac{1}{2} \eta^+_{\underline{I} \underline{J}} F^{\underline{I}} O^{\underline{J}}  &= \bar \psi_a \Gamma^{abc} \nabla_b \psi_c +\frac{1}{96 } F_{a_1a_2a_3a_4} \bar{\psi}_{b} \Gamma^{b} \Gamma^{a_1a_2a_3a_4} \Gamma^c \psi_{c} \nn\\
&\quad  -  \frac1{192} \frac{1}{7!} \varepsilon_{a_1a_2a_3a_4}{}^{b_1\dots b_7}F_{b_1\dots b_7}  \bar \psi_{b_1} \Gamma^{b_1} \Gamma^{a_1\dots a_4} \Gamma^{b_2} \psi_{b_2} + \dots  \ ,  
\end{align}
which would give the correct equation if one had not included the contribution from the level 2 field. It is difficult to imagine which kind of contribution would eliminate it. 

For higher levels $k\ge 3$ one cannot  get the same type of cancellation, because the $J^\ord{k}_a$ factor does not include the $\chi$ fields present in the gauge invariant combination  ${}^{\scalebox{0.6}{$(-\frac{3}{2}+k)$}}\hspace{-0.6mm}F$. They thus cannot recombine into a component of $\cE^{\underline{I}}$ in order to contribute to a term involving only the metric and the 3-form gauge field. Preliminary calculations show that they do not cancel either. Hence it seems that we are still missing a term in the Lagrangian that would cancel all the contributions from level 2 to infinity. At present, we do not have a candidate for such terms.

\subsection{Non-linear supersymmetry transformations}
\label{sec:NLsusy}

Let us now consider the non-linear supersymmetry transformation of the fermion, the linear transformation was given in~\eqref{dpsi}. At the non-linear level, one expects that the partial derivative will again be replaced by the $\widetilde{K}(E_{11})$ covariant derivative, plus possibly additional terms involving the field strength $F^I$ and higher order terms in $\Psi$, such that the supersymmetry transformation may read
\begin{align}
 \dsusy \psi_a  &= G_a{}^{\underline{M}}  \bigl( \mathcal{D}_{\underline{M}}  + \tfrac{1}{2} T^{\underline{\alpha} \underline{N}}{}_{\underline{M}} \mathcal{P}_{\underline{N} \underline{\alpha}} \bigr) \epsilon -\frac14 \eta^+_{\underline{I}\underline{J}} F^{\underline{I}} \,  G^\intercal_a{}^{\underline{J}}  \epsilon + \frac18   : \eta_{\underline{I} \underline{J}} O^{\underline{I}}  G^\intercal_a{}^{\underline{J}} :  \epsilon  \nn\\
&=G_a{}^{\underline{M}}  \bigl( \mathcal{D}_{\underline{M}}  + \tfrac{1}{2} T^{\underline{\alpha} \underline{N}}{}_{\underline{M}} \mathcal{P}_{\underline{N} \underline{\alpha}} \bigr) \epsilon  - \frac18   \bigl(  \eta_{\underline{I} \underline{J}} F^{\underline{I}} G^\intercal_a{}^{\underline{J}} \; + :  \eta^+_{\underline{I} \underline{J}} \widehat{\cE}^{\underline{I}} G^\intercal_a{}^{\underline{J}} : \bigr)  \epsilon \, ,   \end{align}
where $G_a{}^{\underline{M}}$ and $ G^\intercal_a{}^{\underline{J}} $ are the $\widetilde{K}(E_{11})$ invariant constant tensors in the $Spin(1,10)$ Cifford algebra that define \eqref{dpsi} and \eqref{SupercovarianceRep}. The covariant derivative includes the same weight term with the opposite sign as for the vector-spinor \eqref{CovVecSpin}, consistent with the fact that the $\epsilon \Psi$ bilinear does not carry a weight. The regularisation prescription is understood to work as for the Rarita--Schwinger equation, such that the supersymmetry variation is only covariant  under $\widetilde{K}(E_{11})$  modulo a term in the self-duality equation, and the bilinear term $   : \eta_{\underline{I} \underline{J}} O^{\underline{I}}  G^\intercal_a{}^{\underline{J}} : $  in $\Psi\Psi$ is determined to restore $\widetilde{K}(E_{11})$   covariance. This proposed ansatz is not a priori exhaustive, and we expect to miss some terms that would contribute at higher level.

The spinor is related to the eleven-dimensional spinor parameter as
\be 
\epsilon  = e^{-\frac14} \epsilon^{\scalebox{0.5}{11D}} \ ,   
\ee
so that the covariant derivative term in the eleven-dimensional supergravity supersymmetry transformation reads\footnote{Here, $\nabla_n$ denotes the full covariant derivative including both the spin and affine connection such that $\nabla_n e_m{}^a=0$, such that $\nabla_n$ effectively only acts on $\epsilon^{\scalebox{0.5}{11D}}$.}
\bea
\nabla_n \epsilon &=& e^{-\frac14} \partial_n \epsilon^{\scalebox{0.5}{11D}}  + \frac14 \omega_{n\, ab}\Gamma^{ab}e^{-\frac14} \epsilon^{\scalebox{0.5}{11D}} = \partial_n \epsilon + \frac14 \omega_{n\, ab}\Gamma^{ab} \epsilon + \frac{1}{8} g^{pq} \partial_n g_{pq} \epsilon \nn \\
&=&  \cD^\ord{0}_n \epsilon+\frac14 g^{pq} \Bigl( \partial_p g_{qn} - \tfrac12 \partial_n g_{pq} \Bigr) \epsilon+\frac14 \Bigl( F_{np}{}^p|_{11} - \frac1{2} e_{n\, c} F_{ab}{}^c |_{11} \Gamma^{ab} \Bigr) \epsilon \ . 
\eea
The covariant derivatives itself gives 
\be \cD_a \epsilon = \cD^\ord{0}_a \epsilon-\frac{1}{24}  J^\ord{1}_a{}^{b_1b_2b_3} \Gamma_{b_1b_2b_3}\epsilon  +\frac1{4\cdot 6!}J^\ord{2}_a{}^{b_1\dots b_6}\Gamma_{b_1\dots b_6}\epsilon + \dots  \ . \ee
To construct the coupling to the field strength we need the components of the tensor $G^\intercal_a{}^{\underline{I}}$. They can be computed from the definition \eqref{SupercovarianceRep} using \eqref{SusyEM} as 
\bea G^\intercal_{a;b_1b_2b_3b_4} &=& \frac43 \eta_{a[b_1} \Gamma_{b_2b_3b_4]} - \frac16 \Gamma_{ab_1\dots b_4} \ , \nn \\
G^\intercal_{a;b_1b_2}{}^c &=& \frac12 \delta_a^c \Gamma_{b_1b_2} - \eta_{a[b_1} \Gamma_{b_2]}{}^c +\frac29 \Gamma_a \Gamma_{[b_1} \delta_{b_2]}^c \ ,\nn  \\
G^\intercal_{a;b}{}^{c_1c_2c_3} &=& \frac16 \Gamma_{ab}{}^{c_1c_2c_3} + \delta_a^{[c_1} \Gamma_b{}^{c_2c_3]} + \frac23 \eta_{ab} \Gamma^{c_1c_2c_3} - \frac12 \delta_b^{[c_1}\Gamma_a{}^{c_2c_3]} - \frac12 \delta_{ab}^{[c_1c_2} \Gamma^{c_3]} \; , \nn \\
G^\intercal_{a;}{}^{b,c} &=& \delta_a^{(b} \Gamma^{c)} - \frac16 \eta^{bc} \Gamma_a \; , \nn \\
G^\intercal_{a;b}{}^{c_1\dots c_6} &=&- \frac1{18} \varepsilon_b{}^{c_1\dots c_6d_1\dots d_4} \Bigl( \eta_{ad_1} \Gamma_{d_2d_3d_4} - \frac18 \Gamma_{ad_1d_2d_3d_4} \Bigr) + \eta_{ab} \Gamma^{c_1\dots c_6} \; \nn \\
&& \hspace{30mm}- 2 \delta_b^{[c_1} \Gamma_a{}^{c_2\dots c_6]} - 2 \delta_{ab}^{[c_1c_2} \Gamma^{c_3c_4c_5c_6]} \; . \eea 
Using then \eqref{etaplus}, the components of $ G^{\intercal \eta} \equiv \eta^+_{\underline{I}\underline{J}} G^\intercal_a{}^{\underline{J}} $ are given by
\bea 
G^{\intercal \eta}_{a;b_1b_2b_3b_4} &=&\frac1{18} \Bigl( \eta_{a[b_1} \Gamma_{b_2b_3b_4]} - \frac18\Gamma_{ab_1\dots b_4}\Bigr)  \ ,  
\\
G^{\intercal \eta}_{a;b_1b_2}{}^c &=& \frac14 \delta_a^c \Gamma_{b_1b_2} -\frac12 \eta_{a[b_1} \Gamma_{b_2]}{}^c  - \eta_{a[b_1} \delta_{b_2]}^c \ ,
\nn  \\
G^{\intercal \eta}_{a;b}{}^{c_1c_2c_3} &=& \frac1{36} \Gamma_{ab}{}^{c_1c_2c_3} +\frac16 \delta_a^{[c_1} \Gamma_b{}^{c_2c_3]}
 + \frac19 \eta_{ab} \Gamma^{c_1c_2c_3} + \frac16  \delta_b^{[c_1}\Gamma_a{}^{c_2c_3]} - \frac{2}{3} \delta_{ba}^{[c_1c_2} \Gamma^{c_3]} \ , 
\nn \\
G^{\intercal \eta}_{a;b}{}^{c_1\dots c_6} &=&
- \frac1{18\cdot 6!} \varepsilon_b{}^{c_1\dots c_6d_1\dots d_4} \Bigl( \eta_{ad_1} \Gamma_{d_2d_3d_4} 
 - \frac18 \Gamma_{ad_1d_2d_3d_4} \Bigr) +\frac1{6!}\eta_{ab} \Gamma^{c_1\dots c_6} 
\nn\\
&& \hspace{20mm} - \frac{4}{6!} \Gamma_a{}^{[c_1\dots c_5} \delta^{c_6]}_b+ \frac{10}{6!} \delta_a^{[c_1} \Gamma^{c_2c_3c_4c_5} \delta_b^{c_6]}   \ .
\nn 
 \eea 
Putting these components together, one obtains
\bea 
-\frac{1}{4} \eta^+_{\underline{I}\underline{J}} F^{\underline{I}} \,  G^\intercal_a{}^{\underline{J}}   &=&\tfrac14 F_{ab}{}^b - \tfrac{1}{16} F^{b_1b_2}{}_a \Gamma_{b_1b_2} + \tfrac1{8} F_{ab}{}^c \Gamma^b{}_c +\frac{1}{288} \bigl( \Gamma_{a}{}^{b_1b_2b_3b_4} - 8 \delta_a^{b_1} \Gamma^{b_2b_3b_4} \bigr) F_{b_1b_2b_3b_4} 
 \nn\\
&& + \frac{1}{24}  J^\ord{1}_a{}^{b_1b_2b_3} \Gamma_{b_1b_2b_3}  + \frac14   \Bigl( \frac16 \Gamma_{ab_1b_2} J^\ord{1}_c{}^{b_1b_2c}- \frac23 \Gamma^b J^{c}_\ord{1}{}_{abc} \Bigr) \nn\\
&&  -\frac1{4\cdot 6!}J^\ord{2}_a{}^{b_1\dots b_6}\Gamma_{b_1\dots b_6}+ \frac14 \Bigl( \frac4{6!} \Gamma_{ab_1\dots b_5} J^\ord{2}_c{}^{b_1\dots b_5c} - \frac1{3\cdot 4!} \Gamma^{b_1\dots b_4} J_\ord{2}^c{}_{ab_1\dots b_4 c} \Bigr) 
\nn \\
&& -\frac{1}{24^2} \bigl( \Gamma_{a}{}^{b_1b_2b_3b_4} - 8 \delta_a^{[b_1} \Gamma^{b_2b_3b_4} \bigr) \frac{1}{7!} \varepsilon_{b_1\dots b_4}{}^{c_1\dots c_7} F_{c_1\dots c_7}   +\dots  \ ,
\eea
where the dots state for the terms involving the field strength component $F^{\scalebox{0.6}{$(-\frac{3}{2}-k)$}}_{a;A_k} $ for $k\ge 3$. Then we need to compute the term in $G_a{}^{\underline{M}}T^{\underline{\alpha} \underline{N}}{}_{\underline{M}} \mathcal{P}_{\underline{N} \underline{\alpha}} $. For this one first computes that 
\begin{align}
2T^{\underline{\alpha} \underline{N}}{}_{a} \mathcal{P}_{\underline{N} \underline{\alpha}} \big|_{11}  &= e^{\frac{1}{2}} e_a{}^n g^{pq} \Bigl( \partial_p g_{qn} - \tfrac12 \partial_n g_{pq} \Bigr)\; ,  \nn\\
2T^{\underline{\alpha} \underline{N} a_1a_2} \mathcal{P}_{\underline{N} \underline{\alpha}} \big|_{11}&= J^\ord{1}_c{}^{a_1a_2c}\; , \nn\\
2T^{\underline{\alpha} \underline{N} a_1\dots a_5} \mathcal{P}_{\underline{N} \underline{\alpha}} \big|_{11}&= -J^\ord{2}_c{}^{a_1\dots a_5 c} \ .
\end{align}
For establishing these relations, we have used the $K(E_{11})$ transformations of the various components of $\partial_M$ leading to $\partial_m$
\begin{align}
 \delta \partial_m = k_m{}^n \partial_n - \tfrac12 k_n{}^n \partial_m  \; , \quad
\delta \partial^{n_1n_2} =  f^{n_1n_2p} \partial_p  \; , \quad
\delta \partial^{n_1\dots n_5} = - f^{n_1\dots n_5p} \partial_p  \; . 
\end{align}
These formulas extend the level $\ell=1$ transformations given in~\eqref{eq:KE11der}.
Substituting the components of $G_a{}^{\underline{M}}T^{\underline{\alpha} \underline{N}}{}_{\underline{M}} \mathcal{P}_{\underline{N} \underline{\alpha}} $ to the ones of $\partial_{\underline{M}}$ in \eqref{dpsi} one obtains 
\bea  
2G_a{}^{\underline{M}}T^{\underline{\alpha} \underline{N}}{}_{\underline{M}} \mathcal{P}_{\underline{N} \underline{\alpha}} \big|_{11} &=& e^{\frac12} e_a{}^n g^{pq} \Bigl( \partial_p g_{qn} - \tfrac12 \partial_n g_{pq} \Bigr)  + \frac23 \Gamma^b J_\ord{1}^c{}_{abc} - \frac16 \Gamma_{ab_1b_2} J^\ord{1}_c{}^{b_1b_2c} -
\nn\\
&& \qquad + \frac1{3\cdot 4!} \Gamma^{b_1\dots b_4} J_\ord{2}^c{}_{ab_1\dots b_4 c}- \frac4{6!} \Gamma_{ab_1\dots b_5} J^\ord{2}_c{}^{b_1\dots b_5c} +\dots \ ,
\eea
up to terms involving $J^\ord{k}_{A_k}$ for $k\ge 3$. 

Altogether, one obtains eventually that all the non gauge invariant terms cancel out such that 
\bea && \Bigl( G_a{}^{\underline{M}}  \bigl( \mathcal{D}_{\underline{M}}  + \tfrac{1}{2} T^{\underline{\alpha} \underline{N}}{}_{\underline{M}} \mathcal{P}_{\underline{N} \underline{\alpha}} \bigr) \epsilon -\frac14 \eta^+_{\underline{I}\underline{J}} F^{\underline{I}} \,  G^\intercal_a{}^{\underline{J}}  \epsilon  \Bigr) \Big|_{11} \nn \\
&=&   \cD^\ord{0}_a \epsilon+\frac14 e^{\frac12} e_a{}^n g^{pq} \Bigl( \partial_p g_{qn} - \tfrac12 \partial_n g_{pq} \Bigr) \epsilon+\frac14 \Bigl( F_{ab}{}^b - \frac1{2} F_{b_1b_2;a} \Gamma^{b_1b_2} \Bigr) \epsilon  \nn\\
&& +\frac{1}{288} \bigl( \Gamma_{a}{}^{b_1b_2b_3b_4} - 8 \delta_a^{b_1} \Gamma^{b_2b_3b_4} \bigr) F_{b_1b_2b_3b_4} \epsilon -\frac{1}{24^2} \bigl( \Gamma_{a}{}^{b_1b_2b_3b_4} - 8 \delta_a^{[b_1} \Gamma^{b_2b_3b_4} \bigr) \frac{1}{7!} \varepsilon_{b_1\dots b_4}{}^{c_1\dots c_7} F_{c_1\dots c_7} \epsilon \nn\\
&=& \nabla_a \epsilon + \frac{1}{288} \bigl( \Gamma_{a}{}^{b_1b_2b_3b_4} - 8 \delta_a^{b_1} \Gamma^{b_2b_3b_4} \bigr) \Bigl( F_{b_1b_2b_3b_4}- \frac{1}{2\cdot 7!} \varepsilon_{b_1\dots b_4}{}^{c_1\dots c_7} F_{c_1\dots c_7} \Bigr) \epsilon \ . 
\eea
We recover therefore the same situation as for the Rarita--Schwinger equation. Namely, if we had considered naively the level truncation to the level 1 field, we would have recovered the expected supersymmetry transformation of the gravitino potential in eleven-dimensional supergravity. However, the level 2 fields gives an extra contribution that is nonetheless gauge invariant in eleven-dimensional supergravity. One expects similarly that the higher level fields will also give similar contributions, and the absence of $\chi_M{}^{\tilde{\alpha}}$ field in the current components $J^\ord{k}_a$  for  $k\ge 3$, forbids to possibly eliminate them using the duality equations. For the same reason, this proposal for the non-linear supersymmetry variation of the vector-spinor cannot be complete and there is some structure yet to be understood.

The resolution of this problem would permit to understand the notion of generalised  $SL(32)\cong \widetilde{K}(E_{11})/ \mathcal{I}_\epsilon $ holonomy for the full supergravity field equations, generalising the constructions that have been implemented in $E_{7}$ generalised geometry for the $SU(8)$ holonomy \cite{Coimbra:2016ydd}.

\section{Conclusions}

In this paper we have constructed non-linear duality equations that are invariant under $E_{11}$ generalised diffeomorphisms. These equations involve several crucial $E_{11}$ group theoretical properties that are understood thanks to the use  of the tensor hierarchy algebra $\cT(\mf{e}_{11})$. The tensor hierarchy algebra defines a differential complex for fields satisfying the section constraints, and provide in particular a field strength representation that generalises the embedding tensor representation of gauged supergravity. The field strength can only be defined as an $E_{11}$ tensor provided that one considers additional constrained fields  $\chi_{M}{}^{\tilde{\alpha}}$ transforming in an indecomposable representation of $E_{11}$. We have provided strong evidence that a certain algebraic identity between $E_{11}$ structure coefficients holds, thanks to which one can prove that the first order duality equation we propose in this paper is invariant under $E_{11}$ generalised diffeomorphisms.  We find that there is also a formulation of the theory with yet more fields, such that the scalar fields parametrize not only $E_{11} / K(E_{11})$ but an extended non-semi-simple coset $ G(\overline{\mathcal{T}}_0) / K(E_{11})$, together with some additional $\Sigma$ gauge invariance. Within this extended formulation, one can define supersymmetry transformations in a manifestly $K(E_{11})$ covariant form.   

We have computed the first components of the $E_{11}$ self-duality equation~\eqref{DualityEquation} upon branching on  $GL(3)\times E_8 \subset E_{11}$. By choosing a partial solution to the section constraint such that the fields only depend on  $3+248$ coordinates, we recover the $E_8$ exceptional field theory duality equation between the scalar and the vector fields. An infinite chain of duality equations emerges in this way, but one does not recover the whole dynamics without imposing first order equations for the constrained fields. Similar results hold for $E_7$ exceptional field theory, and it would be interesting to analyse the $E_{11}$ equations in their decompositions under the $E_9$ and $E_{10}$ subgroups as well. 
The cosmological $E_{10}$ coset model constructed in~\cite{Damour:2002cu,Damour:2007dt} is different in essence from the $E_{11}$ exceptional field theory considered here in that it is defined in one dimension only rather than using an infinite-dimensional coordinate module like $R(\Lambda_1)$ together with a section constraint. This one direction is considered to be the time direction. In contrast to the framework studied in this paper, space is conjectured to emerge from the infinitely many components of the $E_{10}$ fields through the gradient expansion of the supergravity fields. It would be interesting to compare the $E_{11}$ exceptional field theory equations in the $E_{10}$ decomposition with the $E_{10}$ cosmological model. 

An important open problem of our work is the construction of the non-linear first order field equations of the constrained fields $\chi_M{}^{\tilde\alpha}$. These equations do not follow simply by substituting the duality equations~\eqref{DualityEquation} into the Bianchi identities. This is perfectly analogous to the situation encountered in $E_n$ exceptional field theories in lower dimensions \cite{Hohm:2013uia}. One may try to construct the desired field equations for the constrained fields directly, or from a (pseudo-)action with the desired gauge symmetry.  It is worth noting that the constrained fields $\chi_M{}^{\tilde{\alpha}}$ are expected to be non-zero in any non-trivial supergravity. Moreover, the structure of the $E_{11}$ equation~\eqref{DualityEquation} is such that it does not admit an obvious consistent truncation to a finite-dimensional subgroup. This is due to the fact that one cannot have a non-trivial solution with a finite number of non-vanishing fields as the duality equation automatically relates an infinite series of fields to each other, as we have explained in Section~\ref{sec:dualfields}.

In this paper, we have also studied the supersymmetric extension of $E_{11}$ exceptional field theory by including an unfaithful vector-spinor representation $\Psi$ of ${\widetilde K} (E_{11})$, the double cover of $K(E_{11})$. We have established that the bilinears in $\Psi$ transform in the same ${K}(E_{11})$-representation as the bosonic first-order self-duality equation, up to a suitable quotient. We have defined the supersymmetry transformation rules on all the fields and presented a  ${\widetilde K} (E_{11})$ covariant Rarita--Schwinger equation of motion for the vector-spinor $\Psi$, at the linearised level.  We have also investigated the extent to which these equations can be made non-linear. Terms in the resulting non-linear Rarita--Schwinger equation include those arising from the Pauli couplings present in $D=11$ supergravity, but the results are incomplete.

Another challenge for any $E_n$ exceptional field theory is to find a global interpretation of the infinitely many new coordinates associated with $R(\Lambda_1)$ that are present in the theory. Locally, any solution of the section constraint depends only on finitely many coordinates. Non-trivial global configurations have appeared as non-geometric backgrounds where patching is done with the $E_n$ symmetry group; most work on this subject has been done in the context of double field theory~\cite{Shelton:2005cf,Dabholkar:2005ve,Hull:2006va,Hull:2007zu,Pacheco:2008ps}. These global problems should probably be first addressed for finite-dimensional exceptional groups before tackling $E_{11}$. 

Our work also suggests some interesting group-theoretic identities for $\cT$, $E_{11}$ and $K(E_{11})$ that might be interesting to investigate further. They define the embeddings of various spinor bilinears in the different representations of $E_{11}$: The symmetric bilinear $\epsilon\otimes \epsilon$ appears to arise as a quotient of the $R(\Lambda_1)$ representation of $E_{11}$ while the antisymmetric $\Psi\otimes\Psi$ bilinear appears to arise as a quotient of the field strength representation $\cT_{-1}$. While these embeddings are natural from a physical perspective and have been checked at low levels here, their existence might also entail interesting mathematical consequences.

One of the main promises of the $E_{11}$ exceptional field theory lies in its power to unify all maximal (gauged) supergravities in all dimensions $D\le 11$. It has been found to provide non-linear and consistent reductions of $D=11$ supergravities on nontrivial internal manifolds  to gauged supergravities in lower dimensions~\cite{Berman:2012uy,Musaev:2013rq,Aldazabal:2013mya,Godazgar:2013oba,Lee:2014mla,Hohm:2014qga}. In attempts to go beyond pure (two-derivative) supergravity, exceptional field theory has also been utilised in the analysis of contributions of BPS states to loop corrections in these theories~\cite{Bossard:2015foa,Bossard:2017kfv}. However, the continuous exceptional symmetry cannot directly be used as a tool for classifying generic higher derivative corrections to supergravity because these corrections are expected to generically break the $E_n$ symmetry to a discrete subgroup $E_n(\ints)$~\cite{Hull:1994ys}, with interesting implications for the low-energy effective action~\cite{Green:1997tv,Green:1997as,Pioline:1998mn,Green:1999pu,Green:2005ba,Pioline:2010kb,Green:2010kv,Green:2011vz,Pioline:2015yea,Fleig:2015vky}. For the continuous $E_{11}$, one finds immediately that only a two-derivative Lagrangian can possibly be invariant. Whether $E_{11}(\ints)$ and its automorphic forms~\cite{Fleig:2012xa,Fleig:2013psa} can be used for higher derivative terms remains to be seen.

\subsection*{Acknowledgements}

We acknowledge useful discussions with Daniel Butter, Franz Ciceri, Gianluca Inverso, Ralf K\"ohl, Hermann Nicolai, Jakob Palmkvist, Henning Samt\-leben and Peter West. We thank the Mitchell foundation for support during the Brinsop Court meeting 2018.  AK acknowledges support from the Simons Center for Geometry and Physics, Stony Brook, and from \'Ecole Polytechnique, Paris, where part of this work was carried out. GB and ES thank the Albert Einstein Institute, Potsdam,  for hospitality during the final stages of this work. The work of GB and AK was partially supported by PHC PROCOPE, projet No 37667ZL and DAAD PPP grant 57316852 (XSUGRA). The work of GB was partially supported by the ANR grant Black-dS-String (ANR-16-CE31-0004). The work of ES is supported in part by the NSF grants PHY-1521099 and PHY-1803875.

\appendix

\section{Representations of \texorpdfstring{$\mf e_{11}$}{e11}}
\label{app:e11}

In this appendix, we collect the decomposition of some of the key representations of $\mf e_{11}$ under its $\mf{gl}(11)$ and  $\mf{gl}(3)\oplus \mf{e}_8$ subalgebras.

\subsection{Level decomposition into \texorpdfstring{$\mf{gl}(11)$}{GL11}}

The generators of $\mf{gl}(11)$ are written as $K^m{}n$ with fundamental indices $m,n\in\{1,\ldots,11\}$ and with commutators 
\be
[K^m{}_n, K^p{}_q]= \delta_n^p K^m{}_q - \delta_q^m K^p{}_n\ , \quad m,n,\ldots=0,1,\ldots 10\ .
\label{gl11}
\ee
Defining the level $\ell$ as the eigenvalue of the generator $\frac13 K^m{}_m $, the levels $0\leq \ell\leq 4$ of the $\mf{gl}(11)$ decomposition of the adjoint representation is given in Table~\ref{tab:e11adj}. Similarly, the low lying levels of the $\mf{gl}(11)$ decomposition of the fundamental representation is displayed in Table~\ref{tab:e11l1}, and that of $R(\Lambda_{10})$ representation in Table~\ref{tab:e11l10}. The method of level decomposition is explained for example in~\cite{Damour:2002cu,West:2002jj}.

\renewcommand{\arraystretch}{1.5}
\begin{table}[t!]
\centering
\begin{tabular}{|c|c|c|c|}
\hline
Level $\ell$ & $\mf{sl}(11)$ representation & {Generator}&Potential  \\
\hline
$0$ & $\begin{matrix}(1,0,0,0,0,0,0,0,0,1) \\ (0,0,0,0,0,0,0,0,0,0)\end{matrix}$ & $K^m{}_n$ & $h_m{}^n$\\
\hline
1 & $(0,0,0,0,0,0,0,1,0,0)$ & $E^{n_1n_2n_3}$ & $A_{n_1n_2n_3}$\\
\hline
2 & $(0,0,0,0,1,0,0,0,0,0)$ & $E^{n_1\cdots n_6}$ & $A_{n_1\cdots n_6}$\\
\hline
3 & $(0,0,1,0,0,0,0,0,0,1)$ & $E^{n_1\cdots n_8,m}$ & $h_{n_1\cdots n_8,m}$\\
\hline
4 & 
$\begin{matrix}
(0,1,0,0,0,0,0,1,0,0)\\
(1,0,0,0,0,0,0,0,0,2)\\
(0,0,0,0,0,0,0,0,0,1)
\end{matrix}$
&
$\begin{matrix}
E^{n_1\cdots n_9,p_1p_2p_3}\\
E^{n_1\cdots n_{10},p,q}\\
E^{n_1\cdots n_{11},m}
\end{matrix}$
&
$\begin{matrix}
A_{n_1\cdots n_9,p_1p_2p_3}\\
B_{n_1\cdots n_{10},p,q}\\
C_{n_1\cdots n_{11},m}
\end{matrix}$\\
\hline
\end{tabular}
\caption{\label{tab:e11adj} \textit{Level decomposition of $\mf{e}_{11}$ under its $\mf{gl}(11)$ subalgebra obtained by deleting node $11$ from the Dynkin diagram in Figure~\ref{fig:e11dynk}, up to level $\ell=4$. }}
\end{table}

The commutation relations of $\mf{e}_{11}$ and its action on the $\cT_{-2}$ part of the tensor hierarchy algebra can be summarised in this level decomposition by considering an element of $\cT_{-2}$ given by
\begin{align}
\label{eq:phi11}
\phi^{\widehat\alpha} \bar{t}_{\widehat\alpha}    &= \ldots + \frac{1}{8!} h_-^{n_1\ldots n_8,m} \tilde{F}_{n_1\ldots n_8,m} + \frac1{6!} A_-^{n_1\ldots n_6}\tilde{F}_{n_1\ldots n_6} + \frac1{3!} A_-^{n_1n_2n_3} \tilde{F}_{n_1n_2n_3} + h^+_n{}^m \tilde{K}^n{}_m \nn\\
&\quad + \frac{1}{3!} A^+_{n_1n_2n_3} \tilde{E}^{n_1n_2n_3} + \frac1{6!} A^+_{n_1\ldots n_6} \tilde{E}^{n_1\ldots n_6} + \frac1{8!} h^+_{n_1\ldots n_8,m} \tilde{E}^{n_1\ldots n_8,m} \nn\\
&\quad + \frac{1}{3!\cdot 9!} A^{+}_{n_1\ldots n_9,m_1m_2m_3} \tilde{E}^{n_1\ldots n_9,m_1m_2m_3} \nn\\
&\quad + \frac{1}{8!} X_{n_1\ldots n_9} \tilde{E}^{n_1\ldots n_9} + \frac{1}{2\cdot9!} X_{n_1\ldots n_{10},rs} \tilde{E}^{n_1\ldots n_{10},rs} + \frac{1}{2\cdot 9!} X_{n_1\ldots n_{11},m} \tilde{E}^{n_1\ldots n_{11},m} \nn\\
&\quad + \frac{1}{11!}  Y_{n_1\ldots n_{11},m} \tilde{E}^{n_1\ldots n_{11},m} \ldots \,,
\end{align}
and studying its transformation under $\ell=\pm1$ defined by
\begin{align}
\label{eq:phi11trm}
\delta \phi^{\widehat\alpha} \bar t_{\widehat\alpha} =  \left[ \frac1{3!} e_{n_1n_2n_3} E^{n_1n_2n_3} + \frac1{3!} f^{n_1n_2n_3} F_{n_1n_2n_3},\,  \phi^{\widehat\alpha} \bar{t}_{\widehat\alpha} \right]\,.
\end{align}
In~\eqref{eq:phi11}, we have labelled the generators of the adjoint  at $p=-2$ with a tilde just as in Table~\ref{tab:thaGL11}. The sub- and superscripts $\pm$ on the parameters indicate whether the generator is at level $\ell\geq0$ or $\ell<0$, respectively. The last two lines in~\eqref{eq:phi11} contains the dual of the generators that are not part of $\mf{e}_{11}$ but of the tensor hierarchy algebra, with $X_{\ldots}$ being associated with $\overline{R(\Lambda_2)}$ and $Y_{\ldots}$ being associated with $\overline{R(\Lambda_{10})}$. As these generators only appear for $\ell>0$, we have suppressed the superscript on them.

\begin{table}[t!]
\centering
\begin{tabular}{|c|c|c|c|c|}
\hline
$\ell$ & $\mf{sl}(11)$ representation & Generator & Coordinate& Parameter\\
\hline
$\tfrac32$ 
& $(1,0,0,0,0,0,0,0,0,0)$ & $P_m$ & $x^m$& $\xi^m$\\
\hline
$\tfrac52$ 
& $(0,0,0,0,0,0,0,0,1,0)$ & $Z^{mn}$ & $y_{mn}$ &$\lambda_{mn}$\\  
\hline
$\tfrac72$ 
& $(0,0,0,0,0,1,0,0,0,0)$ & $Z^{n_1\cdots n_5}$ & $y_{n_1\cdots n_5}$&
$\lambda_{n_1\cdots n_5}$\\  
\hline
$\tfrac92$ 
& $\begin{matrix} (0,0,0,1,0,0,0,0,0,1)\\ (0,0,1,0,0,0,0,0,0,0)\end{matrix}$
&  $\begin{matrix} P^{n_1\cdots n_7,m}\\P^{n_1\cdots n_8}\end{matrix}$
& $\begin{matrix}x_{n_1\cdots n_7,m}\\x_{n_1\cdots n_8}\end{matrix}$ & $\begin{matrix}\xi_{n_1\cdots n_7,m}\\
\lambda_{n_1\cdots n_8}\end{matrix}$\\
\hline
\end{tabular}
\caption{\label{tab:e11l1} \textit{Level decomposition of the representation $\overline{R(\Lambda_1)}$
of $\mf{e}_{11}$ under $\mf{gl}(11)$, up to level $\ell=11/2$. }}
\end{table}

\begin{table}[t!]
\centering
\begin{tabular}{|c|c|c|}
\hline
Level $\ell$ & $\mf{sl}(11)$ representation & Generator structure\\[1mm]
\hline
$4$ & 
    (0,0,0,0,0,0,0,0,0,1) & $L^m$ \\ 
\hline
$5$ & 
    (0,0,0,0,0,0,1,0,0,0) & $L^{n_1\ldots n_4}$ \\ 
\hline
$6$ & 
$\begin{matrix} (0,0,0,1,0,0,0,0,0,0) \\ (0,0,0,0,1,0,0,0,0,1) \end{matrix}$
     & $\begin{matrix} L^{n_1\ldots n_7} \\ L^{n_1\ldots n_6,m} \end{matrix}$\\
\hline
\end{tabular}
\caption{\label{tab:e11l10} \textit{Level decomposition of the $\overline{R(\Lambda_{10})}$
representation of $E_{11}$ under $\mf{gl}(11)$.}}
\end{table}

{\allowdisplaybreaks
Performing the $\mf{e}_{11}$ variation~\eqref{eq:phi11trm} one then obtains
\begin{subequations}
\label{E11adjoint}
\begin{align} 
\delta h^+_n{}^m &=\frac{1}{2}e_{np_1p_2} A^{mp_1p_2}_- - \frac{1}{2} f^{mp_1p_2} A^+_{np_1p_2}
 \nn\\ &\quad\, 
- \frac{1}{18} \delta_n^m \scal{e_{p_1p_2p_3}  A_-^{p_1p_2p_3} - f^{p_1p_2p_3} A^+_{p_1p_2p_3}}\ , 
\\
\delta A^+_{n_1n_2n_3} &= -\frac{1}{6} f^{p_1p_2p_3} A^+_{n_1n_2n_3p_1p_2p_3} 
- 3 e_{p[n_1n_2} h^+_{n_3]}{}^{p}\,,
\\
\delta A^+_{n_1\cdots n_6} &= 20 e_{[n_1n_2n_3} A^+_{n_4n_5n_6]} 
-\frac{1}{2} f^{n_7n_8n_9} h^+_{n_1\cdots n_8,n_9}  \ ,
\\
\delta h^+_{n_1\cdots n_8,m} &= 56 e_{\lsharp n_1n_2n_3} A^+_{n_4\cdots n_8,m\rsharp} - \frac{1}{2} f^{p_1p_2p_3} A^+_{p_1\langle n_1\dots n_8,m\rangle p_2p_3}+  \cdots\ , 
\\
\delta X_{n_1\cdots n_9} &= - 28 e_{[n_1n_2n_3} A^+_{n_4\cdots n_9]} - \frac1{18} f^{p_1p_2p_3} A^+_{n_1\dots n_9,p_1p_2p_3} \nn\\
& \qquad -  \tfrac12 f^{p_1p_2p_3}  X_{n_1\ldots n_9p_1,p_2p_3} + f^{p_1p_2p_3}  X_{n_1\ldots n_9p_1p_2,p_3} +\ldots\ ,
\\
\delta X_{n_1\ldots n_{10}, rs} &= 3 \left( 2e_{rs[n_1} X_{n_2\ldots n_{10}]} - 3 e_{r[n_1n_2} X_{n_3\ldots n_{10}] s} +3 e_{s[n_1n_2} X_{n_3\ldots n_{10}] r} \right)+\ldots\ ,
\\
\delta X_{n_1\ldots n_{11},m} &= 11 e_{m[n_1n_2} X_{n_3\ldots n_{11}]}+\ldots\ ,
\\
\delta Y_{n_1\ldots n_{11},m} &= 0 +\ldots \ .
\end{align}
\end{subequations}
Here, we have only given the transformation of the $+$ parameters, the ones for $-$ are obtained by replacing $e_3$ by $f^3$ and changing the sign. The ellipses represent additional terms going into $\mf{e}_{11}$ that will consistently play no role in this paper and that we therefore have not determined.}

We end this subsection by listing the coordinate representation $\overline{R(\Lambda_1)}$ and the (first) section constraint representation $\overline{R(\Lambda_{10})}$ in $\mf{gl}(11)$ decomposition. The coordinate representation of Table~\ref{tab:e11l1} was originally studied in~\cite{West:2003fc,Kleinschmidt:2003jf} and the section constraint representation of Table~\ref{tab:e11l10} in~\cite{Bossard:2017wxl}. 

\subsection{Level decomposition under \texorpdfstring{$\mf{gl}(3)\oplus \mf{e}_8$}{gl3e8}}
\label{app:e8tha}

The $\mf{gl}(3)\oplus \mf{e}_8$ level decomposition of $\mf{e}_{11}$ is obtained by grading the adjoint of $\mf{e}_{11}$ with respect to node $3$ of the Dynkin diagram shown in Figure~\ref{fig:e11dynk}.

We shall label the $\mf{gl}(3)$ generators by $K^\mu{}_\nu$ with fundamental indices $\mu,\nu\in\{1,2,3\}$ and the generators of $\mf{e}_8$ by $t^A$ with $A\in\{1,\ldots,248\}$. Levels $0\leq \ell\leq 2$ of this decomposition are shown in Table~\ref{tab:e11adj8}. The position of the $A$ index on $E_8$ tensors can be changed by using the $E_8$ Killing metric. This decomposition was first given in~\cite{Riccioni:2007au,Bergshoeff:2007qi}.

\renewcommand{\arraystretch}{1.5}
\begin{table}[h!]
\centering
\begin{tabular}{|c|c|c|c|}
\hline
Level $\ell$ & $\mf{sl}(3)\oplus \mf{e}_8$ representation & {Generator}&Potential  \\
\hline
$0$ & $\begin{matrix}(1,1;0,0,0,0,0,0,0,0)  \\ (0,0;0,0,0,0,0,0,0,0) \\ (0,0;1,0,0,0,0,0,0,0)\end{matrix}$ & $\begin{matrix} \\[-3mm]K^\mu{}_\nu\\[-6mm]\\ t^A\end{matrix}$ & $\begin{matrix} \\[-3mm]h_\mu{}^\nu\\[-6mm]\\ \Phi_A\end{matrix}$\\
\hline
1 & $(0,1;1,0,0,0,0,0,0,0)$ & $E^{\mu}_A$ & $A_\mu^A$\\
\hline
2 & $\begin{matrix}(1,0;0,0,0,0,0,0,0,0) \\ (1,0;0,0,0,0,0,0,1,0) \\ (0,2;1,0,0,0,0,0,0,0)\end{matrix} $ & $\begin{matrix} \\[-3mm]E^{\mu\nu}_{AB}\\[-6mm]\\ E^{\mu,\nu}_A\end{matrix}$ & $\begin{matrix} \\[-3mm]B_{\mu\nu}^{AB}\\[-6mm]\\ h_{\mu,\nu}^A\end{matrix}$\\
\hline
\end{tabular}
\caption{\label{tab:e11adj8} \textit{Level decomposition of $\mf{e}_{11}$ under its $\mf{gl}(3)\oplus\mf{e}_8$ subalgebra obtained by deleting node $3$ from the Dynkin diagram in Figure~\ref{fig:e11dynk}, up to level $\ell=3$. The Dynkin labels for the two summands are separated by a semi-colon.}}
\end{table}

In order to list the remaining generators in the indecomposable representation and also the coordinate representation and field strength representation it is more useful to directly consider the tensor hierarchy algebra $\cT(\mf{e}_{11})$ decomposed under $\mf{gl}(3)\oplus \mf{e}_8$. The construction of this algebra is similar to the one performed in the $\mf{gl}(11)$ grading in~\cite{Bossard:2017wxl} and we present only the salient features here.

The local algebra is constructed out of the generators of degree $q=-1,0,1$ in Table~\ref{tab:THA3}. The components of degree $q=0$ are parametrised in the BRST formulation by a bosonic vector superfield  $V_\mu(\vartheta)$ generating the reparametrisation in three Grassmann variables $\vartheta_\mu$, and scalar fermionic superfield $\Phi^A(\vartheta)$ in $\mathfrak{e}_8$. We use $\iota^\mu = \frac{\partial\; }{\partial \vartheta_\mu}$. The components of degree $q=1$ are parametrised by the fermionic superfield $\psi_\mu^A$ and the bosonic superfield $T^{AB}$ in the ${\bf 3875}\oplus {\bf 1}$. The components of degree $q=-1$ are parametrised by the bosonic superfield $S^A$ and the fermionic superfield $\Theta^\mu$. The BRST operator is then 
\bea \delta V_\mu &=& V_\nu \iota^\nu V_\mu + \psi^A_\mu S_A \ , \\
\delta \Phi^A &=& \tfrac12 f_{BC}{}^A \Phi^B \Phi^C + V_\mu \iota^\mu \Phi^A + T^{AB} S_B - f_{BC}{}^A \bigl( \tfrac12 \iota^\mu \psi_\mu^A S^C + \psi_\mu^B \iota^\mu S^C \bigr) + \psi_\mu^A \Theta^\mu \nn \ , \\
\delta S^A &=& V_\mu \iota^\mu S^A + \iota^\mu V_\mu S^A + f_{BC}{}^A \Phi^B S^C \nn\ , \\
\delta \Theta^\mu &=& V_\nu \iota^\nu \Theta^\mu - \iota^\mu V_\nu \Theta^\nu + \iota^\nu V_\nu \Theta^\mu - \iota^\mu \Phi^A S_A \nn\ , \\
\delta \psi_\mu^A &=& V_\nu \iota^\nu \psi_\mu^A + \iota^\nu V_\mu \psi_\nu^A - \iota^\nu V_\nu \psi_\mu^A + f_{BC}{}^A \Phi^B \psi_\mu^C \nn\ , \\
\delta T^{AB} &=& V_\nu \iota^\nu T^{AB}   - \iota^\nu V_\nu T^{AB}  + 2 \Phi^C f_{CD}{}^{(A} T^{B)D} - 2 \iota^\mu \Phi^{(A} \psi_\mu^{B)} - f^{E(A}{}_C f_{ED}{}^{B)} \iota^\mu \Phi^C \psi^D_\mu  \nn\  . 
\eea
One checks indeed that $\delta^2 = 0$ on $V_\mu$ and $\Phi^A$ and vanishes up to terms quadratic in the degree $q=\pm1$ on the components of degree $q=\pm 1$ respectively, showing that this defines a local superalgebra. The tensor hierarchy algebra is defined as the quotient of algebra freely generated from this local algebra by its maximal ideal. The algebra generated at $p=0$ includes by construction $\mathfrak{e}_{11}$, and defining the direct sum over all $q$ for each $p$ one identifies the same $E_{11}$ representations that appear in the tensor hierarchy algebra constructed in~\cite{Bossard:2017wxl} so we conclude that they are indeed the same algebras. One would expect that there is a minimal local algebra, similar to the finite-dimensional construction of~\cite{Palmkvist:2013vya}, which is a subalgebra of both local subalgebras used in the $GL(11)$ covariant construction of~\cite{Bossard:2017wxl} and the $GL(3)\times E_8$ covariant construction presented here. This minimal local algebra would then imply the uniqueness of the tensor hierarchy algebra.

As for~\eqref{eq:phi11}, we parametrise an element at level $p=-2$ as
\begin{align}
\label{eq:phi11E8}
\phi^{\widehat\alpha} \bar{t}_{\widehat\alpha}    &= \ldots + A_A^{-\mu} \tilde{F}_\mu^A + h^+_\mu{}^\nu \tilde{K}^\mu{}_\nu + \Phi_{A} \tilde{t}^A   + A^{+A}_\mu \tilde{E}^\mu_A \nn +\tfrac1{28} \acute{B}_{\mu\nu}^{+AB} \acute{\tilde{E}}_{AB}^{\mu\nu}  +2 B^+_{\mu\nu} \tilde{E}^{\mu\nu} +  h_{\mu\nu}^{+A} \tilde{E}_A^{\mu\nu}  \\
&\hspace{10mm} + X_\mu \tilde{E}^\mu +\tfrac12  X_{\mu\nu}^A \tilde{E}_A^{\mu\nu} +   \ldots \,,
\end{align}
to compute its transformation under $\mathfrak{e}_{11}$ generators at level $\ell=\pm1$ in the $\mf{gl}(3)\oplus \mf{e}_8$ decomposition, defined by
\begin{align}
\label{eq:phi11trmE8}
\delta \phi^{\widehat\alpha} \bar t_{\widehat\alpha} =  \left[e_\mu^A E^{\mu}_A +f^\mu_A F_{\mu}^A ,\,  \phi^{\widehat\alpha} \bar{t}_{\widehat\alpha} \right]\,.
\end{align}
One computes within this level truncation that 
\begin{subequations}
\label{eq:adjointE8} 
\begin{align}
\delta h^+_\mu{}^\nu &= e_\mu^A A_A^{-\nu} - f^\nu_A A^+_\mu{}^A - \delta_\mu^\nu \bigl( e_\sigma^A  A_A^{-\sigma} - f^\sigma_A A^{+A}_\sigma \bigr) \,,\\ 
 \delta \Phi^+_A &= f_{AB}{}^C e_\mu^B A^{-\mu}_C - f_{AB}{}^C f^\mu_C A_\mu^{+B} \,,\\
  \delta A_\mu^{+A} &= - e_\nu^A h^+_\mu{}^\nu + f^{CA}{}_B e_\mu^B \Phi^+_C - f^\nu_B B^+_{\mu\nu}{}^{AB} - f^{AB}{}_C f^\nu_B h^+_{\mu\nu}{}^C \,, \\
    \delta B_{\mu\nu}^{+AB}  &= 28 P^{AB}{}_{CD}  e_{[\mu}^C A^+_{\nu]}{}^D + \tfrac12 \kappa^{AB} \kappa_{CD} e_{[\mu}^C A^+_{\nu]}{}^D   \,, \\
        \delta h_{\mu\nu}^{+A}  &= -  f_{BC}{}^A e_{(\mu}^B A^+_{\nu)}{}^C\,.
\end{align}
\end{subequations}
for the elements in $\mathfrak{e}_{11}$, and 
 \begin{align}
  \delta X_\mu &= f^\nu_A ( X_{\mu\nu}^A + h^+_{\mu\nu}{}^A ) +  e_\mu^A \Phi^+_A \ ,
 \nn\\ 
 \delta  X_{\mu\nu}^A &= - 2 e_{[\mu}^A X_{\nu]} -  f^A{}_{BC} e_{[\mu}^B A^+_{\nu]}{}^C \ , 
\end{align}
 for the elements in $X^{\tilde{\alpha}}$ in  $\overline{R}(\Lambda_2)$. 
 
 The gauge transformations are likewise defined from the parameters 
 \bea  
 \label{DerE8} P^M \partial_M &=& P^\mu \partial_\mu   + F^A \partial_A + \tfrac1{14} \acute{F}_{\mu}^{AB} \acute{\partial}_{AB}^{\mu}  +4 F_{\mu} \partial_\mu + 2 G_\mu^{A} \partial_A^\mu   + \dots \ ,
  \nn \\ 
\xi^M  \bar P_M  &=& \xi^\mu \bar P_\mu    +  \lambda^A \bar F_A  + \tfrac1{14} \acute{\lambda}_{\mu}^{AB} \acute{\bar F}_{AB}^{\mu}  +4 \lambda_{\mu} \bar F^\mu + 2 \xi_\mu^{A} \bar G_A^\mu   + \dots\ ,
 \eea
by 
\be \delta_\xi^+ \phi^{\widehat\alpha} \bar{t}_{\widehat\alpha}  = \partial_M \xi^N \{ P^M , \bar P_N\} \; , \ee
which gives 
 \bea \delta_\xi^+ B^{-\mu\nu}_{AB} &=& -2\partial_{AB}^{[\mu}  \xi^{\nu]} + \dots \ ,
 \nn\\
  \delta_\xi^+  h^{-\mu\nu}_A &=& 2\partial^{(\mu}_A \xi^{\nu)} + \dots \ ,
  \nn\\
  \delta_\xi^+   A^{-\mu}_A &=& \partial_A \xi^\mu  + \partial_{AB}^\mu \lambda^B +f_{AB}{}^C \partial_C^\mu \lambda^B  + \dots \ ,
  \nn\\
  \delta_\xi^+  h^+_\mu{}^\nu &=& \partial_\mu \xi^\nu - \tfrac1{14} \acute{\partial}^\nu_{AB} \acute{\lambda}_\mu^{AB} - 4 \partial^\nu \lambda_\mu + 2 \partial^\nu_A \xi_\mu^A + \delta_\mu^\nu \Bigl( \partial_A \lambda^A+ \tfrac17  \acute{\partial}^\sigma_{AB} \acute{\lambda}_\sigma^{AB}+8 \partial^\sigma \lambda_\sigma   \Bigr) + \dots \ ,
  \nn \\
  \delta_\xi^+  \Phi^+_A &=& - f_{AB}{}^C \partial_C \lambda^B- \tfrac17 f_{AB}{}^C \partial_{CD}^\mu \lambda_\mu^{BD} + 2 f_{AB}{}^C \partial_C^\mu \xi_\mu^B  + \dots \nn\\
  \delta_\xi^+  A^+_\mu{}^A &=& \partial_\mu \lambda^A + \partial_B \lambda^{AB}_\mu - f^{AB}{}_C \partial_B \xi_\mu^C  + \dots \nn\\
  \delta_\xi^+  B_{\mu\nu}^{+AB} &=& 2 \partial_{[\mu} \lambda_{\nu]}^{AB}  + \dots\ ,
   \nn\\
  \delta_\xi^+  h_{\mu\nu}^{+A} &=& 2 \partial_{(\mu} \xi^A_{\nu)}  + \dots
  \label{GaugeStrucCoeff} 
 \eea
 for the $\mathfrak{e}_{11}$ fields and  for the fields in  $\overline{R}(\Lambda_2)$ we obtain
 \begin{align}
  \delta_\xi^+ X_\mu = 2 \partial_A \xi^A_\mu  + \dots \ ,
\hspace{10mm} \delta_\xi^+ X_{\mu\nu}^A = - 2  \partial_{[\mu} \xi^A_{\nu]}  + \dots \ .
 \end{align}

One also defines the following invariant bilinear forms. The Killing--Cartan form expands as 
\begin{multline} \kappa^{\alpha\beta} \Phi^+_\alpha \Phi^+_\beta = h^+_\mu{}^\nu h^+_\nu{}^\mu - \tfrac12 h^+_\mu{}^\mu h^+_\nu{}^\nu + \kappa^{AB} \Phi^+_A \Phi^+_B + 2  A^{-\mu}_A A_\mu^{+A} \\ + \frac1{14} \acute{B}^{-\mu\nu}_{AB} \acute{B}_{\mu\nu}^{+AB} + 4 B^{-\mu\nu} B^+_{\mu\nu} + 2 h^{-\mu\nu}_A h_{\mu\nu}^{+A} + \dots \end{multline}
One can also check the $K(E_{11})$ invariant bilinear form on $R(\Lambda_1)$ and $\mathcal{T}_{-1}$ respectively expand as
\begin{multline} \eta^{AB} \partial_A \partial_B  =\eta^{\mu\nu} \partial_\mu \partial_\nu + \delta^{AB} \partial_A \partial_B  + \frac{1}{14} \delta^{AC} \delta^{BD} \eta_{\mu\nu} \acute{\partial}^\mu_{AB}  \acute{\partial}^\nu_{CD} \\ + 4 \eta_{\mu\nu}\partial^\mu \partial^\nu + 2 \delta^{AB} \eta_{\mu\nu} \partial^\mu_A \partial^\nu_B + \dots  \end{multline}
and 
 \begin{multline} 
 \eta_{IJ} F^I F^J = \tfrac12  \eta^{\mu\rho} \eta^{\nu\lambda}  \eta_{\sigma\kappa} F_{\mu\nu}{}^\sigma F_{\rho\lambda}{}^\kappa -  \eta^{\mu\nu} F_{\mu\sigma}{}^\sigma F_{\nu\rho}{}^\rho + \eta^{\mu\nu} \delta^{AB} F_{\mu A} F_{\nu B} \\ + \tfrac12 \eta^{\mu\sigma} \eta^{\nu\rho} \delta_{AB} F_{\mu\nu}^A F_{\sigma\rho}^B + \eta^{\mu\sigma} \eta^{\nu\rho} F_{\mu;\nu} F_{\rho;\sigma}  \\+ \tfrac1{14} \delta^{AC} \delta^{BD} \acute{F}_{AB} \acute{F}_{CD} + 4 F^2 +  \delta^{AB} ( F_{\mu}{}^\nu_A F_{\nu}{}^\mu_B -F_{\mu}{}^\mu_A F_{\nu}{}^\nu_B ) + \dots  \ ,
 \end{multline}
 with $F_{AB} = \acute{F}_{AB}  + \kappa_{AB} F $ with $ \acute{F}_{AB}$ in the $\bf 3875$. We have also the $E_{11}$ invariant symplectic form 
 \be 
 \Omega_{IJ} F^I G^J = \tfrac12 
 \varepsilon^{\mu\nu\sigma} \bigl( F_{\mu\nu}^A G_{\sigma A} + F_{\mu\nu}{}^\rho G_{\rho;\sigma} - 2 F_{\sigma\rho}{}^\rho G_{\mu;\nu}-G_{\mu\nu}^A F_{\sigma A} -G_{\mu\nu}{}^\rho F_{\rho;\sigma} + 2 G_{\sigma\rho}{}^\rho F_{\mu;\nu}+ \dots  \bigr) \ .
 \ee

\section{Variations of  the fermionic bilinears under \texorpdfstring{$K(E_{11})$}{KE11}}
\label{BilinearVariations} 

In this appendix, we tabulate the variations of various fermionic bilinears under the $\widetilde{K}(E_{11})$ transformations~\eqref{LV}. The $Spin(1,10)$ $\Gamma$-matrices $\Gamma_a$ that we are using satisfy the duality relation
\begin{align}
\label{gammadual}
 \Gamma_{a_1...a_n} = \frac{ (-1)^{\frac{n(n-1)}{2}} }{(11-n)!}\,  \varepsilon_{a_1...a_n}{}^{a_{n+1}...a_{11}}\,   \Gamma_{a_{n+1}...a_{11}} \ .
\end{align}
Using the charge conjugation matrix $\mathcal{C}= i \Gamma^0$ for Majorana spinors, we have that the combinations $\mathcal{C}\Gamma^a$, $\mathcal{C} \Gamma^{a_1a_2}$ and $\mathcal{C}\Gamma^{a_1\ldots a_5}$ are symmetric in their spinor indices, meaning that for anti-commuting spinors
\begin{align}
\bar{\epsilon}_1 \epsilon_2 =  + \bar{\epsilon}_2 \epsilon_1\,,\quad
\bar{\epsilon}_1 \Gamma^a \epsilon_2 =  - \bar{\epsilon}_2 \Gamma^a \epsilon_1\,,\quad
\bar{\epsilon}_1 \Gamma^{a_1a_2} \epsilon_2 =  - \bar{\epsilon}_2 \Gamma^{a_1a_2} \epsilon_1\,, \textrm{etc.}
\end{align}
For calculating the explicit variations below we made use of the gamma package~\cite{Gran:2001yh}.

\subsection{Bilinears of the form \texorpdfstring{$\bar{\epsilon} \psi_a$}{epsia}}

{\allowdisplaybreaks
Under $\widetilde{K}(E_{11})$ as defined infinitesimally in~\eqref{LV} we find
\begin{align}
\delta\left(\beps \psi_a\right) &= \frac16\Lambda_{c_1c_2c_3} \left( -\beps \Gamma^{c_1c_2c_3}\psi_a + \beps\Gamma_a{}^{c_1c_2} \psi^{c_3} -4 \delta_a^{c_1} \beps\Gamma^{c_2} \psi^{c_3}\right)\ ,
\w2
\delta( \beps \Gamma_b \psi_a ) &= \frac16\Lambda_{c_1c_2c_3} \Big(-\beps\Gamma_{ab}{}^{c_1c_2} \psi^{c_3}   -2\,\delta_b^{c_1} \beps\Gamma_a{}^{c_1} \psi^{c_3} - 4\, \delta_a^{c_1} \beps\Gamma_b{}^{c_2} \psi^{c_3} 
\nn\\
&\quad -3\, \delta_b^{c_1} \beps\Gamma^{c_2c_3} \psi_a +\eta_{ab} \beps\Gamma^{c_1c_2} \psi^{c_3} -4  \delta_a^{c_1}\delta_b^{c_2} \beps\psi^{c_3} \Big)\ ,
\w2
\delta( \beps \Gamma_{b_1b_2} \psi_a ) &= \frac16\Lambda_{c_1c_2c_3} \Big( \beps\Gamma_{ab_1b_2}{}^{c_1c_2} \psi^{c_3} - \beps\Gamma_{b_1b_2}{}^{c_1c_2c_3} \psi_a +4\, \delta_{b_1}^{c_1} \beps\Gamma_{b_2a}{}^{c_2} \psi^{c_3}
\nn\\*
& \quad -4\,\delta_a^{c_1} \beps\Gamma_{b_1b_2}{}^{c_2} \psi^{c_3}  -2\, \eta_{ab_1} \beps\Gamma_{b_2}{}^{c_1c_2} \psi^{c_3} + 6\,\delta_{b_1}^{c_1} \delta_{b_2}^{c_2}\,\beps\Gamma^{c_3} \psi_a  
\nn\\*
& \quad -2\, \delta_{b_1}^{c_1} \delta_{b_2}^{c_2}\, \beps\Gamma_a \psi^{c_3} +8\, \delta_{a}^{c_1}\delta_{b_1}^{c_2}\,\beps \Gamma_{b_2} \psi^{c_3}
-4\, \eta_{ab_1} \delta_{b_2}^{c_1}\, \beps\Gamma^{c_2} \psi^{c_3}  \Big)\Big\vert_{[b_1b_2]}\ ,\w2
\delta( \beps \Gamma_{b_1b_2b_3} \psi_a ) &= \frac16\Lambda_{c_1c_2c_3} \Big( -\beps\Gamma_{ab_1b_2b_3}{}^{c_1c_2} \psi^{c_3} + 3\,\eta_{ab_1}\,\beps \Gamma_{b_2b_3}{}^{c_1c_2} \psi^{c_3} -4\,\delta_a^{c_1}\,\beps \Gamma_{b_1b_2b_3}{}^{c_2} \psi^{c_3} 
\nn\\
&\quad -6\, \delta_{b_1}^{c_1}\,\beps \Gamma_{b_2b_3a}{}^{c_2} \psi^{c_3} -9\, \delta_{b_1}^{c_1}\,\beps\Gamma_{b_2b_3}{}^{c_2c_3} \psi_a -6\,\delta_{b_1}^{c_1} \delta_{b_2}^{c_2}\,\beps \Gamma_{b_3a} \psi^{c_3} 
\nn\\
& \quad  - 12\, \delta_a^{c_1} \delta_{b_1}^{c_2}\,\beps \Gamma_{b_2b_3} \psi^{c_3}  -12\, \eta_{ab_1}\,\delta_{b_2}^{c_1}\,\beps \Gamma_{b_3}{}^{c_2} \psi^{c_3} \nn\\
&\quad  -6\,\eta_{ab_1} \delta_{b_2}^{c_1}\delta_{b_3}^{c_3}\, \beps \psi^{c_3} +6\,\delta_{b_1}^{c_1} \delta_{b_2}^{c_2} \delta_{b_3}^{c_3}\, \beps \psi_a \Big)\Big\vert_{[b_1b_2b_3]}\ ,
\w2
\delta( \beps \Gamma_{b_1b_2b_3b_4} \psi_a ) &= \frac16\Lambda_{c_1c_2c_3} \Big(\beps \Gamma_{ab_1b_2b_3b_4}{}^{c_1c_2} \psi^{c_3} -\beps \Gamma_{b_1b_2b_3b_4}{}^{c_1c_2c_3} \psi_a +8\, \delta_{b_1}^{c_1}\, \beps \Gamma_{b_2b_3b_4a}{}^{c_2}\psi^{c_3}
\nn\\*
& \quad -4\, \delta_a^{c_1}\,\beps \Gamma_{b_1b_2b_3b_4}{}^{c_2} \psi^{c_3}  -4\, \eta_{ab_1}\,\beps \Gamma_{b_2b_3b_4}{}^{c_1c_2} \psi^{c_3}-12\,\delta_{b_1}^{c_1} \delta_{b_2}^{c_2}\,\beps \Gamma_{b_3b_4a} \psi^{c_3}
\nn\\*
&  \quad +16\, \delta_a^{c_1} \delta_{b_1}^{c_2}\, \beps \Gamma_{b_2b_3b_4} \psi^{c_3} -24\,\eta_{ab_1}\, \delta_{b_2}^{c_1}\, \beps \Gamma_{b_3b_4}{}^{c_2} \psi^{c_3} 
\nn\\*
& \quad  + 36\, \delta_{b_1}^{c_1} \delta_{b_2}^{c_2}\, \beps \Gamma_{b_3b_4}{}^{c_3} \psi_a +24\, \eta_{ab_1}\,\delta_{b_2}^{c_1} \delta_{b_3}^{c_2}\, \beps \Gamma_{b_4} \psi^{c_3} \Big) \Big\vert_{[b_1b_2b_3b_4]}\ ,
\w2
\delta( \beps \Gamma_{b_1b_2b_3b_4b_5} \psi_a ) &= \frac16\Lambda_{c_1c_2c_3} \Big(-\beps \Gamma_{ab_1b_2b_3b_4b_5}{}^{c_1c_2} \psi^{c_3} - 10\,\delta_{b_1}^{c_1}\,\beps \Gamma_{b_2b_3b_4b_5a}{}^{c_2} \psi^{c_3}\nn\\*
& \quad  -4\,\delta_a^{c_1}\,\beps \Gamma_{b_1b_2b_3b_4b_5}{}^{c_2} \psi^{c_3}
 +5\,\eta_{ab_1}\,\beps \Gamma_{b_2b_3b_4b_5}{}^{c_1c_2} \psi^{c_3}  -15\,\delta_{b_1}^{c_1}\,\beps \Gamma_{b_2b_3b_4b_5}{}^{c_2c_3} \psi_a \nn\\*
& \quad  -20\,\delta_{b_1}^{c_1} \delta_{b_2}^{c_2}\,\beps \Gamma_{b_3b_4b_5a} \psi^{c_3}
 -20\,\delta_a^{c_1}\delta_{b_1}^{c_2}\,\beps \Gamma_{b_2b_3b_4b_5} \psi^{c_3} -40\,\eta_{ab_1} \delta_{b_2}^{c_1} \beps \Gamma_{b_3b_4b_5}{}^{c_2} \psi^{c_3}
\nn\\*
& \quad -60 \eta_{ab_1} \delta_{b_2}^{c_1} \delta_{b_3}^{c_2} \beps \Gamma_{b_4b_5} \psi^{c_3}
+60\, \delta_{b_1}^{c_1} \delta_{b_2}^{c_3} \delta_{b_3}^{c_3}\,\beps \Gamma_{b_4b_5} \psi_a \Big) \Big\vert_{[b_1b_2b_3b_4b_5]}\ .
\end{align}
The bar notation in the above formula denotes (anti-)symmetrizations to be carried out on a tensor expression, e.g.  $T_{ab}|_{[ab]} = T_{[ab]} = \frac12(T_{ab}-T_{ba})$ and so on. The totally antisymmetric, hook symmetric and single traces can easily be obtained from the expressions given above. Furthermore, the variations of $\beps \Gamma_{b_1...b_n} \psi_a$ for $n\ge 6$ can be obtained from~\eqref{gammadual}.
}

\subsection{Bilinears in \texorpdfstring{$\psi_a \psi_b$}{psipsi}}

{\allowdisplaybreaks
The variations of the gravitino bilinears are
\begin{align}
\delta (\bar\psi_a \psi_b) &= -\frac16\,\Lambda_{d_1d_2d_3} \left( \bar\psi_a \Gamma^{d_1d_2d_3} \psi_b -2\bar\psi_a \Gamma_b{}^{d_1d_2} \psi^{d_3} + 8\,\delta_a^{d_1} \,\bar\psi_b\Gamma^{d_2} \psi^{d_3} \right)\Big\vert_{(ab)}\ ,
\w2
\delta (\bar\psi_a \Gamma_c \psi_b) &= \frac16\,\Lambda_{d_1d_2d_3} \Big( -2 \psi_a \Gamma_{bc}{}^{d_1d_2} \psi^{d_3} -3\,\delta_c^{d_1}\,\psi_a \Gamma^{d_2d_3} \psi_b + 8\,\delta_a^{d_1}\,\bar\psi_b \Gamma_c{}^{d_2} \psi^{d_3}
\nn\\
&\quad    -4\, \delta_c^{d_1}\,\bar\psi_a \Gamma_b{}^{d_2} \psi^{d_3}  -2\,\eta_{ac}\,\bar\psi_b \Gamma^{d_1d_2} \psi^{d_3} +8\,\delta_a^{d_1} \delta_c^{d_2}\,\bar\psi_b \psi^{d_3} \Big)\Big\vert_{[ab]}\ ,
\w2
\delta (\bar\psi_a \Gamma_{c_1c_2} \psi_b) &= \frac16\,\Lambda_{d_1d_2d_3} \Big( 2\,\bar\psi_a \Gamma_{bc_1c_2}{}^{d_1d_2} \psi^{d_3} -\bar\psi_a \Gamma_{c_1c_2}{}^{d_1d_2d_3} \psi_b +8\,\delta_a^{d_1}\,\bar\psi_b \Gamma_{c_1c_2}{}^{d_2}\psi^{d_3}
\nn\\*
&\quad  +8\, \delta_{c_1}^{d_1}\,\bar\psi_b \Gamma_{ac_2}{}^{d_2} \psi^{d_3} +4\,\eta_{ac_1}\,\bar\psi_b \Gamma_{c_2}{}^{d_1d_2} \psi^{d_3} +6\, \delta_{c_1}^{d_1}\delta_{c_2}^{d_2}\,\bar\psi_a \Gamma^{d_3} \psi_b
\\*
&\quad -16\,\delta_a^{d_1} \delta_{c_1}^{d_2}\,\bar\psi_b \Gamma_{c_2} \psi^{d_3} -4\, \delta_{c_1}^{d_1}\delta_{c_2}^{d_2}\,\bar\psi_a \Gamma_b \psi^{d_3} +8\,\eta_{ac_1}\,\delta_{c_2}^{d_1}\,\bar\psi_b \Gamma^{d_2} \psi^{d_3} \Big)\Big\vert_{[ab][c_1c_2]}\ ,\nn
\w2
\delta (\bar\psi_a \Gamma_{c_1c_2c_3} \psi_b) &= \frac16\,\Lambda_{d_1d_2d_3} \Big(-2\,\bar\psi_a \Gamma_{bc_1c_2c_3}{}^{d_1d_2} \psi^{d_3}  -9\,\delta_{c_1}^{d_1}\,\bar\psi_a \Gamma_{c_2c_3}{}^{d_2d_3} \psi_b 
\nn\\
&\quad -8\,\delta_a^{d_1}\,\bar\psi_b \Gamma_{c_1c_2c_3}{}^{d_2} \psi^{d_3} -12\,\delta_{c_1}^{d_1}\,\bar\psi_b \Gamma_{ac_2c_3}{}^{d_2} \psi^{d_3} +6\,\eta_{ac_1}\,\bar\psi_b \Gamma_{c_2c_3}{}^{d_1d_2} \psi^{d_3}
 \nn\\
&\quad -24\,\delta_a^{d_1} \delta_{c_1}^{d_2}\,\bar\psi_b \Gamma_{c_2c_3} \psi^{d_3} +12\,\delta_{c_1}^{d_1} \delta_{c_2}^{d_2}\,\bar\psi_b \Gamma_{ac_3} \psi^{d_3}
  -24\,\eta_{ac_1} \delta_{c_2}^{d_1}\,\bar\psi_b \Gamma_{c_3}{}^{d_2} \psi^{d_3} 
  \nn\\
&\quad   +6\,\delta_{c_1}^{d_1}\delta_{c_2}^{d_2}\delta_{c_3}^{d_3}\,\bar\psi_a\psi_b 
 -12\, \eta_{ac_1}\delta_{c_2}^{d_1}\delta_{c_3}^{d_2} \bar\psi_b \psi^{d_3} \Big)\Big\vert_{(ab)[c_1c_2c_3]}\ ,
\w2
\delta (\bar\psi_a \Gamma_{c_1c_2c_3c_4} \psi_b) &= \frac16\,\Lambda_{d_1d_2d_3} \Big(2\,\bar\psi_a \Gamma_{bc_1c_2c_3c_4}{}^{d_1d_2} \psi^{d_3} -\bar\psi_a \Gamma_{c_1c_2c_3c_4}{}^{d_1d_2d_3}\psi_b 
 \nn\\*
 &\quad -8\,\delta_a^{d_1}\,\bar\psi_b \Gamma_{c_1c_2c_3c_4}{}^{d_2} \psi^{d_3}  -16\,\delta_{c_1}^{d_1}\,\bar\psi_b \Gamma_{ac_2c_3c_4}{}^{d_2} \psi^{d_3}  -8\,\eta_{ac_1}\,\bar\psi_b \Gamma_{c_2c_3c_4}{}^{d_1d_2} \psi^{d_3}\nn\\*
&\quad  +36\,\delta_{c_1}^{d_1}\delta_{c_2}^{d_2}\,\bar\psi_a \Gamma_{c_3c_4}{}^{d_3} \psi_b +32\,\delta_a^{d_1}\delta_{c_1}^{d_2}\,\bar\psi_b \Gamma_{c_2c_3c_4} \psi^{d_3} 
-24\,\delta_{c_1}^{d_1}\delta_{c_2}^{d_2}\,\bar\psi_b \Gamma_{ac_3c_4} \psi^{d_3}
\nn\\*
&\quad  
+48\,\eta_{ac_1} \delta_{c_2}^{d_1}\,\bar\psi_b \Gamma_{c_3c_4}{}^{d_2} \psi^{d_3}
 +48\,\eta_{ac_1}\delta_{c_2}^{d_1}\delta_{c_3}^{d_2}\, \bar\psi_b \Gamma_{c_4} \psi^{d_3}
\Big)\Big\vert_{(ab)[c_1c_2c_3c_4]}\ ,
\w2
\delta (\bar\psi_a \Gamma_{c_1c_2c_3c_4c_5} \psi_b) &= \frac16\,\Lambda_{d_1d_2d_3} \Big(2\,\bar\psi_b \Gamma_{ac_1c_2c_3c_4c_5}{}^{d_1d_2} \psi^{d_3}-15\,\delta_{c_1}^{d_1}\,\bar\psi_a \Gamma_{c_2c_3c_4c_5}{}^{d_2d_3} \psi_b
\nn\\*
&\quad +8\,\delta_a^{d_1}\,\bar\psi_b \Gamma_{c_1c_2c_3c_4c_5}{}^{d_2} \psi^{d_3} +20\,\delta_{c_1}^{d_1}\,\bar\psi_b \Gamma_{ac_2c_3c_4c_5}{}^{d_2} \psi^{d_3} 
\nn\\*
&\quad -10\,\eta_{ac_1}\,\bar\psi_b \Gamma_{c_2c_3c_4c_5}{}^{d_1d_2} \psi^{d_3}+40\,\delta_a^{d_1} \delta_{c_1}^{d_2}\,\bar\psi_b \Gamma_{c_2c_3c_4c_5} \psi^{d_3}
\\*
&\quad -40\,\delta_{c_1}^{d_1} \delta_{c_2}^{d_2}\,\bar\psi_b \Gamma_{ac_3c_4c_5} \psi^{d_3}
+80\,\eta_{ac_1}\delta_{c_2}^{d_1}\,\bar\psi_b \Gamma_{c_3c_4c_5}{}^{d_2} \psi^{d_3} 
\nn\\*
&\quad +60\,\delta_{c_1}^{d_1} \delta_{c_2}^{d_2}\delta_{c_3}^{d_3}\,\bar\psi_a \Gamma_{c_4c_5} \psi_b  +120\,\eta_{ac_1}\delta_{c_2}^{d_1} \delta_{c_3}^{d_2}\,\bar\psi_b \Gamma_{c_4c_5} \psi^{d_3}
\Big)\Big\vert_{[ab][c_1c_2c_3c_4c_5]}\ .\nn
\end{align}
Here we have also used the symmetry properties of the Gamma matrices.
}

\section{ Eleven dimensional supergravity }
\label{11Dsugra} 

{\allowdisplaybreaks
Eleven dimensional supergravity equations in our conventions
\footnote{In this section only we shall denote $\psi_a^{11D}$ simply by $\psi_a$, thereby suppressing the $11D$ superscript  in order to avoid clutter in notation. Everywhere else in this paper $\psi_a$ is as defined in \eqref{11DF}. In our conventions $\bar\psi = \psi^\dagger i\Gamma_0,\  \{\Gamma_a,\Gamma_b\} =2\eta_{ab}$ with $\eta_{ab} ={\rm diag}(-,++...+),\  D_m(\omega) \epsilon  = 
\partial_m\epsilon +\frac14 \omega_m{}^{ab} \Gamma_{ab}\epsilon$, $\Gamma^{a_1...a_{11}} = -\epsilon^{a_1...a_{11}}$ and $R= e^m_a e^n_b R_{mn}{}^{ab}$. } 
are given by 
\bea
{\cal L}= e R(\omega)-\frac{1}{48} e F_{mnpq} F^{mnpq} - \frac{1}{144^2} \varepsilon^{m_1...m_{11}} F_{m_1...m_4} F_{m_5...m_8} A_{m_9m_{10}m_{11}}
\nn\\
+ e\bar\psi_m \Gamma^{mnp} D_n (\frac{\omega+\hat\omega}{2}) \psi_p  + \frac{1}{192} e \bar\psi_{[r} \Gamma^r\Gamma^{mnpq} \Gamma^s \psi_{s]}
 \left(F_{mnpq} + \widehat F_{mnpq} \right)\ ,
\label{CSJ}
\eea  
where
\bea
\widehat\omega_{mab} &=& \omega^{(0)}_{m ab} -\frac14 (\bar\psi_m\Gamma_a\psi_b -\bar\psi_m\Gamma_b\psi_a 
+\bar\psi_a\Gamma_m\psi_b )\ ,
\nn\\
\omega_{m ab} &=& \widehat\omega_{m ab} -\frac{1}{8}\bar\psi_n \Gamma_{m ab}{}^{np} \psi_p\ ,
\nn\\
F_{mnpq} &=& 4 \partial_{[m} A_{npq]}\ ,\qquad \widehat F_{mnpq} = F_{mnpq} -3 \bar\hat\psi_{[m} \Gamma_{np} \psi_{q]}\ ,
\eea
and $ \omega^{(0)}_{m ab} $ is the spin connection without torsion. The local supersymmetry transformations are
\bea
\delta e_m{}^a &=& -\frac12 \bar\epsilon \Gamma^a \psi_m\ ,\qquad \delta A_{mnp} = \frac32 \bar\epsilon\Gamma_{[mn} \psi_{p]}\ ,
\nn\w2
\delta\psi_m &=& D_m (\widehat\omega)\epsilon +\frac{1}{288} \left(\Gamma_m{}^{npqr} -8\delta_m^n \Gamma^{pqr} \right) \widehat F_{npqr}\ .
\eea

For the purposes of this paper, it is of interest to identify the quartic fermion terms coming from different sources. These are 
\be
{\cal L}_{\psi^4} = { \cal L}^{\rm EH} _{\psi^4} + { \cal L}^{\rm RS} _{\psi^4} + {\cal L}^{\rm Pauli}_{\psi^4}\ ,
\ee
where
\bea
e^{-1}  { \cal L}^{\rm EH} _{\psi^4}  &=& \frac{1}{16} \bar\psi_c \Gamma_a\psi_b \left( \bar\psi^c \Gamma^a\psi^b -2\bar\psi^b \Gamma^c \psi^a\right) -\frac14 \bar{\psi}_b \Gamma^b \psi_a \bar\psi_c \Gamma^c \psi^a
 \nn\\
 &&-\frac{1}{64} \bar\psi_d\Gamma^{cabde}\psi_e \left( \bar\psi_f\Gamma^{cabfg}\psi_g  + 4 \bar\psi^c \Gamma^a\psi^b \right)\ ,
 \w2
e^{-1}  { \cal L}^{\rm RS} _{\psi^4} &=&- \frac{1}{8} \bar\psi_c \Gamma_a\psi_b \left( \bar\psi^c \Gamma^a\psi^b -2\bar\psi^b \Gamma^c \psi^a\right) +\frac12 \bar{\psi}_b \Gamma^b \psi_a \bar\psi_c \Gamma^c \psi^a
 \nn\\
 &&+ \frac{1}{64} \bar\psi_d\Gamma^{cabde}\psi_e \left( \bar\psi_f\Gamma^{cabfg}\psi_g  + 6 \bar\psi^c \Gamma^a\psi^b \right)\ ,
 \w2
e^{-1} {\cal L}^{\rm Pauli}_{\psi^4} &=& -\frac{1}{64} \bar\psi_{[a}\Gamma_{bc} \psi_{d]} \left( \bar\psi_e \Gamma^{abcdef}\psi_f +12 \bar\psi^a \Gamma^{bc} \psi^d\right)\ .
\eea
 Summing these up yields
\bea
e^{-1} {\cal L}_{\psi^4} &=& -\frac{1}{32} \bar\psi^c \Gamma^a \psi^b \left( 2\bar\psi_c\Gamma_a\psi_b -  4\bar\psi_b\Gamma_c\psi_a - 8\eta_{ca} \bar\psi_d\Gamma^d\psi_b - \bar\psi_d\Gamma^{abcde}\psi_e  \right) 
\nn\\
&& -\frac{1}{64} \bar\psi_{[a} \Gamma_{bc} \psi_{d]} \left( \bar\psi_e \Gamma^{abcdef} \psi_f +12  \bar\psi^a \Gamma^{bc} \psi^d \right)\ .
\eea
}

\section{ The gauge parameter representation }
\label{GaugeParaSpinor} 

The gauge algebra suggests an embedding of the bilinear spinor in the gauge parameter representation. Recalling \eqref{eq:KE11der},  we want to check this embedding. The representation satisfies 
\bea 
\delta \left(\bar \epsilon\, \Gamma^a \epsilon \right) &=& - \tfrac12 \Lambda^{abc} \bar \epsilon\,  \Gamma_{bc} \epsilon\ ,
 \nn\w2
\delta \left( \bar \epsilon\,  \Gamma^{ab} \epsilon \right) &=& - \tfrac16 \Lambda_{cde} \bar \epsilon\,  \Gamma^{abcde} \epsilon + \Lambda^{abc} \bar \epsilon\,  \Gamma_c \epsilon \ ,
\w2
\delta \left(\bar \epsilon\,  \Gamma^{a_1a_2a_3a_4a_5} \epsilon\right)  &=& 10 \Lambda^{[a_1a_2a_3} \bar \epsilon\,  \Gamma^{a_4a_5]}  \epsilon - \tfrac1{48} \Lambda^{b_1b_2[a_1} \varepsilon_{b_1b_2}{}^{a_2a_3a_4a_5]c_1c_2c_3c_4c_5} \bar \epsilon\,  \Gamma_{c_1c_2c_3c_4c_5} \epsilon \ .
\nn
\eea
At this level truncation one finds a consistent embedding with 
\bea \label{GaugeParameter} 
\xi^a &=& \bar \epsilon\,  \Gamma^a \epsilon \ ,
\nn\\
\lambda_{ab} &=& - \bar \epsilon\,  \Gamma_{ab} \epsilon \ ,
\nn\\
\lambda_{a_1a_2a_3a_4a_5} &=& \bar \epsilon\,  \Gamma_{a_1a_2a_3a_4a_5} \epsilon \ ,
\nn\\
\xi_{a_1\dots a_7,b} &=& - \frac{7}{5!} \eta_{b[a_1} \varepsilon_{a_2\dots a_7]c_1c_2c_3c_4c_5} \bar \epsilon\,  \Gamma^{c_1c_2c_3c_4c_5} \epsilon \ , 
\nn\\
\lambda_{a_1\dots a_8} &=& 0 \ ,
\nn\\
\lambda_{a_1\dots a_8,b_1b_2b_3} &=& -42 \eta_{b_1[a_1}\eta_{|b_2|a_2} \eta_{|b_3|a_3}\bar \epsilon\,  \Gamma_{a_4a_5a_6a_7a_8]} \epsilon \ ,
\nn \\
\xi_{a_1\dots a_{10},b} &=& \frac72 \delta_{b[a_1} \varepsilon_{a_2\dots a_{10}]c_1c_2} \bar \epsilon\,  \Gamma^{c_1c_2} \epsilon\ , 
\nn\\
\xi_{a_1\dots a_{9},b,c} &=& - \frac{21}{20}\bigl( \eta^{bc} \varepsilon^{a_1\dots a_9d_1d_2} + \eta^{b[a_1}  \varepsilon^{a_2\dots a_9]cd_1d_2}+ \eta^{c[a_1}  \varepsilon^{a_2\dots a_9]bd_1d_2} \bigr)  \bar \epsilon\,  \Gamma^{d_1d_2} \epsilon \ ,
\nn\\
\xi^\prime_{a_1\dots a_{10},b} &=& 0 \ , 
\eea
and two the other parameters $\xi_{9,2}$ and $\lambda_{11}$ at this level vanish. One checks indeed that $\xi^a$, $\lambda_{a_1a_2}$ and $\lambda_{a_1a_2a_3a_4a_5}$ transform as in \eqref{eq:KE11der}, while $\xi_{a_1\dots a_7,b}$ and $\lambda_{a_1\dots a_8}$ do as well provided on defines  the additional variation 
\bea 
\delta \xi_{a_1\dots a_7,b} &=&-\frac{105}{8} \left(  \Lambda_{[ a_1a_2a_3} \lambda_{a_4a_5a_6a_7]b} + \Lambda_{b[ a_1a_2} \lambda_{a_3a_4a_5a_6a_7]} \right) - 4 \Lambda^{c_1c_2c_3} \lambda_{c_1\langle  a_1\dots a_7,b\rangle c_2c_3}
\nn\\
&& + \frac16 \Lambda^{c_1c_2c_3} \xi_{c_1c_2c_3\langle a_1\dots a_7,b\rangle} + \frac12 \Lambda^{c_1c_2c_3} \xi_{c_1c_2\langle a_1\dots a_7,b\rangle,c_3} \ ,
\nn\\
 \delta \lambda_{a_1\dots a_8} &=&7  \Lambda_{[ a_1a_2a_3} \lambda_{a_4a_5a_6a_7a_8]}  + \tfrac16  \Lambda^{c_1c_2c_3} \lambda_{ a_1\dots a_8,c_1c_2c_3}  + \tfrac12 \Lambda^{c_1c_2c_3} \xi^\prime_{a_1\dots a_8c_1c_2,c_3}\ . 
 \eea
In particular one has 
\bea 
\delta \xi^{a_1\dots a_7,b}  &=& - \frac{7}{12} \eta^{b[a_1} \varepsilon^{a_2\dots a_7]c_1c_2c_3d_1d_2} \Lambda_{c_1c_2c_3}\bar \epsilon\,  \Gamma_{d_1d_2} \epsilon - 105 \eta^{b[a_1} \Lambda^{a_2a_3}{}_c \bar \epsilon\,  \Gamma^{a_4a_5a_6a_7]c} \epsilon 
\nn \\
&=&- \frac{7}{12} \eta^{b[a_1} \varepsilon^{a_2\dots a_7]c_1c_2c_3d_1d_2} \Lambda_{c_1c_2c_3}\bar \epsilon\,  \Gamma_{d_1d_2} \epsilon - 168 \Lambda_{c_1c_2c_3}  \eta^{\langle b,[a_1} \eta^{|c_1|a_2} \eta^{|c_2|a_3}   \bar \epsilon\,  \Gamma^{a_4a_5a_6a_7]c_3} \epsilon 
\nn\\
&& - \frac{105}{8} \bigl(  \Lambda^{[ a_1a_2a_3} \bar \epsilon\,  \Gamma^{a_4a_5a_6a_7]b} \epsilon + \Lambda^{b[ a_1a_2}  \bar \epsilon\, \Gamma^{a_3a_4a_5a_6a_7]} \epsilon \bigr) \ .  
\eea

Alternatively, one can consider the submodule parametrised by the combination
\bea \tilde{\xi}_{a_1\dots a_7,b} &=& {\xi}_{a_1\dots a_7,b} + \frac{7}{5!} \eta_{b[a_1} \varepsilon_{a_2\dots a_7]c_1\dots c_5} \lambda^{c_1\dots c_5} \ ,
\\
\tilde{\lambda}_{a_1\dots a_8,b_1b_2b_3} &=& {\lambda}_{a_1\dots a_8,b_1b_2b_3} +42 \eta_{b_1[a_1}\eta_{|b_2|a_2} \eta_{|b_3|a_3} \lambda_{a_4a_5a_6a_7a_8]} \ ,
\nn\\
\tilde{\xi}_{a_1\dots a_{10},b} &=&\xi_{a_1\dots a_{10},b}- \frac72 \delta_{b[a_1} \varepsilon_{a_2\dots a_{10}]c_1c_2}  \lambda^{c_1c_2}  \ , 
\nn\\
\tilde{\xi}_{a_1\dots a_{9},b,c} &=&{\xi}_{a_1\dots a_{9},b,c}  + \frac{21}{20}\bigl( \eta^{bc} \varepsilon^{a_1\dots a_9d_1d_2} + \eta^{b[a_1}  \varepsilon^{a_2\dots a_9]cd_1d_2}+ \eta^{c[a_1}  \varepsilon^{a_2\dots a_9]bd_1d_2} \bigr) \lambda^{d_1d_2} \ ,
\nn
\eea
and $\lambda_{a_1\dots a_8} $ and $\xi^\prime_{a_1\dots a_{10},b}$ and check that they vary into each others. One has for instance 
\bea 
\delta \tilde{\xi}_{a_1\dots a_7,b} &=& - 4 \Lambda^{c_1c_2c_3} \tilde{\lambda}_{c_1\langle  a_1\dots a_7,b\rangle c_2c_3}  + \frac{7}{2\cdot 5!} \eta_{b[a_1} \varepsilon_{a_2\dots a_7]c_1\dots c_5} \Lambda_{c_6c_7d} \tilde{ \xi}^{c_1\dots c_7,d} 
\\
&& + \frac16 \Lambda^{c_1c_2c_3} \tilde{\xi}_{c_1c_2c_3\langle a_1\dots a_7,b\rangle} +\! \frac12 \Lambda^{c_1c_2c_3} \tilde{\xi}_{c_1c_2\langle a_1\dots a_7,b\rangle,c_3}  +\! \frac{7}{6!}\eta_{b[a_1} \varepsilon_{a_2\dots a_7]c_1\dots c_5} \Lambda_{c_6c_7c_8} \lambda^{c_1\dots c_8} ,
\nn  \\
\delta \lambda_{a_1\dots a_8} &=&  \frac16  \Lambda^{c_1c_2c_3} \tilde{\lambda}_{ a_1\dots a_8,c_1c_2c_3}  + \frac12 \Lambda^{c_1c_2c_3} \xi^\prime_{a_1\dots a_8c_1c_2,c_3}\ ,
\nn
\eea 
up to terms in $\xi_{9,2}$ and $\lambda_{11}$ that we did not compute and are part of the invariant subspace. 

The above results  exhibit to this level truncation that there is an invariant subspace in  $R(\Lambda_1)$ such that the associated quotient is the symmetric $SL(32)$ tensor representation of  $K(E_{11})$ obtained from the symmetric bilinear. A complete proof seems out of reach.

\section{\texorpdfstring{$\widetilde{K}(E_{11})$}{K(E11)} fermions under \texorpdfstring{$Spin(1,3) \times SU(8)$}{Spin(1,3)xSU(8)}} 
\label{E7susy} 

In this appendix, we perform the decomposition of the $\widetilde{K}(E_{11})$ spinors under $Spin(1,3) \times SU(8)$ associated with exceptional field theory in four external dimensions. We use this to probe the action of generalised diffeomorphisms discussed in~\eqref{eq:liePsi} and compare with results available in the literature for the supersymmetric $E_7$ exceptional field theory~\cite{Godazgar:2014nqa,Butter:2018bkl}.

The $K(\mathfrak{e}_{11})$ level 1 variation in the $\mf{gl}(4)\oplus \mf{e}_7$ decomposition  is defined from the generators 
\be
 \Lambda^{ij a} ( E_{a\, ij}  + \eta_{ab} F^b{}_{ij} ) - \Lambda_{ij a} ( E^{a\, ij} + \eta^{ab} F_b{}^{ij} )\; , 
 \ee
such that the Fermi fields transform as follows\footnote{We introduce the real parameter $z$ to compare with other conventions.}
\bea
\label{ke11E7}  \delta \psi_a^i &=& 2 i  \Lambda^{ijb} \gamma_{[a} \psi_{b] j}  +\frac{z i }{2} \Lambda_{jkb} \gamma^b \gamma_a \chi^{ijk} \ , \CR
\delta \chi^{ijk} &=&  \frac{3}{z}  i \Lambda^{[ij|a} \psi_a^{k]} + \frac{i}{12} \varepsilon^{ijklpqrs} \Lambda_{lp a} \gamma^a \chi_{qrs}\  .  
\eea
One can check the closure of the $K(\mathfrak{e}_{11})$ algebra by treating $\Lambda^{ij a}$ as a Grassmann odd parameter such that the commutator is the square of the variation
\bea 
\delta^2 \psi_a^i &=&  i z \varepsilon_{a}{}^{bcd} \gamma_d  \bigl( \Lambda^{[ij}{}_b \Lambda^{kl]}{}_c + \frac{1}{12} \varepsilon^{ijklpqrs} \Lambda_{pq b} \Lambda_{rs c} \bigr) \chi_{jkl} \CR
 &&+ \Bigl( \Lambda^{ik [b} \Lambda_{jk}{}^{c]} - \tfrac18 \delta^i_j \Lambda^{kl [b} \Lambda_{kl}{}^{c]} \Bigr) \bigl( 4 \gamma_{ab} \psi_c^j  + \gamma_{bc} \psi_a^j \bigr) + \frac12 \Lambda^{kl(b} \Lambda_{kl}{}^{c)} \bigl( - \gamma_b \gamma_a \psi_c^i + \tfrac14 \eta_{bc} \psi_a^i \bigr) \CR
 && +\Bigl( \Lambda^{ik b} \Lambda_{jkb} - \tfrac18 \delta^i_j \Lambda^{kl b} \Lambda_{klb} \Bigr)  \psi_a^j + \frac12 \Lambda^{kl[b} \Lambda_{kl}{}^{c]} \bigl( \eta_{ab} \psi_c^i + \tfrac14 \gamma_{bc}  \psi_a^i \bigr) 
 \eea
where the last line reproduces an $\mathfrak{so}(1,3) \oplus \mathfrak{su}(8)=K(\mf{gl}(4)\oplus \mf{e}_7)$ transformation, while the others appear at level 2 in $K(\mathfrak{e}_{11})$. Indeed, level decomposition predicts at level 2~\cite{Riccioni:2007au,Bergshoeff:2007qi}: a rank 2 antisymmetric tensor of $SL(4)$ in the ${\bf 133}$ of $E_7$ that branches under $SU(8)$ as a complex self-dual rank 2 antisymmetric tensor and a anti-hermitian traceless tensor, and a symmetric tensor of $SL(4)$ in the singlet representation of $E_7$. Similarly 
\bea 
\delta^2 \chi^{ijk}  &=&  \frac{6}{z}  \bigl( \Lambda^{[ij}{}_a \Lambda^{kl]}{}_b +\frac{1}{12} \varepsilon^{ijklpqrs} \Lambda_{pq a} \Lambda_{rs b} \bigr) \gamma^{[a} \psi^{b]}_{l} \CR
& &-3  \Bigl( \Lambda^{[i|p [b} \Lambda_{lp}{}^{c]} - \tfrac18 \delta^{i|}_l \Lambda^{pq [b} \Lambda_{pq}{}^{c]} \Bigr) \gamma_{bc} \chi^{jk]l}  - \frac18 \Lambda^{pqb} \Lambda_{pqb} \chi^{ijk}  \CR
 && +3 \Bigl( \Lambda^{[i|p b} \Lambda_{lpb} - \tfrac18 \delta^{[i}_l \Lambda^{pq b} \Lambda_{pqb} \Bigr) \chi^{jk]l}  + \frac18 \Lambda^{pq[b} \Lambda_{pq}{}^{c]} \gamma_{bc} \chi^{ijk}
 \eea
where the last line reproduces a $\mathfrak{so}(1,3) \oplus \mathfrak{su}(8)$ transformation, while the others appear at level 2 in $K(\mathfrak{e}_{11})$. This shows that the action~\eqref{ke11E7} of $K(\mf{e}_{11})$ on the fermions closes in this decomposition.

We next work out the general formula~\eqref{eq:liePsi} for the generalised diffeomorphism on the fermions with the partial solution of the section constraint corresponding to the $GL(4)\times E_7$ subgroup. In this decomposition one defines the $K(E_{11})$ spinor as $\Psi = ( e^{\frac14} e_a{}^\mu \psi_\mu^i ,  e^{\frac14} \chi_{ijk})$, and the compensating transformation $X$ of~\eqref{eq:kmpg} has only one parameter $\Lambda^{ija}$ at level 1   given by 
\be 
X^{ij a} =  -2 V^{A ij}e_\mu{}^a  \partial_A \xi^\mu \; . 
\ee
One obtains in this way from~\eqref{ke11E7} the external diffeomorphism 
\begin{align}
\delta_\xi  \psi^i_\mu &= \xi^\nu \partial_\nu \psi_\mu^i + \partial_\mu \xi^\nu \psi_\nu^i 
+4i V^{A  ij} \partial_A \xi^\nu \gamma_{a} e_{[\mu}{}^a  \psi_{\nu]}^j - z i V^{A}{}_{ij} e_\mu{}^a e_{\nu}{}^b  \partial_A \xi^\nu  \gamma_b \gamma_a \chi^{ijk}\,,\nn\\
\delta_\xi  \chi^{ijk} &= \xi^\mu \partial_\mu  \chi^{ijk}  -\frac{6}{z} i  V^{A  [ij}  \partial_A \xi^\mu  \psi_{\mu}{}^{k]} - \frac{i}{6} \varepsilon^{ijklpqrs}   V^{A}{}_{lp}    e_{\mu a} \partial_A \xi^\mu  \gamma^a \chi_{qrs}\,.
\end{align}
If we put $z =  \frac1{\sqrt{2}}$ this is in complete agreement with~\cite[Eq.~(3.1)]{Butter:2018bkl}.\footnote{Note that the appearence of covariant derivatives in  \cite{Butter:2018bkl} is  because they take a gauge parameter $\xi^M = ( \xi^\mu , \xi^\mu A_\mu^A , \dots )$ whereas we take $\xi^M = ( \xi^\mu , 0 , \dots )$.}

The spinor representation~\eqref{LV2} of $\widetilde{K}(E_{11})$ in the present decomposition becomes
\be 
\delta \epsilon^i =- i  \Lambda^{ij a} \gamma_a \epsilon_j \; . 
\ee
It closes according to 
\begin{align} 
\delta^2 \epsilon^i &= \frac{1}{8} \Lambda^{kl a} \Lambda_{kl }{}^{b} \gamma_{ab}  \epsilon^i + \Bigl( \Lambda^{ik a} \Lambda_{jka} - \tfrac18 \delta^i_j \Lambda^{kl a} \Lambda_{kla} \Bigr)  \epsilon^j  \nn \\ 
& \qquad +  \frac{1}{8} \Lambda^{kl a} \Lambda_{kl a} \epsilon^i  + \Bigl( \Lambda^{ik a} \Lambda_{jk}{}^b - \tfrac18 \delta^i_j \Lambda^{kl a} \Lambda_{kl}{}^b \Bigr) \gamma_{ab}  \epsilon^j \; , 
\end{align}
where the second line corresponds to level 2 generators.

Generalising~\eqref{Xi}, one obtains moreover for the components in $\mathfrak{e}_{11}\ominus K(\mathfrak{e}_{11})$ 
\bea \label{susyVarE7} 
\Xi_{ab} &=&  \overline{ \epsilon}_i \gamma_{(a} \psi_{b)}^i  + \overline{ \epsilon}^i \gamma_{(a} \psi_{b)i }  \ , \CR
\Xi^{ijkl} &=& \overline{\epsilon}^{[i} \chi^{jkl]} + \frac1{24} \varepsilon^{ijklpqrs} \overline{\epsilon}_p \chi_{qrs} \ , \CR
\Xi^{ij}_a &=& \overline{\epsilon}^{[i}  \psi^{j]}_a + \frac{z}{2} \overline{\epsilon}_k \gamma_a \chi^{ijk} \ , \CR
\Xi^{ijkl}_{ab} &=&  \overline{\epsilon}^{[i} \gamma_{ab} \chi^{jkl]} - \frac{1}{24} \varepsilon^{ijklpqrs} \overline{\epsilon}_p \gamma_{ab} \chi_{qrs}  \ , \CR
\Xi_{ab}{}^{\hspace{-1mm} i}{}_j &=&  \overline{\epsilon}^i \gamma_{[a} \psi_{b] j} + \overline{\epsilon}_j \gamma_{[a} \psi_{b]}^i- \frac{1}{8} \delta_j^i \bigl(\overline{\epsilon}^k \gamma_{[a} \psi_{b] k} +\overline{\epsilon}_k \gamma_{[a} \psi_{b]}^k \bigr)  \ , \CR
\Xi^\prime_{a,b} &=& i \,  \overline{ \epsilon}_i \gamma_{(a} \psi_{b)}^i - i  \, \overline{ \epsilon}^i \gamma_{(a} \psi_{b)i } \; , 
\eea
that transform under $K(\mf{e}_{11})$ as
\bea 
\delta \Xi_{ab} &=&  -2 i \Lambda_{ij (a} \Xi_{b)}^{ij} + i \eta_{ab} \Lambda_{ij}{}^c \Xi^{ij}_c +2 i \Lambda^{ij}{}_{(a} \Xi_{b) ij} -i \eta_{ab} \Lambda^{ij c} \Xi_{c ij }\ ,  \\
\delta \Xi^{ijkl}  &=& - \frac{3 i }{z} \Bigl(  \Lambda^{[ij|a}  \Xi^{kl]}_a  - \frac{1}{24} \varepsilon^{ijklpqrs}  \Lambda_{pq}{}^{a}  \Xi_{rs a} \Bigr) \  , \CR
\delta \Xi_a^{ij} &=& i z   \Lambda_{kl a} \Xi^{ijkl}  + \frac{i}{4} \Lambda^{ij b} \Xi_{ab} -  i z \Lambda_{kl}{}^b \Xi_{ab}^{ijkl} + 2 i \Lambda^{k[i} \Xi_{ab}{}^{\hspace{-0.4mm} j]}{}_k + \frac{1}{4} \Lambda^{ij b} \left(\Xi^\prime_{a,b} +\Upsilon_{ab}\right)\; . \nn  
 \eea
The extra bilinear $ \Upsilon_{ab} $ from the $R(\Lambda_2)$ module is
\be 
\Upsilon_{ab} =  i  \overline{ \epsilon}_i \gamma_{[a} \psi_{b]}^i -i   \overline{ \epsilon}^i \gamma_{[a} \psi_{b]i }  \ . 
\ee
One checks that it transforms into a vector in the ${\bf 28}$ that also belongs to $R(\Lambda_2)$ and not into the components  in $\mathfrak{e}_{11}\ominus K(\mathfrak{e}_{11})$, as required by the indecomposable structure of the module:
\be 
\delta \Upsilon_{ab} = i \Lambda^{ij c} \Upsilon_{abc\, ij} - i  \Lambda_{ij}{}^c \Upsilon_{abc}^{\hspace{2mm} ij} \ , 
\ee
with 
\be 
\Upsilon_{abc\, ij} = 3 i \overline{\epsilon}_{[i|} \gamma_{[ab} \psi_{c]| j]} +\frac{z }{2} \varepsilon_{abcd} \overline{\epsilon}^k \gamma^d \chi_{ijk} \; .
 \ee
This is indeed consistent with the supersymmetry transformations of $E_7$ exceptional field theory \cite[Eq. (3.32)]{Godazgar:2014nqa} with
\bea 
\dsusy h_{ab} &=& \Xi_{ab}  \ , \CR
\dsusy \phi^{ijkl} &=& 2 z \Xi^{ijkl} \ , \CR
\dsusy A_a^{ij} &=& -  i z   \Xi^{ij}_a \ , \CR
\dsusy ( B_{ab}^{ijkl} ,B_{ab}{}^{\hspace{-1mm} i}{}_j)  &=&-2 i  ( z \, \Xi_{ab}^{ijkl} ,\Xi_{ab}{}^{\hspace{-1mm} i}{}_j)  \ . 
 \eea
Similarly as for $D=11$ in~\eqref{ExtendedFields} one can combine the symmetric dual graviton $\tilde{h}_{a,b}$ and the anti-symmetric field $\tilde{h}_{ab}$ of the indecomposable module $R(\Lambda_2)$ into a single reducible field $\tilde{h}_{a;b}$ with supersymmetry transformation
\begin{align}
\dsusy \tilde{h}_{a;b} = \Xi^\prime_{a,b} +\Upsilon_{ab} = i \,  \overline{ \epsilon}_i \gamma_{a} \psi_{b}^i  -  i  \, \overline{ \epsilon}^i \gamma_{a} \psi_{bi } \,.
\end{align}

Moreover, one checks that the fermion bilinears $O^{\underline{I}}$ introduced in Section~\ref{sec:psiO} become
\bea 
O_{ab}^{ij} &=& \bar \psi_{[a}^{[i} \psi_{b]}^{j]} + \frac{i}{2} \varepsilon_{ab}{}^{cd} \bar \psi_{c}^{[i} \psi_{d}^{j]} + \frac z2 \bar \psi_{c k} \gamma_{ab} \gamma^c \chi^{ijk} + \frac{z^2}{72} \varepsilon^{ijklpqrs} \bar \chi_{klp} \gamma_{ab} \chi_{qrs}\; ,  \CR
O_{ab}{}^c &=& - \bar \psi_{[a}^i \gamma^c \psi_{b]i} -  \bar \psi^{k c} \gamma_{[a} \psi_{b] k}-  \bar \psi^{c}_k \gamma_{[a} \psi_{b]}^k- \frac{i z^2}{6} \varepsilon_{ab}{}^{cd} \bar \chi^{ijk} \gamma_{d} \chi_{ijk} - i \varepsilon_{ab}{}^{ef}  ( \bar \psi^{k c} \gamma_{e} \psi_{f k} - \bar \psi^{c}_k \gamma_{e} \psi_{f}^k)  \; ,   \CR
O_{a}^{ijkl} &=& \bar \psi_a^{[i} \chi^{jkl]} + \frac1{24} \varepsilon^{ijklpqrs} \bar \psi_{a p} \chi_{qrs}- \frac{i}{2} \varepsilon_{a}{}^{bcd} \bigl( \bar \psi_{b}^{[i} \gamma_{cd}  \chi^{jkl]} - \tfrac1{24} \varepsilon^{ijklpqrs} \bar \psi_{b p} \gamma_{cd}  \chi_{qrs} \bigr)   \; , \CR
O_{abc}{}^{\hspace{-1mm} i}{}_j &=&6 \bar \psi^i_{[a} \gamma_b \psi_{c] j} - \frac 34 \delta^i_j \bar \psi^k_{[a} \gamma_b \psi_{c]k} + i z^2 \varepsilon_{abcd} \bigl( \bar \chi^{ikl} \gamma^d \chi_{jkl} - \tfrac18 \delta^i_j \bar \chi^{klp} \gamma^d \chi_{klp} \bigr) \; .
\eea
For these one finds the $K(\mathfrak{e}_{11})$ variation 
\begin{multline} 
\label{eq:OE7}
\delta O_{ab}^{ij} = - \frac{i}{4} \Lambda^{ij}{}_c ( O_{ab}{}^c + \tfrac{i}{2} \varepsilon_{ab}{}^{ef} O_{ef}{}^c ) + i \Lambda^{k[i|c} ( O_{abc}{}^{\hspace{-1mm} j]}{}_k + \tfrac{i}{2} \varepsilon_{ab}{}^{ef} O_{efc}{}^{\hspace{-1mm} j]}{}_k) \\
- 2 i z \Lambda_{kl [ a} O_{b]}^{ijkl} + z \varepsilon_{ab}{}^{cd} \Lambda_{kl c} O_{d}^{ijkl} \; .  
\end{multline}
The absence of undesirable representations determines the relative coefficients in $O_{ab}^{ij}$, which in turn reproduces the supercovariantisation of the twisted self-duality equation in $\mathcal{N}=8$ supergravity \cite{deWit:1982bul}.  The expressions in~\eqref{eq:OE7} have been fixed by requiring that $O_{ab}{}^c$ is real, $O_{abc}{}^{\hspace{-1mm} i}{}_j$ Hermitian traceless, and $O_{a}^{ijkl}$ complex self-dual. The first two terms in $O_a^{ijkl}$ give the supercovariantisation of the scalar field momentum $P_a^{ijkl}$ \cite{Cremmer:1979up,deWit:1982bul}, while the last two terms represent the supercovariantisation of the 3-form field strength $H_{abc}^{ijkl}$ \cite{Butter:2018bkl}. Similarly, $ O_{abc}{}^{\hspace{-1mm} i}{}_j$ is the supercovariantisation of the 3-form field strength $H_{abc}{}^{\hspace{-1mm} i}{}_j$ \cite{Butter:2018bkl}. This is consistent with the duality equation relating the scalar field current and the 3-form field strength in the adjoint representation. The first four terms in $O_{ab}{}^c$ reproduce the supercovariantisation of the spin connection \cite{deWit:1982bul}, while the last two terms define the supercovariantisation of the dual-graviton field strength, consistently with the dual-graviton supersymmetry transformation $\Xi_{ab}^\prime$ in \eqref{susyVarE7}.

\end{document}